\pdfoutput=1
\documentclass[12pt,a4paper]{article}

\usepackage{ifthen} 
\newboolean{pdflatex}
\setboolean{pdflatex}{true} 

\newboolean{articletitles}
\setboolean{articletitles}{true} 

\newboolean{uprightparticles}
\setboolean{uprightparticles}{false} 

\newboolean{inbibliography}
\setboolean{inbibliography}{false} 

\usepackage{microtype}
\usepackage{lineno}  
\usepackage{xspace} 
\usepackage{caption} 

\usepackage{graphicx}  
\usepackage{color}
\usepackage{colortbl}
\graphicspath{{./figs/}} 

\usepackage{amsmath} 
\usepackage{amssymb}
\usepackage{amsfonts}
\usepackage{upgreek} 

\newcommand*\patchAmsMathEnvironmentForLineno[1]{%
\expandafter\let\csname old#1\expandafter\endcsname\csname #1\endcsname
\expandafter\let\csname oldend#1\expandafter\endcsname\csname
end#1\endcsname
 \renewenvironment{#1}%
   {\linenomath\csname old#1\endcsname}%
   {\csname oldend#1\endcsname\endlinenomath}%
}
\newcommand*\patchBothAmsMathEnvironmentsForLineno[1]{%
  \patchAmsMathEnvironmentForLineno{#1}%
  \patchAmsMathEnvironmentForLineno{#1*}%
}
\AtBeginDocument{%
\patchBothAmsMathEnvironmentsForLineno{equation}%
\patchBothAmsMathEnvironmentsForLineno{align}%
\patchBothAmsMathEnvironmentsForLineno{flalign}%
\patchBothAmsMathEnvironmentsForLineno{alignat}%
\patchBothAmsMathEnvironmentsForLineno{gather}%
\patchBothAmsMathEnvironmentsForLineno{multline}%
\patchBothAmsMathEnvironmentsForLineno{eqnarray}%
}

\usepackage{hyperref}    
\usepackage[all]{hypcap} 

\def\lhcb {\mbox{LHCb}\xspace}

\def\babar  {\mbox{BaBar}\xspace}

\def\MagUp {\mbox{\em Mag\kern -0.05em Up}\xspace}

\ifthenelse{\boolean{uprightparticles}}%
{

 \def\Pmu         {\ensuremath{\upmu}\xspace}

 \def\Ppi         {\ensuremath{\uppi}\xspace}

 \def\Ppsi        {\ensuremath{\uppsi}\xspace}

 \def\PDelta      {\ensuremath{\Delta}\xspace}                 
 \def\PXi      {\ensuremath{\Xi}\xspace}                 
 \def\PLambda      {\ensuremath{\Lambda}\xspace}                 
 \def\PSigma      {\ensuremath{\Sigma}\xspace}                 
 \def\POmega      {\ensuremath{\Omega}\xspace}                 
 \def\PUpsilon      {\ensuremath{\Upsilon}\xspace}                 
 
 \def\PB      {\ensuremath{\mathrm{B}}\xspace}                 
                  
 \def\PD      {\ensuremath{\mathrm{D}}\xspace}

 \def\PJ      {\ensuremath{\mathrm{J}}\xspace}                 
 \def\PK      {\ensuremath{\mathrm{K}}\xspace}

 \def\Pb      {\ensuremath{\mathrm{b}}\xspace}                 
 \def\Pc      {\ensuremath{\mathrm{c}}\xspace}

 \def\Pi      {\ensuremath{\mathrm{i}}\xspace}

 \def\Ps      {\ensuremath{\mathrm{s}}\xspace}

}
{

 \def\Pmu         {\ensuremath{\mu}\xspace}

 \def\Ppi         {\ensuremath{\pi}\xspace}

 \def\Ppsi        {\ensuremath{\psi}\xspace}                 
                  
 \mathchardef\PDelta="7101
 \mathchardef\PXi="7104
 \mathchardef\PLambda="7103
 \mathchardef\PSigma="7106
 \mathchardef\POmega="710A
 \mathchardef\PUpsilon="7107
                  
 \def\PB      {\ensuremath{B}\xspace}                 
                  
 \def\PD      {\ensuremath{D}\xspace}

 \def\PJ      {\ensuremath{J}\xspace}                 
 \def\PK      {\ensuremath{K}\xspace}

 \def\Pb      {\ensuremath{b}\xspace}                 
 \def\Pc      {\ensuremath{c}\xspace}

 \def\Pi      {\ensuremath{i}\xspace}

 \def\Ps      {\ensuremath{s}\xspace}

}

\makeatletter
\ifcase \@ptsize \relax
  \newcommand{\miniscule}{\@setfontsize\miniscule{4}{5}}
\or
  \newcommand{\miniscule}{\@setfontsize\miniscule{5}{6}}
\or
  \newcommand{\miniscule}{\@setfontsize\miniscule{5}{6}}
\fi
\makeatother

\DeclareRobustCommand{\optbar}[1]{\shortstack{{\miniscule (\rule[.5ex]{1.25em}{.18mm})}
  \\ [-.7ex] $#1$}}

\def\mup        {{\ensuremath{\Pmu^+}}\xspace}
\def\mun        {{\ensuremath{\Pmu^-}}\xspace} 
\def\mumu       {{\ensuremath{\Pmu^+\Pmu^-}}\xspace}

\def\squark    {{\ensuremath{\Ps}}\xspace}

\def\cquark    {{\ensuremath{\Pc}}\xspace}
\def\cquarkbar {{\ensuremath{\overline \cquark}}\xspace}
\def\ccbar     {{\ensuremath{\cquark\cquarkbar}}\xspace}
\def\bquark    {{\ensuremath{\Pb}}\xspace}
\def\bquarkbar {{\ensuremath{\overline \bquark}}\xspace}

\def\pion   {{\ensuremath{\Ppi}}\xspace}

\def\pip    {{\ensuremath{\pion^+}}\xspace}
\def\pim    {{\ensuremath{\pion^-}}\xspace}

\def\kaon    {{\ensuremath{\PK}}\xspace}
  \def\Kbar    {{\kern 0.2em\overline{\kern -0.2em \PK}{}}\xspace}

\def\KorKbar    {\kern 0.18em\optbar{\kern -0.18em K}{}\xspace}

\def\Kp      {{\ensuremath{\kaon^+}}\xspace}
\def\Km      {{\ensuremath{\kaon^-}}\xspace}

\def\Kstarz  {{\ensuremath{\kaon^{*0}}}\xspace}
\def\Kstarzb {{\ensuremath{\Kbar{}^{*0}}}\xspace}
\def\Kstar   {{\ensuremath{\kaon^*}}\xspace}

  \def\Dbar    {{\kern 0.2em\overline{\kern -0.2em \PD}{}}\xspace}

\def\DorDbar    {\kern 0.18em\optbar{\kern -0.18em D}{}\xspace}

\def\B       {{\ensuremath{\PB}}\xspace}
\def\Bbar    {{\ensuremath{\kern 0.18em\overline{\kern -0.18em \PB}{}}}\xspace}

\def\BorBbar    {\kern 0.18em\optbar{\kern -0.18em B}{}\xspace}
\def\Bz      {{\ensuremath{\B^0}}\xspace}
\def\Bzb     {{\ensuremath{\Bbar{}^0}}\xspace}
\def\Bu      {{\ensuremath{\B^+}}\xspace}

\def\Bd      {{\ensuremath{\B^0}}\xspace}
\def\Bs      {{\ensuremath{\B^0_\squark}}\xspace}
\def\Bsb     {{\ensuremath{\Bbar{}^0_\squark}}\xspace}
\def\Bdb     {{\ensuremath{\Bbar{}^0}}\xspace}

\def\jpsi     {{\ensuremath{{\PJ\mskip -3mu/\mskip -2mu\Ppsi\mskip 2mu}}}\xspace}
\def\psitwos  {{\ensuremath{\Ppsi{(2S)}}}\xspace}

  \def\Y#1S{\ensuremath{\PUpsilon{(#1S)}}\xspace}

\def\Lz          {{\ensuremath{\PLambda}}\xspace}
\def\Lbar        {{\ensuremath{\kern 0.1em\overline{\kern -0.1em\PLambda}}}\xspace}
\def\LorLbar    {\kern 0.18em\optbar{\kern -0.18em \PLambda}{}\xspace}

\def\Lb      {{\ensuremath{\Lz^0_\bquark}}\xspace}

\newcommand{\decay}[2]{\ensuremath{#1\!\to #2}\xspace}         

\def\to                 {\ensuremath{\rightarrow}\xspace}

\def\qsq       {{\ensuremath{q^2}}\xspace}

\def\eps   {{\ensuremath{\varepsilon}}\xspace}

\def\CP                {{\ensuremath{C\!P}}\xspace}

\def\AT#1     {\ensuremath{A_{\mathrm{T}}^{#1}}\xspace}           

\def\C#1      {\ensuremath{\mathcal{C}_{#1}}\xspace}                       
\def\Cp#1     {\ensuremath{\mathcal{C}_{#1}^{'}}\xspace}                    
\def\Ceff#1   {\ensuremath{\mathcal{C}_{#1}^{\mathrm{(eff)}}}\xspace}        
\def\Cpeff#1  {\ensuremath{\mathcal{C}_{#1}^{'\mathrm{(eff)}}}\xspace}       
\def\Ope#1    {\ensuremath{\mathcal{O}_{#1}}\xspace}                       
\def\Opep#1   {\ensuremath{\mathcal{O}_{#1}^{'}}\xspace}                    

\newcommand{\tev}{\ifthenelse{\boolean{inbibliography}}{\ensuremath{~T\kern -0.05em eV}\xspace}{\ensuremath{\mathrm{\,Te\kern -0.1em V}}}\xspace}
\newcommand{\gev}{\ensuremath{\mathrm{\,Ge\kern -0.1em V}}\xspace}
\newcommand{\mev}{\ensuremath{\mathrm{\,Me\kern -0.1em V}}\xspace}
\newcommand{\kev}{\ensuremath{\mathrm{\,ke\kern -0.1em V}}\xspace}
\newcommand{\ev}{\ensuremath{\mathrm{\,e\kern -0.1em V}}\xspace}
\newcommand{\gevc}{\ensuremath{{\mathrm{\,Ge\kern -0.1em V\!/}c}}\xspace}
\newcommand{\mevc}{\ensuremath{{\mathrm{\,Me\kern -0.1em V\!/}c}}\xspace}
\newcommand{\gevcc}{\ensuremath{{\mathrm{\,Ge\kern -0.1em V\!/}c^2}}\xspace}
\newcommand{\gevgevcccc}{\ensuremath{{\mathrm{\,Ge\kern -0.1em V^2\!/}c^4}}\xspace}
\newcommand{\mevcc}{\ensuremath{{\mathrm{\,Me\kern -0.1em V\!/}c^2}}\xspace}

\def\mum  {\ensuremath{{\,\upmu\rm m}}\xspace}

\def\invfb   {\ensuremath{\mbox{\,fb}^{-1}}\xspace}

\newcommand{\chisq}{\ensuremath{\chi^2}\xspace}

\def\deriv {\ensuremath{\mathrm{d}}}

\def\gsim{{~\raise.15em\hbox{$>$}\kern-.85em
          \lower.35em\hbox{$\sim$}~}\xspace}
\def\lsim{{~\raise.15em\hbox{$<$}\kern-.85em
          \lower.35em\hbox{$\sim$}~}\xspace}

\def\ptot       {\mbox{$p$}\xspace}
\def\pt         {\mbox{$p_{\rm T}$}\xspace}

\def\evtgen     {\mbox{\textsc{EvtGen}}\xspace}

\def\geant      {\mbox{\textsc{Geant4}}\xspace}

\def\photos     {\mbox{\textsc{Photos}}\xspace}

\def\pythia     {\mbox{\textsc{Pythia}}\xspace}

\def\tell1  {TELL1\xspace}
\def\ukl1   {UKL1\xspace}

\newcommand{\eg}{\mbox{\itshape e.g.}\xspace}
\newcommand{\ie}{\mbox{\itshape i.e.}\xspace}

\usepackage{cite} 
\usepackage{mciteplus}

\def\thetal {\ensuremath{\theta_{l}}\xspace}
\def\thetak {\ensuremath{\theta_{K}}\xspace}

\newcommand{\comment}[1]{}

\definecolor{darkred}{rgb}{0.6,0.0,0.0}
\definecolor{darkgreen}{rgb}{0.0,0.5,0.0}
\definecolor{lightgreen}{rgb}{0.75,1.0,0.75}
\definecolor{lightred}{rgb}{1.0,0.75,0.75}
\definecolor{lightblue}{rgb}{0.75,0.75,1.0}
\definecolor{darkblue}{RGB}{100,100,200}
\definecolor{verylightblue}{rgb}{0.9,0.9,1.0}
\definecolor{verylightred}{rgb}{1.0,0.9,0.9}
\definecolor{lightgray}{rgb}{0.9,0.9,0.9}
\definecolor{verylightgray}{rgb}{0.95,0.95,0.95}
\definecolor{darkgray}{rgb}{0.75,0.75,0.75}
\definecolor{orange}{rgb}{1.0,0.75,0.0}

\usepackage{longtable} 
\usepackage{multirow}
\usepackage{afterpage}

\begin{document}

\renewcommand{\thefootnote}{\fnsymbol{footnote}}
\setcounter{footnote}{1}


\begin{titlepage}
\pagenumbering{roman}

\vspace*{-1.5cm}
\centerline{\large EUROPEAN ORGANIZATION FOR NUCLEAR RESEARCH (CERN)}
\vspace*{1.5cm}
\hspace*{-0.5cm}
\begin{tabular*}{\linewidth}{lc@{\extracolsep{\fill}}r}
\ifthenelse{\boolean{pdflatex}}
{\vspace*{-2.7cm}\mbox{\!\!\!\includegraphics[width=.14\textwidth]{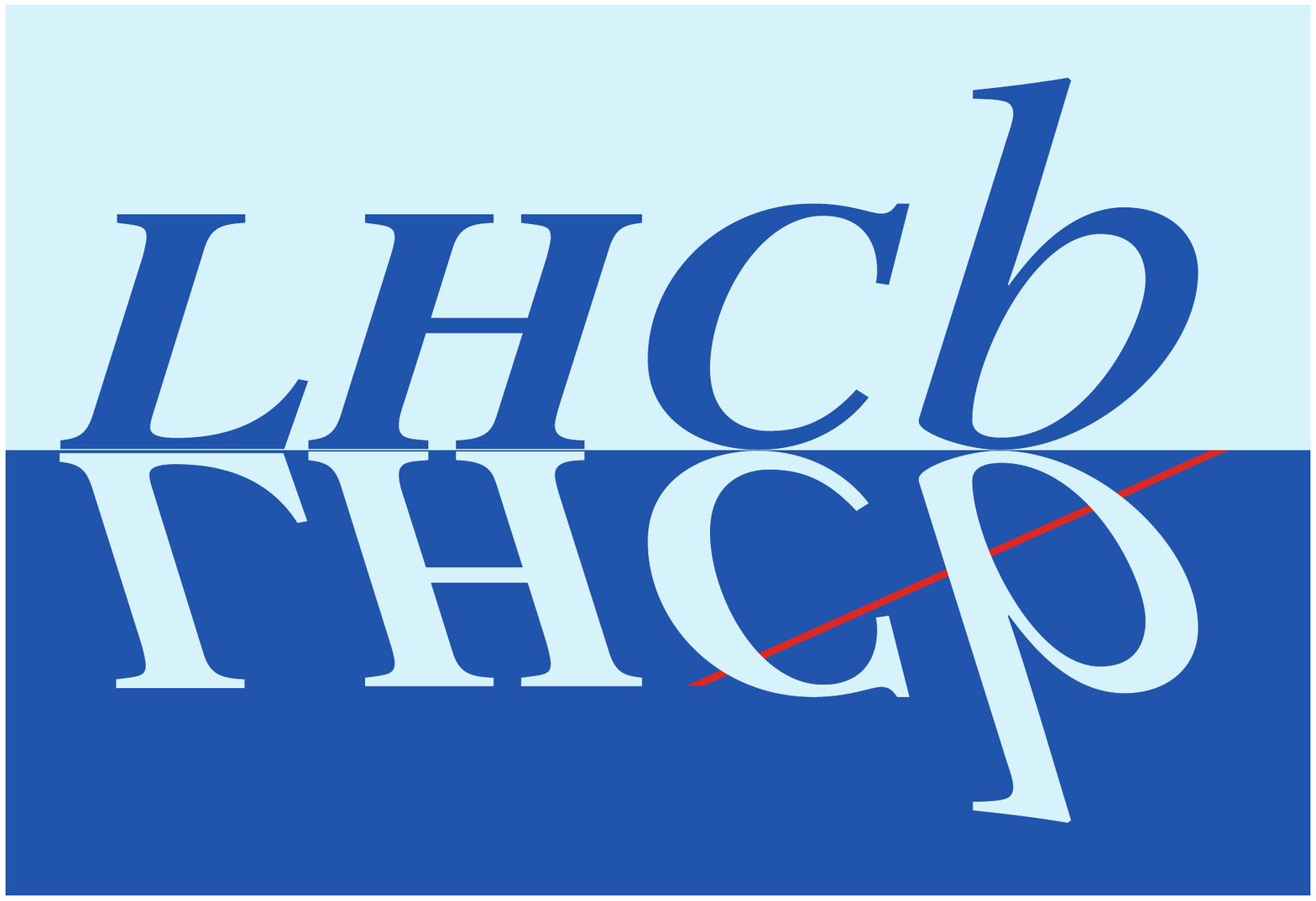}} & &}%
{\vspace*{-1.2cm}\mbox{\!\!\!\includegraphics[width=.12\textwidth]{lhcb-logo.eps}} & &}%
\\
 & & CERN-PH-EP-2015-314 \\  
 & & LHCb-PAPER-2015-051 \\  
 & & December 14, 2015 \\ 
 & & \\
\end{tabular*}

\vspace*{1.2cm}

{\bf\boldmath\huge
  \begin{center}
      Angular analysis of the $\decay{\Bd}{\Kstarz\mup\mun}$ decay using $3\invfb$ of integrated luminosity 
\end{center}
}

\vspace*{0.8cm}

\begin{center}
The LHCb collaboration\footnote{Authors are listed at the end of this paper.}
\end{center}

\vspace{\fill}

\begin{abstract}
  \noindent
  An angular analysis of the
  $B^{0}\rightarrow K^{*0}(\to\Kp\pim)\mu^{+}\mu^{-}$ decay is
  presented. The dataset corresponds to an integrated luminosity of
  $3.0{\mbox{\,fb}^{-1}}$ of $pp$ collision data collected at the LHCb
  experiment. The complete angular information from the decay is used to determine 
  $C\!P$-averaged observables and $C\!P$ asymmetries, taking account of
  possible contamination from decays with the $\Kp\pim$ system in an S-wave configuration.  The angular
  observables and their correlations are reported in bins of $q^2$,
  the invariant mass squared of the dimuon system.  The observables
  are determined both from an unbinned maximum likelihood fit and
  by using the principal moments of the angular
  distribution. In addition, by fitting for $q^2$-dependent decay amplitudes in the region
  $1.1<q^{2}<6.0\mathrm{\,Ge\kern -0.1em V}^{2}/c^{4}$, the zero-crossing points of several
  angular observables are computed.
A global fit is performed to the complete set of $C\!P$-averaged observables obtained from the maximum likelihood fit. 
This fit indicates differences with predictions based on the Standard Model at the level of 3.4~standard deviations.
These differences could be explained by contributions from physics beyond the Standard Model, or by an unexpectedly large hadronic effect that is not accounted for in the Standard Model predictions. 
 
\end{abstract}

\vspace*{1.0cm}

\begin{center}
  Published in JHEP {\bf 02} (2016) 104
\end{center}

\vspace{\fill}

{\footnotesize 
\centerline{\copyright~CERN on behalf of the \lhcb collaboration, licence \href{http://creativecommons.org/licenses/by/4.0/}{CC-BY-4.0}.}}
\vspace*{2mm}

\end{titlepage}

\newpage
\setcounter{page}{2}
\mbox{~}

\cleardoublepage


\renewcommand{\thefootnote}{\arabic{footnote}}
\setcounter{footnote}{0}



\pagestyle{plain} 
\setcounter{page}{1}
\pagenumbering{arabic}


%

\sloppy

\section{Introduction}
\label{sec:introduction}

The decay $\decay{\Bd}{\Kstarz\mumu}$
proceeds via a $\bquark \to \squark$ quark flavour-changing neutral current (FCNC) transition.
In the Standard Model (SM) the decay is therefore forbidden at tree level and occurs, at lowest order,  via electroweak penguin and box processes. 
In extensions of the SM, new particles may enter in competing processes and can significantly change the branching fraction of the decay and 
the angular distribution of the final-state particles.
Angular observables are of particular interest, since theoretical predictions of such observables tend to be less affected by the hadronic uncertainties associated with the $\Bz \to \Kstarz$ transition. 
Throughout this paper \Kstarz is used to refer to the $\Kstar(892)^0$ resonance. 

The LHCb collaboration previously determined a set of angular observables in the $\decay{\Bd}{\Kstarz\mumu}$ decay,
\footnote{The inclusion of charge conjugate processes is implied throughout this paper, unless otherwise noted.} 
using data collected during 2011, corresponding to an integrated luminosity of $1.0\invfb$~\cite{LHCb-PAPER-2013-019}.
Different subsets of these angular observables have also been measured by the \babar, Belle, CDF and CMS collaborations~\cite{Aubert:2006vb,Lees:2015ymt,:2009zv,Aaltonen:2011ja,Chatrchyan:2013cda,Khachatryan:2015isa} and all of these measurements are in good agreement with SM predictions. 
The LHCb collaboration has also used the 2011 dataset to determine an alternative set of angular observables that have reduced theoretical uncertainties~\cite{LHCb-PAPER-2013-037}.
In contrast to the previous analyses, these observables cannot be extracted from single angle distributions. 
This second LHCb analysis found a local deviation with respect to the SM prediction in one observable, $P_5^\prime$, with a significance corresponding to 3.7~standard deviations. Possible interpretations of this discrepancy and the consistency of all of the measurements of $\bquark\to\squark$ transitions have been widely discussed in the literature~\cite{Descotes-Genon:2013wba,Altmannshofer:2013foa,Beaujean:2013soa,Hurth:2013ssa,Jager:2012uw,Descotes-Genon:2014uoa,Lyon:2014hpa,Altmannshofer:2014cfa,Crivellin:2015mga,Gauld:2013qja,Altmannshofer:2014rta,Mahmoudi:2014mja, Datta:2013kja}. 
 
The present paper describes an updated angular analysis of the $\decay{\Bd}{\Kstarz\mumu}$ decay, using the LHCb Run 1 data sample, corresponding to an integrated luminosity of $3.0\invfb$. 
The data were recorded in $pp$ collisions at centre-of-mass energies of $7$ and $8\tev$ during 2011 and 2012, respectively.
All previous analyses of the $\decay{\Bd}{\Kstarz\mumu}$ decay have extracted only part of the information available, by fitting simplified forms of the angular distribution. 
This paper presents a complete set of observables for the first time, based on the full angular distribution.
The simultaneous determination of these observables allows correlations between the measured quantities to be computed, enabling the use of the results in global fits to theoretical models.
This is critical to understand whether SM dynamics are sufficient to explain the above discrepancy, or if extensions to the SM are necessary. 

The structure of this paper is as follows. In Sec.~\ref{sec:ang}, the angular distribution and observables for the $\decay{\Bd}{\Kstarz\mumu}$ decay are presented. Section~\ref{sec:Detector} describes the experimental setup. The reconstruction and selection of the \decay{\Bz}{\Kstarz\mumu} candidates and sources of background are presented in Sec.~\ref{sec:selection}. 
The method used to correct the angular distribution for experimental effects is detailed in Sec.~\ref{sec:acceptance} and the parameterisation of the mass distribution is described in Sec.~\ref{sec:mass}. 
The determination of the angular observables is detailed in Sec.~\ref{sec:angular}, and Sec.~\ref{sec:systematics} discusses sources of  systematic uncertainty. 
Results are given in Sec.~\ref{sec:Results} and the compatibility with predictions based on the Standard Model is discussed in Sec.~\ref{sec:discussion}. 
Finally, conclusions are presented in Sec.~\ref{sec:conclusions}.

\section{Angular distribution and observables}
\label{sec:ang}

The final state of the decay $\decay{\Bd}{\Kstarz\mumu}$ can be described by $q^2$,
the invariant mass squared of the dimuon system, and three decay angles $\vec{\Omega}=(\cos\thetal,\cos\thetak,\phi)$. 
The angle between the $\mup$ ($\mun$) and the direction opposite to that of the $\Bd$ ($\Bdb$) in the rest frame of the dimuon system is denoted $\thetal$.
In this analysis, the $\Kstarz$ meson is reconstructed through the decay $\Kstarz\to\Kp\pim$. 
The angle between the direction of the $\Kp$ ($\Km$) and the $\Bd$ ($\Bdb$) in the rest frame of the $\Kstarz$ ($\Kstarzb$) system is denoted $\thetak$. 
The angle between the plane defined by the dimuon pair and the plane defined by the kaon and pion in the $\Bd$ ($\Bdb$) rest frame is denoted $\phi$. 
More details of the angular basis adopted in this analysis are given in Appendix A of Ref.~\cite{LHCb-PAPER-2013-019}.

The differential decay rates of \decay{\Bzb}{\Kstarzb\mumu} and \decay{\Bz}{\Kstarz\mumu} decays, in terms of \qsq and the three angles, are given by
\begin{align}
\begin{split}
\frac{\deriv^{4}\Gamma[\decay{\Bzb}{\Kstarzb\mumu}]}{\deriv\qsq\,\deriv\vec{\Omega}} = & \frac{9}{32\pi} \sum\limits_{i}^{} I_{i}(\qsq) f_{i}(\vec{\Omega})~~\text{and} \\
\frac{\deriv^{4}\bar{\Gamma}[\decay{\Bz}{\Kstarz\mumu}]}{\deriv\qsq\,\deriv\vec{\Omega}} = & \frac{9}{32\pi} \sum\limits_{i}^{} \bar{I}_{i}(\qsq) f_{i}(\vec{\Omega})~,
\end{split}
\end{align} 
where $\Gamma$ ($\bar{\Gamma}$) refers to decays involving a \bquark (\bquarkbar) quark and hence a \Bzb (\Bz) meson,
the terms $f_{i}(\vec{\Omega})$ are formed from combinations of spherical harmonics and the $I_{i}$ ($\bar{I}_{i}$) are $q^2$-dependent angular observables.
The $I_{i}$ can be expressed as bilinear combinations of six complex decay amplitudes, ${\cal A}_{0,\parallel,\perp}^{\rm L,R}$,
which correspond to the different transversity states of the \Kstarz meson and the different (left- and right-handed) chiralities of the dimuon system.
An additional suffix $s$ or $c$ is conventionally added to some of the $I_i$ terms to indicate that they have a $\sin^2\thetak$ or $\cos^2\thetak$ dependence.
When \qsq is sufficiently large ($\qsq \gsim 1\gevgevcccc$), the muons can be considered massless.
The list of the angular terms and observables that remain in this massless limit is given in Table~\ref{tab:angular}.

Following the notation of Ref.~\cite{Altmannshofer:2008dz}, \qsq-dependent 
\CP\ averages, $S_{i}$, and \CP\ asymmetries, $A_{i}$, can be defined as
\begin{align}
\begin{split}
  S_i &= \left( I_{i} + \bar{I}_{i} \right)\Big/\left(\frac{\deriv\Gamma}{\deriv\qsq} + \frac{\deriv\bar{\Gamma}}{\deriv\qsq}\right) ~\text{and}\\
  A_i &= \left( I_{i} - \bar{I}_{i} \right)\Big/\left(\frac{\deriv\Gamma}{\deriv\qsq} + \frac{\deriv\bar{\Gamma}}{\deriv\qsq}\right) .
\end{split}
\end{align}
In the massless limit, the \CP-averaged observables $S_{1(s,c)}$ and $S_{2({s,c})}$ obey the relations $S_{1s}=3S_{2s}$, $S_{1c}=-S_{2c}$ and $\tfrac{3}{4}(2S_{1s}+S_{1c})-\tfrac{1}{4}(2S_{2s}+S_{2c})=1$ (see for example Ref.~\cite{Altmannshofer:2008dz}).
These relationships reduce the number of independent \CP-averaged observables from eleven to eight.
The relations between the observables also hold to a good approximation for $\qsq < 1\gev^{2}/c^{4}$ and are therefore adopted for the full \qsq range.
The $S_{1c}$ observable corresponds to the fraction of longitudinal polarisation of the $\Kstarz$ meson  and is therefore more commonly referred to as $F_{\rm L}$, with
\begin{align}
F_{\rm L} = S_{1c} = \frac{|{\cal A}_{0}^{\rm L}|^{2} + |{\cal A}_{0}^{\rm R}|^{2}}{|{\cal A}_{0}^{\rm L}|^{2} + |{\cal A}_{0}^{\rm R}|^{2} + |{\cal A}_{\parallel}^{\rm L}|^{2} + |{\cal A}_{\parallel}^{\rm R}|^{2} + |{\cal A}_{\perp}^{\rm L}|^{2} + |{\cal A}_{\perp}^{\rm R}|^{2}}\,.
\end{align}
It is also conventional to replace $S_{6s}$ by the forward-backward asymmetry of the dimuon system $A_{\rm FB}$, with $A_{\rm FB}=\tfrac{3}{4}S_{6s}$. 
The \CP-averaged angular distribution of the \mbox{\decay{\Bz}{\Kstarz\mumu}} decay can then be written as 
\begin{equation}
\begin{split}
\frac{1}{\deriv(\Gamma+\bar{\Gamma})/\deriv q^2}\,\frac{\deriv^4(\Gamma+\bar{\Gamma})}{\deriv\qsq\,\deriv\vec{\Omega}} =
\frac{9}{32\pi} \Big[
 & \tfrac{3}{4} (1-{F_{\rm L}})\sin^2\thetak + {F_{\rm L}}\cos^2\thetak \\
\phantom{\Big[}+& \tfrac{1}{4}(1-{F_{\rm L}})\sin^2\thetak\cos 2\thetal\\
\phantom{\Big[}-& {F_{\rm L}} \cos^2\thetak\cos 2\thetal + {S_3}\sin^2\thetak \sin^2\thetal \cos 2\phi\\
\phantom{\Big[}+& {S_4} \sin 2\thetak \sin 2\thetal \cos\phi + {S_5}\sin 2\thetak \sin \thetal \cos \phi\\
\phantom{\Big[}+& \tfrac{4}{3} {A_{\rm FB}} \sin^2\thetak \cos\thetal + {S_7} \sin 2\thetak \sin\thetal \sin\phi\\
+& {S_8} \sin 2\thetak \sin 2\thetal \sin\phi + {S_9}\sin^2\thetak \sin^2\thetal \sin 2\phi  \Big]\,.
\end{split}
\label{eq:pdfpwave}
\end{equation}

Additional sets of observables, for which the leading $\Bz \to \Kstarz$ form-factor uncertainties cancel, can be built from $F_{\rm L}$ and $S_{3}$--$S_{9}$. Examples of such {\it optimised} observables include the transverse asymmetry $A_{\rm T}^{(2)}$~\cite{Kruger:2005ep}, where $A_{\rm T}^{(2)} = 2 S_3/(1 - F_{\rm L})$, and the $P^{(\prime)}_{i}$ series of observables~\cite{DescotesGenon:2012zf}. 
In this paper the notation used is
\begin{align}
\begin{aligned}
P_1 & = \frac{2\,S_3}{(1 - F_{\rm L})} = A_{\rm T}^{(2)}\, , \\ 
P_2 & = \frac{2}{3} \frac{A_{\rm FB}}{( 1- F_{\rm L} )} \, ,\\
P_3 & = \frac{-S_9}{(1 - F_{\rm L})}\, , \\
P^\prime_{4,5,8} & = \frac{S_{4,5,8}}{\sqrt{F_{\rm L}(1 - F_{\rm L})}}\, ,\\ 
P^\prime_{6} & = \frac{S_{7}}{\sqrt{F_{\rm L}(1 - F_{\rm L})}}\, .
\end{aligned}
\end{align}
The definition of the $P'_{i}$ observables differs from that of Ref.~\cite{DescotesGenon:2012zf}, but is consistent with the notation used in the LHCb analysis of Ref.~\cite{LHCb-PAPER-2013-037}.

In addition to the resonant P-wave \Kstarz contribution to the $\Kp\pim\mumu$ final state, the $\Kp\pim$ system can also be in an S-wave configuration.
The addition of an S-wave component introduces two new complex amplitudes, ${\cal A}_{\rm S}^{\rm L,R}$, and results in the six additional angular terms that are given in the lower part of Table~\ref{tab:angular}. 
In the analyses described in Refs.~\cite{LHCb-PAPER-2013-019,LHCb-PAPER-2013-037} the S-wave contribution, which is expected to be approximately 5\%, was treated as a systematic uncertainty. 
The presence of a $\Kp\pim$ system in an S-wave configuration modifies the angular distribution to
\begin{equation}
\begin{split}
\left.\frac{1}{\deriv(\Gamma+\bar{\Gamma})/\deriv q^2}\,\frac{\deriv^4(\Gamma+\bar{\Gamma})}{\deriv\qsq\,\deriv\vec{\Omega}}\right|_{{\rm S}+{\rm P}} ~=~ &
(1-F_{\rm S}) \left.\frac{1}{\deriv(\Gamma+\bar{\Gamma})/\deriv q^2}\,\frac{\deriv^4(\Gamma+\bar{\Gamma})}{\deriv\qsq\,\deriv\vec{\Omega}}\right|_{\rm P}\\
& + \frac{3}{16\pi} F_{\rm S}\sin^{2}\theta_{l}  \\
& + \frac{9}{32\pi} ( S_{11}  + S_{13} \cos 2\theta_{l} ) \cos\theta_{K}  \\
& + \frac{9}{32\pi} ( S_{14} \sin 2\theta_{l} + S_{15} \sin\theta_{l} ) \sin\theta_{K}\cos\phi  \\
& + \frac{9}{32\pi} ( S_{16} \sin\theta_{l} + S_{17} \sin 2\theta_{l} ) \sin\theta_{K}\sin\phi \, ,
\end{split}
\label{eq:pdfswave}
\end{equation}
where $F_{\rm S}$ denotes the S-wave fraction,
\begin{align}
F_{\rm S} &=  \frac{|{\cal A}_{\rm S}^{\rm L}|^{2} + |{\cal A}_{\rm S}^{\rm R}|^{2}}{|{\cal A}_{\rm S}^{\rm L}|^{2} + |{\cal A}_{\rm S}^{\rm R}|^{2} + |{\cal A}_{0}^{\rm L}|^{2} + |{\cal A}_{0}^{\rm R}|^{2} + |{\cal A}_{\parallel}^{\rm L}|^{2} + |{\cal A}_{\parallel}^{\rm R}|^{2} + |{\cal A}_{\perp}^{\rm L}|^{2} + |{\cal A}_{\perp}^{\rm R}|^{2} }  \, ,
\end{align} 
and the terms $S_{11}$, $S_{13}$--$S_{17}$ arise from interference between the S- and P-wave amplitudes. 
Note that $F_{\rm S}$ replaces the terms $S_{10}$ and $S_{12}$, with $F_{\rm S} = 3 S_{10} =-3 S_{12}$.
Throughout this paper, $F_{\rm S}$ and the interference terms between the S- and P-wave are treated as nuisance parameters.

Due to the flavour specific final state of the decay, the \CP asymmetries $A_i$ can be determined from differences in the angular distributions between \Bz\ and \Bzb\ decays. 

\begin{table}[!hb]
\caption{
Angular observables $I_j$ and their corresponding angular terms for dimuon masses that are much larger than twice the muon mass. 
The terms in the lower part of the table arise from the $\Kp\pim$ S-wave contribution to the  $\Kp\pim\mumu$ final state.  
The $\bar{I}_i$ coefficients are obtained by making the substitution ${\cal A} \to \bar{{\cal A}}$, \ie by complex conjugation of the weak phases in the amplitudes. 
\label{tab:angular}
} 
\centering 
{\setlength{\extrarowheight}{6pt}%
\begin{tabular}{c|l|l}
$i$ & $I_{i}$ & $f_{i}$ \\
\hline
$1s$ & \,$\frac{3}{4}  \left[ |{\cal A}_{\parallel}^{\rm L}|^{2} + |{\cal A}_{\perp}^{\rm L}|^{2}  + |{\cal A}_{\parallel}^{\rm R}|^{2} + |{\cal A}_{\perp}^{\rm R}|^{2}\right]$ & $\sin^{2}\theta_{K}$ \\
$1c$ &  $|{\cal A}_{0}^{\rm L}|^{2} +  |{\cal A}_{0}^{\rm R}|^{2}$ & $\cos^{2}\theta_{K}$ \\ 
$2s$ & $\frac{1}{4}  \left[ |{\cal A}_{\parallel}^{\rm L}|^{2} + |{\cal A}_{\perp}^{\rm L}|^{2}  + |{\cal A}_{\parallel}^{\rm R}|^{2} + |{\cal A}_{\perp}^{\rm R}|^{2}\right]$ & $\sin^{2}\theta_{K}\cos 2\theta_{l}$ \\
$2c$ &  $- |{\cal A}_{0}^{\rm L}|^{2} -  |{\cal A}_{0}^{\rm R}|^{2} $ & $\cos^{2}\theta_{K}\cos 2\theta_{l}$ \\
3 & $\frac{1}{2}   \left[ |{\cal A}_{\perp}^{\rm L}|^{2} - |{\cal A}_{\parallel}^{\rm L}|^{2}  + |{\cal A}_{\perp}^{\rm R}|^{2} - |{\cal A}_{\parallel}^{\rm R}|^{2} \right]$  & $\sin^{2}\theta_{K} \sin^{2}\theta_{l} \cos 2\phi$ \\
4 & $\sqrt{\frac{1}{2}}   {\rm Re}({\cal A}_{0}^{\rm L} {\cal A}_{\parallel}^{{\rm L}\ast} + {\cal A}_{0}^{\rm R} {\cal A}_{\parallel}^{{\rm R}\ast}) $  & $\sin 2\theta_{K} \sin 2\theta_{l} \cos \phi$ \\
5 &  $\sqrt{2}  {\rm Re}({\cal A}_{0}^{\rm L} {\cal A}_{\perp}^{{\rm L}\ast} - {\cal A}_{0}^{\rm R} {\cal A}_{\perp}^{{\rm R}\ast})$ &  $\sin 2\theta_{K} \sin \theta_{l} \cos \phi$ \\
$6s$ &  $2   {\rm Re}({\cal A}_{\parallel}^{\rm L} {\cal A}_{\perp}^{{\rm L}\ast} - {\cal A}_{\parallel}^{\rm R} {\cal A}_{\perp}^{{\rm R}\ast})$ &  $\sin^{2}\theta_{K} \cos \theta_{l}$ \\ 
7 &  $\sqrt{2}  {\rm Im}({\cal A}_{0}^{\rm L} {\cal A}_{\parallel}^{{\rm L}\ast} - {\cal A}_{0}^{\rm R} {\cal A}_{\parallel}^{{\rm R}\ast})$ &  $\sin 2\theta_{K} \sin \theta_{l} \sin \phi$ \\
8 &  $\sqrt{\frac{1}{2}}  {\rm Im}({\cal A}_{0}^{\rm L} {\cal A}_{\perp}^{{\rm L}\ast} + {\cal A}_{0}^{\rm R} {\cal A}_{\perp}^{{\rm R}\ast})$&  $\sin 2\theta_{K} \sin 2\theta_{l} \sin \phi$ \\
9 &  ${\rm Im}({\cal A}_{\parallel}^{{\rm L}\ast} {\cal A}_{\perp}^{\rm L} + {\cal A}_{\parallel}^{{\rm R}\ast} {\cal A}_{\perp}^{\rm R}) \phantom{\sqrt{\frac{1}{1}}}$ &  $\sin^{2}\theta_{K} \sin^{2}\theta_{l} \sin 2\phi$ \\ 
\hline 
10 & $\frac{1}{3}  \left[ |{\cal A}_{\rm S}^{\rm L}|^{2} + |{\cal A}_{\rm S}^{\rm R}|^{2}\right]$ & 1 \\  
11 & $\sqrt{\frac{4}{3}}  {\rm Re}({\cal A}_{\rm S}^{\rm L} {\cal A}_{0}^{{\rm L}\ast} + {\cal A}_{\rm S}^{\rm R} {\cal A}_{0}^{{\rm R}\ast})$ & $\cos\theta_{K}$ \\
12 & $-\frac{1}{3}  \left[ |{\cal A}_{\rm S}^{\rm L}|^{2} + |{\cal A}_{\rm S}^{\rm R}|^{2}\right]$ & $\cos 2\theta_{l}$ \\
13 & $-\sqrt{\frac{4}{3}}  {\rm Re}({\cal A}_{\rm S}^{\rm L} {\cal A}_{0}^{{\rm L}\ast} + {\cal A}_{\rm S}^{\rm R} {\cal A}_{0}^{{\rm R}\ast})$ & $\cos\theta_{K} \cos 2\theta_{l}$ \\
14 & $\sqrt{\frac{2}{3}}  {\rm Re}({\cal A}_{\rm S}^{\rm L} {\cal A}_{\parallel}^{{\rm L}\ast} + {\cal A}_{\rm S}^{\rm R} {\cal A}_{\parallel}^{{\rm R}\ast})$ & $\sin \theta_{K} \sin 2\theta_{l} \cos\phi$ \\
15 & $\sqrt{\frac{8}{3}}  {\rm Re}({\cal A}_{\rm S}^{\rm L} {\cal A}_{\perp}^{{\rm L}\ast} - {\cal A}_{\rm S}^{\rm R} {\cal A}_{\perp}^{{\rm R}\ast})$ & $\sin \theta_{K} \sin \theta_{l} \cos\phi$ \\
16 & $\sqrt{\frac{8}{3}}  {\rm Im}({\cal A}_{\rm S}^{\rm L} {\cal A}_{\parallel}^{{\rm L}\ast} - {\cal A}_{\rm S}^{\rm R} {\cal A}_{\perp}^{{\rm R}\ast})$ & $\sin \theta_{K} \sin \theta_{l} \sin\phi$ \\
17 & $\sqrt{\frac{2}{3}}  {\rm Im}({\cal A}_{\rm S}^{\rm L} {\cal A}_{\perp}^{{\rm L}\ast} + {\cal A}_{\rm S}^{\rm R} {\cal A}_{\perp}^{{\rm R}\ast})$ & $\sin \theta_{K} \sin 2\theta_{l} \sin \phi$ \\
\end{tabular}
}
\end{table}

In this analysis, three separate techniques are used to study the angular distribution:
\begin{enumerate}
\item{An unbinned maximum likelihood fit is used to determine the \CP-averaged observables $F_{\rm L}$, $A_{\rm FB}$, and $S_{3}$--$S_{9}$,
  as well as the \CP asymmetries $A_{3}$--$A_{9}$, averaged over bins of $q^2$.
In addition, the $P^{(\prime)}_{i}$ observables are determined by reparameterising the likelihood fit.
  The data are analysed in \qsq bins of approximately $2\gev^{2}/c^{4}$ width and also in wider $1.1 < \qsq < 6.0\gev^{2}/c^{4}$ and $15.0 < \qsq < 19.0\gev^{2}/c^{4}$ bins for which there are particularly precise theoretical predictions.
  The unbinned maximum likelihood fit is described in Sec.~\ref{sec:likelihood}.
}
\item{The same observables are also determined using principal angular moments. 
  This so-called {\it method of moments} gives an approximately 15\% less precise determination of the observables than the likelihood fit but is particularly robust for low signal yields and does not require a complex angular fit~\cite{Beaujean:2015xea}. 
  This allows the observables to be determined in approximately $1\gev^{2}/c^{4}$ wide $q^2$ bins, which gives additional shape information that is useful in regions where the observables vary rapidly with \qsq. 
  The method is described in Sec.~\ref{sec:moments}. 
}
\clearpage
\item{
Finally, the observables $S_{4}$, $S_{5}$ and $A_{\rm FB}$ vary as a function of \qsq and are known to change sign in the SM.  
By fitting for  the  decay amplitudes as a function of \qsq, the \qsq values at which these observables cross zero can be determined. 
At leading order these zero-crossing points are free from $\Bz \to \Kstarz$ form-factor uncertainties and consequently provide a precision test of the SM~\cite{Ali:1999mm, Kumar:2014bna}.
The method is applied in the range $1.1 < \qsq < 6.0\gev^{2}/c^{4}$ and is described in Sec.~\ref{sec:amplitudes}.
 }  
\end{enumerate}
The three methods are complementary, but their results are correlated and cannot be combined. 
Method~1 is the most precise and is therefore used to compare to the SM predictions.
The \qsq bins used for the likelihood fit of the angular observables and the method of moments are given in Tables~\ref{tab:results:likelihood:averages} and \ref{tab:results:moments:averages} of Appendix~\ref{sec:appendix:results}, respectively.

\section{Detector and simulation}
\label{sec:Detector}

The \lhcb detector~\cite{Alves:2008zz,LHCb-DP-2014-002} is a single-arm forward
spectrometer, covering the \mbox{pseudorapidity} range $2<\eta <5$,
designed for the study of particles containing \bquark or \cquark quarks. 
The detector includes a high-precision tracking system
consisting of a silicon-strip vertex detector surrounding the $pp$
interaction region, a large-area silicon-strip detector located
upstream of a dipole magnet with a bending power of about
$4{\rm\,Tm}$, and three stations of silicon-strip detectors and straw
drift tubes placed downstream of the magnet.
The tracking system provides a measurement of momentum, \ptot, of charged particles with
a relative uncertainty that varies from 0.5\% at low momentum to 1.0\% at 200\gevc.
The minimum distance of a track to a primary $pp$ interaction vertex (PV), the impact parameter, is measured with a resolution of $(15+29/\pt)\mum$,
where \pt is the component of the momentum transverse to the beam, in \gevc.
Different types of charged hadrons are distinguished using information
from two ring-imaging Cherenkov (RICH) detectors. 
Photons, electrons and hadrons are identified by a calorimeter system consisting of
scintillating-pad and preshower detectors, an electromagnetic
calorimeter and a hadronic calorimeter. Muons are identified by a
system composed of alternating layers of iron and multiwire
proportional chambers.
The online event selection is performed by a trigger, 
which consists of a hardware stage, based on information from the calorimeter and muon
systems, followed by a software stage, which applies a full event
reconstruction~\cite{LHCb-DP-2012-004}. 

Simulated signal events are used to determine the impact of the detector geometry, trigger, reconstruction and candidate selection on the angular distribution of the signal. 
In addition,  simulated samples are used to estimate the contribution of possible background processes. 
In the simulation, $pp$ collisions are generated using \pythia~\cite{Sjostrand:2006za,*Sjostrand:2007gs} with a specific \lhcb 
configuration~\cite{LHCb-PROC-2010-056}.  Decays of hadronic particles
are described by \evtgen~\cite{Lange:2001uf}, in which final-state
radiation is generated using \photos~\cite{Golonka:2005pn}. 
The interaction of the generated particles with the detector, and its response,
are implemented using the \geant
toolkit~\cite{Allison:2006ve, *Agostinelli:2002hh} as described in Ref.~\cite{LHCb-PROC-2011-006}.
Data-driven corrections are applied to the simulation to account for a small level of mismodelling of the detector occupancy, $\Bd$ momentum and $\Bd$ vertex quality.
Similarly, the simulated particle identification (PID) performance is corrected to match that determined from control samples selected from the data.

\section{Selection of signal candidates}
\label{sec:selection}

The $\decay{\Bd}{\Kstarz\mumu}$ signal candidates are required to pass a hardware trigger,
which selects events containing at least one muon with $\pt>1.48\gevc$ in the 7\tev data or $\pt>1.76\gevc$ in the 8\tev data.
In the subsequent software trigger, at least one of the final-state particles is required to have both
$\pt>0.8\gevc$ and impact parameter larger than $100\mum$ with respect to all PVs in the event. 
Finally, the tracks of two or more of the final-state particles are required to form a vertex that is significantly displaced from any PV.

Signal candidates are formed from a pair of oppositely charged tracks that are identified as muons, combined with a \Kstarz meson candidate. The \Kstarz candidate is formed from two charged tracks that are identified as a kaon and a pion, respectively.
The four tracks of the final-state particles are required to have a significant impact parameter with respect to all PVs in the event. 
The tracks are then fitted to a common vertex,
which is required to be of good quality.
The impact parameter of the $\Bd$ candidate with respect to one of the PVs is required to be small and the vertex of the \Bz candidate is required to be significantly displaced from the same PV.
The angle $\theta_{\rm DIRA}$ between the reconstructed $\Bd$ momentum and the vector connecting the PV to the reconstructed $\Bd$ decay vertex is required to be small.  
Candidates are required to have reconstructed $\Bz$ invariant mass, $m(\Kp\pim\mumu)$, in the range $5170<m(\Kp\pim\mumu)<5700\mevcc$.  
Finally, the reconstructed mass of the $\Kp\pim$ system, $m(\Kp\pim)$, is required to be in the range $796<m(\Kp\pim)<996\mevcc$. 

Background formed by combining particles from different \bquark- and \cquark-hadron decays (referred to as {\it combinatorial} background) is further reduced using a boosted decision tree~(BDT)~\cite{Breiman,AdaBoost}, which is trained using data.
As a proxy for the signal decay, $\decay{\Bd}{\jpsi\Kstarz}$ decays are used to train the BDT,  
where the \jpsi is reconstructed through its decay into \mumu.
Candidates from the upper mass sideband $5350<m(\Kp\pim\mumu)<7000\mevcc$ are used as a proxy for the background. 
As input variables, the BDT uses the reconstructed $\Bd$ lifetime and vertex fit quality,
the momentum and transverse momentum of the $\Bd$ candidate, $\cos\theta_{\rm DIRA}$, 
particle identification information from the RICH detectors and the muon system, 
as well as variables describing the isolation of the final state tracks~\cite{LHCb-PAPER-2011-004}. 
To best exploit the data available for training, the $k$-folding technique~\cite{Blum:1999:BHB:307400.307439} is employed with $k=10$. 
At the chosen working point, the BDT has a background rejection of $97\%$ and a signal efficiency of $85\%$. 
The signal efficiency and background rejection of the BDT is uniform in $m(\Kp\pim\mumu)$ and  $m(\Kp\pim)$. 
The distortion induced in \qsq and the angular distributions is discussed in Sec.~\ref{sec:acceptance}.

\begin{figure}
  \centering
  \includegraphics[width=12cm]{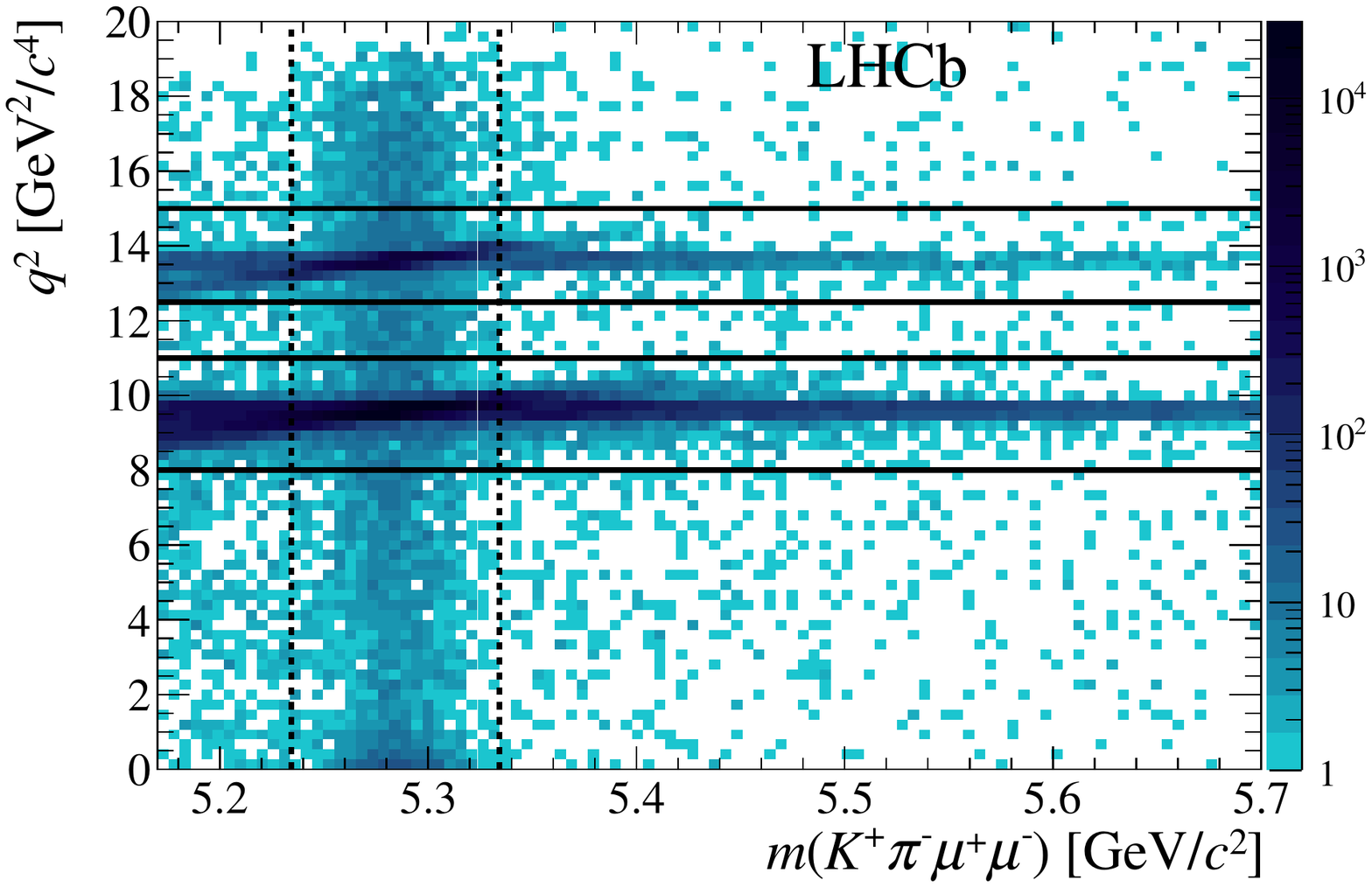}
  \caption{Invariant mass of the $\Kp\pim\mumu$ system versus $q^2$.
    The decay $\decay{\Bd}{\Kstarz\mumu}$ is clearly visible inside the dashed vertical lines.
    The horizontal lines denote the charmonium regions, where the tree-level decays $\decay{\Bd}{\jpsi\Kstarz}$ and $\decay{\Bd}{\psitwos\Kstarz}$ dominate. These candidates are excluded from the analysis. \label{fig:mbvsq2}}
\end{figure}

The $\Kp\pim\mumu$ invariant mass versus $q^2$ for candidates that pass the full selection is shown in Fig.~\ref{fig:mbvsq2}. 
The $\decay{\Bd}{\Kstarz\mumu}$ signal candidates are clearly visible as a vertical band.
The contributions from the decays $\decay{\Bd}{\jpsi\Kstarz}$ and $\decay{\Bd}{\psitwos\Kstarz}$,
which proceed through tree-level $b \to \ccbar s$ transitions, have a dimuon mass consistent with the known \jpsi or \psitwos meson mass and $m(\Kp\pim\mumu)$ consistent with that of the known \Bz meson mass.
The horizontal bands are formed from combinatorial background comprising a genuine \jpsi or \psitwos meson and a \Kstarz candidate selected from elsewhere in the event.

\subsection{Background composition}
\label{sec:backgrounds}

In addition to combinatorial background, there are several sources of background that accumulate in $m(\Kp\pim\mumu)$ and can potentially mimic the signal decay if they are mis-reconstructed in the detector. 
These are referred to as {\it peaking} backgrounds.
Contamination from peaking backgrounds is estimated using samples of simulated events.

The tree-level decays $\decay{\Bd}{\jpsi\Kstarz}$ and $\decay{\Bd}{\psitwos\Kstarz}$
dominate in the regions $8.0<q^2<11.0\gevgevcccc$ and $12.5<q^2<15.0\gevgevcccc$, respectively, and these $q^2$ regions are therefore excluded from the analysis of the signal decay. 
However, these decays can still form a source of background if the $\mun$ ($\mup$) is misidentified as a $\pim$ ($\Kp$) and the $\pim$ ($\Kp$) is misidentified as a $\mun$ ($\mup$). 
To remove this background, candidates are rejected if the \pim (\Kp) satisfies the muon identification criteria and  the mass of the $\pim\mup$ ($\Kp\mun$) system, when the \pim (\Kp) is assigned the muon mass, is consistent with that of a \jpsi or \psitwos meson.
Possible pollution from $\decay{\Bd}{\Kstarz\phi(\to\mumu)}$ decays is removed by excluding from the analysis the $q^2$ region $0.98<q^2<1.10\gevgevcccc$. 

The decay $\decay{\Lb}{p\Km\mumu}$, which can proceed via \eg the $\PLambda(1520)$ resonance, can be a source of peaking background if the proton is misidentified as a pion.
This background is suppressed by rejecting candidates where the pion is not unambiguously identified by the RICH detectors and which have a mass close to the known \Lb mass, when the pion is assigned the proton mass.
Similarly, $\decay{\Lb}{p\Km\mumu}$ backgrounds with double misidentification of the hadrons,
\ie~where the proton is misidentified as a kaon and the kaon is misidentified as a pion, are suppressed using PID information. 

The decay $\decay{\Bs}{\phi(\to\Kp\Km)\mumu}$ can mimic the signal decay if one of the kaons is misidentified as a pion.
This background is suppressed by requiring stringent PID criteria if, after assigning the kaon mass to the pion candidate, the reconstructed invariant masses of the \Bz and \Kstarz candidates are consistent with the known $\Bs$ and $\phi$ masses.

The decay $\decay{\Bu}{\Kp\mumu}$ can form a background if a
low momentum pion from elsewhere in the event is added to form a four-particle final state.
The resulting invariant mass $m(\Kp\pim\mumu)$  will be larger than the known \Bz mass but can contribute to the upper mass-sideband. Such decays can therefore distort the estimate of the angular distribution of the residual background, which is assessed from this sideband. This background is suppressed by removing candidates with $5220<m(\Kp\mumu)<5340\mevcc$. 
It is also possible to have backgrounds from \decay{B^{0,+}}{K^{\ast 0,+}\mumu} decays, where the pion from the \Kstar meson is replaced by another pion from the rest of the event. 
This background does not peak in the signal region and is considered as part of the combinatorial background.  
Finally, \decay{\Bz}{\Kstarz\mumu} decays can form a background to \decay{\Bzb}{\Kstarzb\mumu} decays (and vice versa) if the $\Kp$ (\Km) is misidentified as the $\pip$ (\pim) and the $\pim$ (\pip) is misidentified as the $\Km$ (\Kp). 
These misidentified decays are referred to as {\it signal swaps} and are suppressed using PID information.

After all vetoes are applied, the largest peaking background contribution is from
$\Lb\to p\Km\mumu$ decays. The residual background from these decays is expected to be at a level of $(1.0\pm0.4)\%$ of the signal yield. 
The next largest backgrounds are $\decay{\Bd}{\Kstarz\mumu}$ signal swaps at $(0.64\pm0.06)\%$,  
misidentified $\decay{\Bs}{\phi\mumu}$ events at $(0.33\pm0.12)\%$ and
$\decay{\Bd}{\jpsi\Kstarz}$ decays with double misidentification at $(0.05\pm0.05)\%$ of the signal yield.
All of the sources of peaking background are sufficiently small such that they are neglected in the angular analysis  but are considered further as sources of systematic uncertainty. 
The background from \bquark-hadron decays where two hadrons are misidentified as muons is negligible. 
The largest residual background is combinatorial in nature and varies smoothly with $m(\Kp\pim\mumu)$, $m(\Kp\pim)$ and the decay angles.

\section{Angular acceptance}
\label{sec:acceptance}

The triggering, reconstruction and selection criteria distort the distributions of the decay angles $\thetal$, $\thetak$ and $\phi$, as well as the $q^2$ distribution, giving rise to so-called {\it acceptance effects}.  
The dominant acceptance effects come from momentum and impact parameter requirements. 
In particular, the implicit momentum threshold that is required for tracks to traverse the magnetic spectrometer removes low momentum particles.
In contrast to the previous LHCb  analyses~\cite{LHCb-PAPER-2013-019,LHCb-PAPER-2013-037},
the acceptance is not assumed to factorise in the three decay angles. 
Instead, the efficiency is parameterised in four dimensions, according to
\begin{align}
  \eps(\cos\thetal,\cos\thetak,\phi,q^2) &= \sum_{ijmn} c_{ijmn} L_i(\cos\thetal) L_j(\cos\thetak) L_m(\phi) L_n(q^2),
\label{eq:acceptance}
\end{align}
where the terms $L_h(x)$ denote Legendre polynomials of order $h$ and the observables are rescaled to the range $-1 < x < +1$ when evaluating the polynomial.  
For $\cos\thetal$, $\cos\thetak$ and $\phi$, the sum in Eq.~(\ref{eq:acceptance}) encompasses $L_h(x)$ up to fourth, fifth and sixth order, respectively.
The $q^2$ parameterisation comprises $L_h(x)$ up to fifth order. 
The coefficients $c_{ijmn}$ are determined using a principal moment analysis of simulated three-body $\decay{\Bd}{\Kstarz\mumu}$ phase-space decays. 
As the efficiency is parameterised in terms of all of the relevant kinematic variables needed to describe the decay, it does not depend on the model used in the simulation.

\begin{figure}[htb]
\begin{center}
\includegraphics[width=0.48\linewidth]{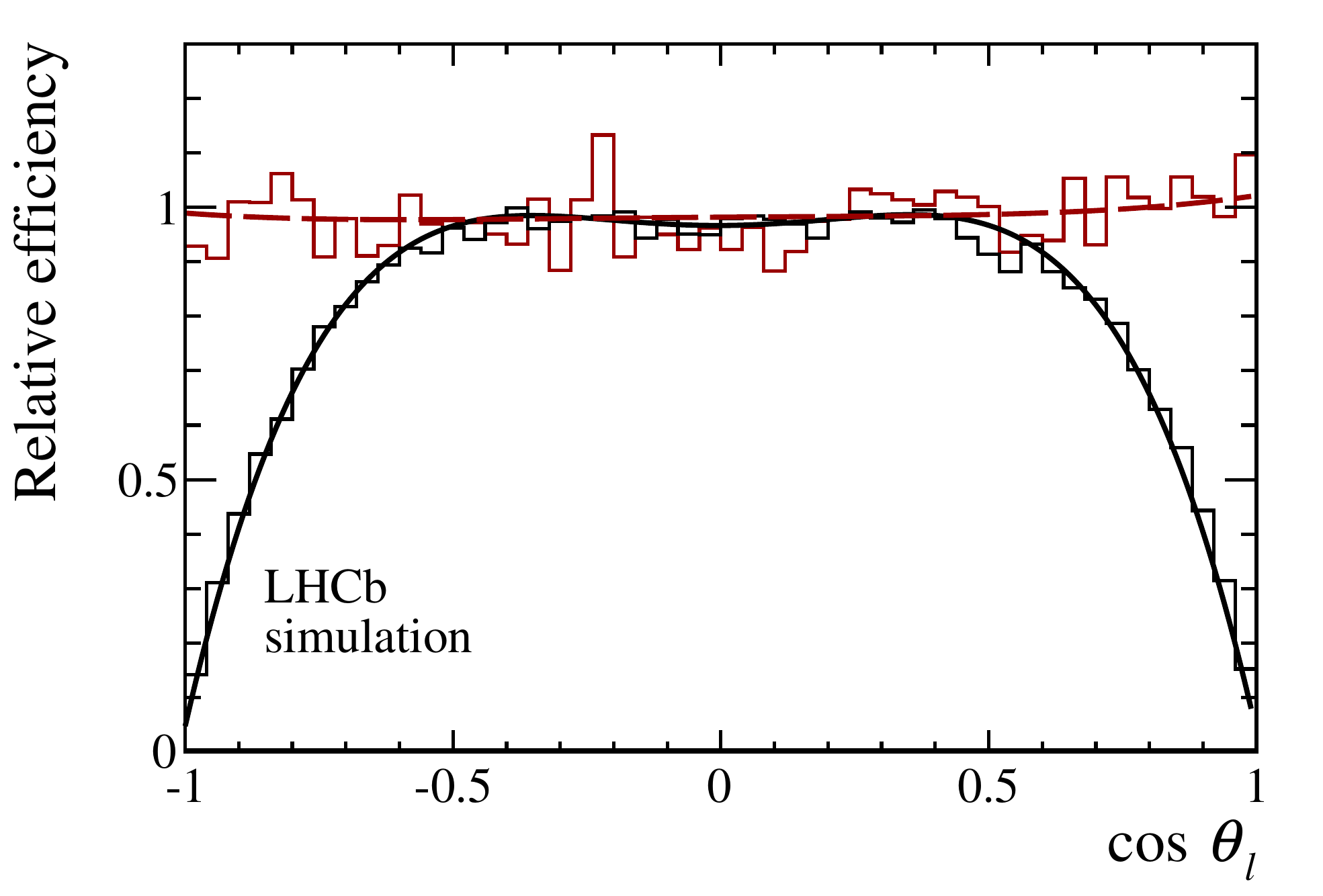} 
\includegraphics[width=0.48\linewidth]{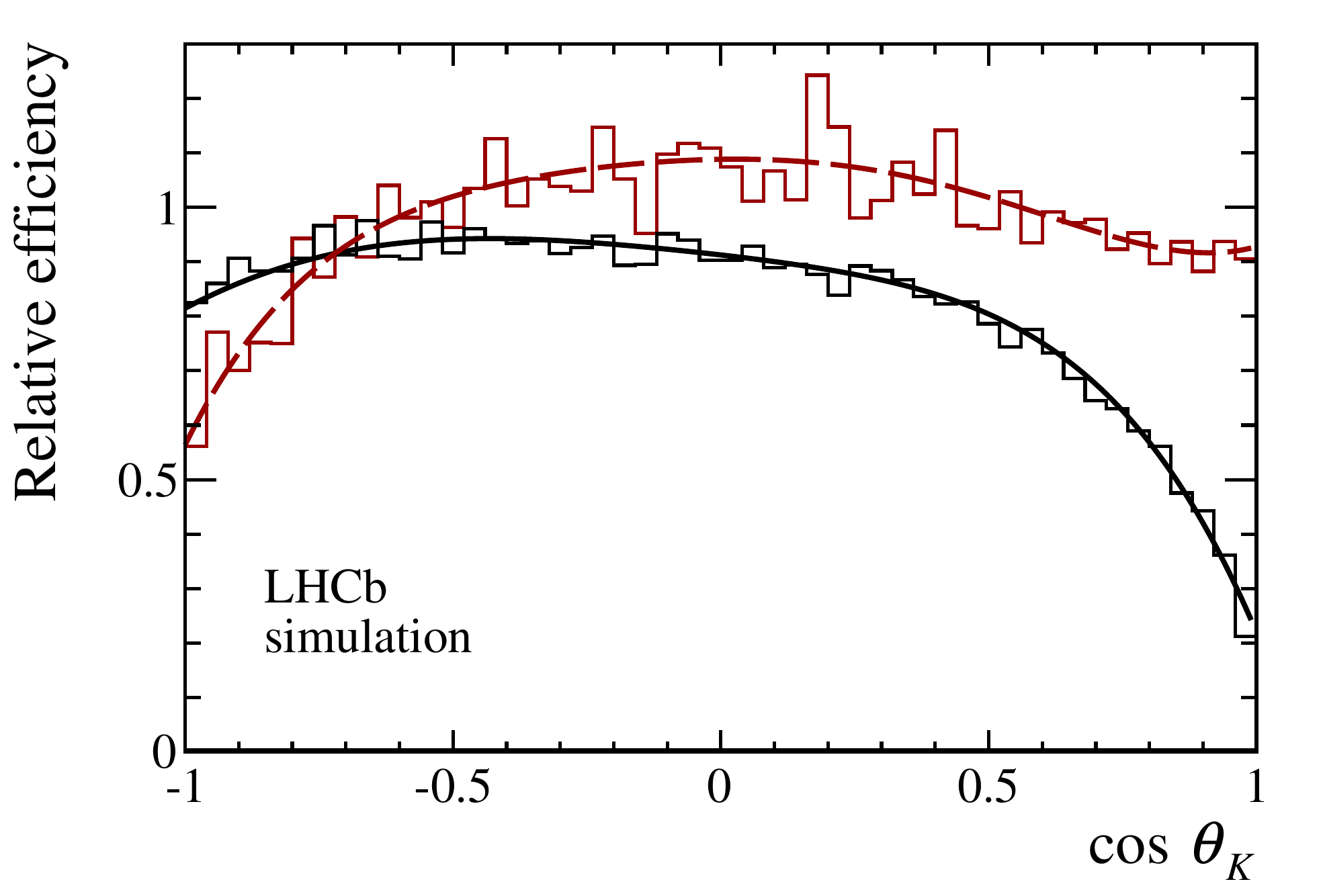}  \\ 
\includegraphics[width=0.48\linewidth]{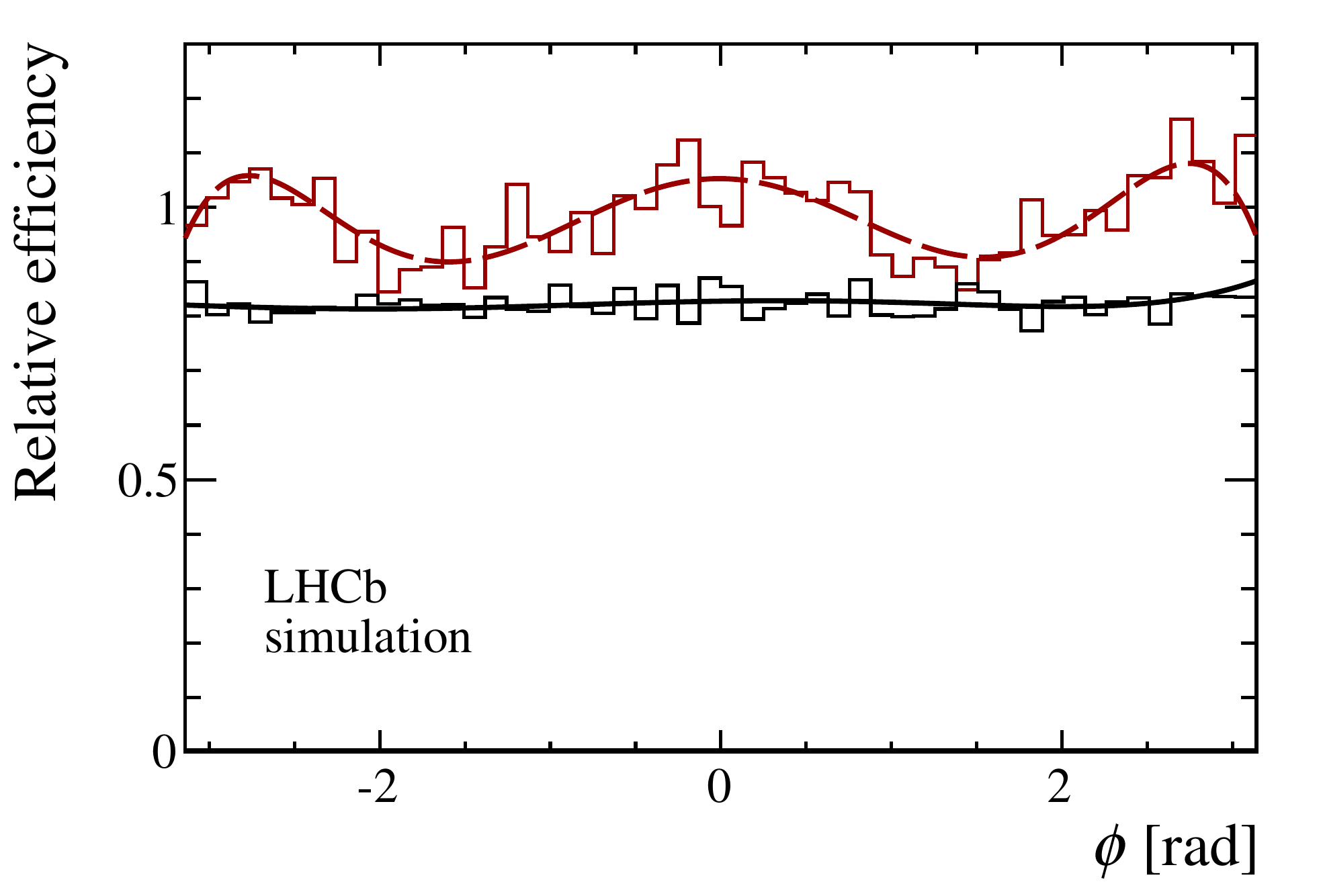} 
\includegraphics[width=0.48\linewidth]{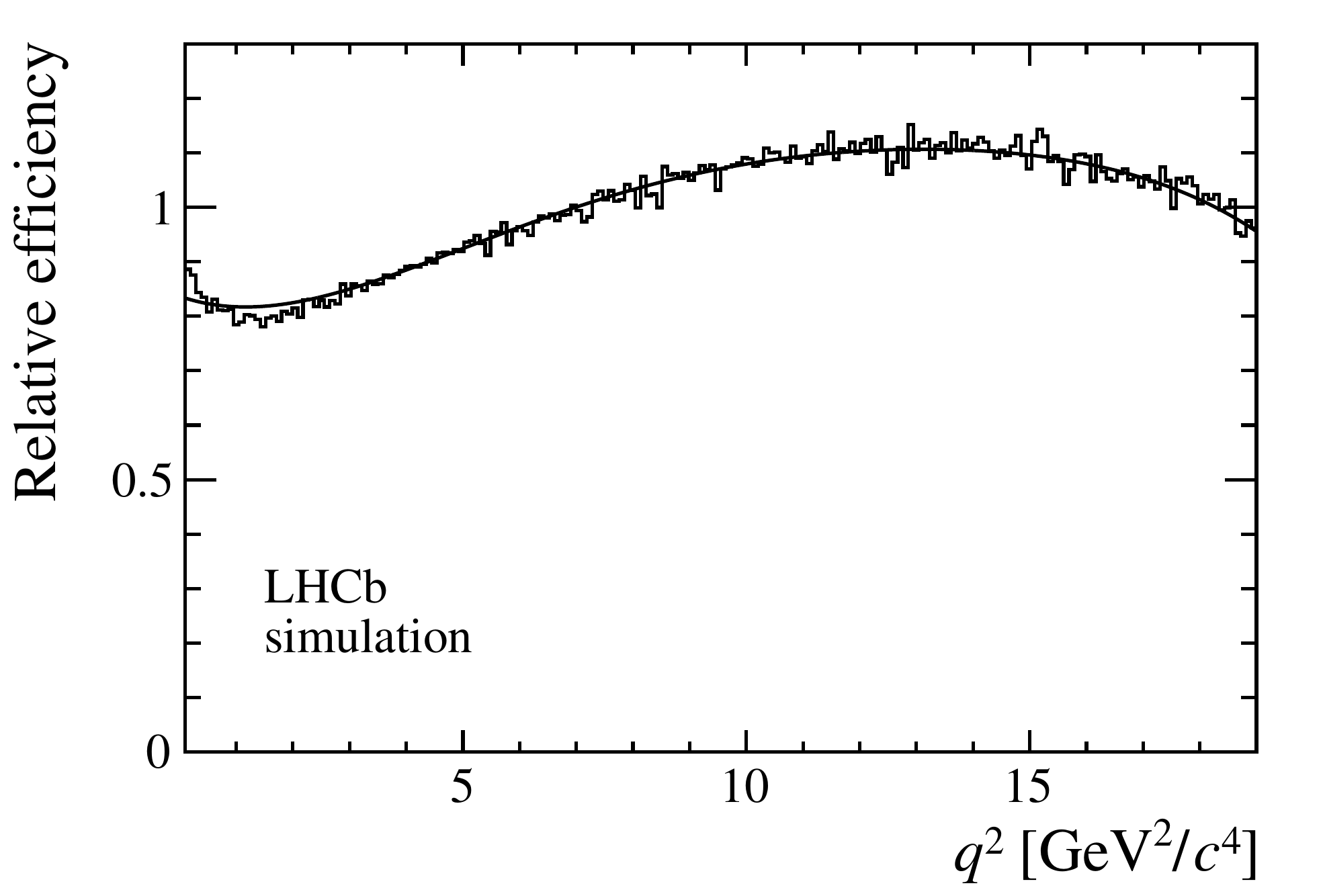} 
\end{center}
\caption{ 
Relative efficiency in $\cos\thetal$, $\cos\thetak$, $\phi$ and \qsq, as determined from a principal moment analysis of simulated three-body $\decay{\Bd}{\Kstarz\mumu}$ phase-space decays. 
The efficiency as a function of  $\cos\thetal$, $\cos\thetak$ and $\phi$ is shown for the regions $0.1 < \qsq < 0.98\gev^{2}/c^{4}$ (black solid line) and $18.0 < \qsq <19.0 \gev^{2}/c^{4}$ (red dashed line). 
The efficiency as a function of \qsq is shown after integrating over the decay angles.
The histograms indicate the distribution of the simulated three-body $\decay{\Bd}{\Kstarz\mumu}$ phase-space decays used to determine the acceptance. 
\label{fig:acceptance} 
} 
\end{figure}

The angular acceptance in $\cos\thetal$, $\cos\thetak$ and $\phi$ is shown for $0.10 < \qsq < 0.98\gev^{2}/c^{4}$ and $18.0 < \qsq < 19.0 \gev^{2}/c^{4}$ in Fig.~\ref{fig:acceptance}. 
The acceptance varies smoothly as a function of \qsq between these extremes.
The acceptance as a function of \qsq, after integrating over the decay angles, is also shown.
The description of the angular acceptance is cross-checked, for $\qsq = m^{2}(\jpsi)$, using the decay $\decay{\Bd}{\jpsi\Kstarz}$. 
This decay can be selected in the data with background contamination below 1\% and the angular structure has been determined by measurements made by the \babar, Belle and  \lhcb  collaborations~\cite{Aubert:2007hz, Chilikin:2014bkk, Aaij:2013cma}. 
With the acceptance correction derived using the above method, the $\decay{\Bd}{\jpsi\Kstarz}$ angular observables obtained from the LHCb data are in good agreement with these previous measurements. 
The angular fit of the \decay{\Bz}{\jpsi\Kstarz} data is shown in Fig.~\ref{fig:supp:jpsikstar} of Appendix~\ref{sec:projections}. 

\section[The $\Kp\pim\mumu$ mass distribution]{The \boldmath{$\Kp\pim\mumu$} and \boldmath{$\Kp\pim$} mass distributions} 
\label{sec:mass} 

\begin{figure}
  \centering
  \includegraphics[width=0.48\linewidth]{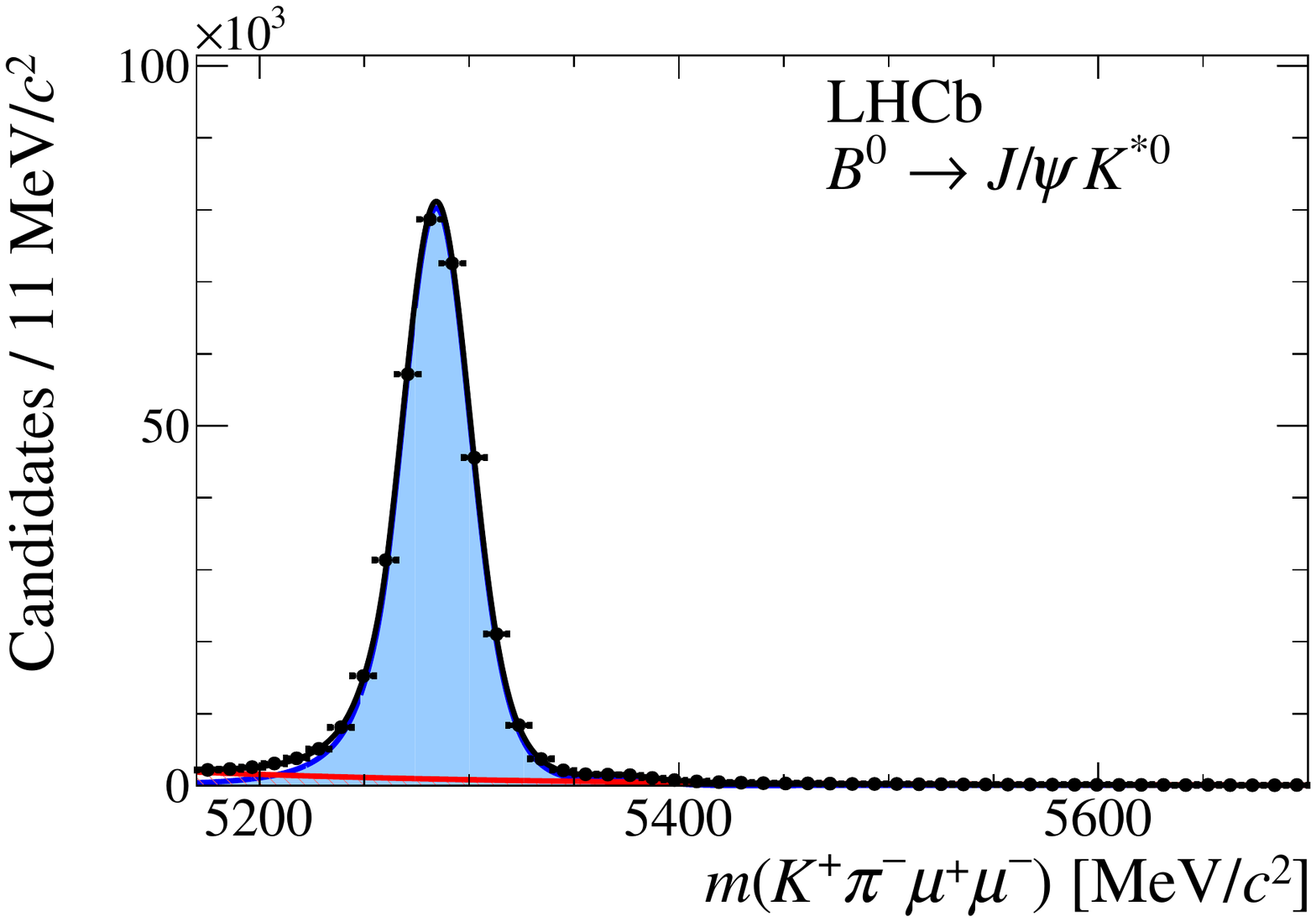}
  \includegraphics[width=0.48\linewidth]{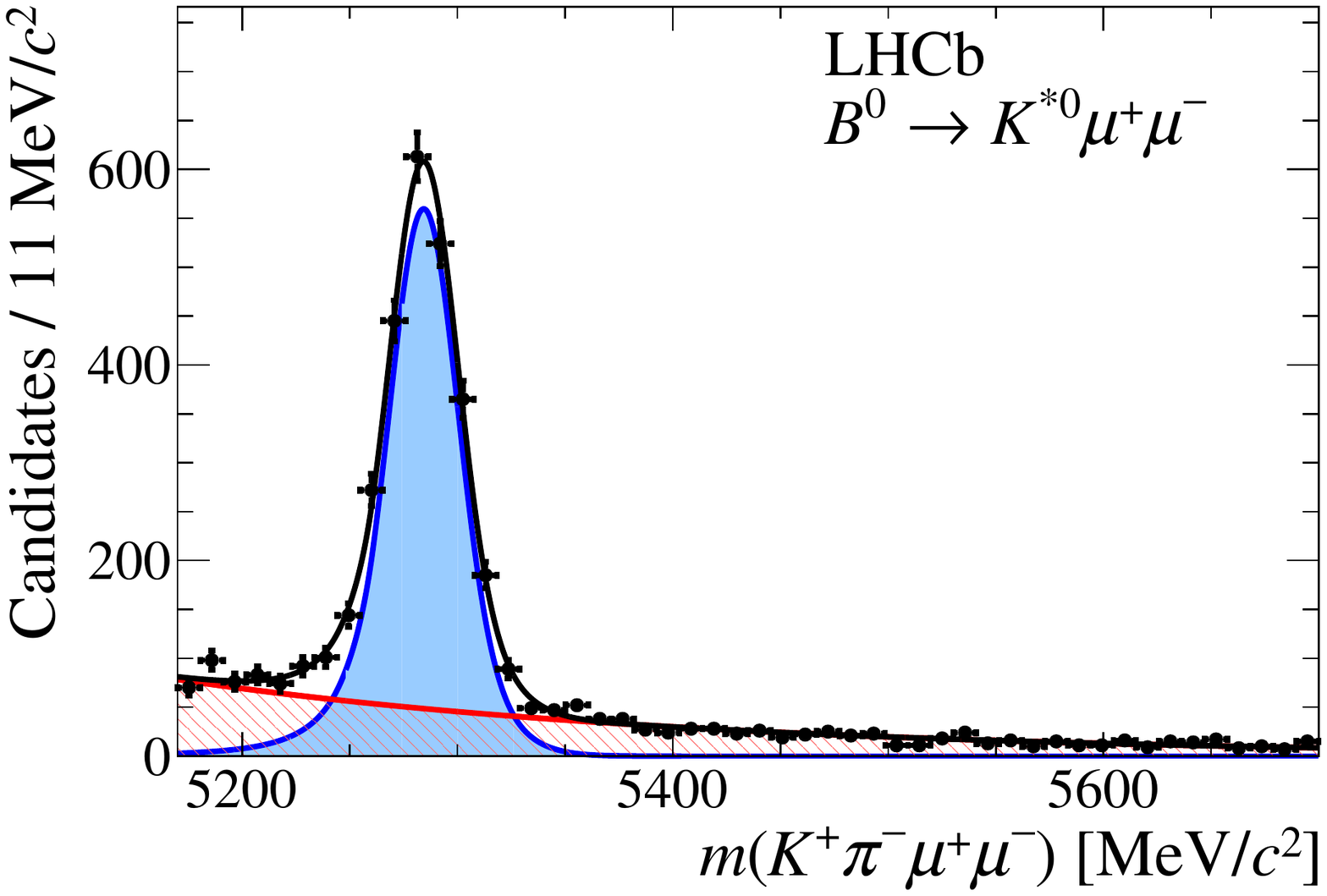}
  \caption{Invariant mass $m({\Kp\pim\mumu})$ for (left) the control decay $\decay{\Bd}{\jpsi\Kstarz}$ and (right) the signal decay $\decay{\Bd}{\Kstarz\mumu}$, integrated over the full $q^2$ range (see text). 
    Overlaid are the projections of the total fitted distribution (black line) and the signal and background components.
   The signal is shown by the blue shaded area and the background  by the red hatched area.
\label{fig:massfit}}
\end{figure}

The ${\Kp\pim\mumu}$ invariant mass is used to discriminate between signal and background. 
The distribution of the signal candidates is modelled using the sum of two Gaussian functions with a common mean, each with a power-law tail on the low-mass side.
The parameters describing the signal mass-shape are determined from a fit to the $\decay{\Bd}{\jpsi\Kstarz}$ decay in the data, as shown in Fig.~\ref{fig:massfit}, and are subsequently fixed when fitting the \decay{\Bz}{\Kstarz\mumu} candidates.
In samples of simulated \decay{\Bz}{\Kstarz\mumu} decays, the mass resolution is observed to vary with \qsq by 2--8\%. A scale factor is therefore taken from the simulation and is used to correct the width of the Gaussian functions in the different \qsq bins. 
A component is included in the fit to account for \decay{\Bsb}{\jpsi\Kstarz} decays, which are at a level of 0.8\% of the \decay{\Bz}{\jpsi\Kstarz} signal yield~\cite{LHCb-PAPER-2012-014}.
However, the \decay{\Bsb}{\Kstarz\mumu} decay is neglected when fitting \decay{\Bz}{\Kstarz\mumu} candidates.
Combinatorial background is described well by a single exponential distribution in $m(\Kp\pim\mumu)$. 
The $\decay{\Bz}{\Kstarz\mumu}$ signal yield integrated over the $q^2$ ranges $0.1 < \qsq < 8.0\gev^{2}/c^{4}$, $11.0 < \qsq < 12.5\gev^{2}/c^{4}$ and $15.0 < \qsq < 19.0\gev^{2}/c^{4}$ is determined to be $2398\pm 57$. 
The signal yield in the range $1.1 < \qsq < 6.0\gev^{2}/c^{4}$ is $624\pm30$.

As detailed in Secs.~\ref{sec:likelihood} and \ref{sec:moments}, the likelihood fit and the method of moments use additional information from the $m(\Kp\pim)$ distribution to constrain the fraction of $\Kp\pim$ S-wave present in the data.
To describe this distribution, the $\Kstarz$ signal component is modelled using a relativistic Breit-Wigner function for the P-wave component and the LASS parameterisation~\cite{Aston1988493} for the S-wave component. 
The combinatorial background is described by a linear function in $m(\Kp\pim)$. 
There is no evidence for a \Kstarz component in the background $m(\Kp\pim)$ distribution.
When fitting for \Kstarz decay amplitudes, $m(\Kp\pim)$ is integrated over, as detailed in Sec.~\ref{sec:amplitudes}. 

\section{Angular analysis of the decay} 
\label{sec:angular}

The three methods used to determine the \CP-averaged angular observables, \CP asymmetries and the zero-crossing points of $S_4$, $S_5$ and $A_{\rm FB}$ are detailed below. 
Section~\ref{sec:likelihood} describes the determination of the observables in bins of \qsq using a maximum likelihood fit. 
Section~\ref{sec:moments} discusses the determination of the same set of observables using a principal moment analysis. 
Finally, Sec.~\ref{sec:amplitudes} describes a fit to the angular and \qsq distribution of the decay, parameterised in terms of the decay amplitudes rather than the observables.  This fit is used to determine the zero-crossing points of $S_4$, $S_5$ and $A_{\rm FB}$.

\subsection{Determination of angular observables with a likelihood fit} 
\label{sec:likelihood}

In each $q^2$ bin, an unbinned maximum likelihood fit to $m(\Kp\pim\mumu)$
and the three decay angles $\cos\thetal$, $\cos\thetak$ and $\phi$ is used to determine the angular observables introduced in Sec.~\ref{sec:ang}.
The angular distribution of the signal is described using Eq.~(\ref{eq:pdfswave}). 
The background angular distribution is modelled with second order polynomials in $\cos\thetal$, $\cos\thetak$ and $\phi$, the parameters of which are left free in the fit. 
The angular distribution is assumed to factorise in the three decay angles. 
This assumption has been validated in the upper mass sideband.

In order to describe the signal angular distribution, the
angular acceptance discussed in Sec.~\ref{sec:acceptance} must be accounted for.  
The acceptance is treated in one of two ways, depending on the \qsq range being fitted.  
In the narrow $q^2$ bins, the acceptance is treated as being constant across each bin and is included in the fit by multiplying Eq.~(\ref{eq:pdfswave}) by the acceptance function evaluated at the bin centre.  
In the wider $1.1 < \qsq < 6.0\gev^{2}/c^{4}$ and $15.0 < \qsq < 19.0\gev^{2}/c^{4}$ bins, the shape of the acceptance can vary significantly across the bin.  
In this case, the candidates are weighted in the likelihood fit by the inverse of their efficiency.  
The event weights are scaled such that this pseudo-likelihood fit has confidence intervals with the correct coverage.

\begin{figure}[!ht]
  \centering
  \includegraphics[width=7cm]{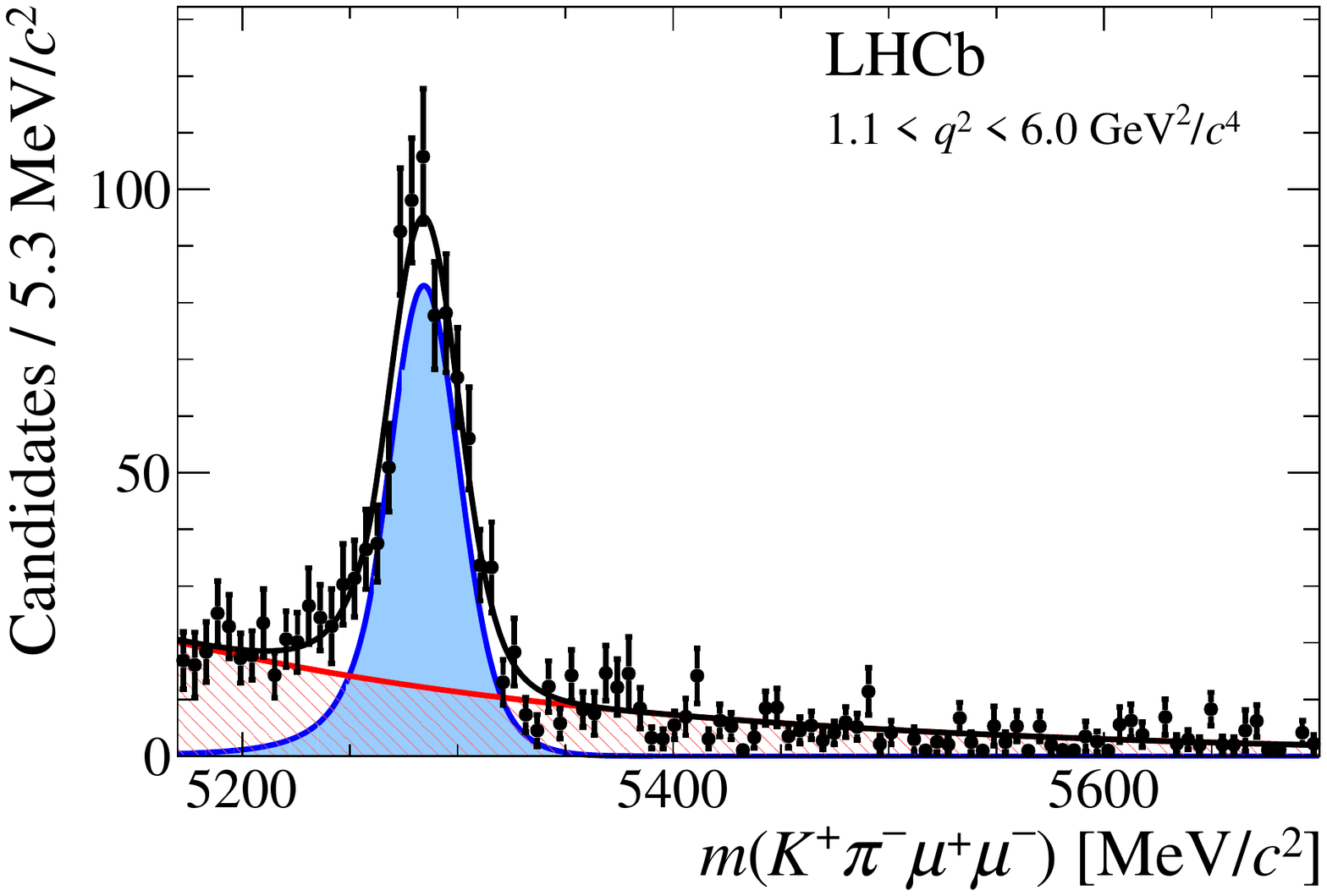}
  \includegraphics[width=7cm]{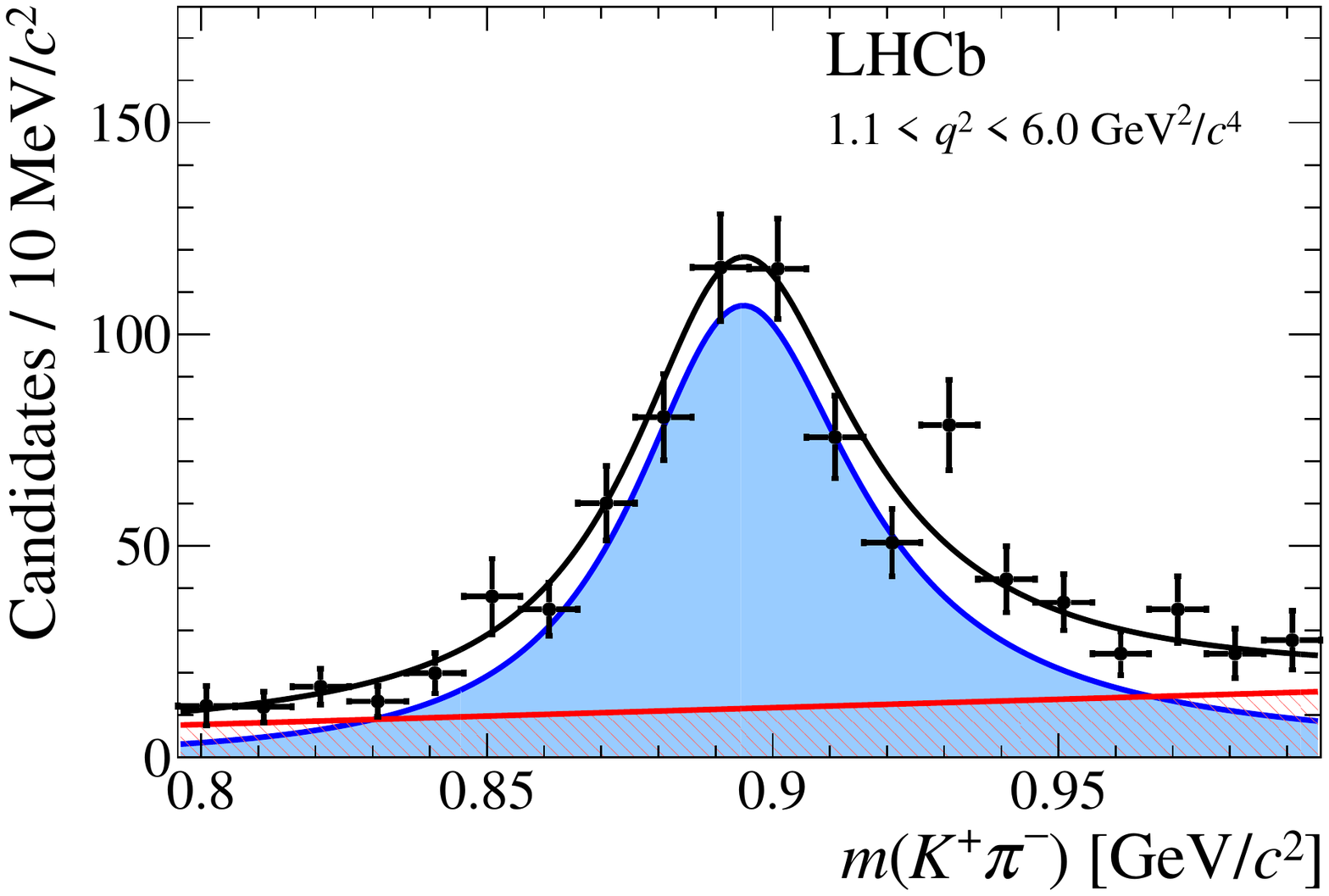}\\
  \includegraphics[width=7cm]{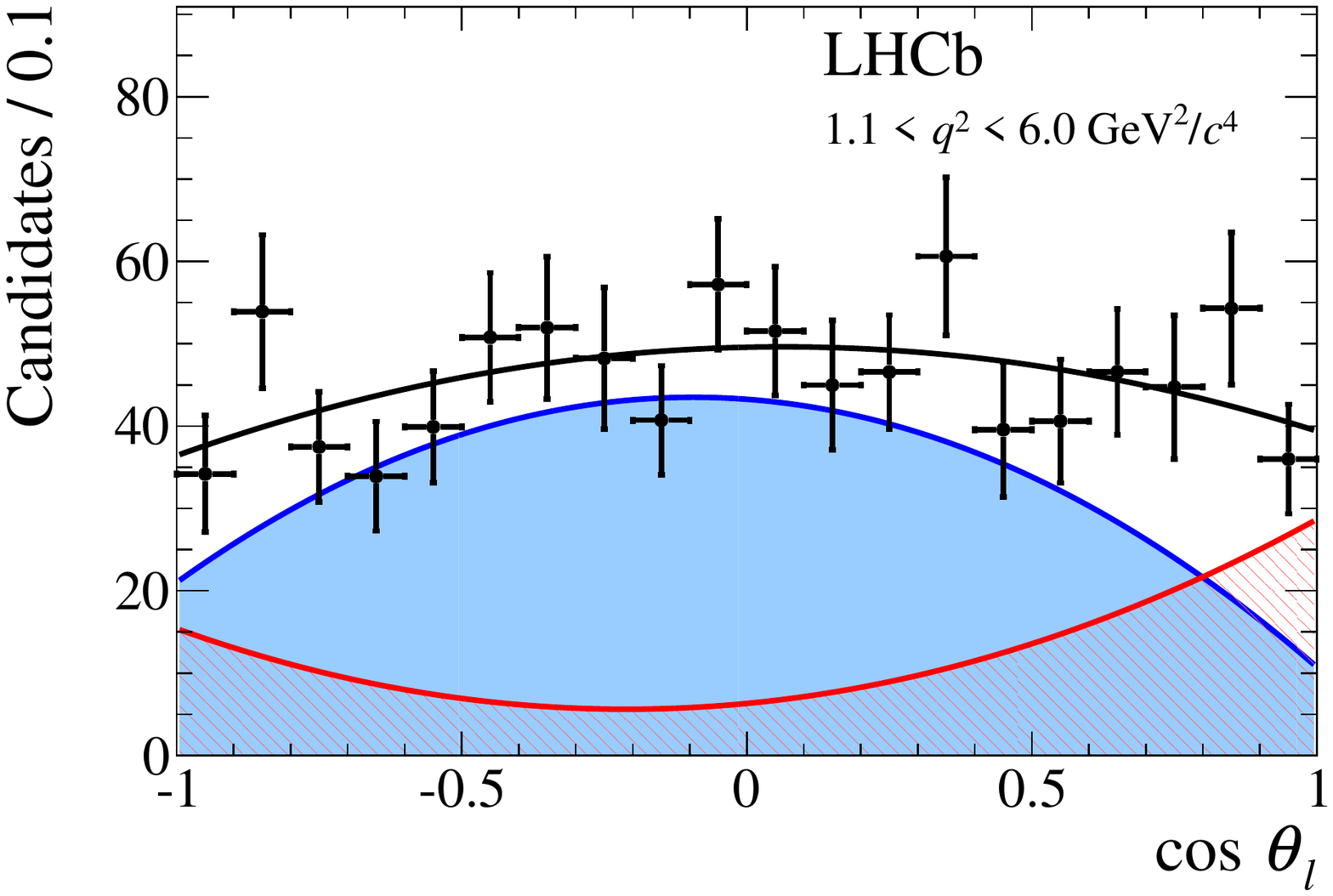}
  \includegraphics[width=7cm]{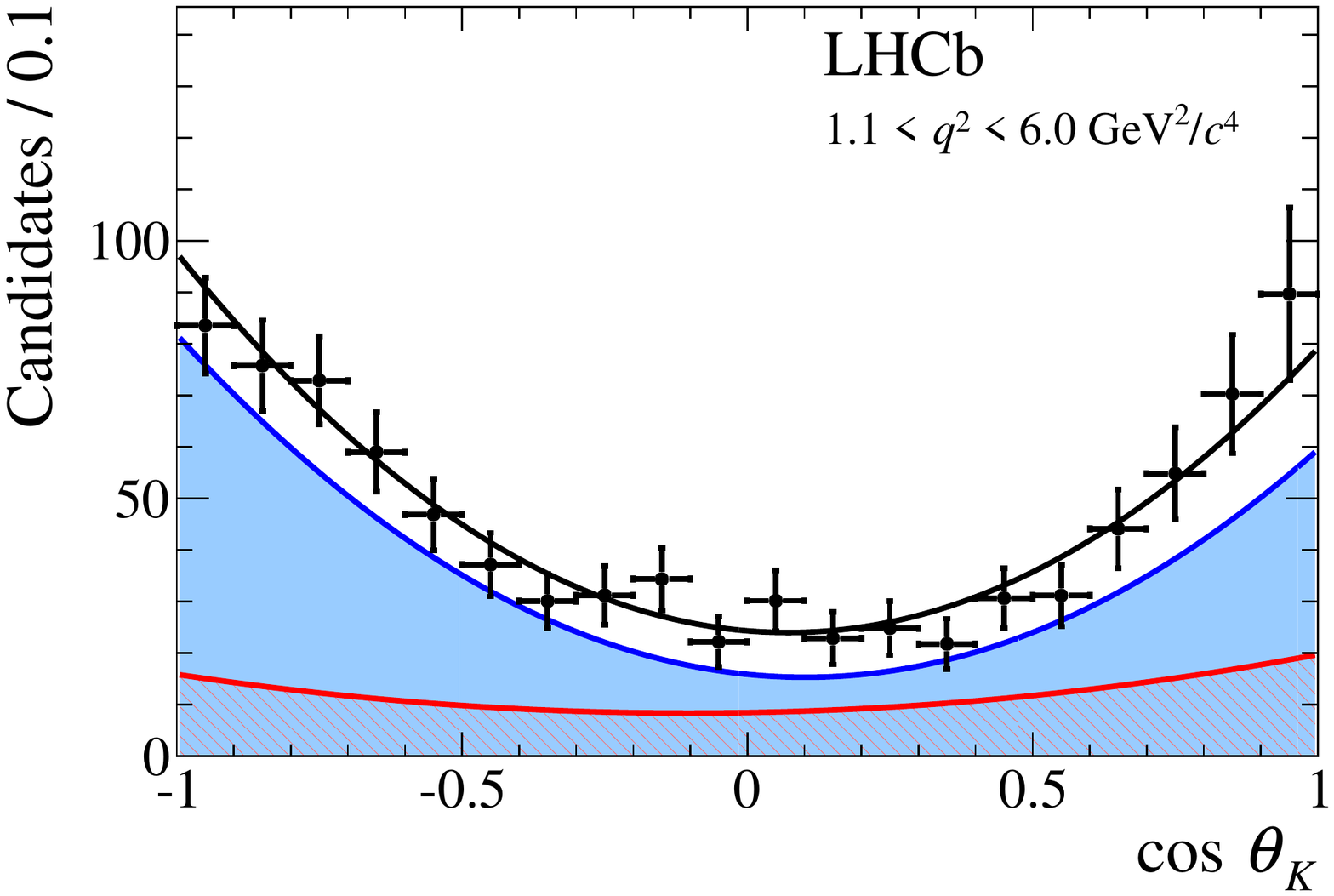}\\
  \includegraphics[width=7cm]{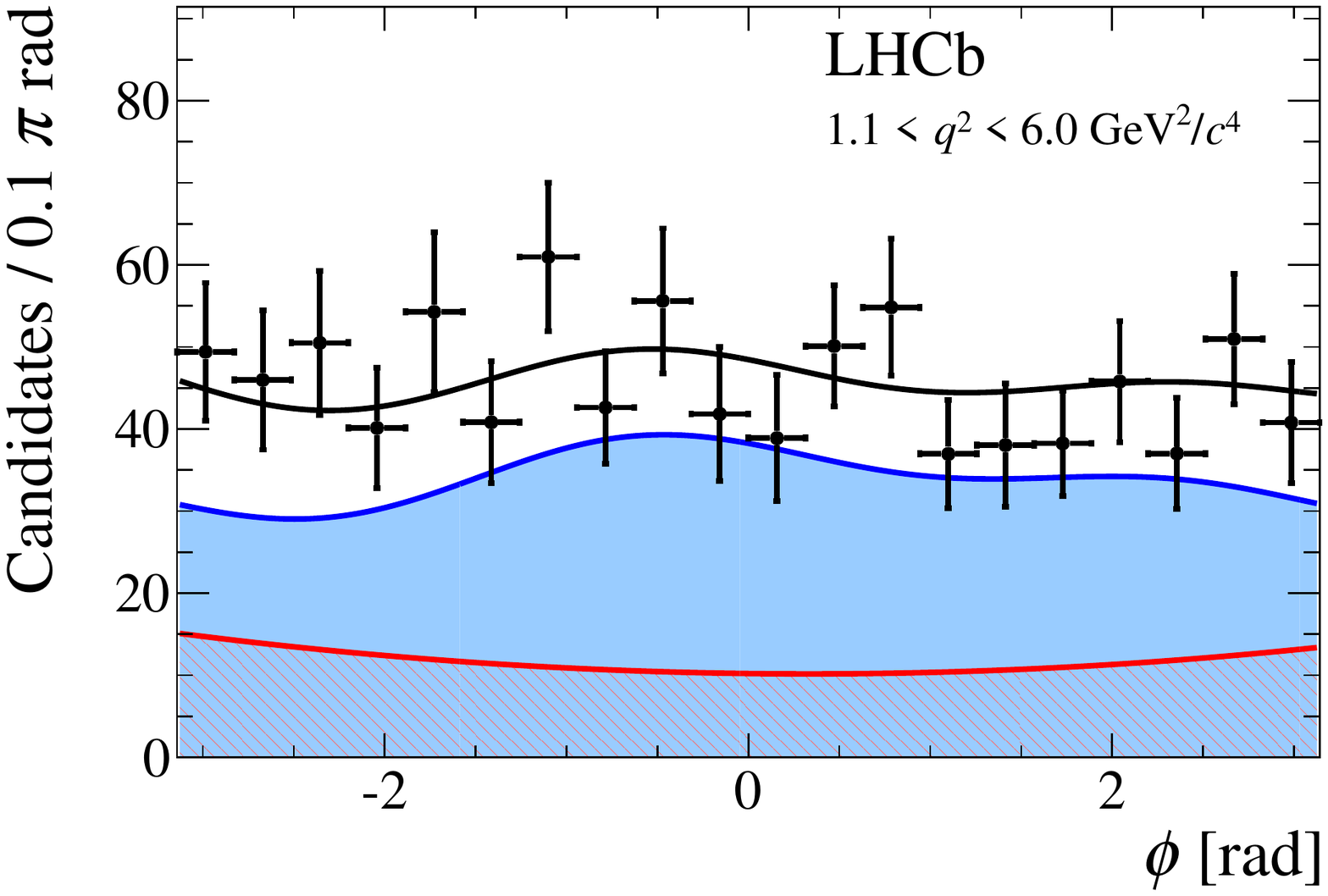}  
  \caption{Angular and mass distributions for $1.1<q^2<6.0\gevgevcccc$.
    The distributions of $m(\Kp\pim)$ and the three decay angles are given for candidates in the signal mass window $\pm50\mevcc$ around the known $\Bd$ mass. 
The candidates have been weighted to account for the acceptance.
    Overlaid are the projections of the total fitted distribution (black line) and its different components.
      The signal is shown by the blue shaded area and the background by the red hatched area. \\ \vspace{2cm}
\label{fig:projections}
  }
\end{figure}

The $\Kp\pim\mumu$ invariant mass is included in the fit to separate signal from background.
The signal and background mass distributions are parameterised as described in Sec.~\ref{sec:mass}.
In order to better constrain the S-wave fraction, a simultaneous fit of the $m(\Kp\pim)$ distribution is performed using the parameterisation described in Sec.~\ref{sec:mass}.
The signal fraction and $F_{\rm S}$ are common parameters in the simultaneous fits to the $m(\Kp\pim)$ distribution and to the angular and $m(\Kp\pim\mumu)$ distributions.
Figure~\ref{fig:projections} shows the projections of the fitted
probability density function on the angular and mass distributions for
the $1.1<q^2<6.0\gevgevcccc$ $q^2$~bin.  
Good agreement of the fitted function with the data is observed.  
Projections for the other \qsq bins are provided in Appendix~\ref{sec:projections}.

The $P_i^{(\prime)}$ observables introduced in Sec.~\ref{sec:ang} are determined by reparameterising
Eq.~(\ref{eq:pdfpwave}) using a basis comprising $F^{}_{\rm L}$, $P^{}_{1,2,3}$ and $P_{4,5,6,8}^{\prime}$.  
The \CP asymmetries are determined by modifying the angular convention, introducing a relative sign between the angular terms $f_3(\vec{\Omega})$--$f_9(\vec{\Omega})$ for \Bz\ and \Bzb\ decays, such that Eq.~(\ref{eq:pdfpwave}) is given in terms of $F_{L}$ and the \CP asymmetries $A_{3}$--$A_{9}$. 
The \Bz\ or \Bzb\ flavour is determined from the charge of the final-state kaon.

To ensure correct coverage for the uncertainties of the angular
observables, the Feldman-Cousins method~\cite{1998PhRvD..57.3873F} is
used with nuisance parameters treated according to the plug-in method~\cite{plugin}. 
Angular observables are considered one at a time, with the other angular observables treated as nuisance parameters.
The nuisance parameters also include the signal fraction, the background parameters, $F_{\rm S}$ and the angular terms that arise from interference between the S- and P-wave.

\subsection{Determination of angular observables using the method of moments}
\label{sec:moments}

The angular observables are also determined using a principal moment analysis of the angular distribution, without making any angular fit to the data~\cite{Beaujean:2015xea,Gratrex:2015hna}.
As a continuous function of \qsq, the moments are given by
\begin{align}
M_{i}(\qsq) = \int \left(\frac{1}{\deriv(\Gamma+\bar{\Gamma})/\deriv q^2}\right)\frac{\deriv^4(\Gamma+\bar{\Gamma})}{\deriv\qsq\,\deriv\vec{\Omega}}  f_{i}(\vec{\Omega})  \deriv\vec{\Omega}\, .
\label{eq:moments:qsqdependent}
\end{align}
The average $M_i(\qsq)$ in a bin of \qsq is estimated by replacing the integral in Eq.~\ref{eq:moments:qsqdependent} with a sum over the candidates in the dataset. 
The angular acceptance is accounted for by weighting the candidates in the sum, 
\begin{equation}
 \widehat{M}_i = \frac{1}{\sum_{e} w_{e}} \sum_{e} w_{e} f_{i}(\vec{\Omega}_{e}) \, . 
\label{eq:momentsdefinition}
\end{equation}
The sum is evaluated for candidates within $\pm 50\mevcc$ of the \Bz mass.
The weight, $w_{e}$, is the reciprocal of the candidate's efficiency and is computed as described in Sec.~\ref{sec:acceptance}.
The mass window contains more than 95\% of the signal candidates.
This sum is also computed for candidates with $5350<m(\Kp\pim\mumu)< 5700\mevcc$ and the resulting value of $\widehat{M}_i$ is used to subtract the background contribution from the $\pm 50\mevcc$ window. 
The functions $f_i$ are given in Table~\ref{tab:angular}.
Due to their dependence on the spherical harmonics, most of the angular terms are orthogonal. For $f_{i = 3\text{--}9}$,
\begin{align}
\int f_i(\vec{\Omega})f_j(\vec{\Omega}) \deriv\vec{\Omega} = \lambda_i \delta_{ij}~,
\end{align}
such that the moments give the \CP-averaged observables $S_{3}$ to $S_{9}$ with a coefficient, $\lambda_i$, that takes into account the normalisation. 
In the limit of massless muons, the moments are related to the observables by the expressions
\begin{align}
M_i= \begin{cases}
\frac{8}{25} (1-F_{\rm S}) S_{i} & \text{if}\quad  i=3,4,8,9,\\
\frac{2}{5} (1-F_{\rm S}) S_{i}  &\text{if}\quad i=5,6s,7,\\
\frac{2}{5}(1-F_{\rm S}) (2-F_{\rm L}) + \frac{2}{3}F_{\rm S} &\text{if}\quad i= {1s},\\
\end{cases}
\label{eq:mom_ave_swave}
\end{align}
where, as noted previously, $A_{\rm FB} = \tfrac{3}{4} S_{6s}$.
The relevant signal and background yields and the S-wave fraction $F_{\rm S}$ are determined 
from a two-dimensional extended unbinned maximum likelihood fit to the $m(\Kp\pim\mumu)$ and $m(\Kp\pim)$ distributions, 
using the shapes described in Sec.~\ref{sec:mass}. 

The statistical uncertainties of the angular moments are estimated using a bootstrapping technique~\cite{Efron:1979}.  
Confidence intervals are defined such that they include the 16$^{\text{th}}$--84$^{\text{th}}$ percentiles of the bootstrap distribution of the observables. 
When computing the $P_i^{(\prime)}$ observables, bootstrapped data with unphysical $F_{\rm L}$ ($F_{\rm L} < 0$ or $F_{\rm L} > 1$) are added at $\pm \infty$ to ensure that the resulting intervals do not undercover.
As in the likelihood method, the \CP asymmetries are determined by flipping the sign of the relevant \Bz angular terms.
The resulting moments are then used to determine the \CP asymmetries by substituting $A_{i}$ for $S_i$ in  Eq.~(\ref{eq:mom_ave_swave}).

In the moment analysis, an additional angular observable that is not present in the massless limit is determined. 
This observable is sensitive to large new scalar or tensor contributions to the decay and is associated with a new forward-backward asymmetry of the dimuon system, $f_{6c}(\vec{\Omega}) = \cos^{2}\thetak \cos\thetal$~\cite{Altmannshofer:2008dz,Beaujean:2015gba}. 
The corresponding observable is highly correlated to $A_{\rm FB}$ but can be determined from the moments $M_{6c}$ and $M_{6s}$, using $S_{6c} = 8 M_{6c} - 2 M_{6s}$.

\subsection{Determination of zero-crossing points using the decay amplitudes}
\label{sec:amplitudes}

In the $1.1 < \qsq < 6.0\gev^{2}/c^{4}$ region, it is also possible to determine the amplitudes ${\cal A}_{0,\parallel,\perp}^{\rm L, R}$, appearing in Table~\ref{tab:angular}, using a smoothly varying \qsq-dependent parameterisation.  
For $\qsq \gsim 6.0\gev^{2}/c^{4}$, resonant \ccbar states make a simple parameterisation of the \qsq-dependence impossible. 
A similar problem exists below 1.1~\gevgevcccc due to the presence of light resonances. 

The amplitudes  ${\cal A}_{0,\parallel,\perp}^{\rm L, R}$ are complex functions of \qsq and therefore, at each point in \qsq, the decay \decay{\Bd} {\Kstarz\mup\mun} is described by twelve real degrees of freedom.  
Several symmetries leave the angular distribution of the final-state particles unchanged~\cite{Egede:2010zc}. 
These symmetries allow four components of the amplitudes to be set to zero. 
This simplification results in eight independent degrees of freedom which completely describe the $\decay{\Bd}{\Kstarz\mup\mun}$ decay. 
In this paper, the choice 
\begin{align}
{\rm Re}({\cal A}_{0}^{\rm R}) =  {\rm Im}({\cal A}_{0}^{\rm R}) = {\rm Im}({\cal A}_{0}^{\rm L}) = {\rm Im}({\cal A}_{\perp}^{\rm R}) = 0
\end{align}
is made and the P-wave amplitudes are expressed using the form 
\begin{align} {\cal A}_{i = 0,\parallel,\perp}^{\rm L, R}(\qsq) = \alpha_i^{\rm L, R} + \beta_i^{\rm L, R} \qsq + \frac{\gamma_i^{\rm L, R}}{\qsq}~,
\end{align} 
where $\alpha_i^{\rm L, R}$, $\beta_i^{\rm L, R}$ and $\gamma_i^{\rm L,R}$ are complex coefficients.
The choice of which amplitude components to fix to zero and the form of the parameterisation are motivated in Ref.~\cite{Egede:2015kha}.

In the \qsq range considered, the S-wave amplitudes are expected to vary slowly with \qsq~\cite{Wang:2012swave}. 
To simplify the fit, these amplitudes are therefore assumed to be constant in \qsq and are described with a single complex parameter. 
The systematic uncertainty related to this approximation is negligible.  
After applying the symmetry constraints, the \Bz and \Bzb decays are each described by 24 real parameters for the P-wave amplitudes and four real parameters for the S-wave amplitudes. 
With the 3\invfb dataset, it is not possible to determine the parameters describing both the \Bz and \Bzb decays separately. 
It is therefore assumed that \CP symmetry holds in the decay such that the amplitudes describing the \Bz and \Bzb decays are identical.

An unbinned maximum likelihood fit to the distributions of $m(\Kp\pim\mumu)$, $\cos\thetal$, $\cos\thetak$, $\phi$ and \qsq is used to determine the amplitude parameters. 
The integral of the angular distribution is required to be consistent with the number of signal candidates in the fit.
For simplicity,  $m(\Kp\pim)$  is not included in the fit. 
The variation of the amplitudes with  $m(\Kp\pim)$ is accounted for by replacing products of amplitudes ${\cal A}_{i}^{\rm L,R}  {\cal A}_j^{*\,{\rm L,R}}$ with 
\begin{align}
 {\cal A}_{i}^{\rm L,R} {\cal A}^{*\,{\rm L,R}}_{j} \int g_i(m({\Kp\pim})) g^*_j(m({\Kp\pim})) \deriv m({\Kp\pim}) \, ,
\end{align}
where $g_{i}(m({\Kp\pim}))$ describes the variation of the amplitude ${\cal A}_{i}^{\rm L,R}$ with $m(\Kp\pim)$. The same models are used for the S- and P-wave lineshapes as in Secs.~\ref{sec:likelihood} and \ref{sec:moments}.
The acceptance as a function of $\cos\thetal$, $\cos\thetak$, $\phi$ and \qsq (as described in Sec.~\ref{sec:acceptance}) is included in the amplitude fit. 
The combinatorial background is parameterised by a linear function in \qsq. The background angular distribution is assumed to be independent of \qsq and is described by the product of three second-order polynomials. 
The background model and its factorisation in the decay angles and \qsq is checked using candidates in the $m(\Kp\pim\mumu)$ sideband.  

\begin{figure}[!h]
  \centering
  \includegraphics[width=0.45\linewidth]{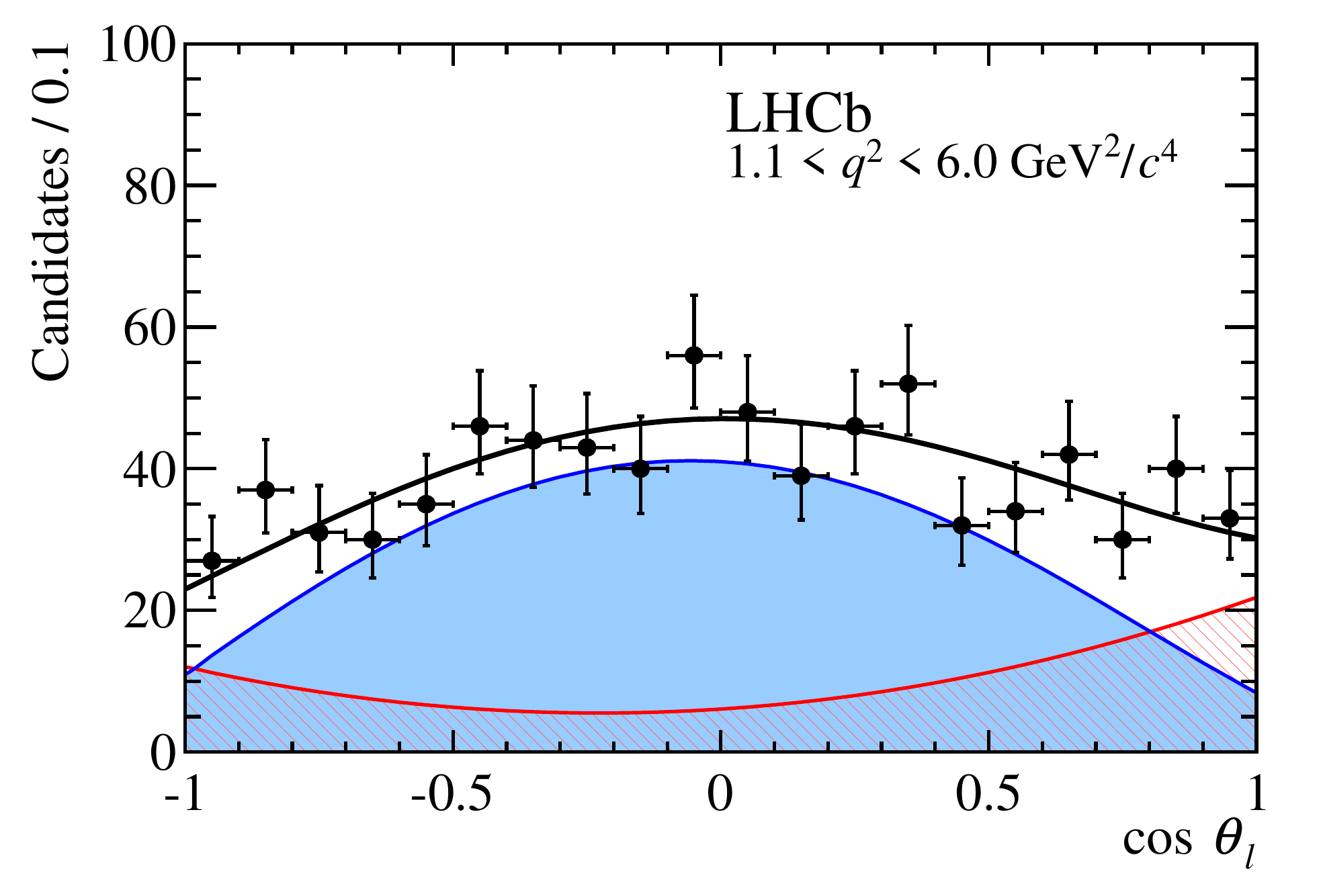}
  \includegraphics[width=0.45\linewidth]{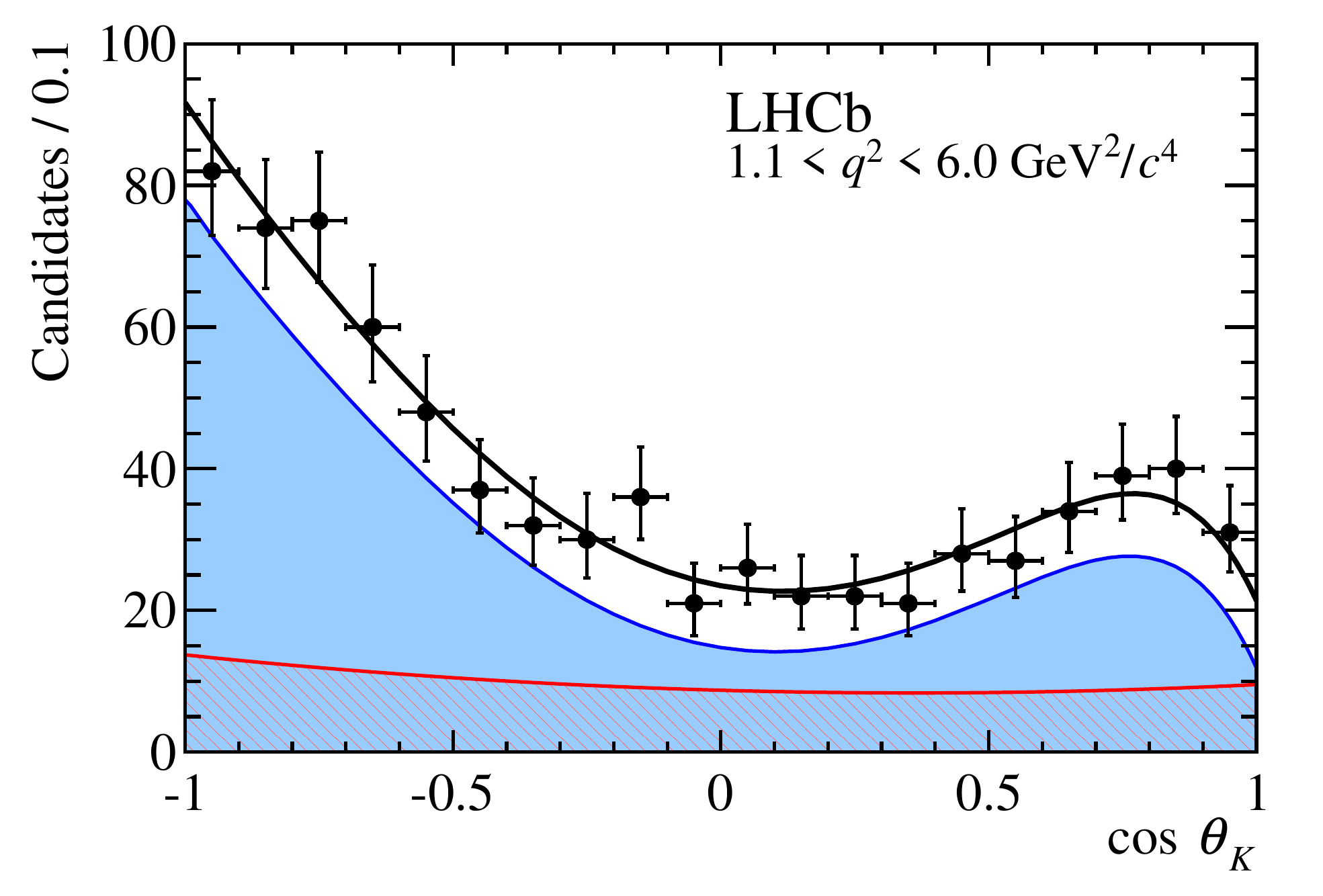} \\ 
  \includegraphics[width=0.45\linewidth]{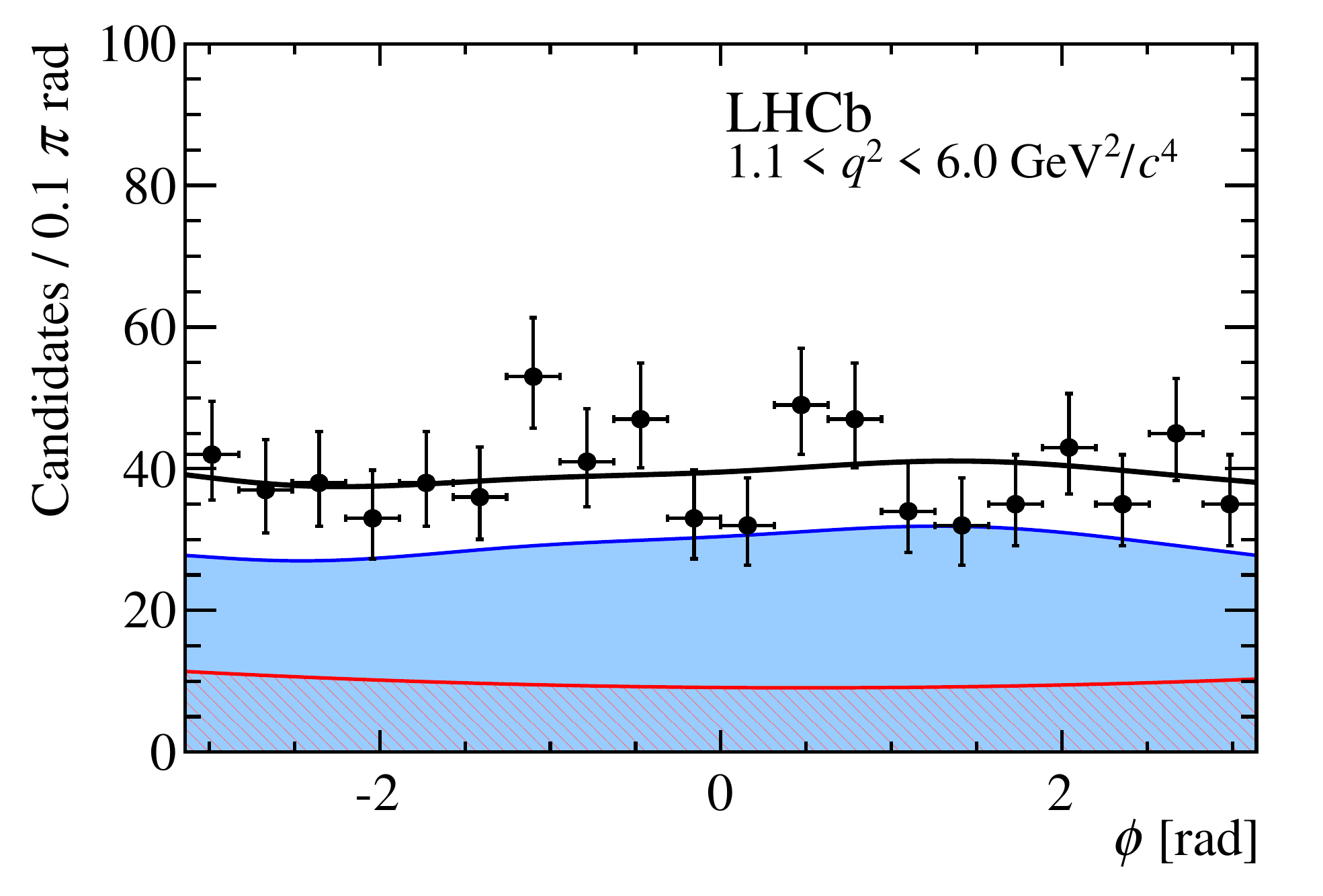}  
  \includegraphics[width=0.45\linewidth]{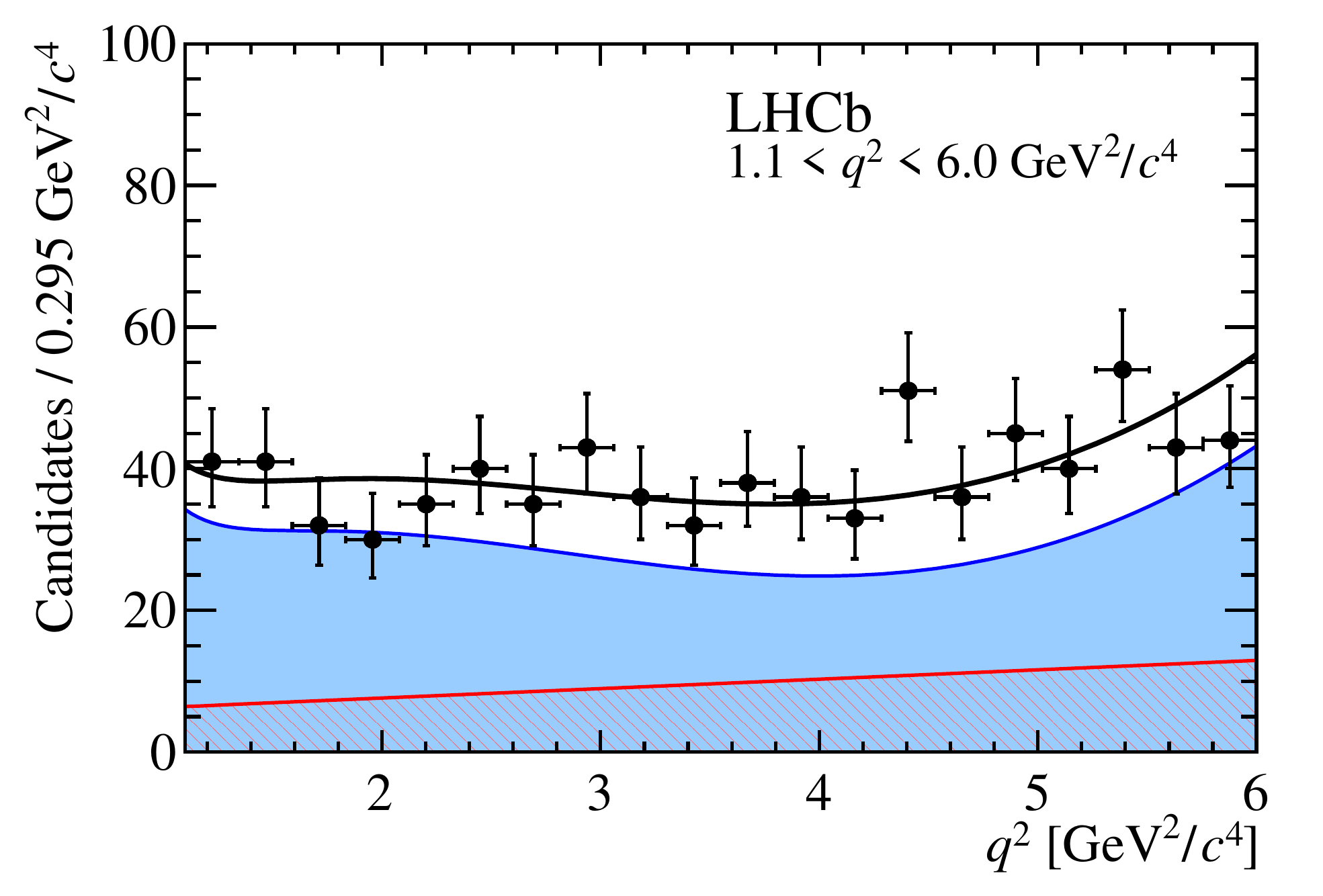}
  \caption{Angular and \qsq distribution of candidates overlaid by the result of the amplitude fit.
    The distribution of candidates in \qsq and the three decay angles is given in a $\pm50\mevcc$ window around the known $\Bz$ mass. 
    Overlaid are the projections of the total fitted distribution (black line) and its different components.
     The signal is shown by the blue shaded area and the background  by the red hatched area.
\label{fig:projections_amps}
}
\end{figure}

Figure~\ref{fig:projections_amps} shows the projections of the fitted probability density functions on the angular and \qsq distributions of the candidates. 
In contrast to Fig.~\ref{fig:projections}, the effect of the selection efficiency on the angles and \qsq is included in the signal distribution. 
The figure therefore shows the distribution of the candidates rather than the candidates weighted by the inverse of their selection efficiency.
Good agreement of the fitted function and the data is observed.

The amplitude parameters are used to construct observables as continuous functions of \qsq.  
The observables $S_4$, $S_5$ and $A_{\rm FB}$ have zero-crossing points and these are determined by solving a quartic equation.
The different solutions of this equation are separable based on the sign of the slope of the observable in the vicinity of the zero-crossing point.
Only zero-crossing points in the range $1.1 < \qsq < 6.0\gev^{2}/c^{4}$ with a local slope consistent with the data above $6.0\gev^{2}/c^{4}$ are retained. 

The large number of parameters floating in the fit, coupled with the limited number of signal candidates present in the dataset, results in a non-parabolic likelihood surface.
Therefore,  as in the determination of the angular moments, the statistical uncertainties of the \qsq-dependent observables and their corresponding zero-crossing points are determined using a bootstrapping technique~\cite{Efron:1979}. 
The statistical coverage of the resulting intervals is checked using simulated events and is found to be correct for the observables $S_4$, $S_5$ and $A_{\rm FB}$. 
Despite the coverage being correct, approximately $10\%$ of the bootstrapped datasets result in no zero-crossing point with the correct slope in the \qsq range $1.1 < \qsq < 6.0\gev^{2}/c^{4}$. 
In these cases, the zero-crossing point is added to the bootstrap distribution at $\pm\infty$ to ensure that the method does not undercover. 
The determination of the \qsq-dependent amplitudes in principle allows the full observable basis to be determined. However, pseudoexperiments indicate that a larger dataset is required in order to guarantee the correct coverage of the uncertainties on the observables other than $S_4$, $S_5$ and $A_{\rm FB}$.

\section{Systematic uncertainties}
\label{sec:systematics}

Effects that can alter the mass or angular distribution of either the signal or background candidates are sources of systematic uncertainty.
The various sources of systematic uncertainty are discussed in detail below and are summarised in Table~\ref{tab:systematics}.
In general, the systematic uncertainties are significantly smaller than the statistical uncertainties.

\begin{table}[ht]
\caption{Summary of the different sources of systematic uncertainty on the angular observables. 
Upper limits or typical ranges are quoted for the different groups of observables.  
The column labelled $q^2_0$ corresponds to the zero-crossing points of $S_4$, $S_5$ and $A_{\rm FB}$.
\label{tab:systematics}}
\begin{center}
\scalebox{1.00}{
\setlength\extrarowheight{2pt}
\begin{tabular}{r|ccccc}
Source                                             & $F_{\rm L}$            & $S_3$--$S_9$            &  $A_3$--$A_9$ & $P^{}_1$--$P'_{8}$ & $q^2_{0}$~\gevgevcccc     \\
\hline
Acceptance stat. uncertainty           & $<0.01$              & $<0.01$                    &  $<0.01$          &   $<0.01$           &  $\phantom{<}0.01$                             \\
Acceptance polynomial order                           & $<0.01$              & $<0.02$                    &  $<0.02$          &   $<0.04$           & $0.01$--$0.03$               \\
Data-simulation differences             & $0.01$--$0.02$  & $<0.01$                    & $<0.01$          &  $<0.01$            &   $<0.02$                         \\
Acceptance variation with $q^2$                      & $<0.01$              & $<0.01$                    &  $<0.01$          &   $<0.01$            & --                                     \\
$m(\Kp\pim)$ model                             & $<0.01$              & $<0.01$                    &   $<0.01$         &    $<0.03$   &  $<0.01$                           \\
Background model                          & $<0.01$              & $<0.01$                    &  $<0.01$          &  $<0.02$             &   $0.01$--$0.05$             \\
Peaking backgrounds                       & $<0.01$              & $<0.01$                    &  $<0.01$          &   $<0.01$            &  $0.01$--$0.04$              \\
$m(\Kp\pim\mumu)$ model                   & $<0.01$              & $<0.01$                    &   $<0.01$         &    $<0.02$           &  $<0.01$                           \\
Det. and prod. asymmetries                        & --                        & --                             & $<0.01$           &   $<0.02$     & --                                     \\   
\end{tabular}
}
\end{center}
\end{table}

The size of each systematic uncertainty is estimated using pseudoexperiments in which one or more parameters are varied. 
The angular observables are determined from these pseudoexperiments using the nominal model and the systematically varied model. 
For each observable, in each \qsq region, the systematic uncertainty is then taken as the average of the difference between the two models.
The pseudoexperiments are generated with signal yields many times larger than that of the data, in order to render statistical fluctuations negligible.

The main systematic effects associated with the signal modelling arise from the estimate of the angular acceptance. 
Four separate sources of systematic uncertainty are considered: 
the statistical uncertainty on the acceptance correction resulting from the limited size of the simulation sample from which it is determined; 
an uncertainty associated with the parameterisation that is used to describe the acceptance function; 
an uncertainty arising from residual data-simulation differences; 
and, for the likelihood fit of the angular observables in narrow \qsq bins, an uncertainty associated with evaluating the acceptance at a fixed point in \qsq. 

The statistical uncertainty on the acceptance function is evaluated using
pseudoexperiments that are generated by coherently fluctuating the acceptance parameters according to the covariance matrix for the angular moments of the acceptance function.  
To evaluate the uncertainty associated with the particular choice of order for the polynomials used to describe the acceptance function, pseudoexperiments are produced in which the polynomial order is simultaneously increased by two in \qsq and in each of the angles.

After the \Bz momentum spectrum, detector occupancy and PID performance of the simulation are corrected to match the data, there is very good agreement between the properties of simulated and genuine \decay{\Bz}{\jpsi\Kstarz} decays. 
There are, however, some small remaining differences in the momentum and transverse momentum spectra of the reconstructed pion that can affect the determination of the acceptance correction. 
A new acceptance correction is derived after re-weighting the simulated phase-space sample to account for the observed differences. 
A more conservative variation has also been considered in which an acceptance correction is derived without any of the data-simulation corrections applied.
The larger of the variations observed is added as a systematic uncertainty. 

When determining the angular observables in the narrow \qsq\ bins with the maximum likelihood fit, 
the acceptance is evaluated using the \qsq value of the bin centre.
Pseudoexperiments are generated to assess the bias generated by this choice, using instead the value of \qsq of the left- or right-hand bin boundary.

Possible contributions from the tails of higher mass \Kstar states in the $796 < m(\Kp\pim) < 996\mevcc$ window are also considered.
Simulation studies indicate that any bias arising from these states is negligible compared to the statistical uncertainty on the angular observables. 

For the background modelling, two sources of systematic uncertainty are considered. 
The first source is associated with the choice of second-order polynomials to model the background angular distribution in the fits of the angular observables and the \qsq-dependent decay amplitudes. 
It is not possible to fit a more complex model to the data because of the small number of background candidates. 
Therefore, to test the model, the BDT requirement is relaxed and the background candidates are fitted with a fourth-order polynomial in each of the three angles. 
This shape is used when generating the pseudoexperiments.
The second source is associated with the fit to the \qsq-dependent decay amplitudes. In this case, the \qsq dependence of the background model is modified from a linear function to a third-order polynomial.

Systematic uncertainties are  assessed for the different sources of peaking background that are neglected in the analysis.  
As detailed in Sec.~\ref{sec:backgrounds}, the most important backgrounds are those from \decay{\Lb}{p\Km\mumu} and \decay{\Bs}{\phi\mumu} decays, where a kaon or proton is misidentified as pion; and \decay{\Bz}{\Kstarz\mumu} decays, where the kaon and pion are both misidentified.  
Taking the angular distribution of the background from simulated events, pseudoexperiments are generated with these backgrounds included, and the angular observables determined as if the background were not present.  
Pseudoexperiments are also generated in which the angular distribution of the \decay{\Bs}{\phi\mumu} and \decay{\Lb}{p\Km\mumu} decays are taken from data. These decays are selected by removing PID information from the BDT and inverting the background vetoes.

Systematic uncertainties are also assessed for the signal mass-modelling in $m(\Kp\pim\mumu)$ and $m(\Kp\pim)$.
To assess the model of $m(\Kp\pim\mumu)$, a fit is performed to \decay{\Bz}{\jpsi\Kstarz} data using the sum of two Gaussian distributions without the power law tails.
To assess the modelling of $m(\Kp\pim)$, pseudoexperiments are produced by systematically varying the S- and P-wave line-shape parameters.
For the S-wave, the LASS line-shape is also exchanged for the sum of resonant $K^{*}_0(800)^{0}$ (sometimes referred to as the $\kappa$ resonance) and $K^*_{0}(1430)^{0}$ contributions.

For the fit to the \qsq-dependent decay amplitudes, an additional uncertainty is assigned for the choice of the \qsq parameterisation of the S-wave components. 
As described in Sec.~\ref{sec:amplitudes}, the S-wave amplitudes are taken to be constant in \qsq. 
Motivated by Ref.~\cite{Wang:2012swave}, a systematic variation is considered by assuming that the S-wave amplitudes ${\cal A}_{\rm S}^{\rm L,R}$ have the same \qsq dependence as the longitudinal P-wave amplitudes ${\cal A}_{0}^{\rm L,R}$. 

The measured \CP asymmetries can be biased due to detection and production asymmetries. 
The \Bd production asymmetry is measured to be less than 1\%~\cite{LHCb-PAPER-2014-042,LHCb-PAPER-2014-053}. 
The effect of this asymmetry is further suppressed due to \Bz-\Bzb mixing.  
The kaon detection asymmetry was measured in Ref.~\cite{LHCb-PAPER-2014-013}.
In contrast to the other sources of systematic uncertainty, the shift due to the detection and production asymmetries is calculated directly without generating pseudoexperiments.
The systematic uncertainty on the angular observables $A_i$ due to production and detection asymmetries is found to be less than $0.01$.  
The effect of these asymmetries on the \CP-averaged observables is negligible.  

In the \qsq-bin $0.10 < \qsq < 0.98\gev^{2}/c^{4}$, the muon mass-squared is comparable to \qsq and the  relations between $S_{1(s,c)}$, $S_{2(s,c)}$ and $F_{\rm L}$ (see Sec.~\ref{sec:ang})  are only approximate.  The assumption that these relations hold has no impact on the measured values of $S_3$--$S_9$ or $A_3$--$A_9$ but results in a biased estimate of $F_{\rm L}$ and hence of the $P^{(\prime)}_i$ observables. In pseudoexperiments based on the SM, this bias is typically at the level of $0.02$.  This can be accounted for in the SM predictions for this \qsq-bin and hence is not considered as a source of systematic uncertainty.

For $F_{\rm L}$ and $A_{\rm FB}$, the largest source of systematic uncertainty comes from the data-simulation comparison of the pion momenta.  
The systematic uncertainty assigned to this effect is at the level of $0.01 - 0.02$, depending on the \qsq bin.  
This uncertainty constitutes up to 30\% of the statistical uncertainty on $F_{\rm L}$ and 20\% of the statistical uncertainty on $A_{\rm FB}$.  
For $S_5$ and $A_5$, the largest source of systematic uncertainty comes from the choice of polynomial order for the angular acceptance.  
If polynomials two orders higher are used, a variation of $\sim0.01$ is observed. 
For the remaining \CP-averaged and \CP-asymmetric observables, the uncertainties arising from the data-simulation comparison and the acceptance are small. 
However, there are three other non-negligible sources of systematic uncertainty.  
Throughout the full \qsq range, peaking backgrounds introduce a systematic uncertainty at the level of $0.01$ or less.  
For the likelihood fit of the angular observables, in the first two \qsq bins (where the acceptance changes most rapidly), the uncertainty arising from using the bin centre, as opposed to a bin edge, is at the level of $0.01$ or less.  
Finally, at high \qsq, the statistical precision on the acceptance correction leads to a systematic uncertainty at the level of $0.01$ or less.
For the $P_i^{(')}$ observables, the situation is more complex and the systematic uncertainty is shared more evenly between the different sources (see Table 2).
The dominant sources of systematic uncertainty can all be reduced in future analyses with larger datasets.

Propagating the above sources of systematic uncertainty to the zero-crossing points yields uncertainties at the level of $0.07\gevgevcccc$ for $S_4$, $0.02\gevgevcccc$ for $S_5$ and $0.03\gevgevcccc$ for $A_{\rm FB}$. These uncertainties are negligible  compared to the statistical uncertainties.

\section{Results} 
\label{sec:Results}

\begin{figure}[h]
\begin{center}
\includegraphics[width=0.47\linewidth]{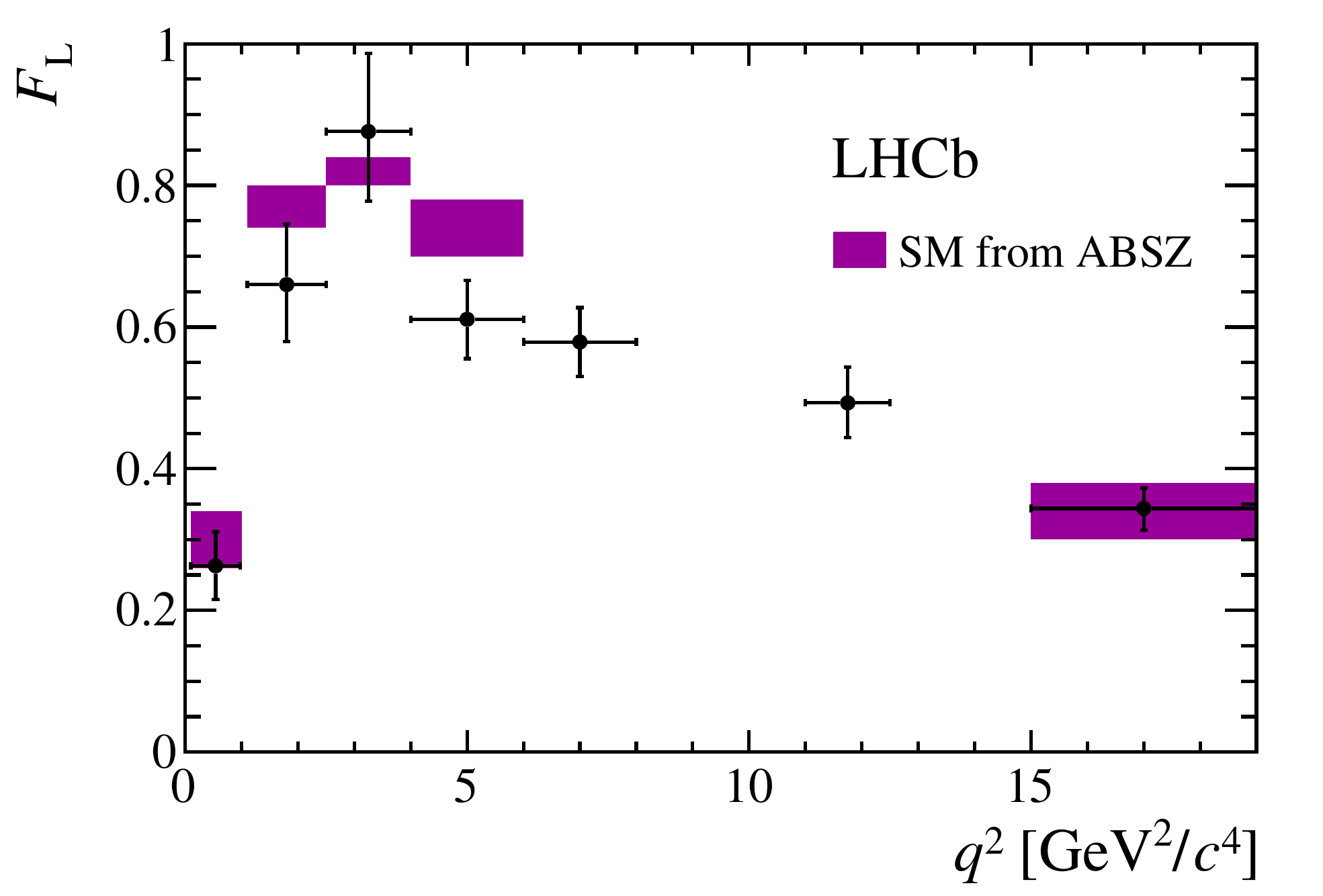} 
\includegraphics[width=0.47\linewidth]{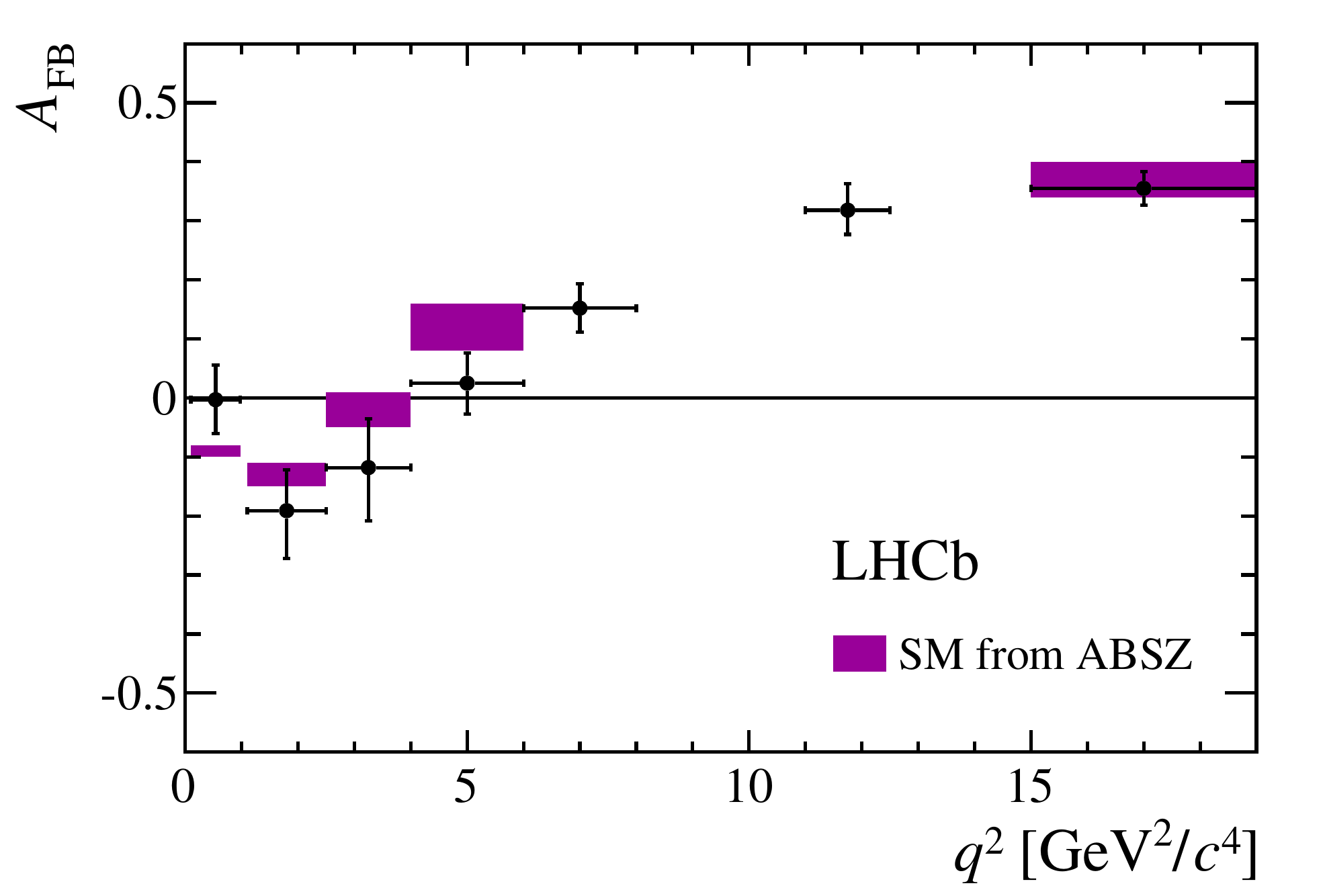} \\
\includegraphics[width=0.47\linewidth]{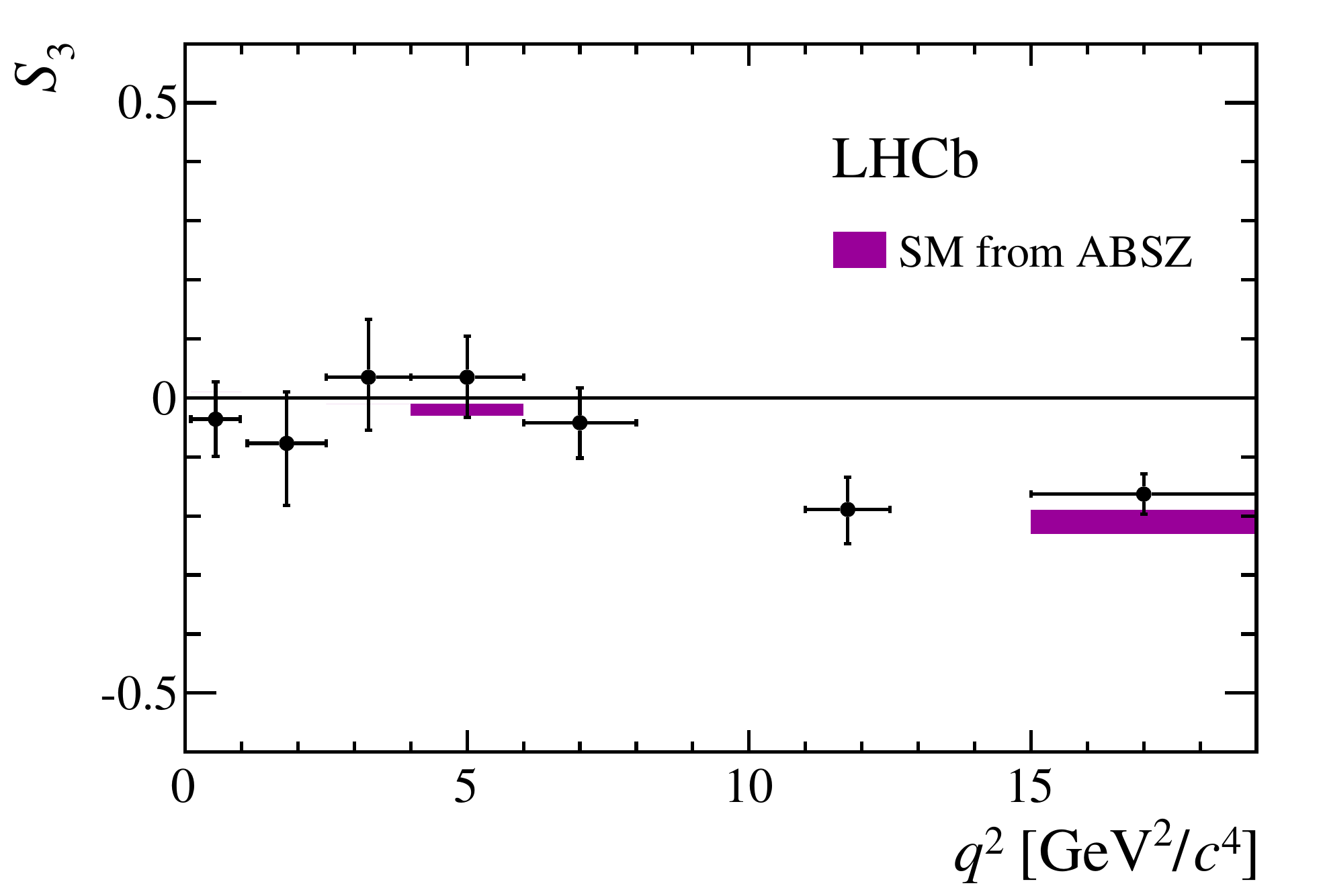} 
\includegraphics[width=0.47\linewidth]{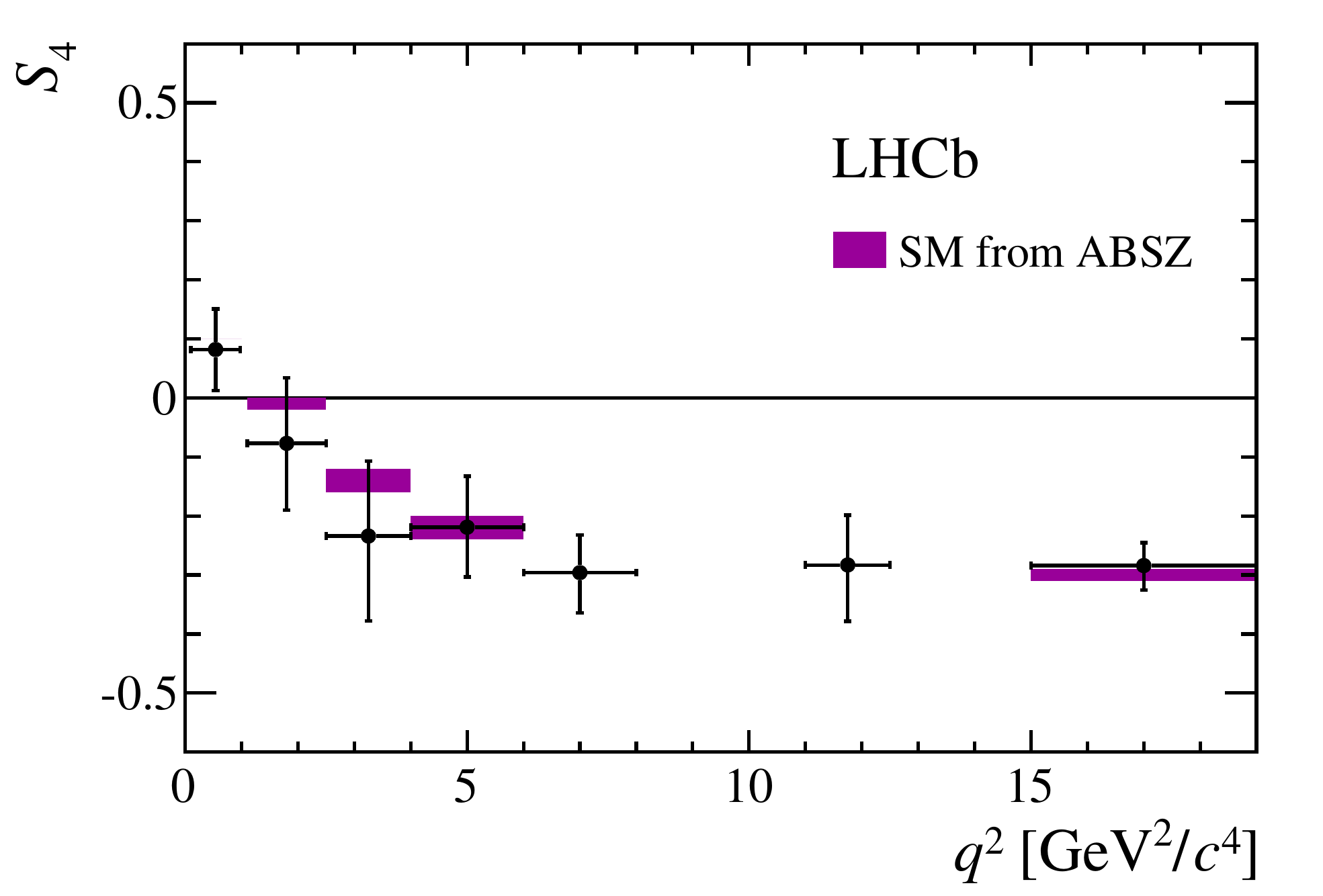} \\
\includegraphics[width=0.47\linewidth]{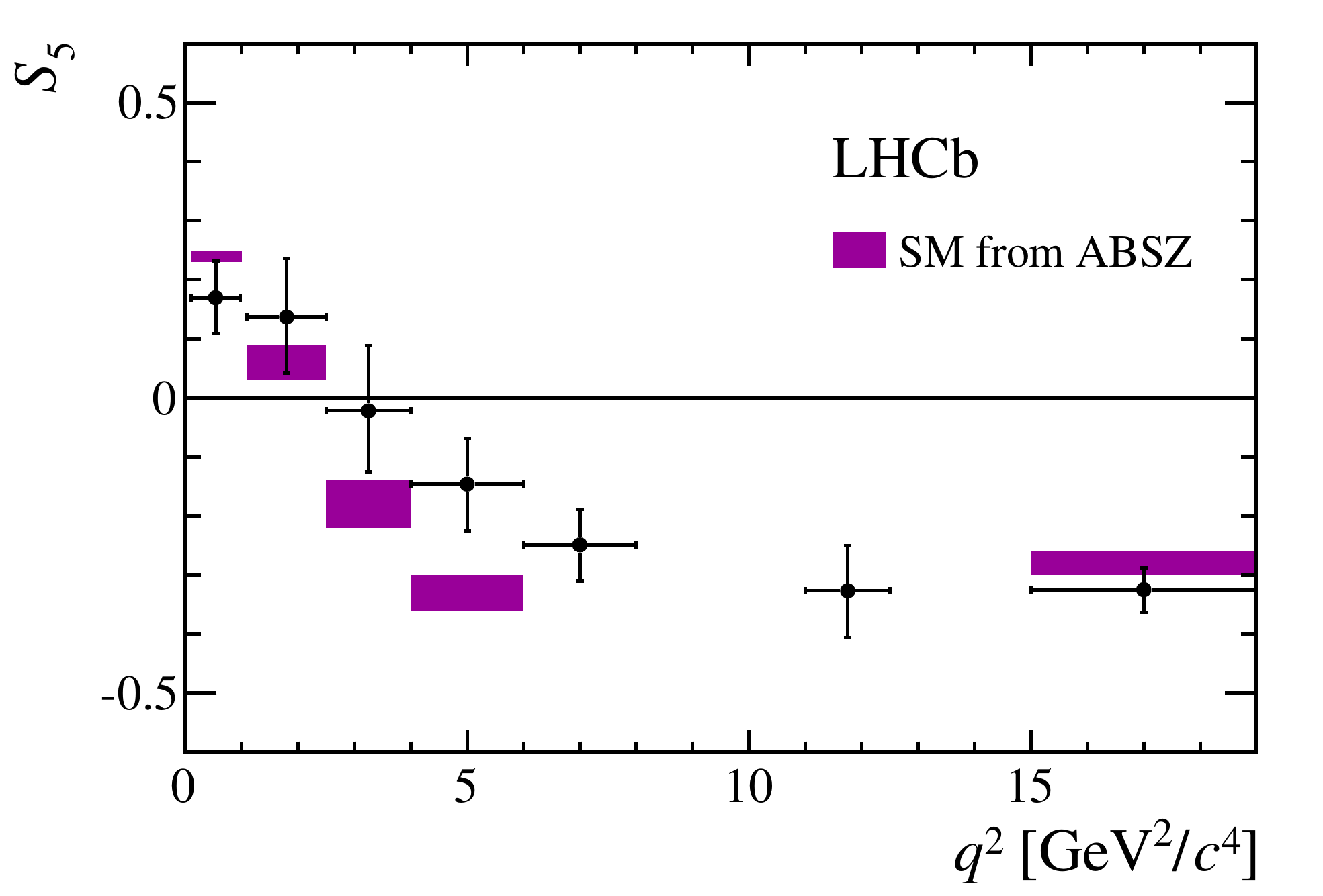} 
\includegraphics[width=0.47\linewidth]{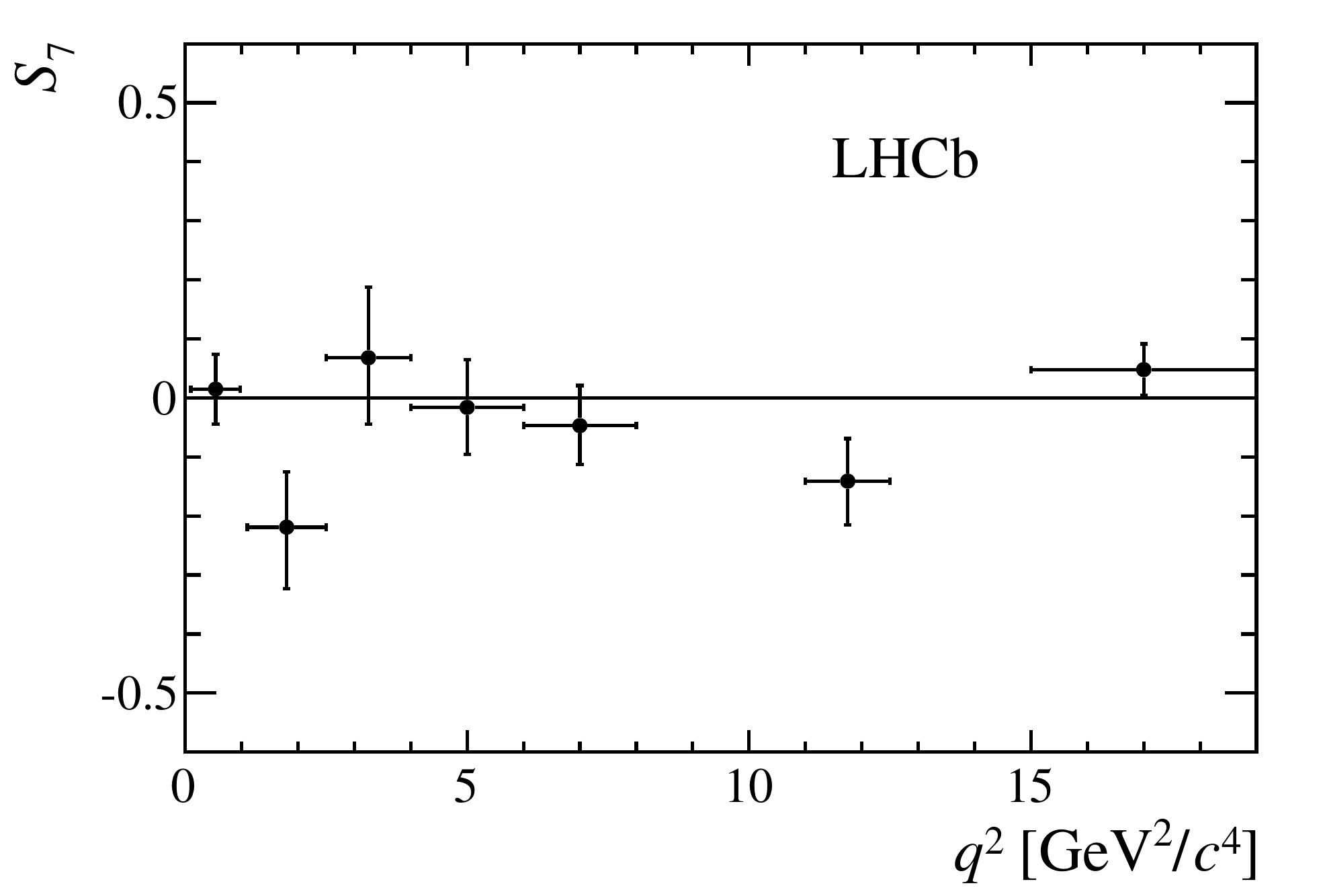} \\
\includegraphics[width=0.47\linewidth]{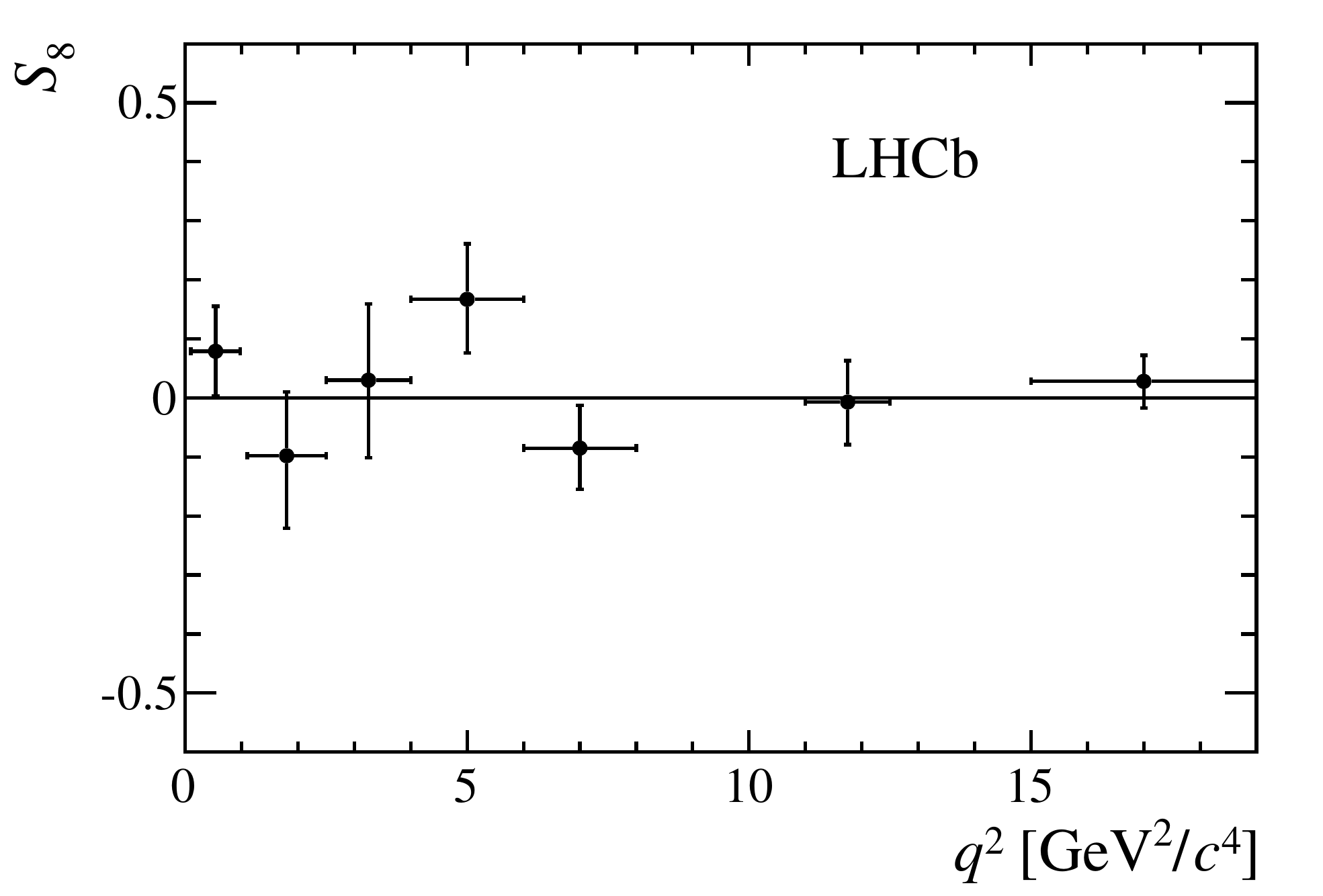} 
\includegraphics[width=0.47\linewidth]{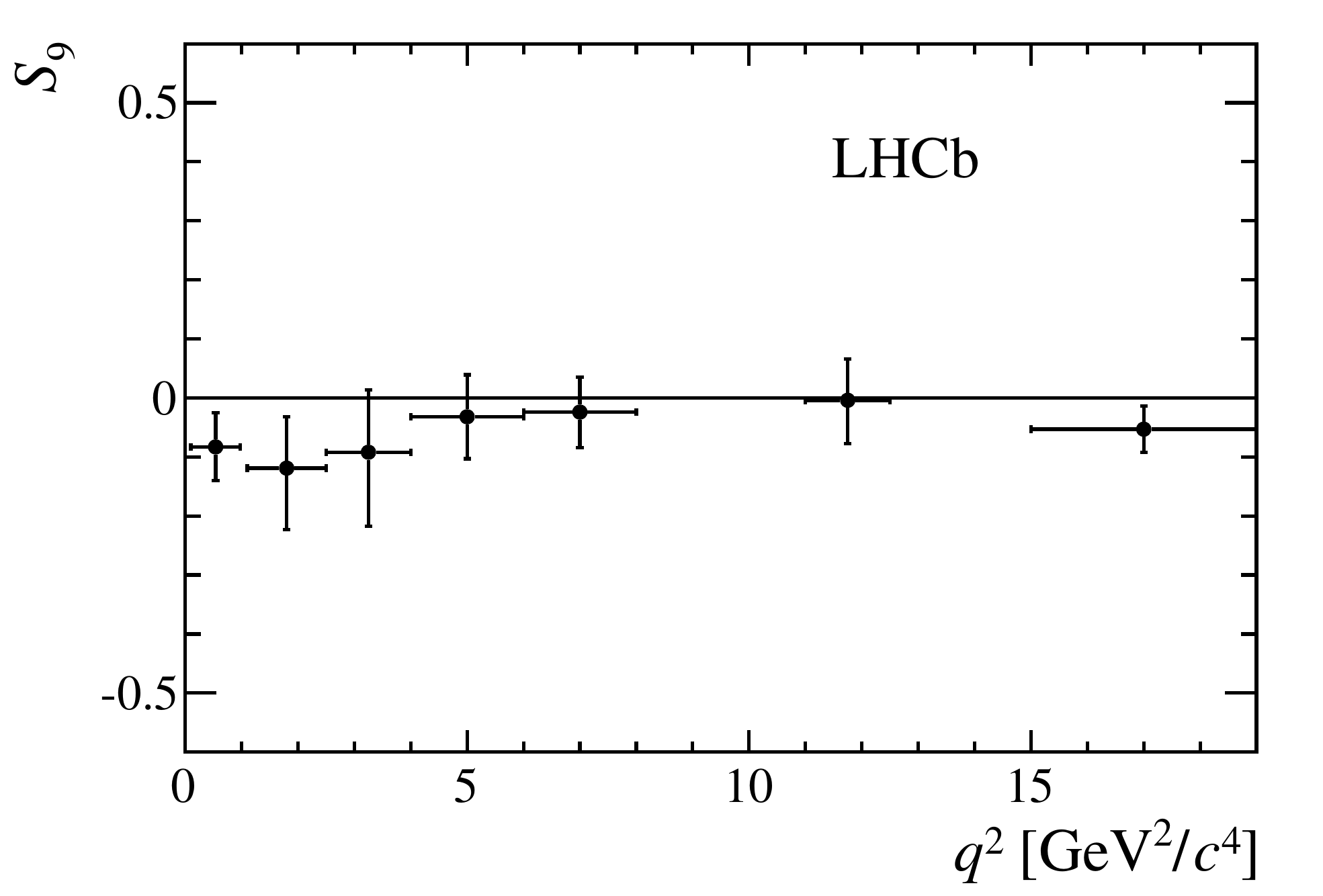} \\
\end{center}
\caption{\label{fig:results:Si} The \CP-averaged observables in bins of $q^2$, determined from a maximum likelihood fit to the data. The shaded boxes show the SM predictions based on the prescription of  Ref.~\cite{Altmannshofer:2014rta}.} 
\end{figure}

\begin{figure}[h]
\begin{center}
\includegraphics[width=0.47\linewidth]{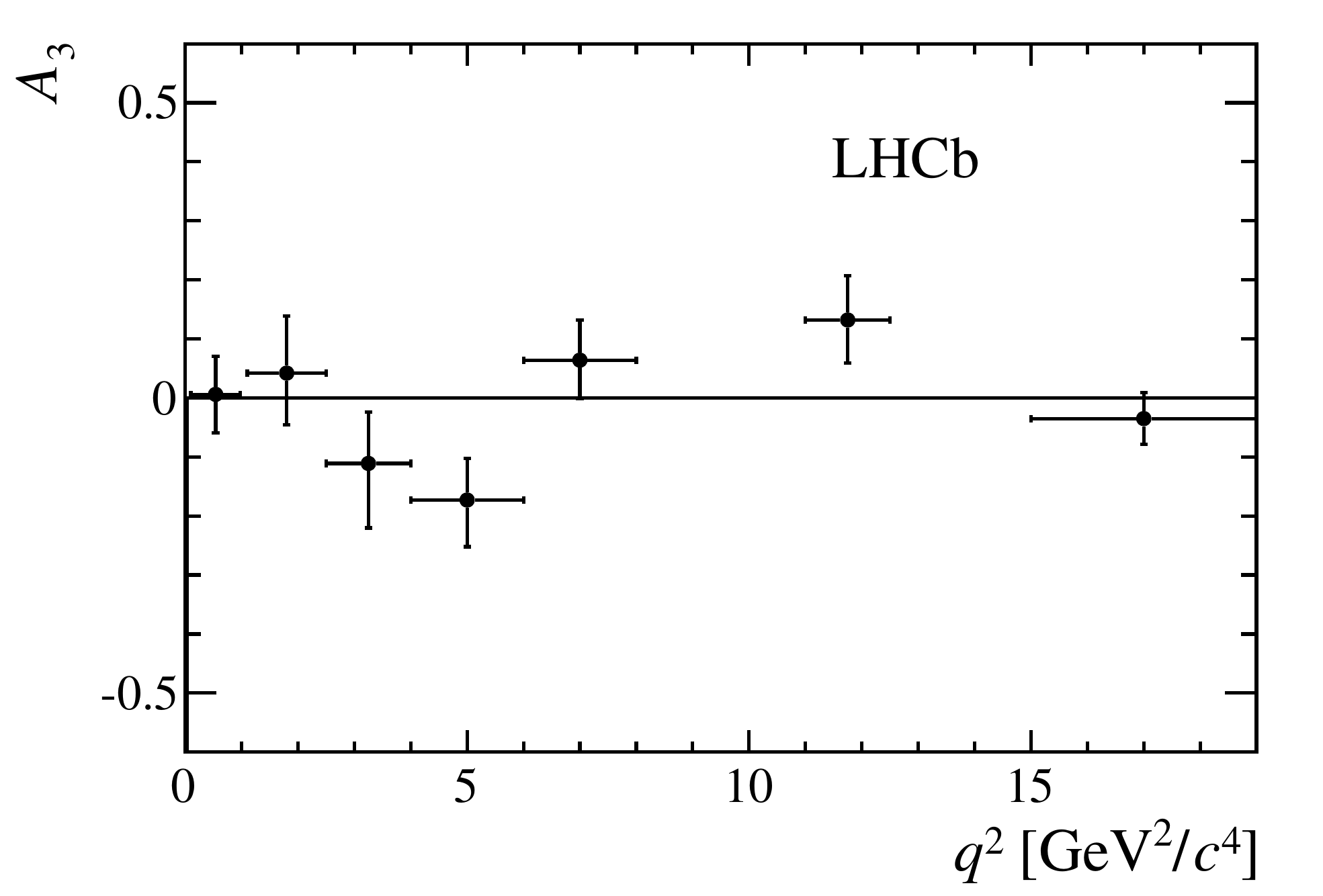}
\includegraphics[width=0.47\linewidth]{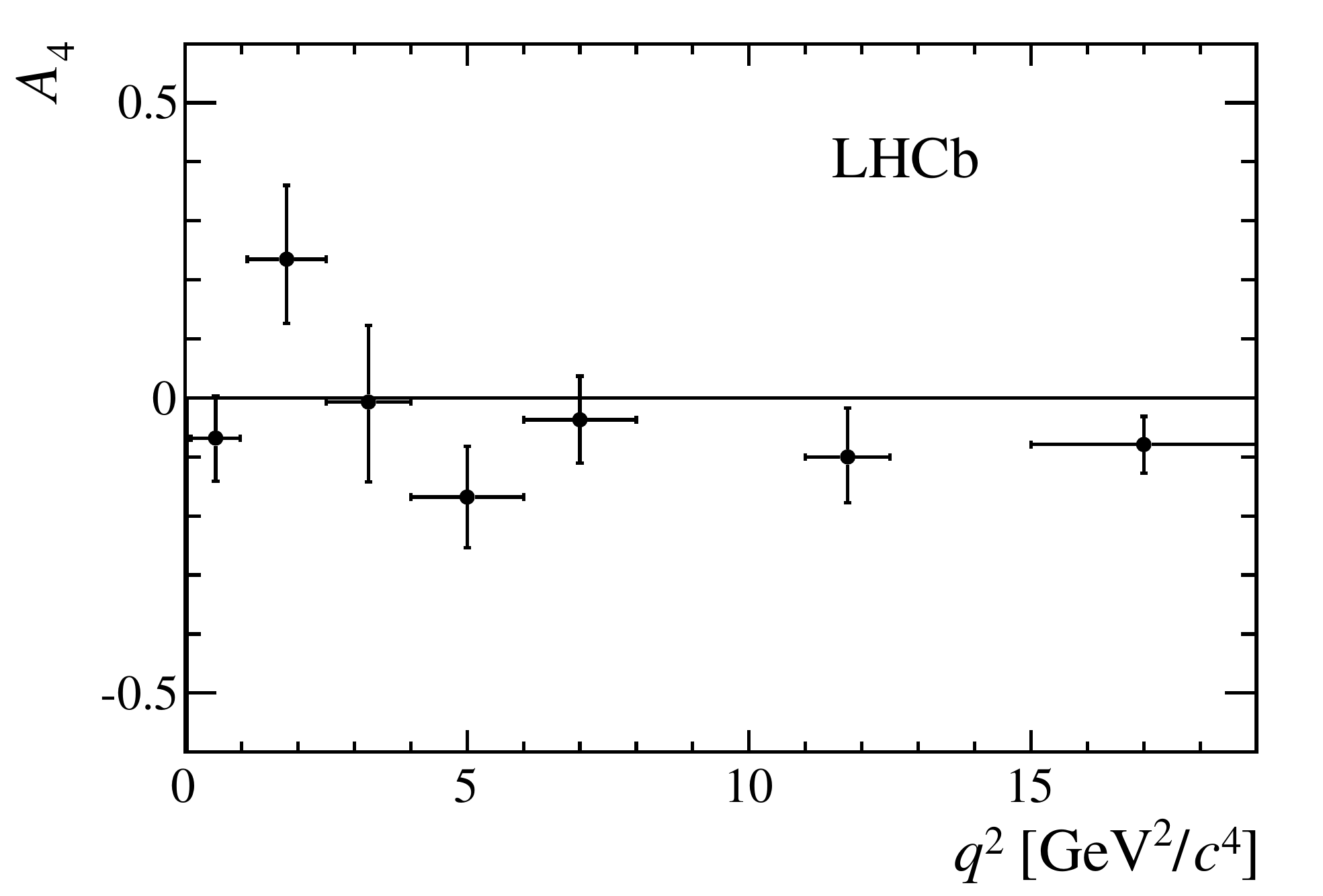}  \\
\includegraphics[width=0.47\linewidth]{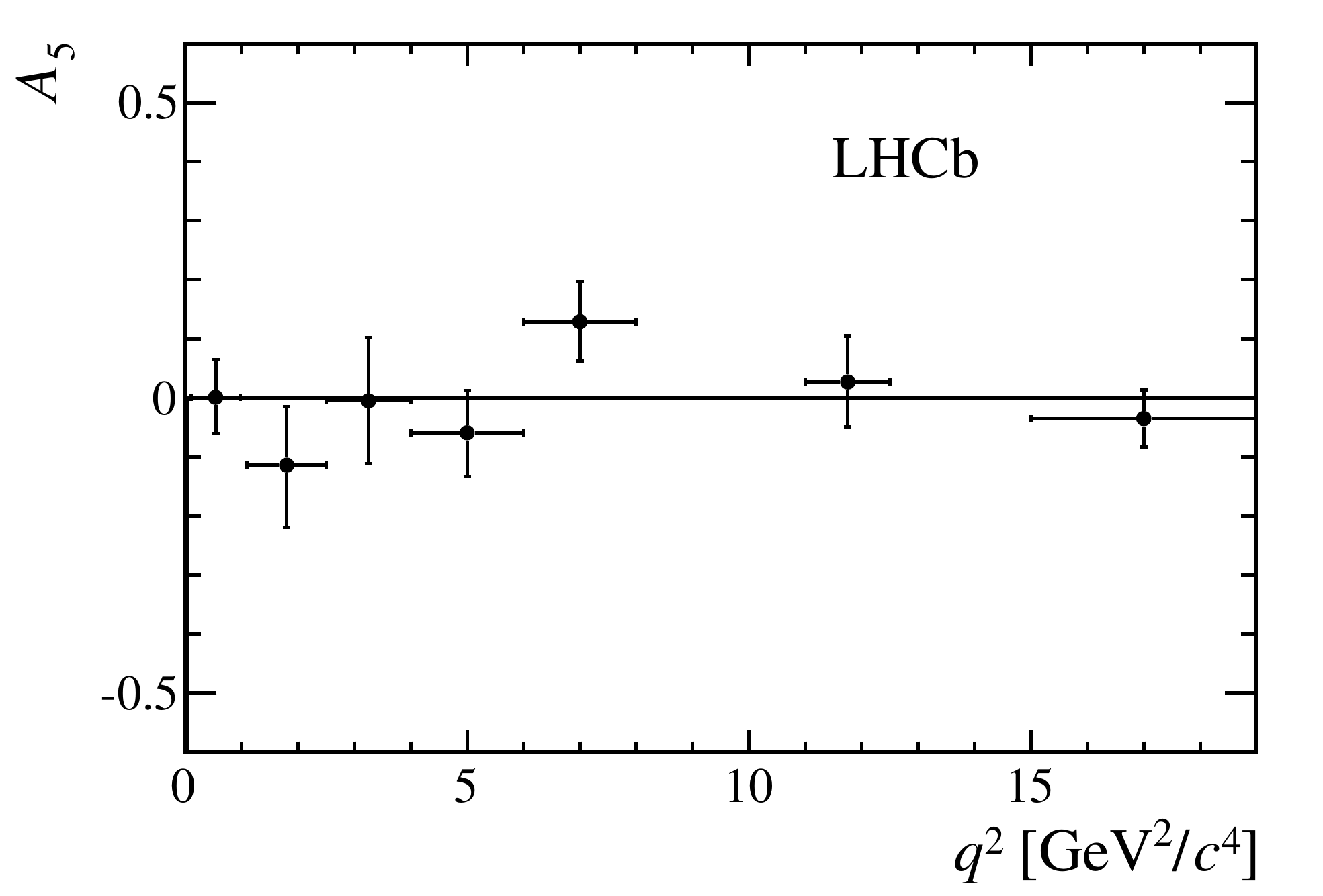} 
\includegraphics[width=0.47\linewidth]{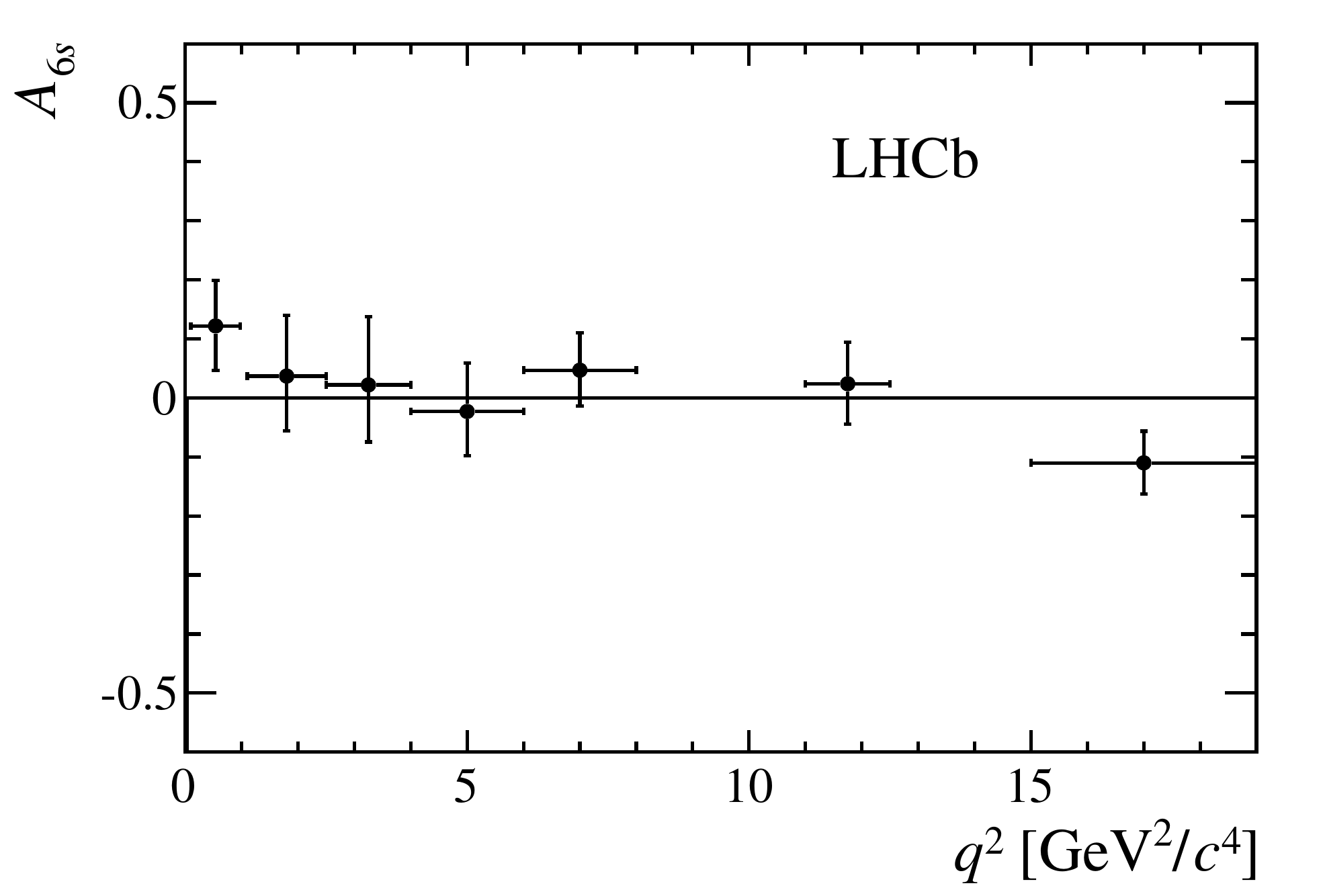} \\
\includegraphics[width=0.47\linewidth]{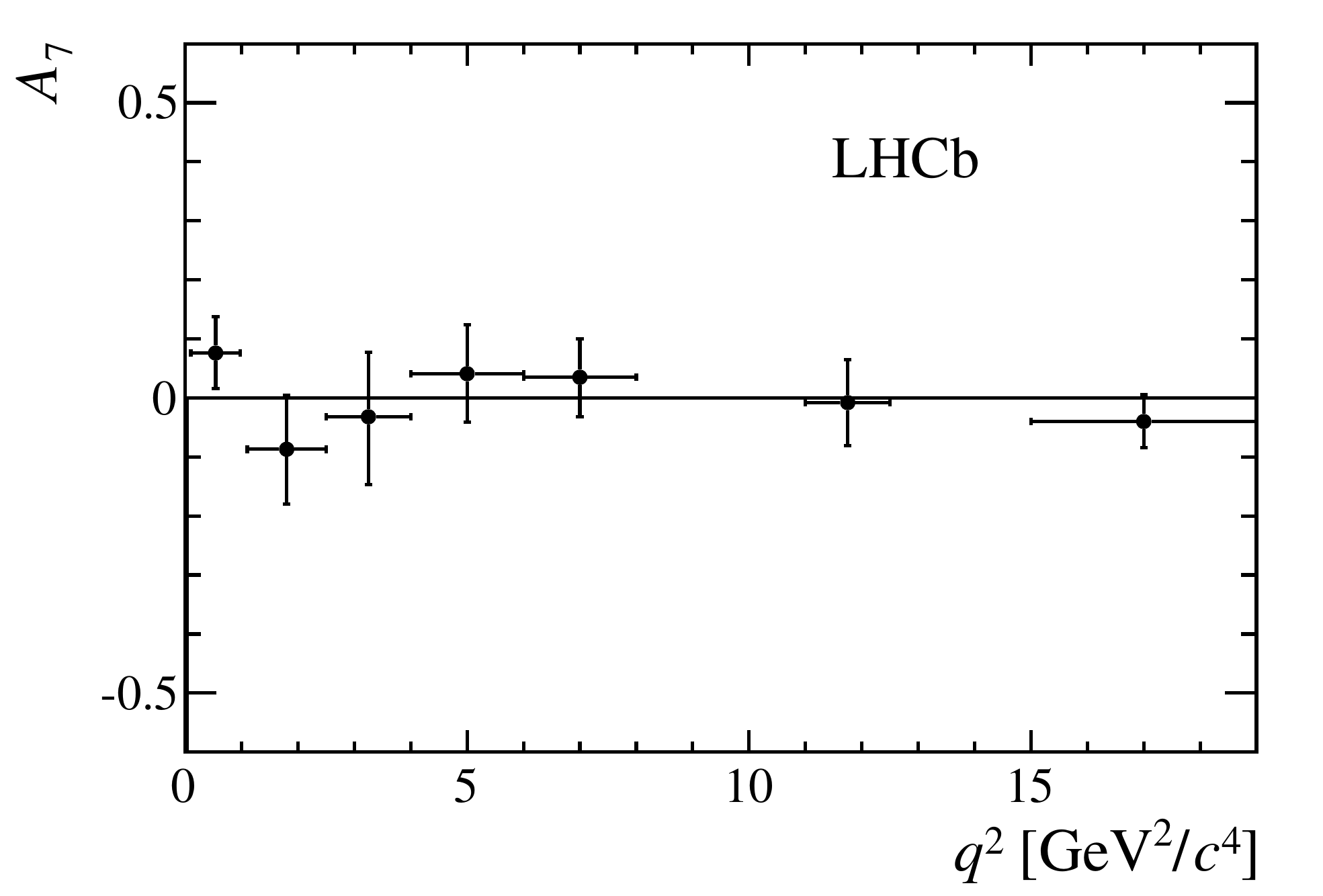} 
\includegraphics[width=0.47\linewidth]{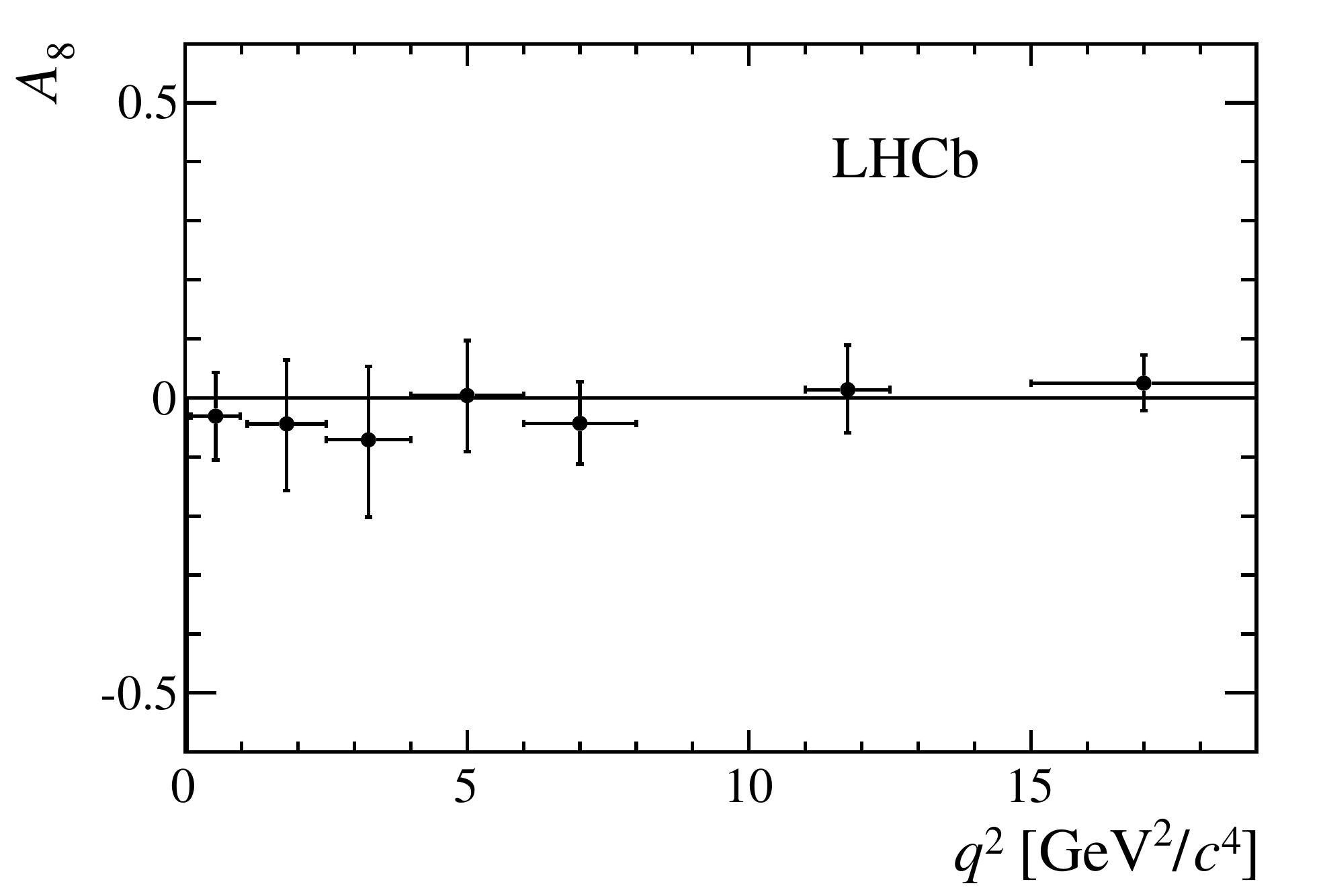} \\
\includegraphics[width=0.47\linewidth]{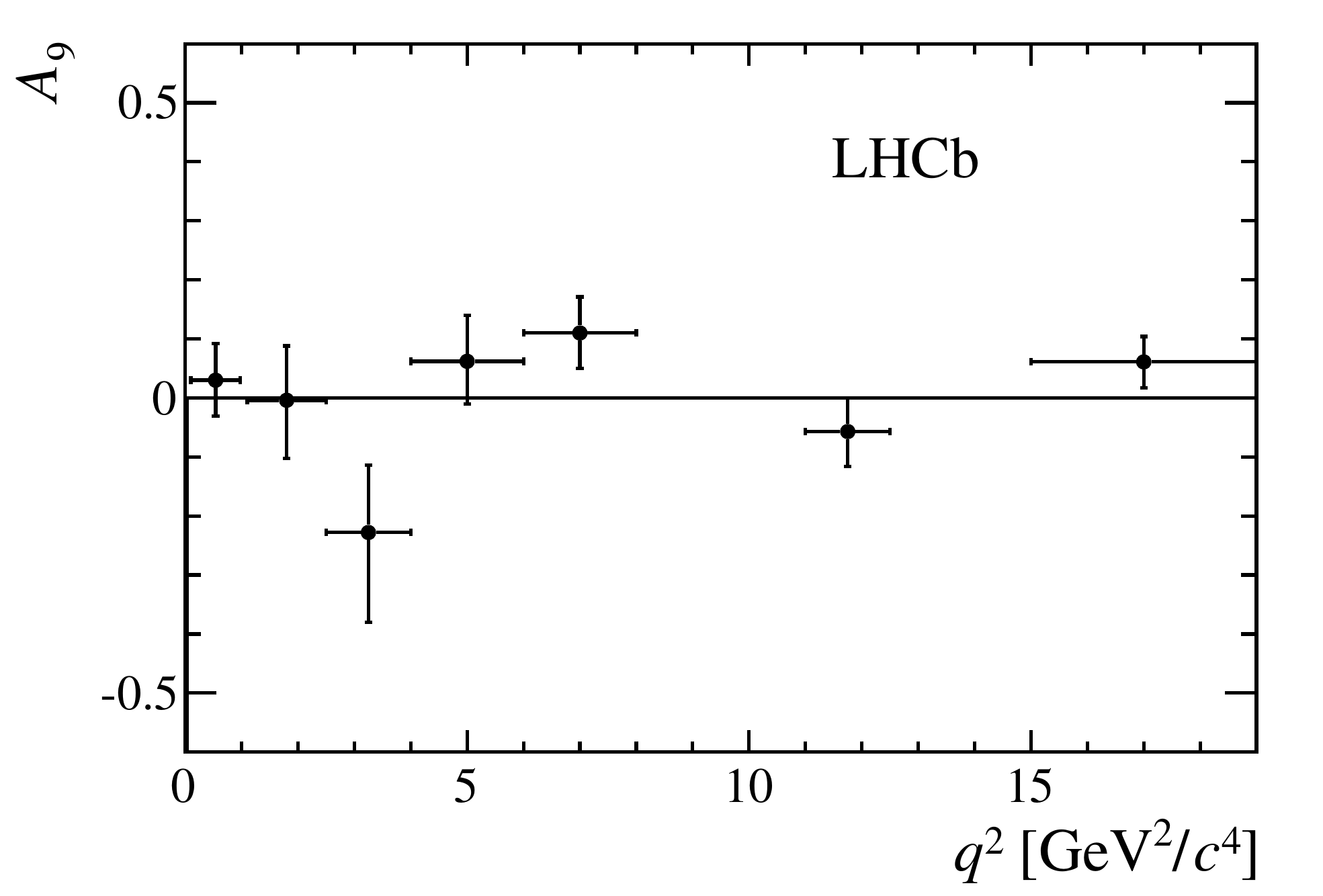} 
\end{center}
\caption{\label{fig:results:Ai} The \CP-asymmetric observables in bins of $q^2$, determined from a maximum likelihood fit to the data.} 
\end{figure}

The \CP-averaged observables that are obtained from the likelihood fits are shown together with the SM  predictions in Fig.~\ref{fig:results:Si}. The \CP asymmetries are shown in Fig.~\ref{fig:results:Ai}. 
The SM predictions  are based on the prescription of Ref.~\cite{Altmannshofer:2014rta}.
In contrast to the alternative SM predictions that are available in Refs.~\cite{Bobeth:2010wg,Jager:2014rwa,Das:2012kz,Horgan:2013pva,Mahmoudi:2014mja,Hambrock:2013zya,Hurth:2014vma}, these predictions update the calculations from Ref.~\cite{Ball:2004rg} to account for the known correlation between the different form factors~\cite{Straub:2015ica}. 
Light-cone sum rule predictions, which are valid in the low-\qsq region, are also combined with lattice determinations at high \qsq~\cite{Horgan:2013hoa,Horgan:2015vla} to yield more precise determinations of the form factors over the full \qsq range. 
The predictions are made in the regions $0.1<q^2<6.0\gev^{2}/c^{4}$ and $15.0<q^2<19.0\gev^{2}/c^{4}$. 
No predictions are made for the region close to the narrow $\ccbar$ resonances, the \jpsi and \psitwos, where many of the assumptions that go into the SM predictions are thought to be invalid. 
Reference~\cite{Altmannshofer:2014rta} does not make predictions for the $S_{7, 8, 9}$ and the $A_{i}$ observables. 
These observables are all expected to be close to zero in the SM.

\begin{figure}[h]
\begin{center}
\includegraphics[width=0.47\linewidth]{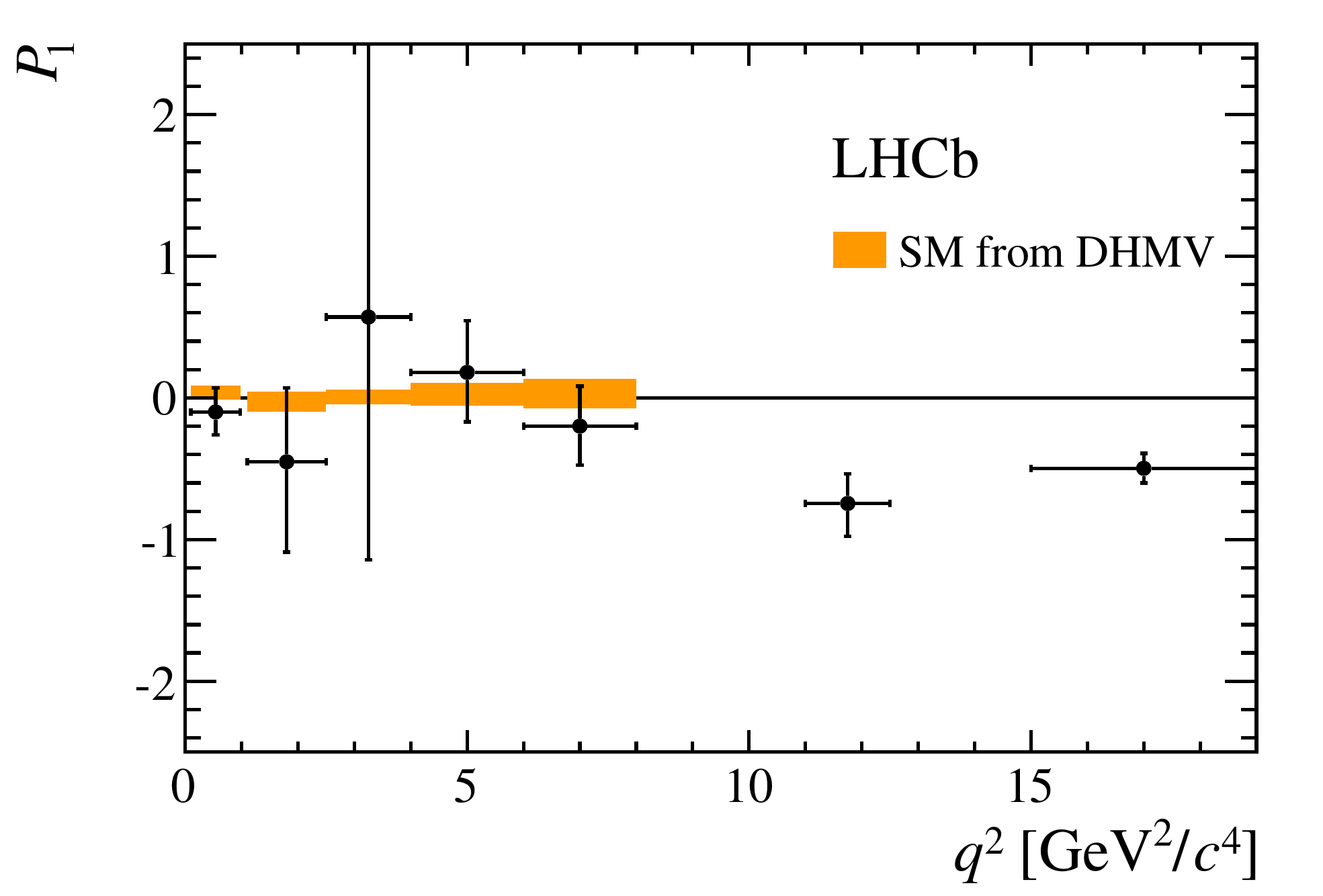}
\includegraphics[width=0.47\linewidth]{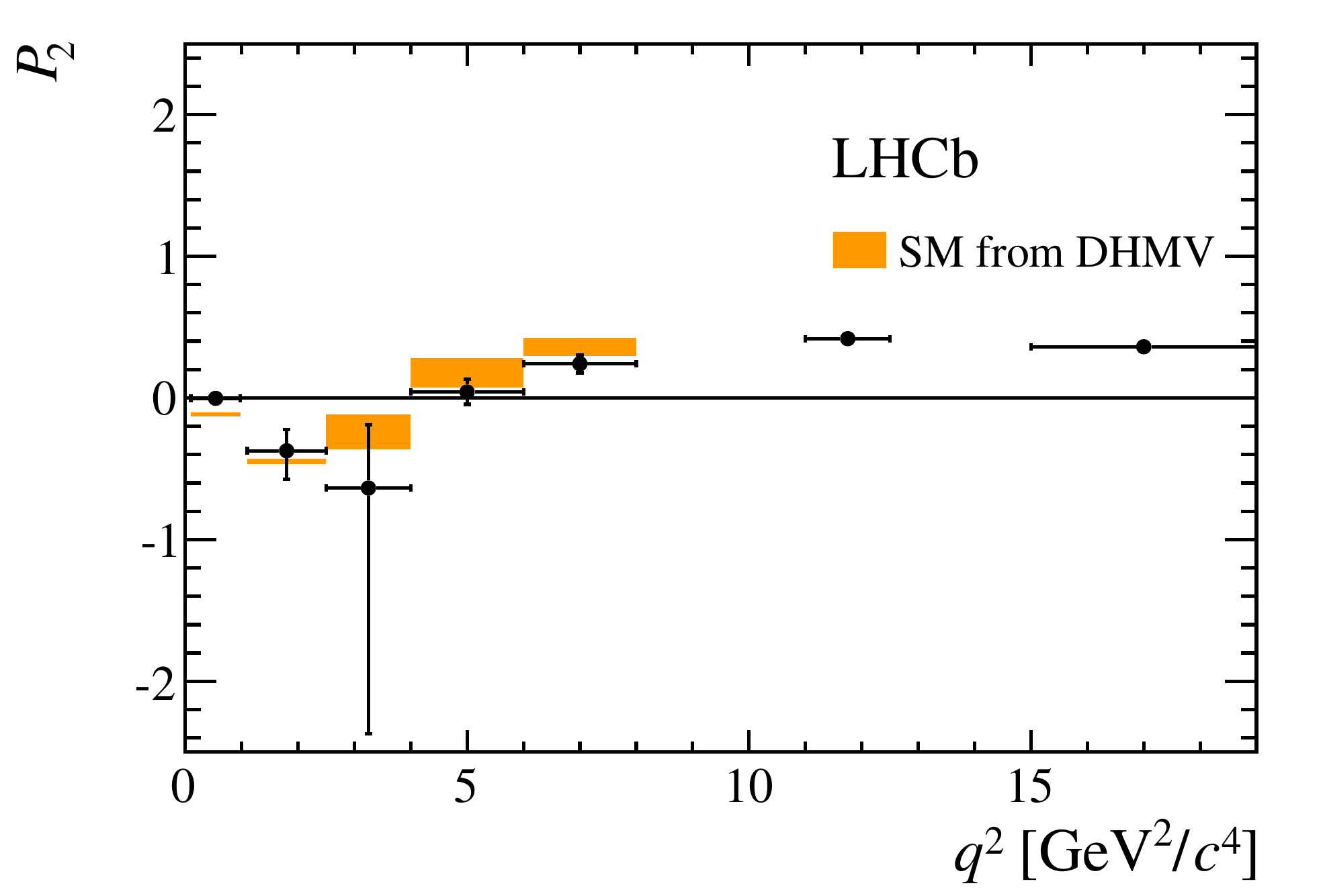}  \\
\includegraphics[width=0.47\linewidth]{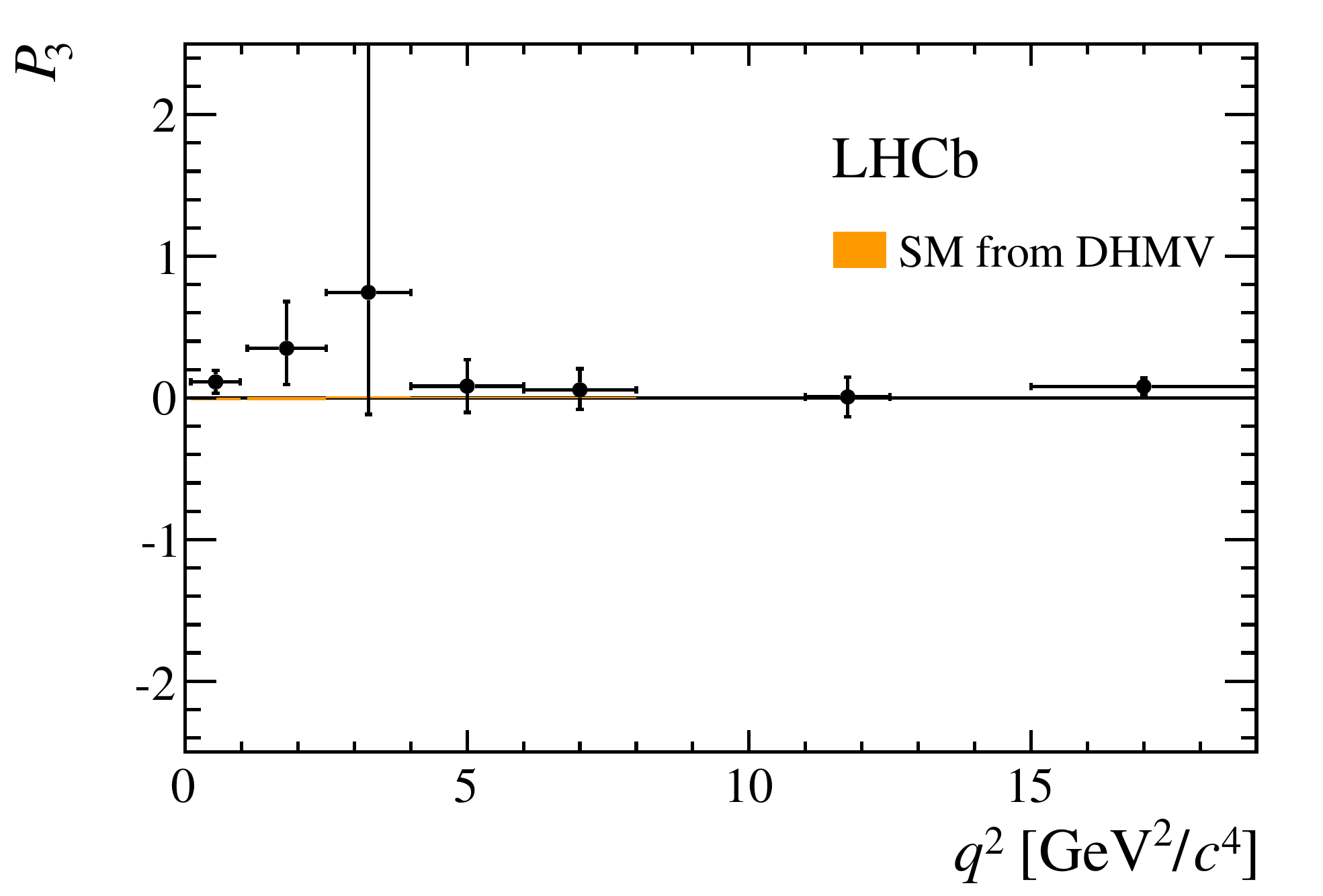} 
\includegraphics[width=0.47\linewidth]{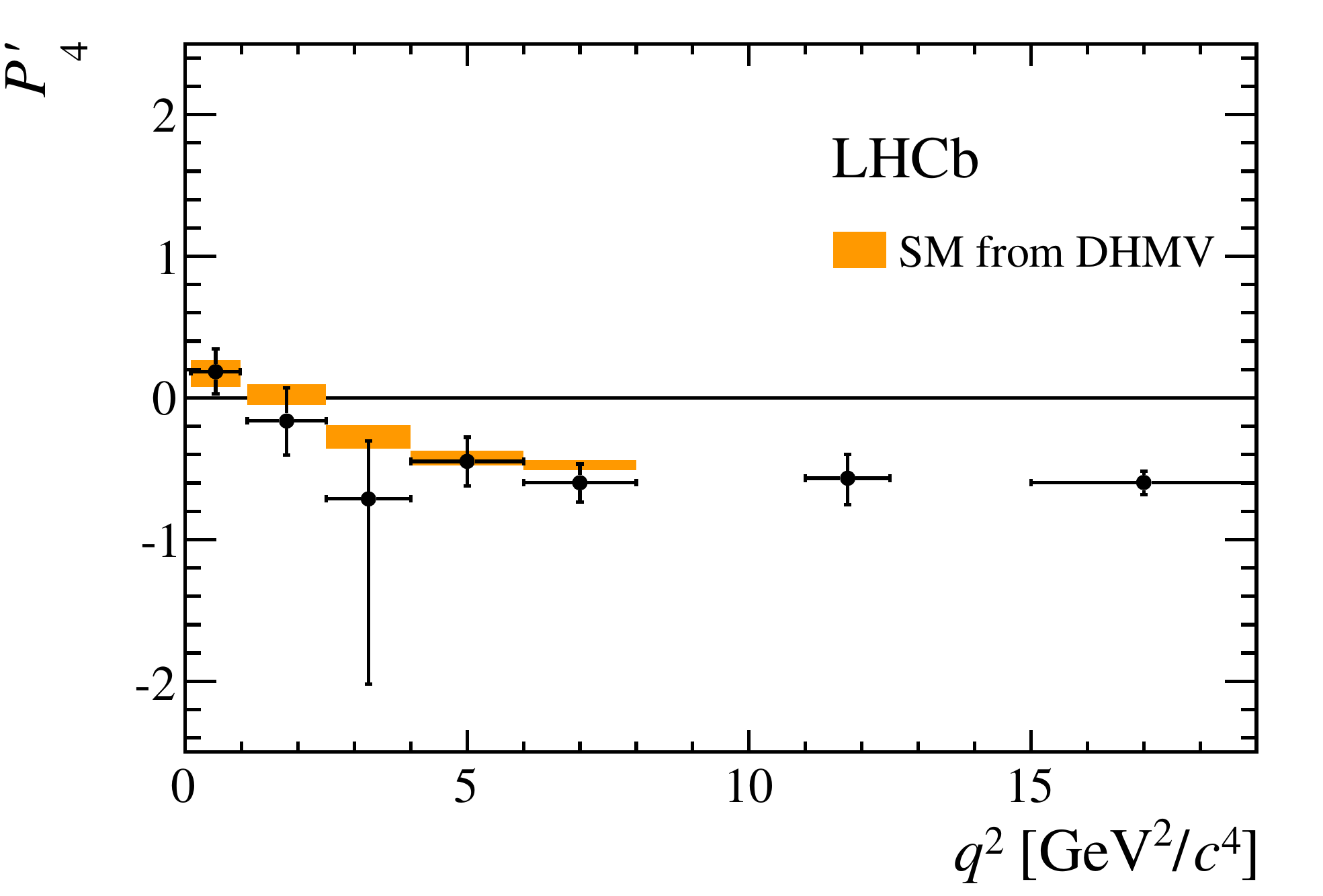} \\
\includegraphics[width=0.47\linewidth]{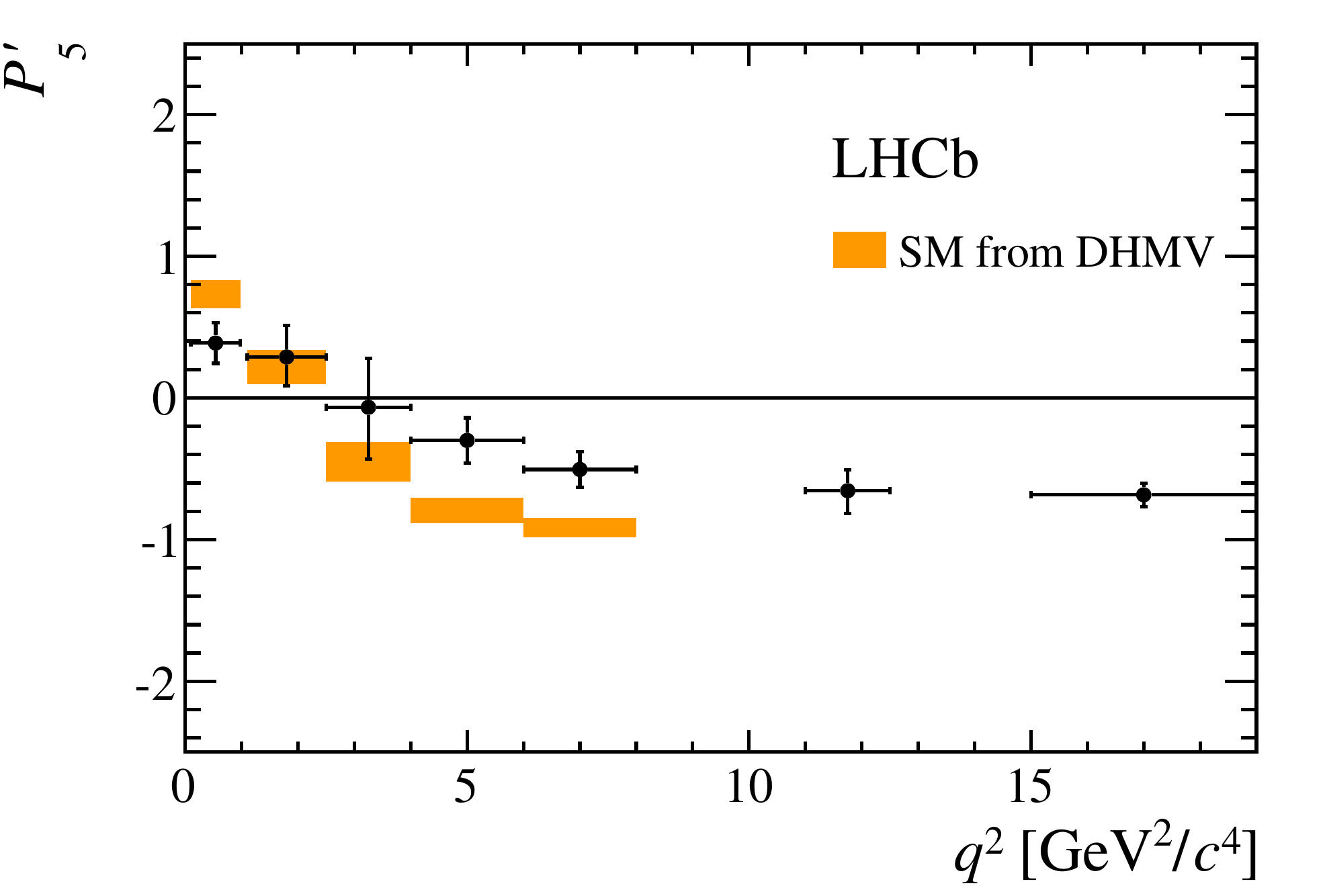} 
\includegraphics[width=0.47\linewidth]{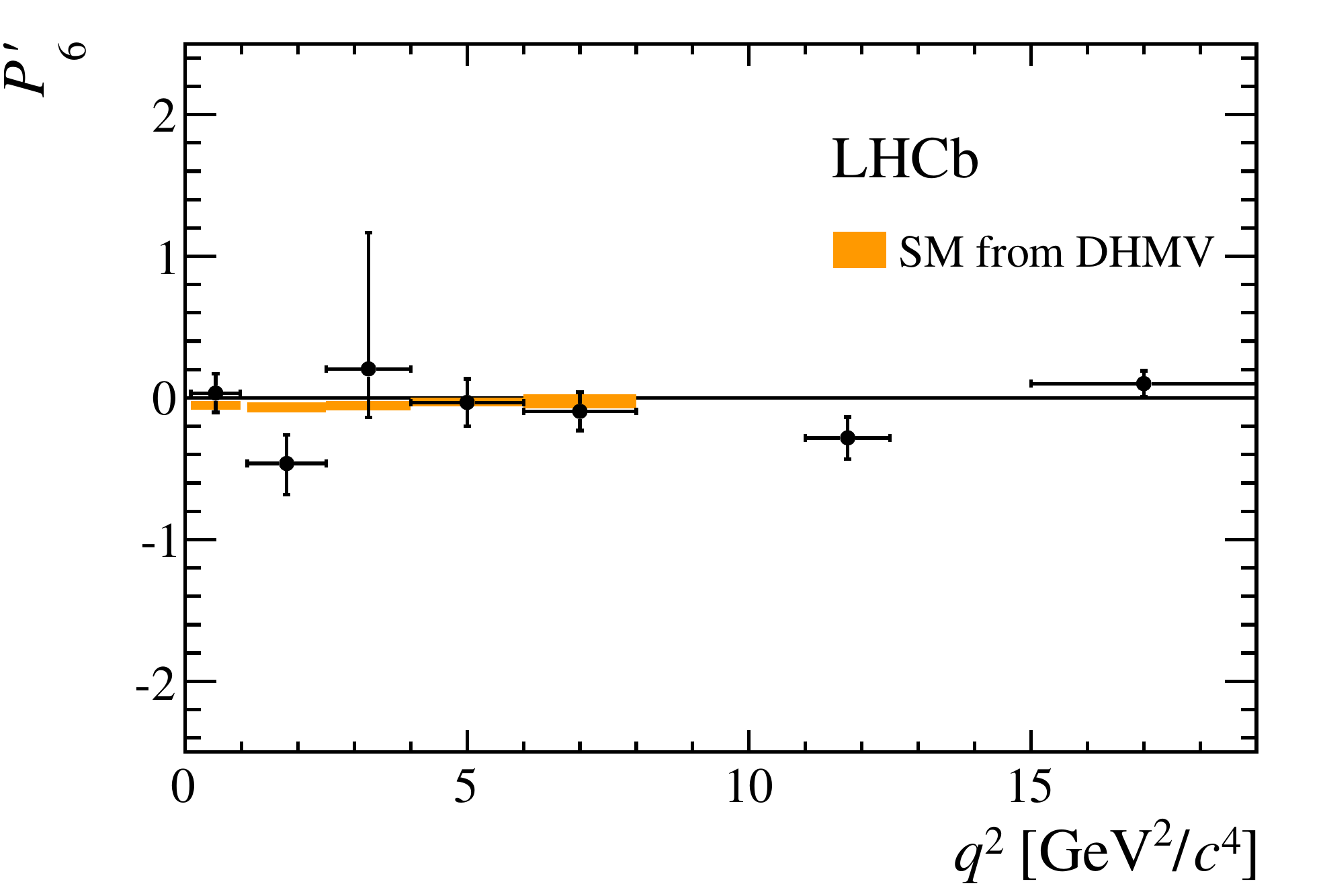} \\
\includegraphics[width=0.47\linewidth]{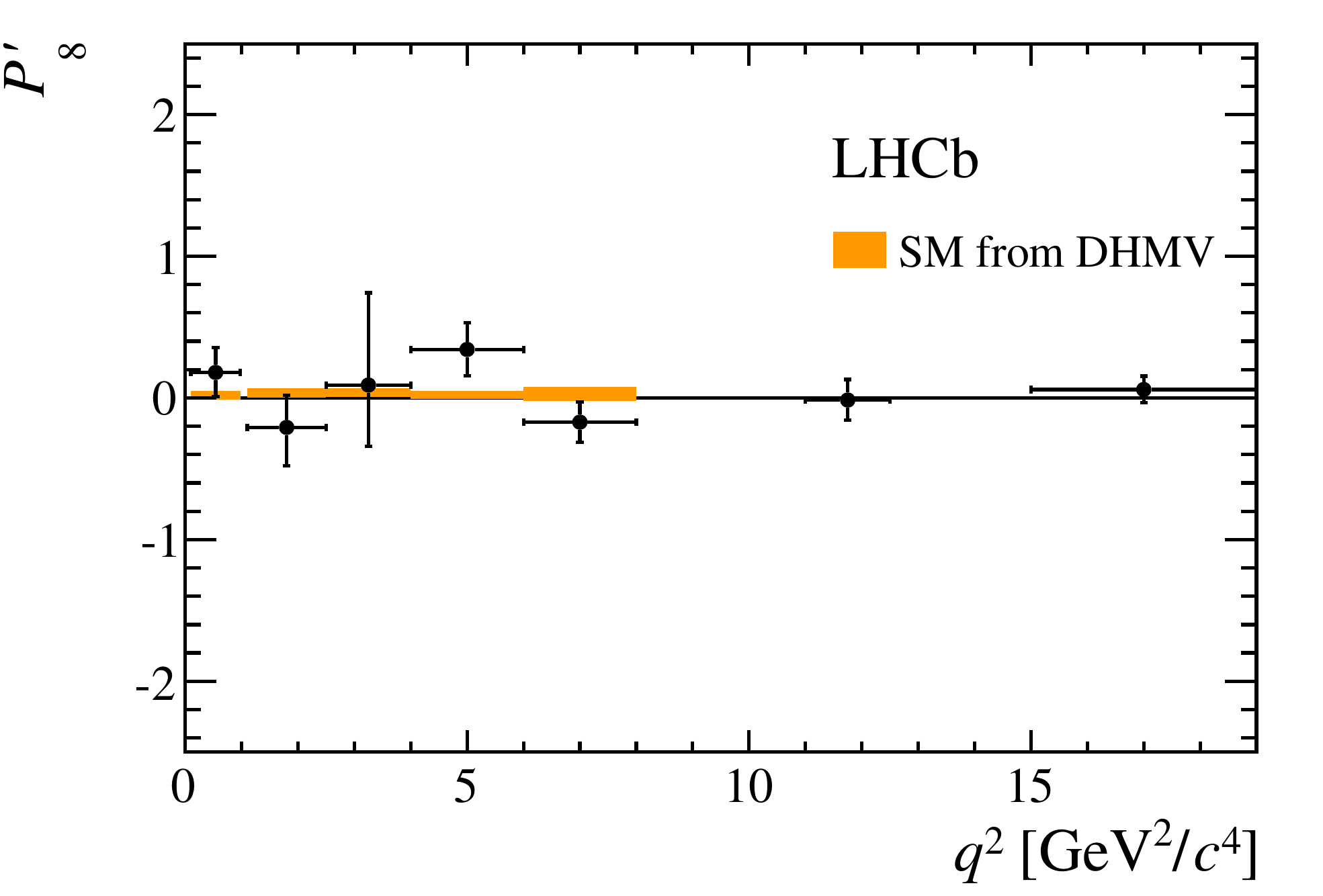} 
\end{center}
\caption{\label{fig:results:Pi} The optimised angular observables in bins of $q^2$, determined from a maximum likelihood fit to the data. The shaded boxes show the SM prediction taken from Ref.~\cite{Descotes-Genon:2014uoa}.} 
\end{figure}

The results of the fits for the optimised angular observables are shown together with their SM predictions in Fig.~\ref{fig:results:Pi}. 
For the $P^{(\prime)}_{i}$ observables, predictions from Ref.~\cite{Descotes-Genon:2014uoa} are shown using form factors from Ref.~\cite{Khodjamirian:2010vf}.
The SM predictions are restricted to the \qsq range $\qsq < 8.0\gev^{2}/c^4$. 
The variation in the size of the uncertainties on the $P^{(\prime)}_i$ observables arises from their  dependence on $F_{\rm L}$. When $F_{\rm L}$ is large, the $(1-F_{\rm L})$ term in the definition of the observables gives a large uncertainty that can be significantly asymmetric.

The results of the likelihood fit for the angular observables are given in Tables~\ref{tab:results:likelihood:widebins}--\ref{tab:results:likelihood:optimised} of Appendix~\ref{sec:appendix:results}, with the statistical and systematic uncertainties separated. 
In general, the correlations between the observables are small. 
The most notable exceptions are the correlations between $A_{\rm FB}$ and $F_{\rm L}$, which can be as large as $60\%$, and the correlations between the different $P^{(\prime)}_i$ observables in the range $2.5 < \qsq < 4.0\gev^2/c^4$. 
The correlations between $A_{\rm FB}$ and $F_{\rm L}$ arise from the requirement that the differential decay rate in Eq.~(\ref{eq:pdfpwave}) be positive across the entire phase space. 
The correlations between the $P^{(\prime)}_i$ observables originate from their common dependence on $F_{\rm L}$. 
The correlation matrices for all of the \qsq bins are available in Appendices~\ref{sec:appendix:likelihood:correlation},~\ref{sec:appendix:correlation:asymmetries} and \ref{sec:appendix:correlation:optimised}. 
The values of $F_{\rm S}$ obtained from the fits are consistent with the S-wave contribution  of approximately $5\%$ observed in \decay{\Bz}{\jpsi\Kstarz} data~\cite{Aubert:2007hz, Chilikin:2014bkk, Aaij:2013cma}.
Considering the observables individually, the results appear largely in agreement with the SM predictions.  
The exception to this is the observable $S_5$ and the related observable $P'_5$.
Small differences can also be seen in the measured $A_{\rm FB}$ distribution, where 
the data lie systematically below the SM predictions in the range $1.1 < \qsq < 6.0\gev^2/c^4$.
No significant \CP asymmetry is seen.

The discrepancy in $P'_5$ confirms the result of the previous LHCb analysis~\cite{LHCb-PAPER-2013-037}, where a difference was seen between the data and the SM predictions in the \qsq range  $4.30 < \qsq < 8.68\gev^2/c^4$.
In the present analysis, a deviation from the SM prediction is observed in each of the $4.0<q^2<6.0\gev^{2}/c^{4}$ and $6.0<q^2<8.0\gev^{2}/c^{4}$ bins at a level of 2.8 and 3.0 standard deviations, respectively. The SM predictions for the optimised observables that are used in this analysis are taken from Ref.~\cite{Descotes-Genon:2014uoa}. The predictions are an update of the SM calculation from Ref.~\cite{Descotes-Genon:2013vna}, which was used to compare the previous LHCb $P^{(\prime)}_i$ measurements to the SM.

\begin{figure}[h]
\begin{center}
\includegraphics[width=0.47\linewidth]{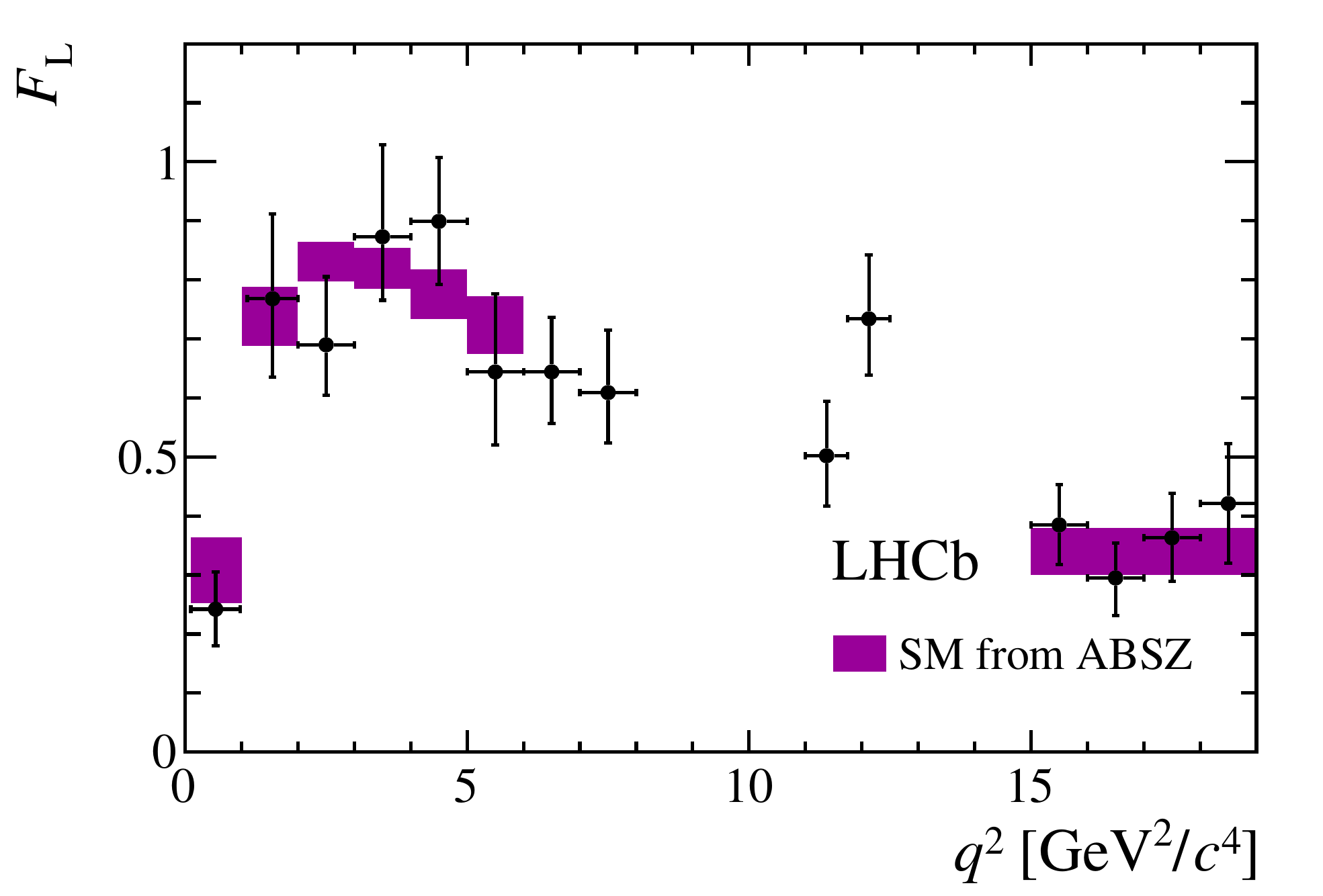} 
\includegraphics[width=0.47\linewidth]{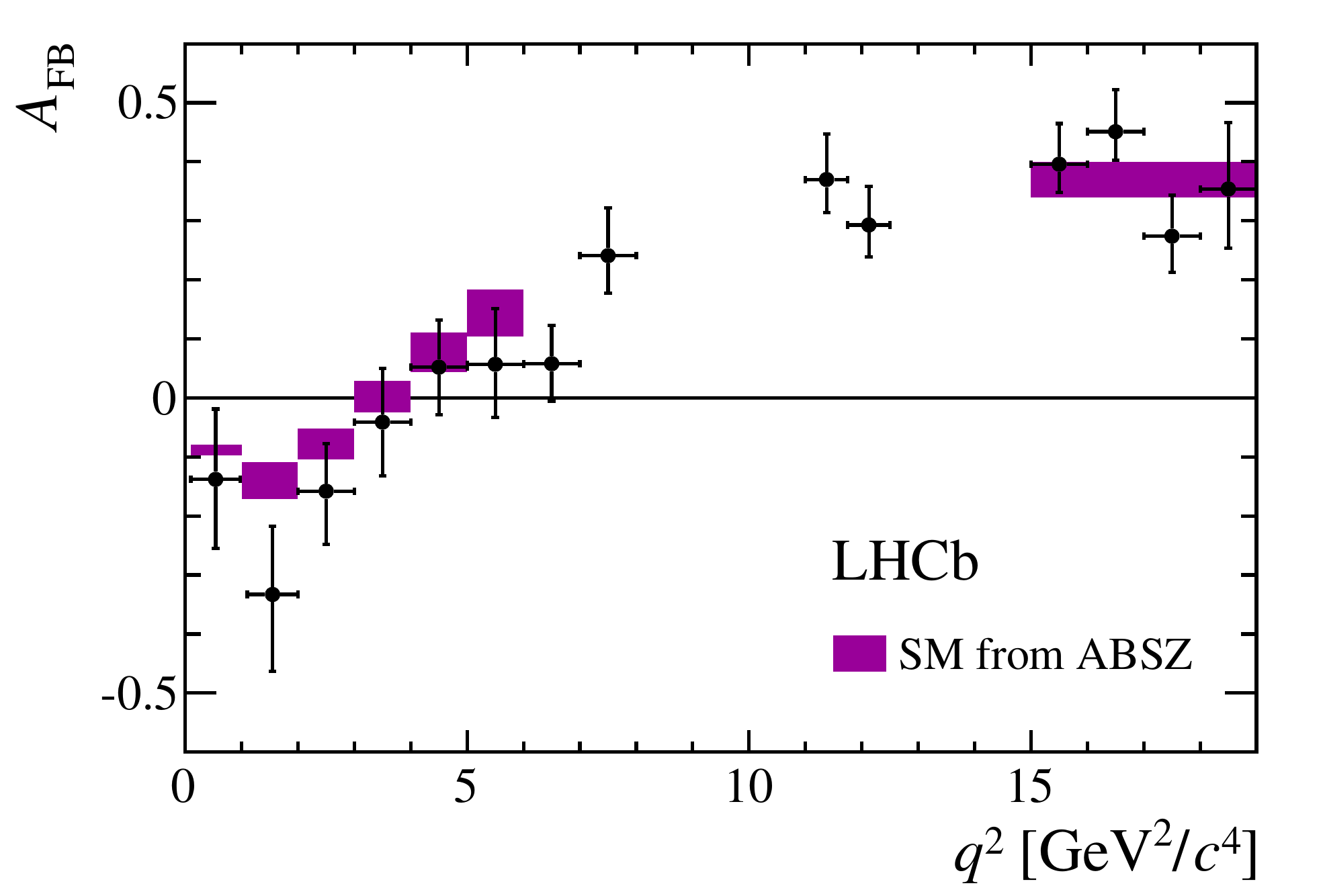} \\
\includegraphics[width=0.47\linewidth]{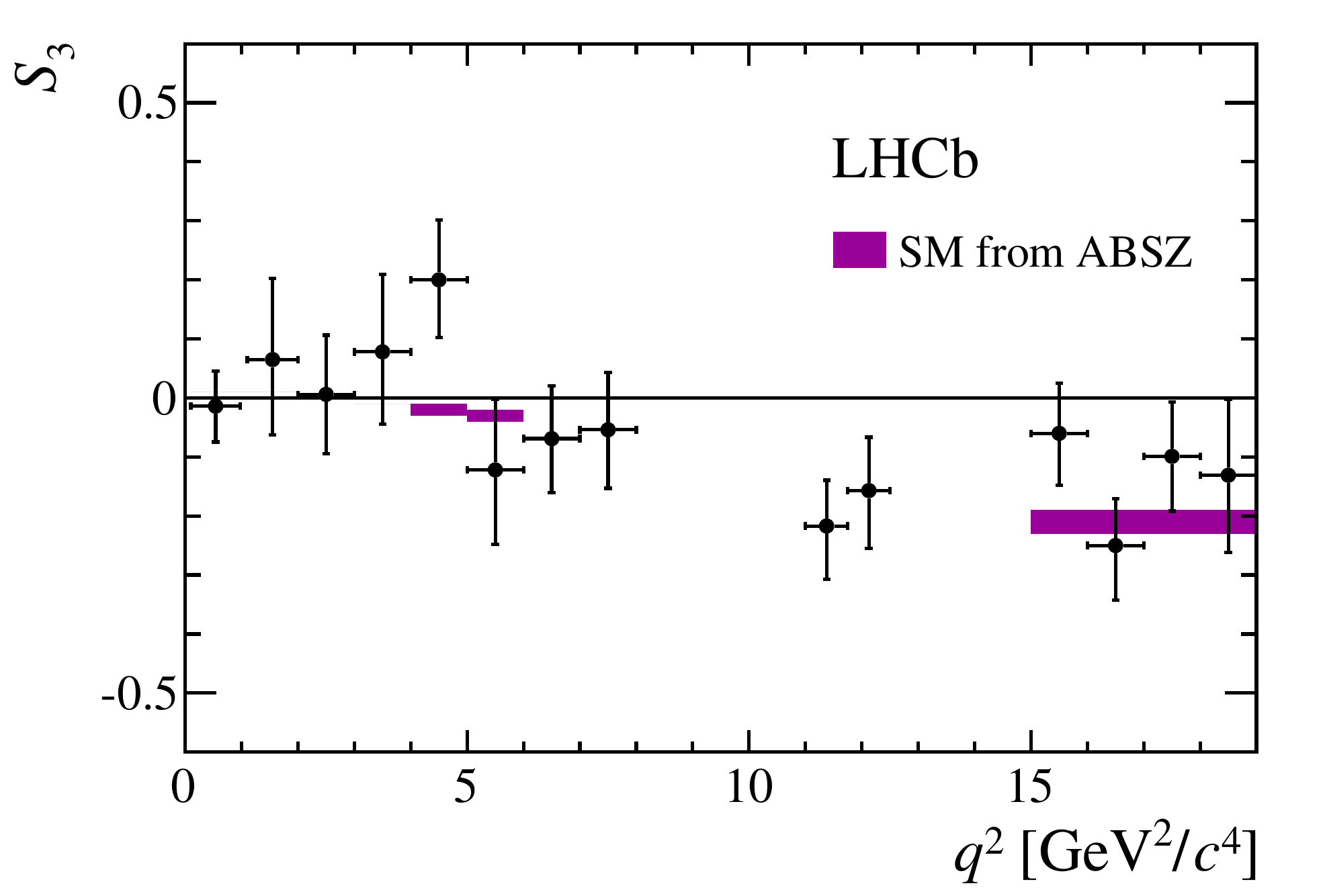} 
\includegraphics[width=0.47\linewidth]{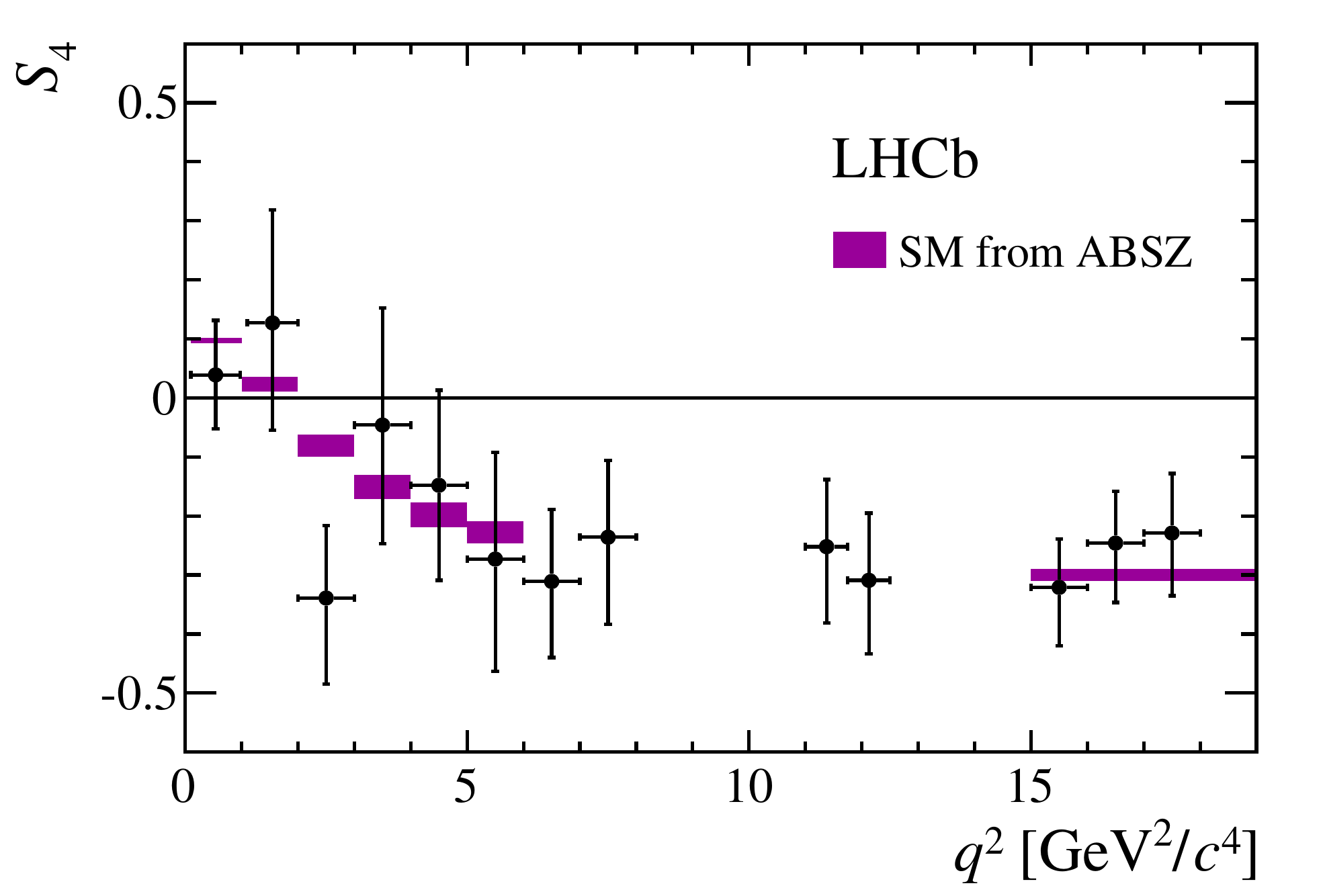} \\
\includegraphics[width=0.47\linewidth]{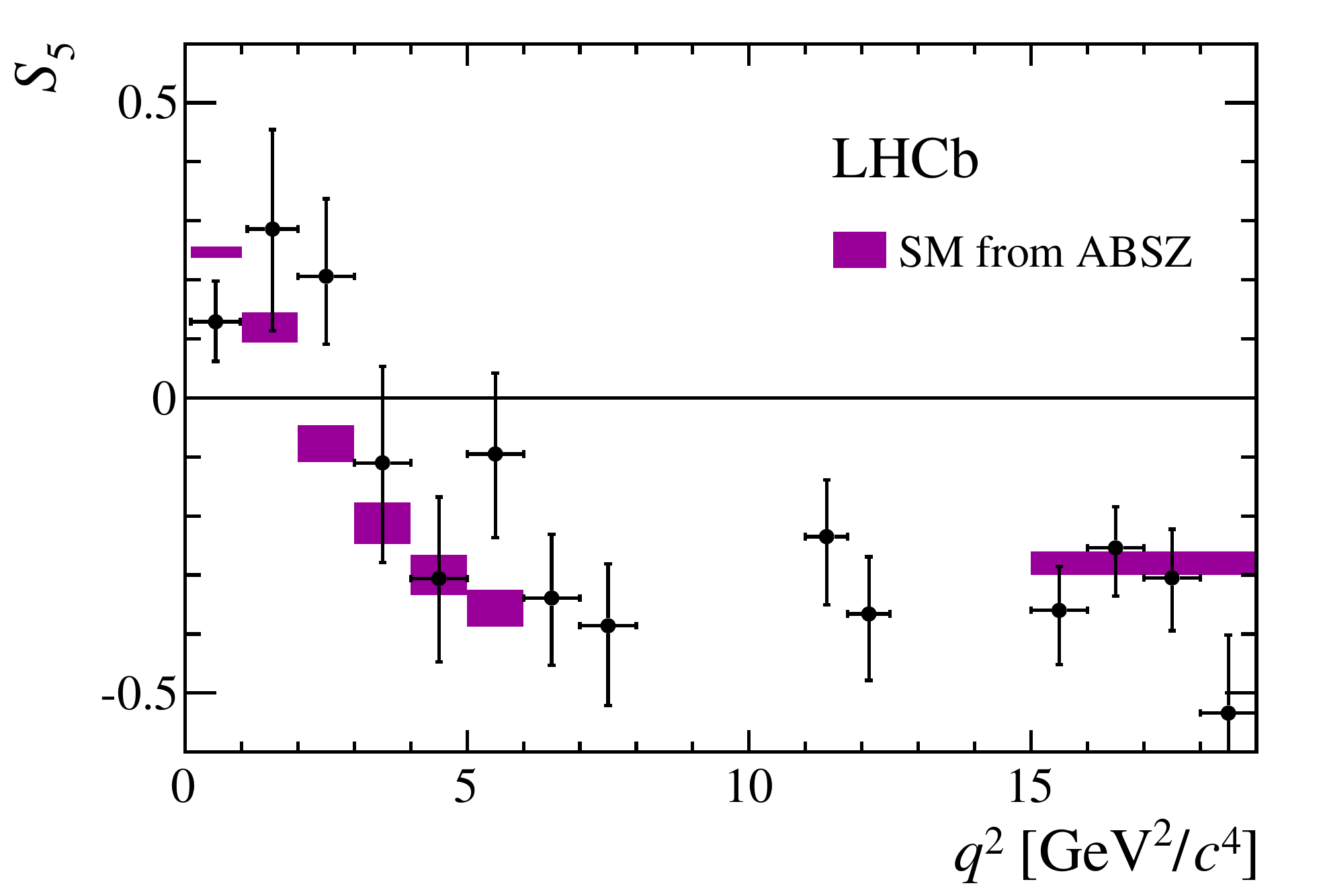} 
\includegraphics[width=0.47\linewidth]{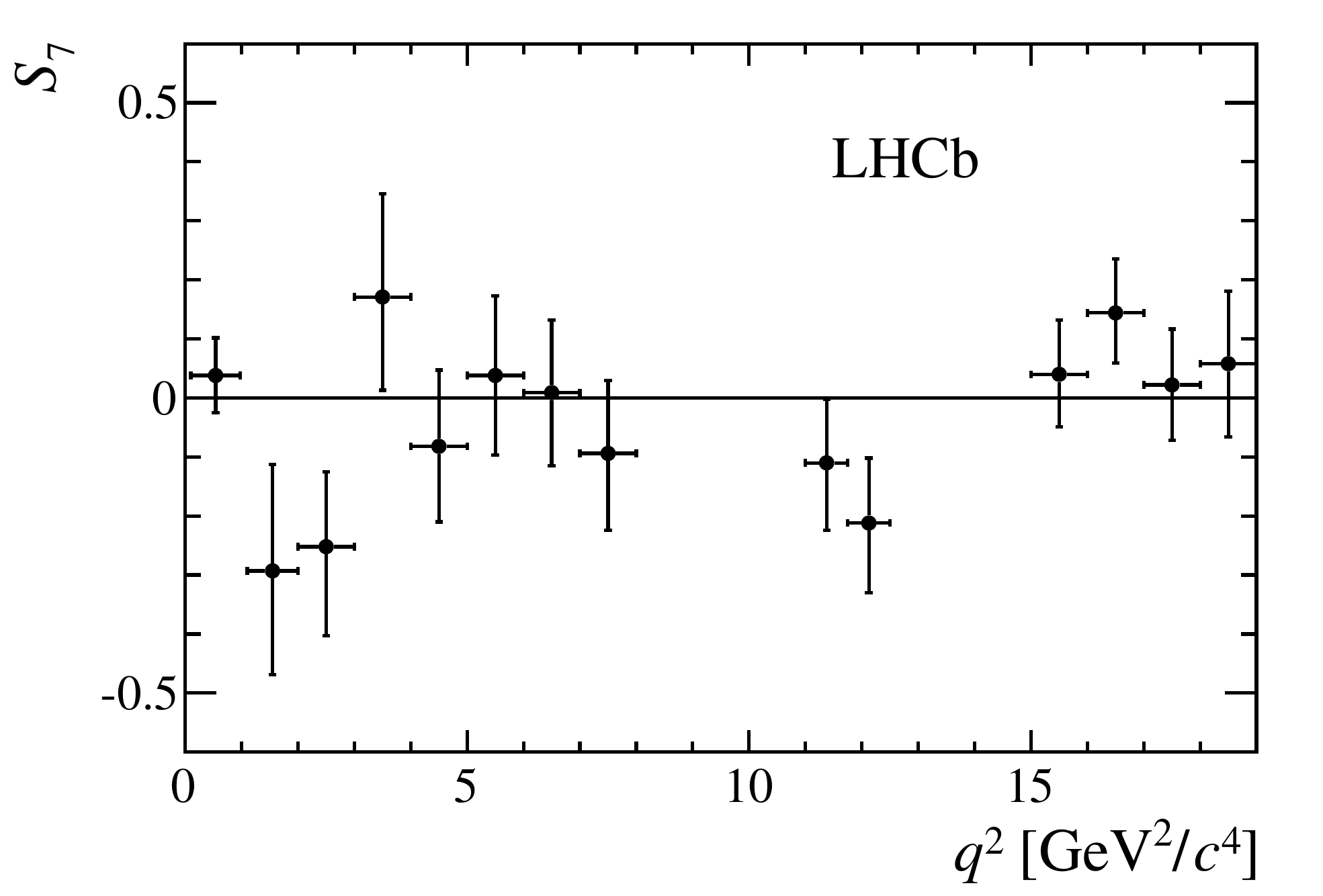} \\
\includegraphics[width=0.47\linewidth]{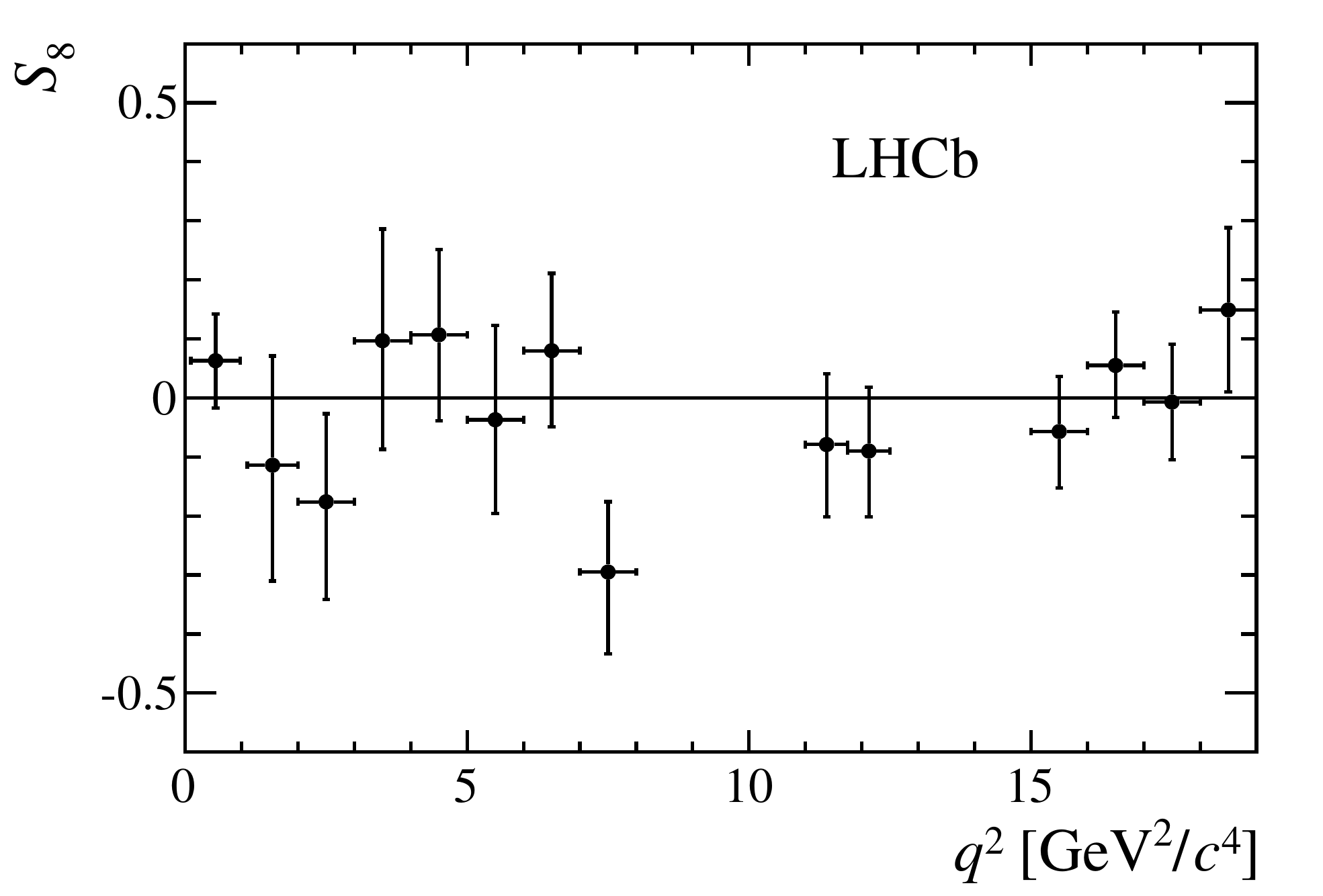} 
\includegraphics[width=0.47\linewidth]{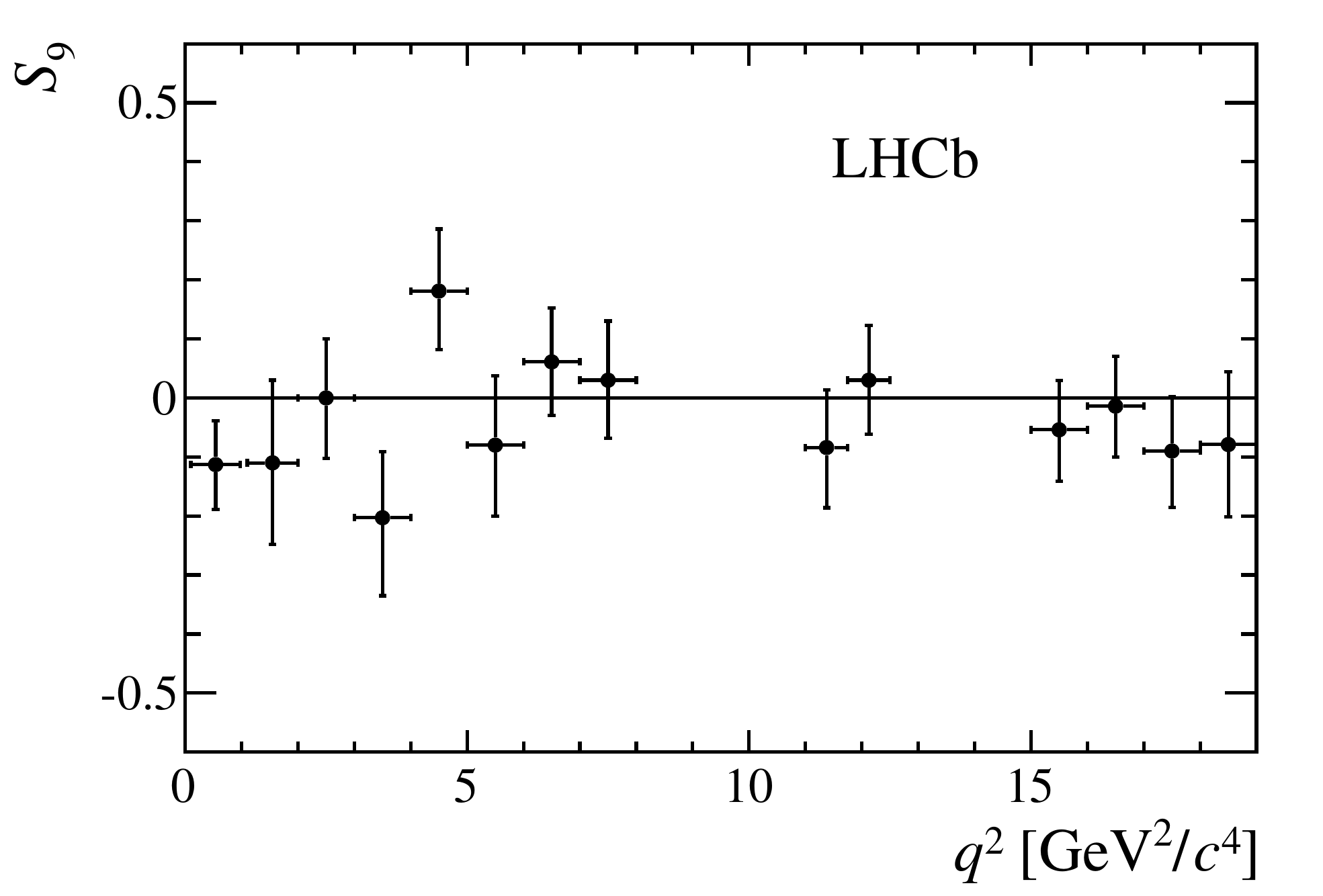} \\
\end{center}
\caption{\label{fig:results:moments:Si} The \CP-averaged observables in bins of $q^2$, determined from a moment analysis of the data.  The shaded boxes show the SM predictions based on the prescription of  Ref.~\cite{Altmannshofer:2014rta}.} 
\end{figure} 

\begin{figure}[h]
\begin{center}
\includegraphics[width=0.47\linewidth]{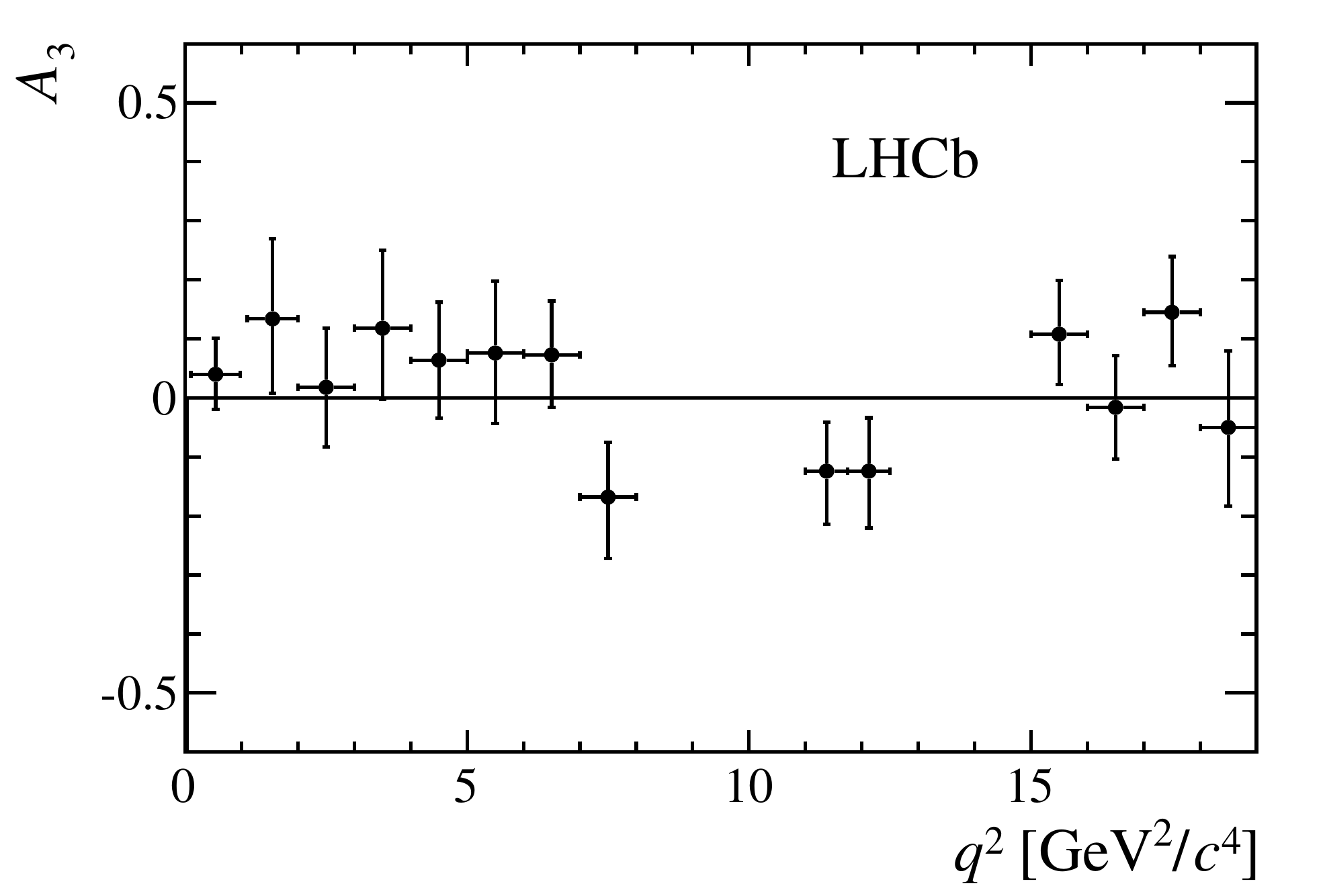}
\includegraphics[width=0.47\linewidth]{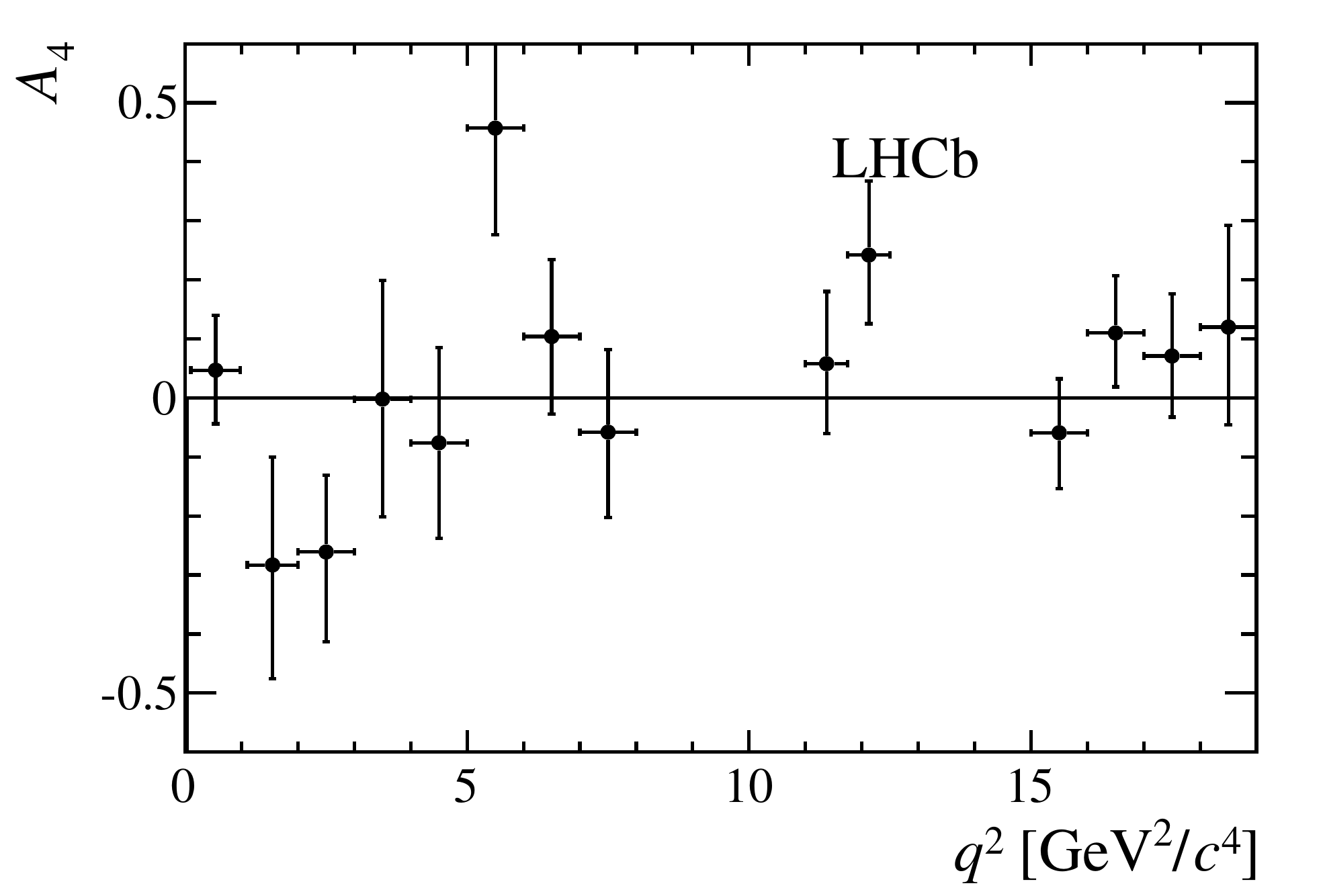}  \\
\includegraphics[width=0.47\linewidth]{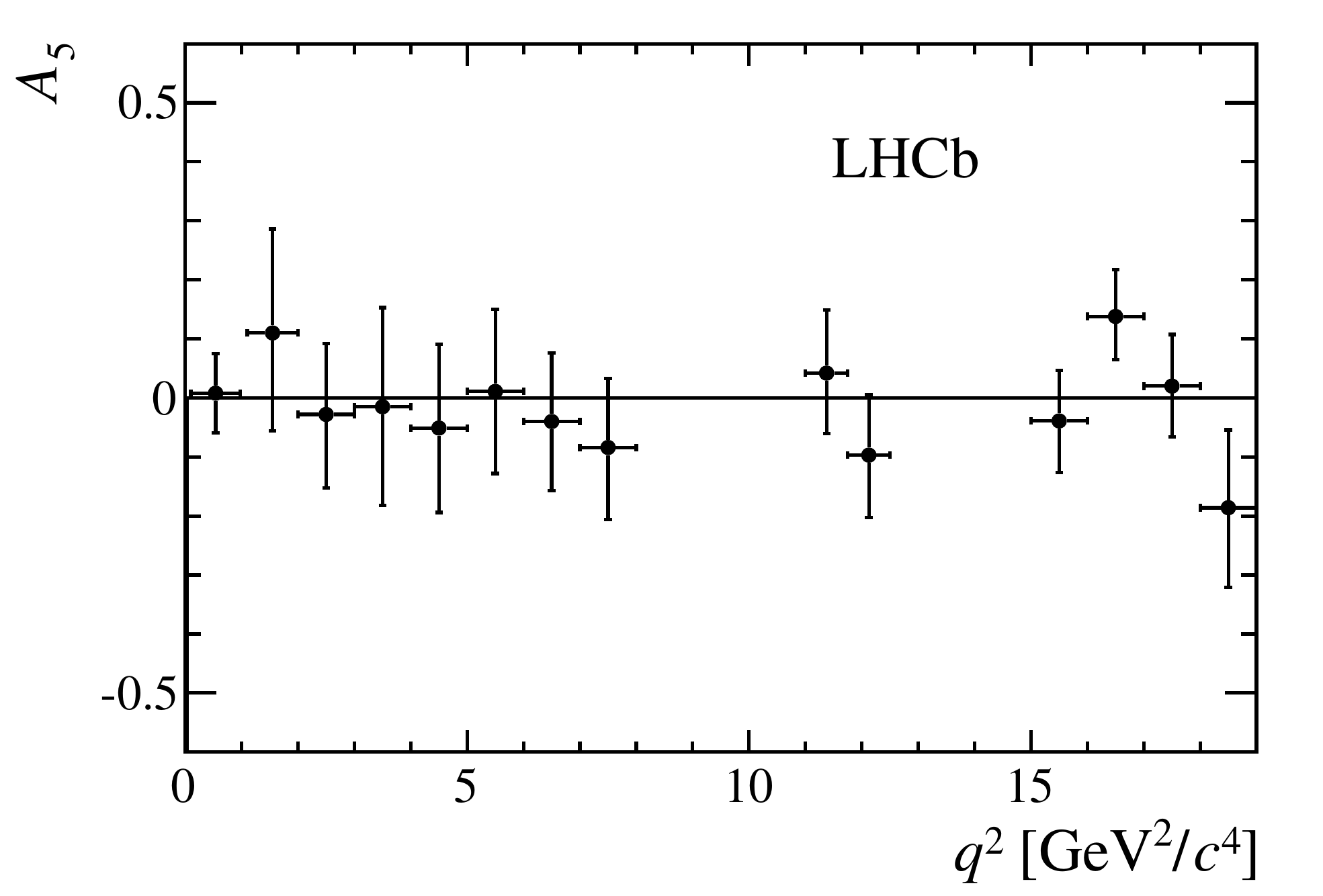} 
\includegraphics[width=0.47\linewidth]{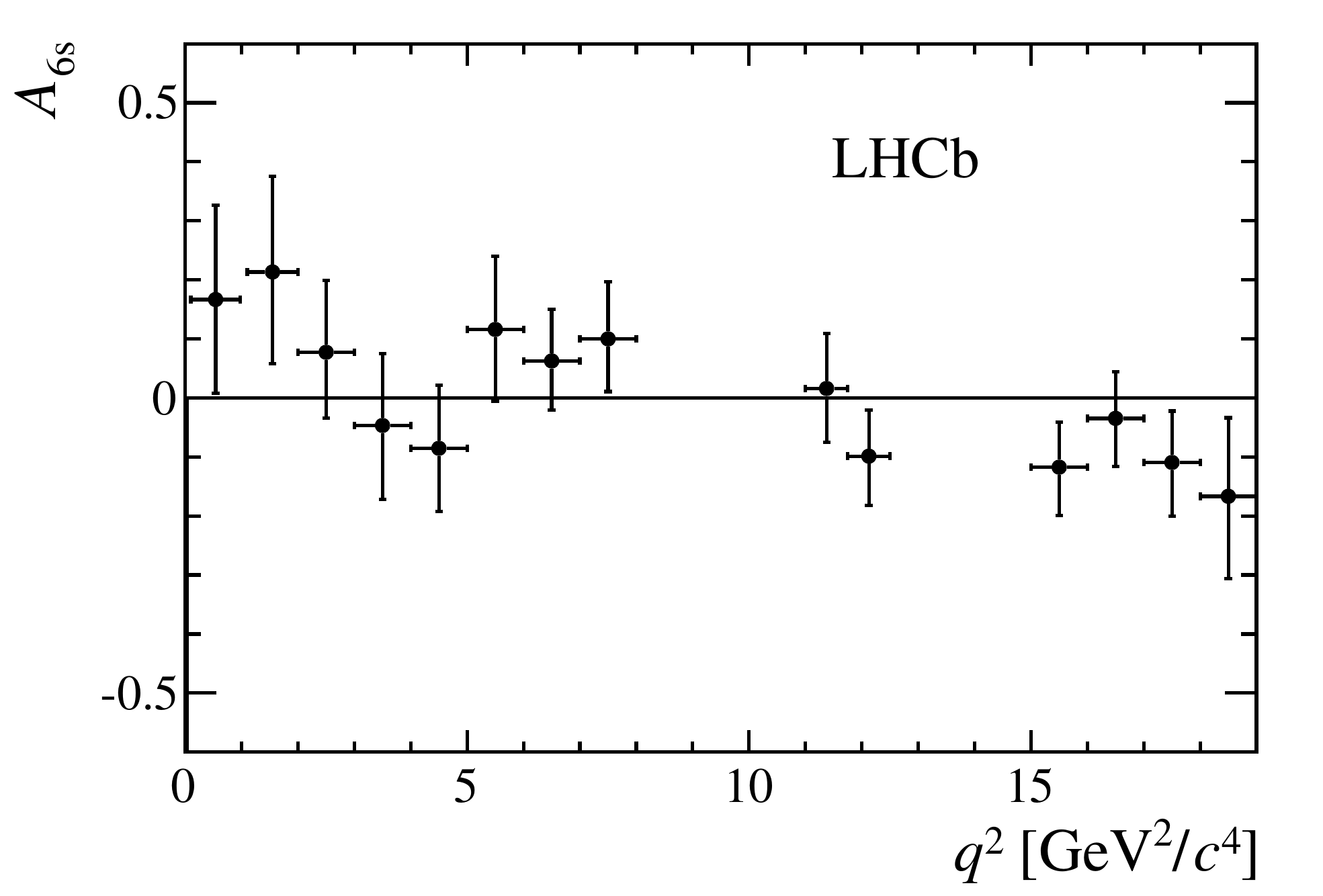} \\
\includegraphics[width=0.47\linewidth]{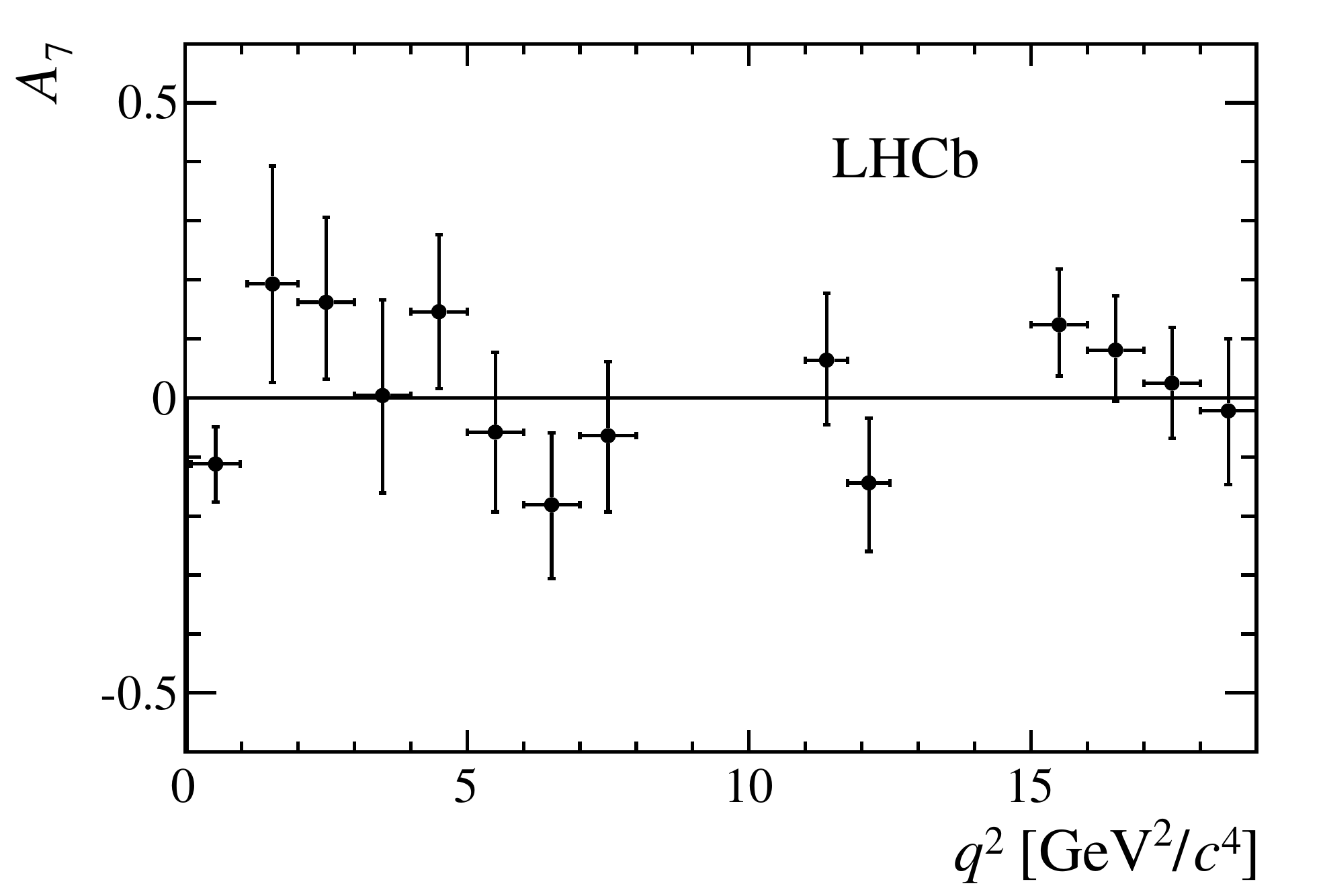} 
\includegraphics[width=0.47\linewidth]{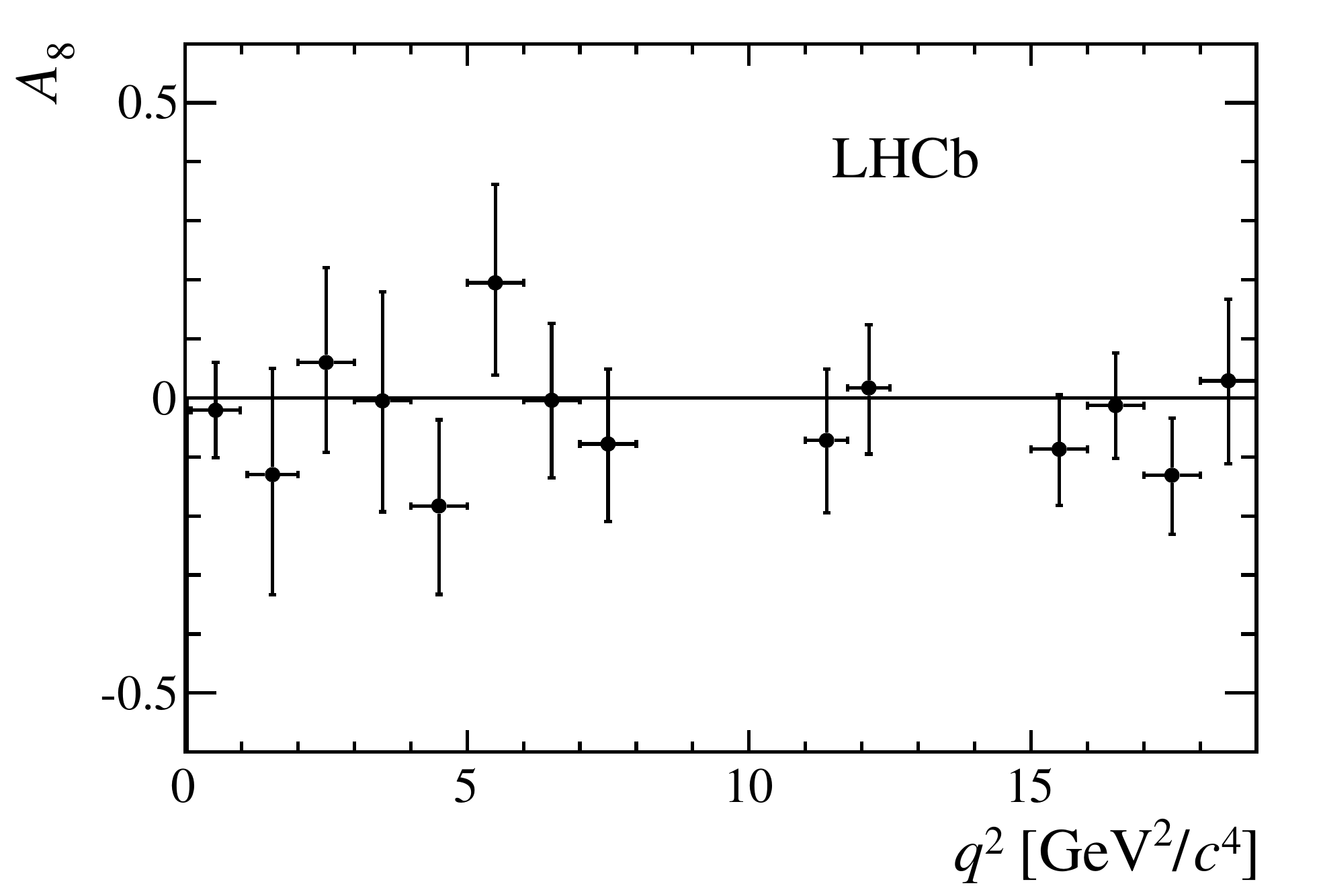} \\
\includegraphics[width=0.47\linewidth]{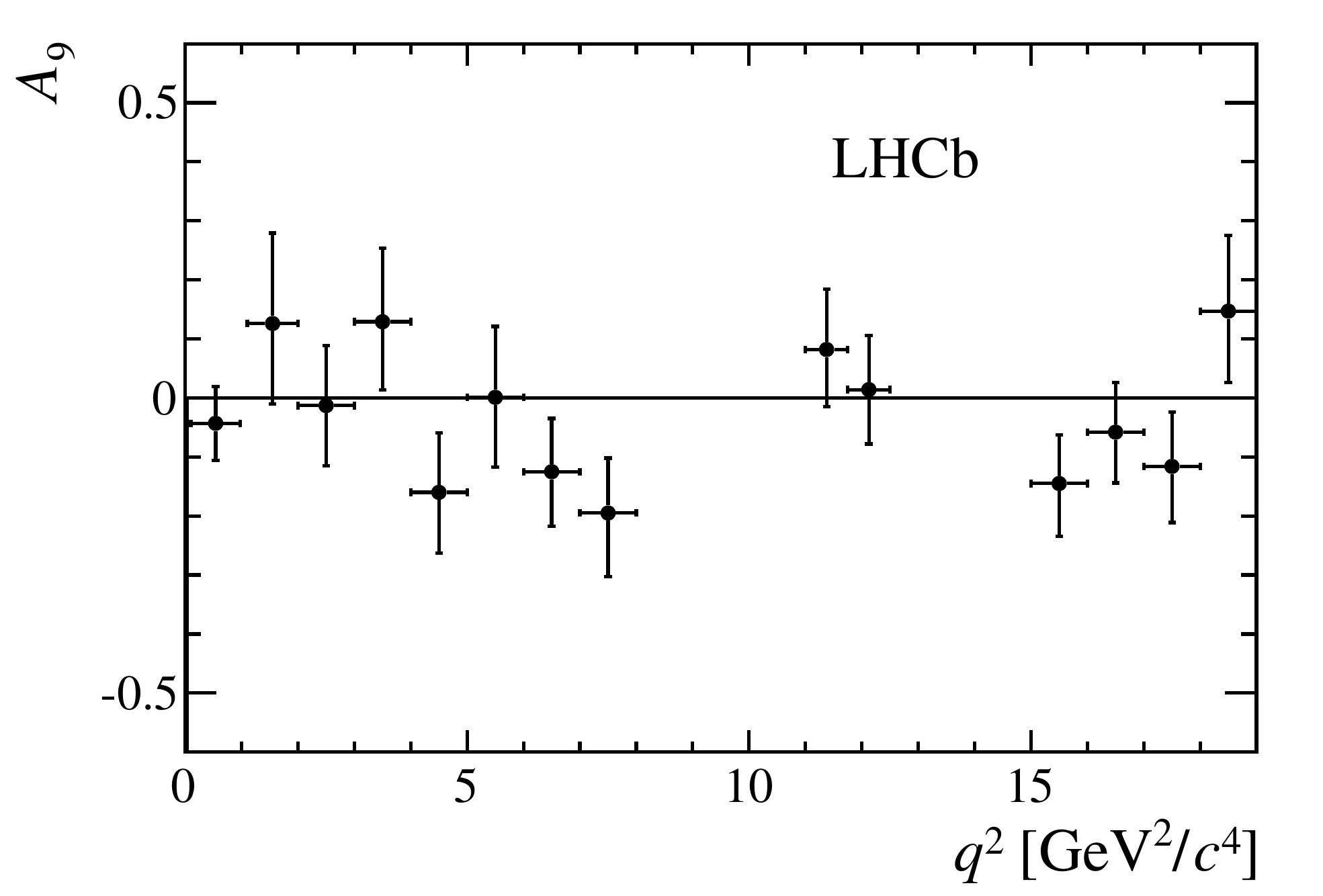} 
\end{center}
\caption{\label{fig:results:moments:Ai} The \CP-asymmetric observables in bins of $q^2$, determined from a moment analysis of the data.} 
\end{figure} 

\begin{figure}[h]
\begin{center}
\includegraphics[width=0.47\linewidth]{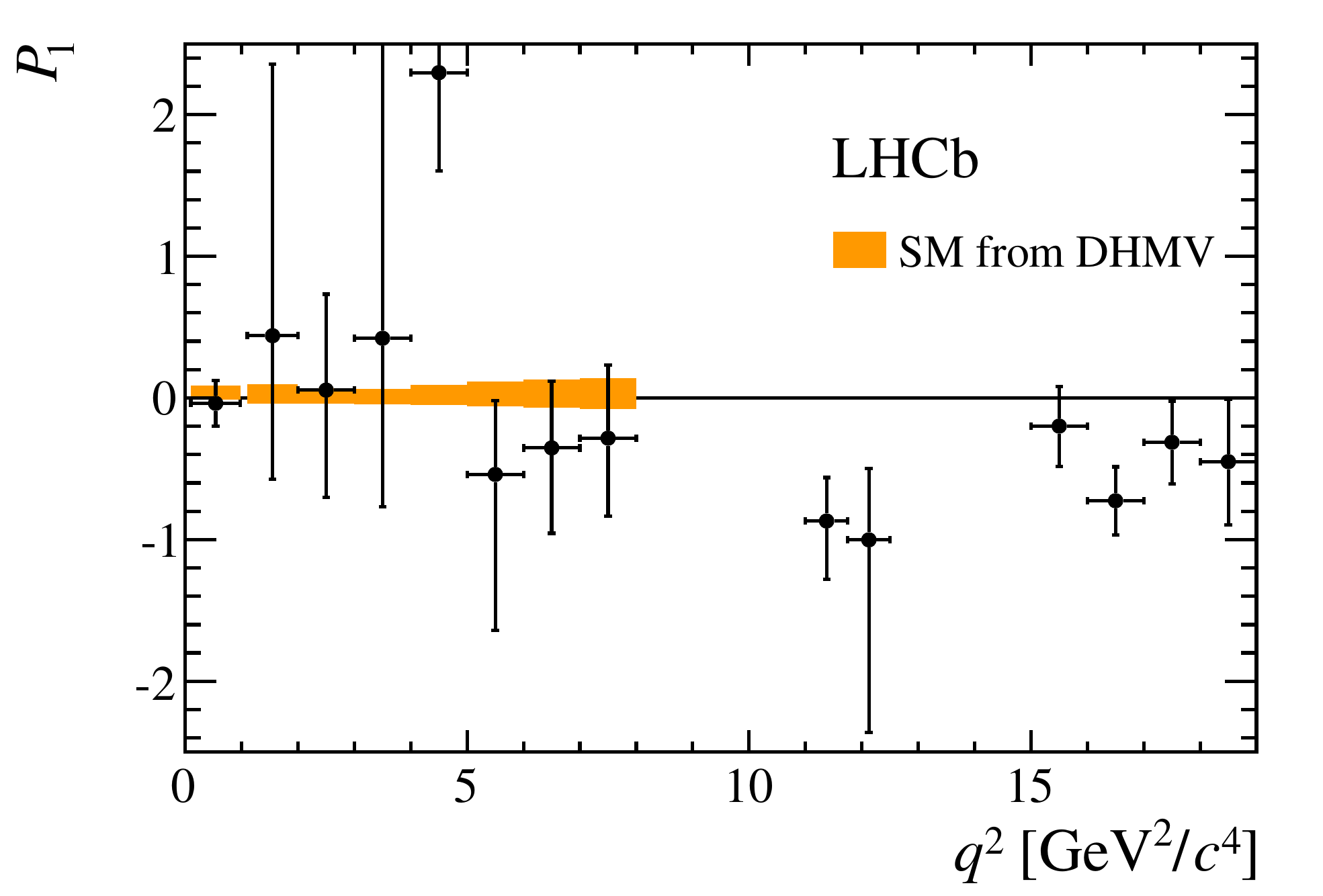}
\includegraphics[width=0.47\linewidth]{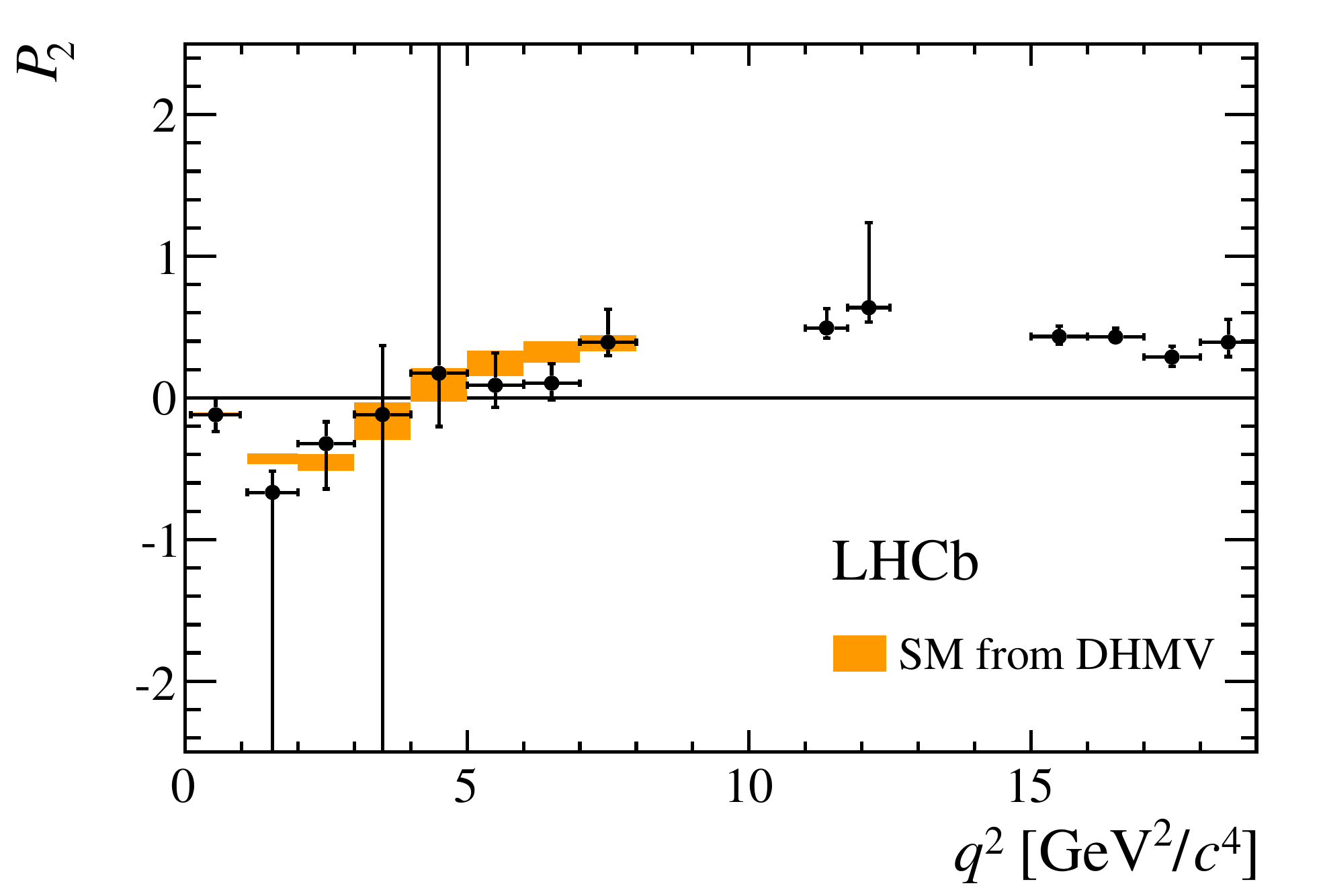}  \\
\includegraphics[width=0.47\linewidth]{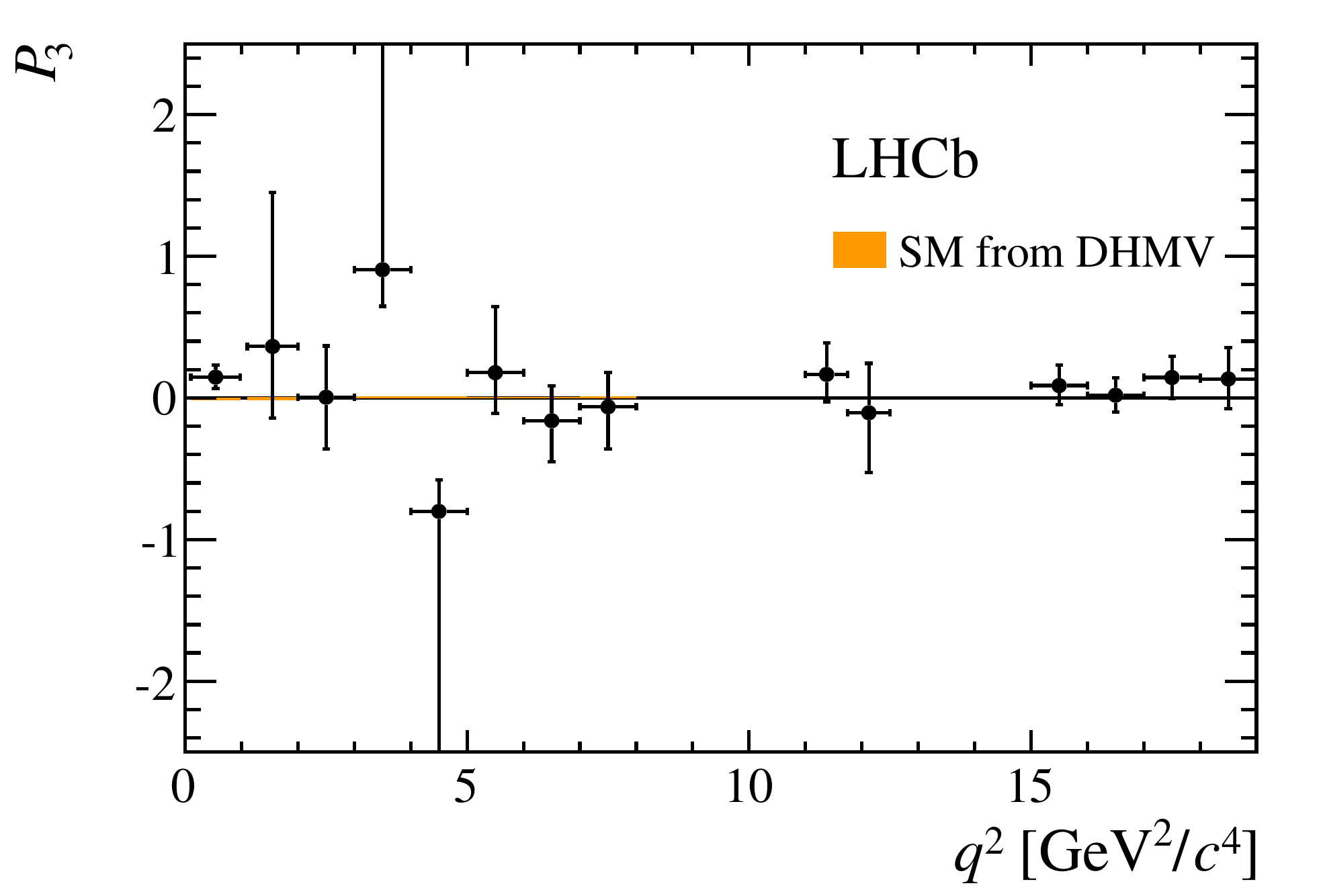} 
\includegraphics[width=0.47\linewidth]{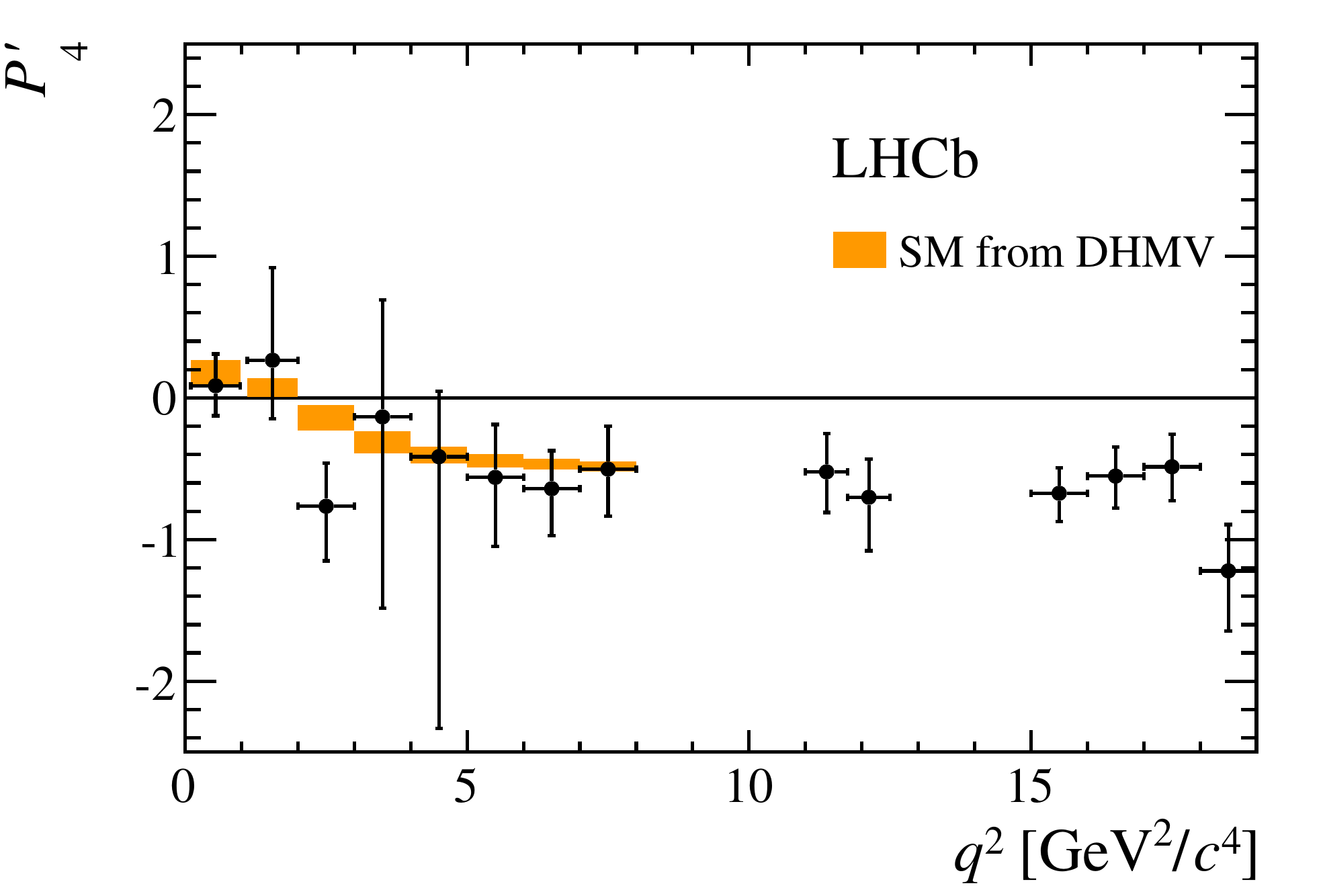} \\
\includegraphics[width=0.47\linewidth]{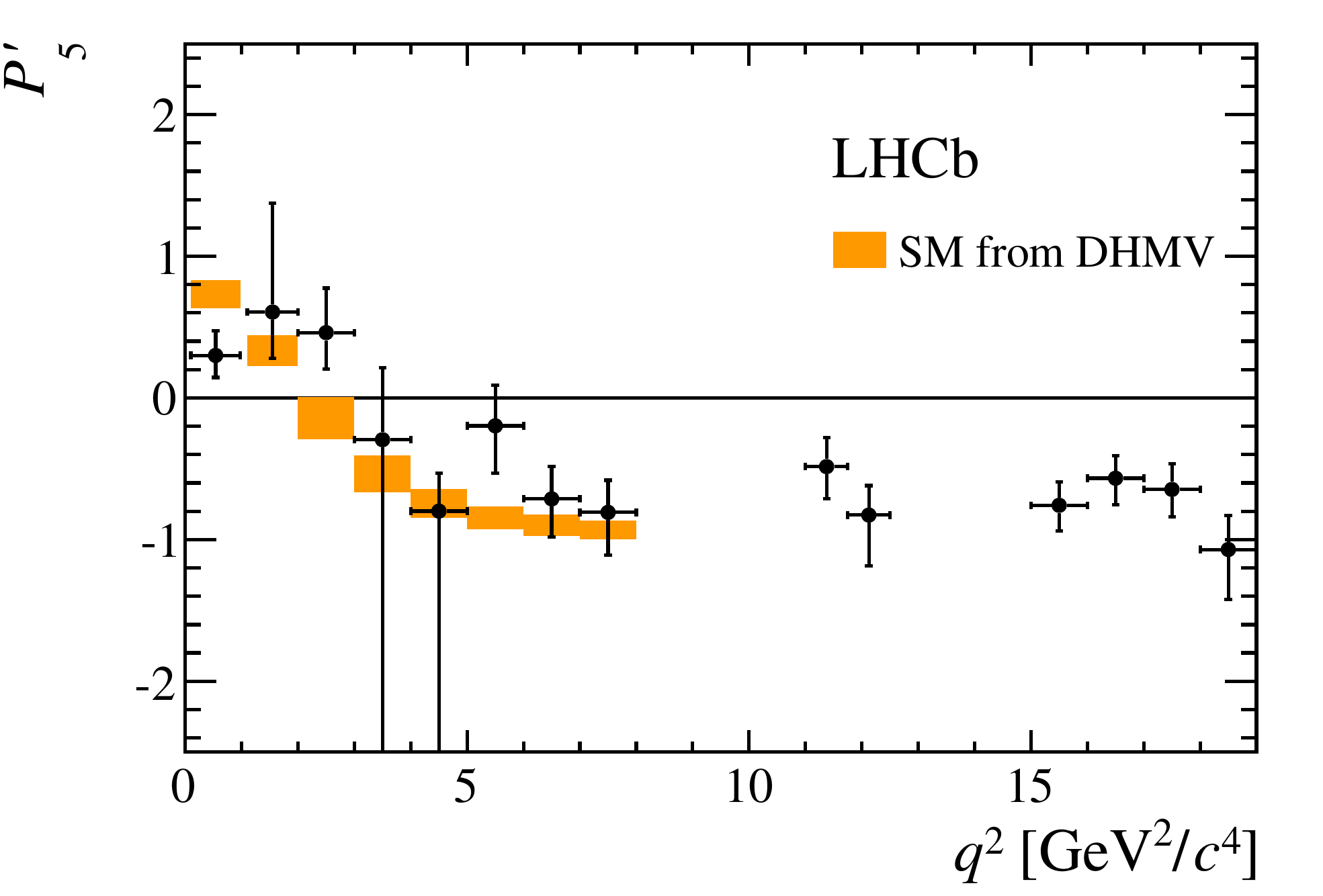} 
\includegraphics[width=0.47\linewidth]{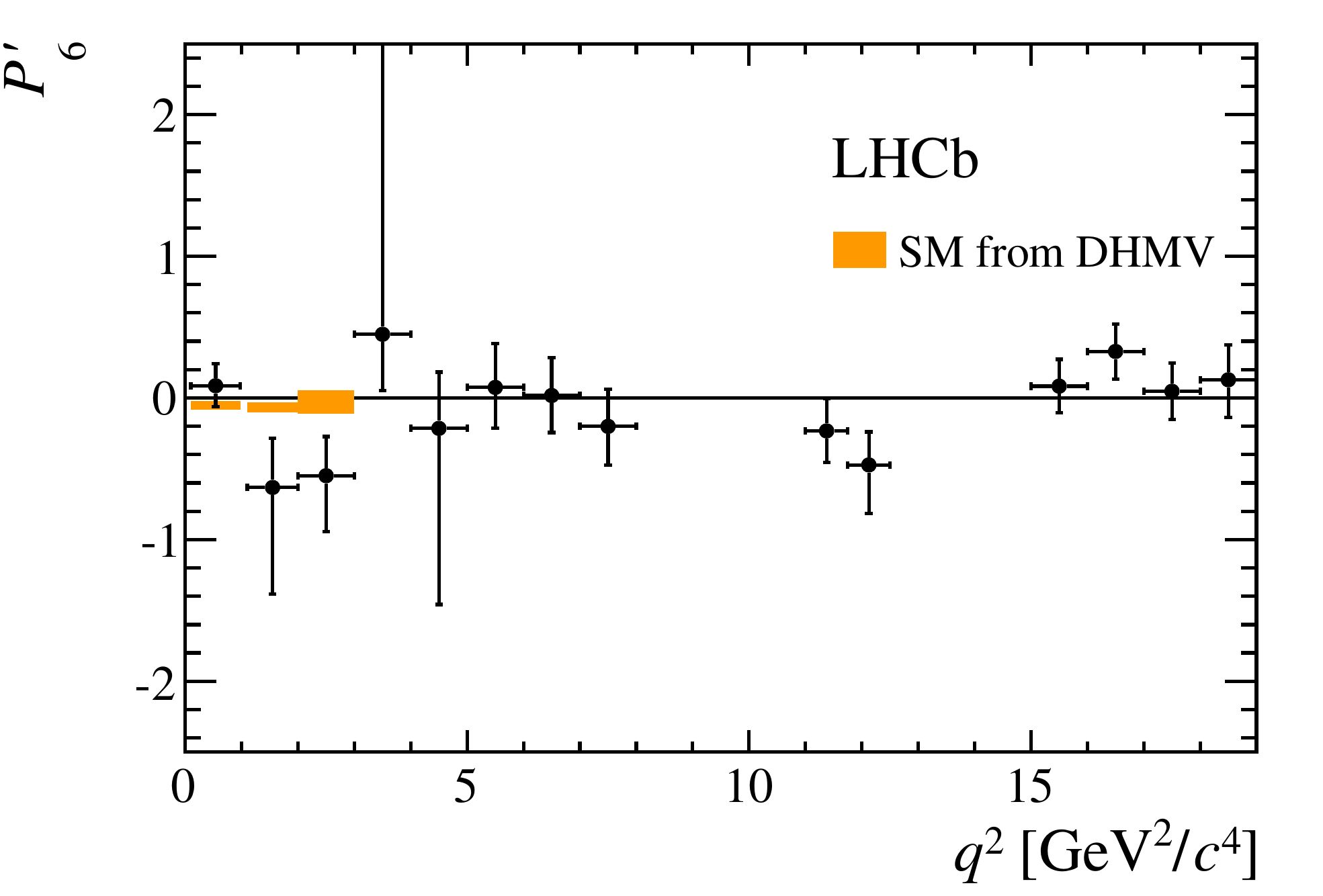} \\
\includegraphics[width=0.47\linewidth]{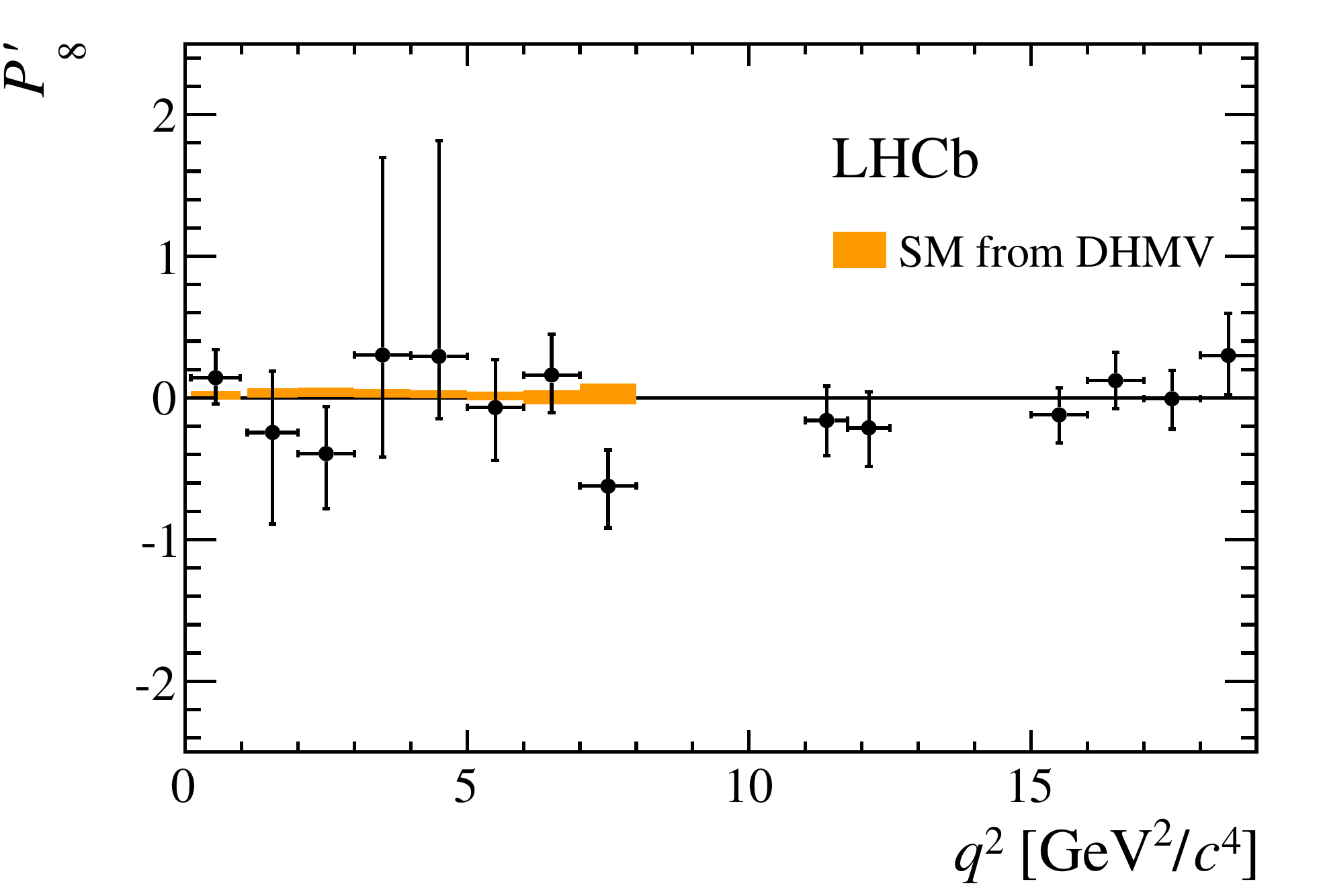} 
\end{center}
\caption{\label{fig:results:moments:Pi} The optimised angular observables in bins of $q^2$, determined from a moment analysis of the data. The shaded boxes show the SM predictions taken from Ref.~\cite{Descotes-Genon:2014uoa}.} 
\end{figure}

\begin{figure}[!htb]
\centering 
\includegraphics[width=0.6\linewidth]{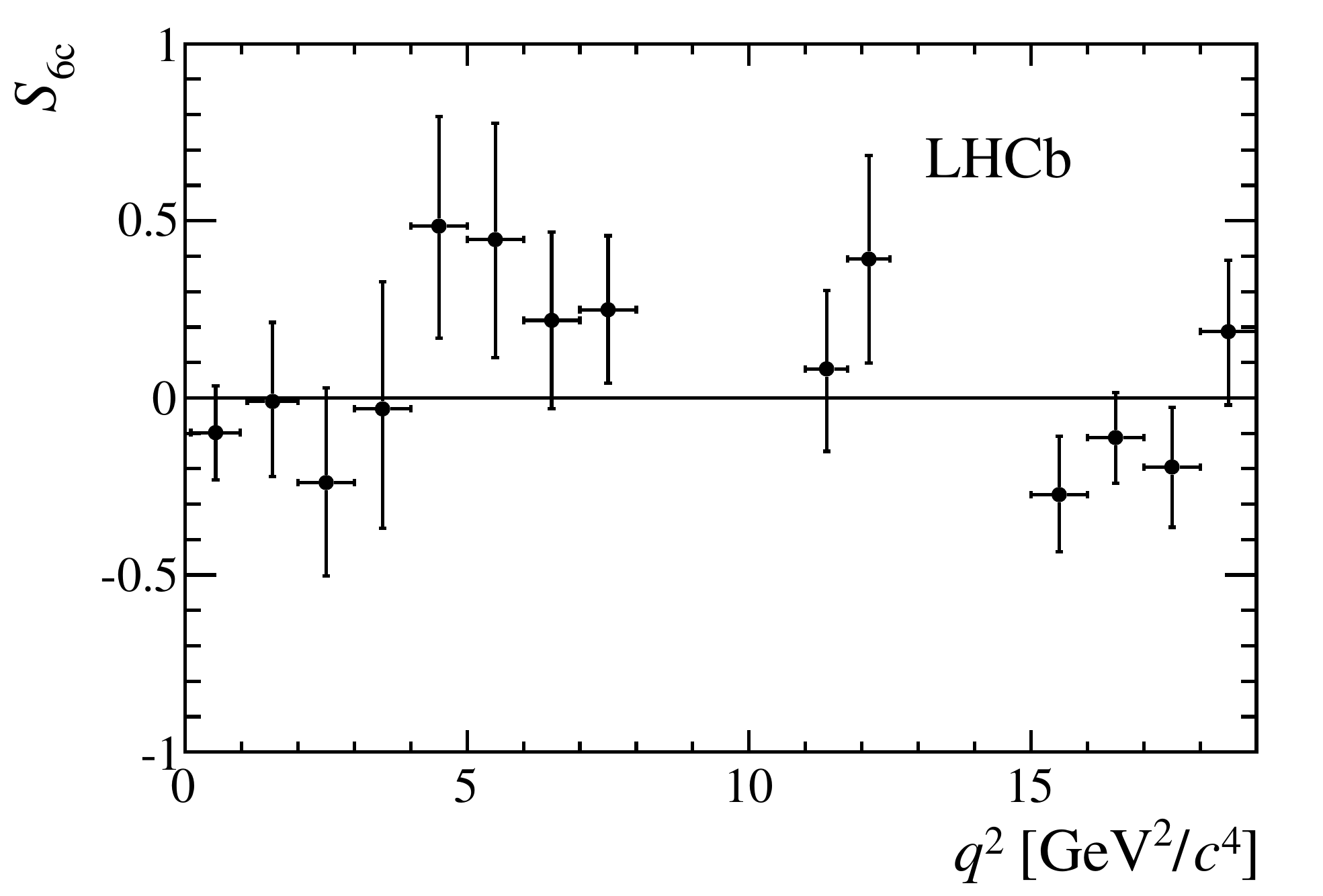} 
\caption{The observable $S_{6c}$ in bins of $q^2$, as determined from a moment analysis of the data.
 \label{fig:results:S6c}
} 
\end{figure}

The results of the moment analysis are shown in Figs.~\ref{fig:results:moments:Si},~ \ref{fig:results:moments:Ai} and \ref{fig:results:moments:Pi} and given in Tables~\ref{tab:results:moments:averages},~\ref{tab:results:moments:asymmetries} and \ref{tab:results:moments:optimised} of Appendix~\ref{sec:appendix:results}. 
The same behaviour is seen as in the likelihood fit, where some differences are observed between the SM predictions and the data in $S_5$ (and $P'_5$) at low values of \qsq. 
The observable $S_{6c}$ is also included in Table~\ref{tab:results:moments:averages} and shown in Fig.~\ref{fig:results:S6c}. 
\clearpage
\noindent This observable is consistent with zero, as expected in the SM.
As a cross-check, the observables have also been determined by a moment analysis in the approximately $2\gev^{2}/c^{4}$ \qsq bins used in the likelihood fit of the angular observables. 
The differences between the central values of the two methods are compatible with those expected from pseudoexperiments.
The correlation matrices for all of the \qsq bins are available in Appendices~\ref{sec:appendix:bootstrap:correlation}--\ref{sec:appendix:bootstrap:correlation:optimised}.

\begin{figure}[!htb]
\begin{center}
\includegraphics[width=0.48\linewidth]{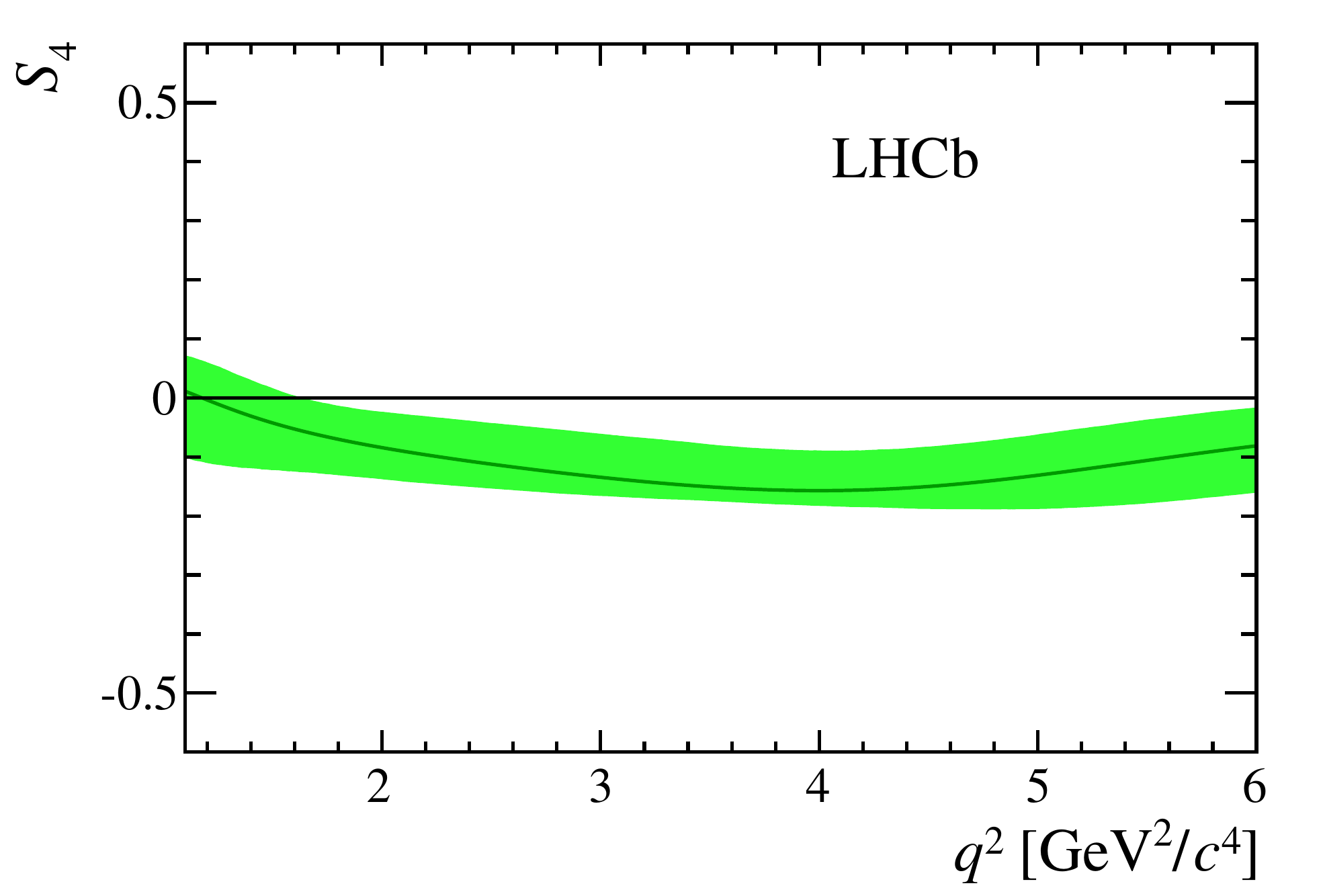} 
\includegraphics[width=0.48\linewidth]{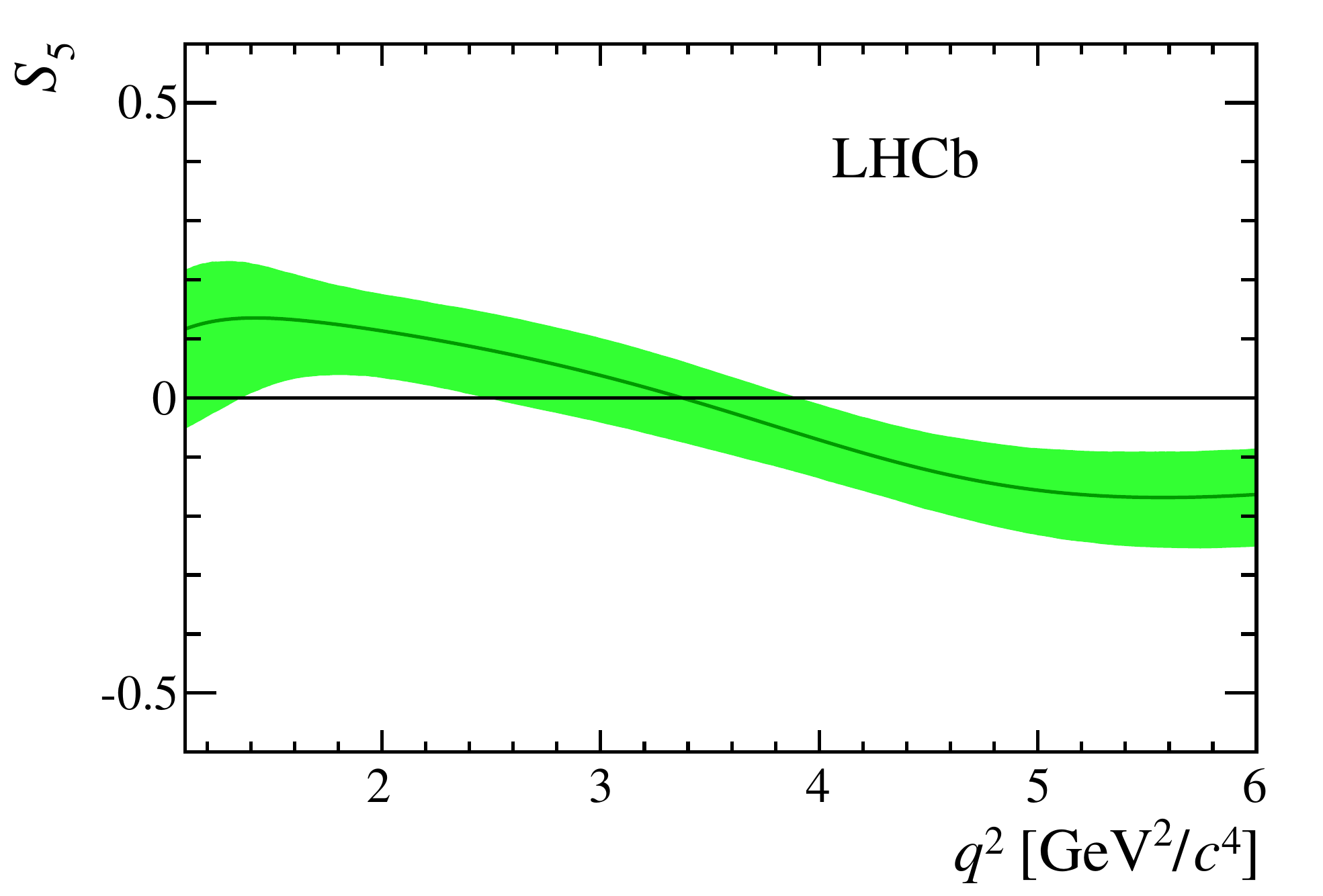}  \\ 
\includegraphics[width=0.48\linewidth]{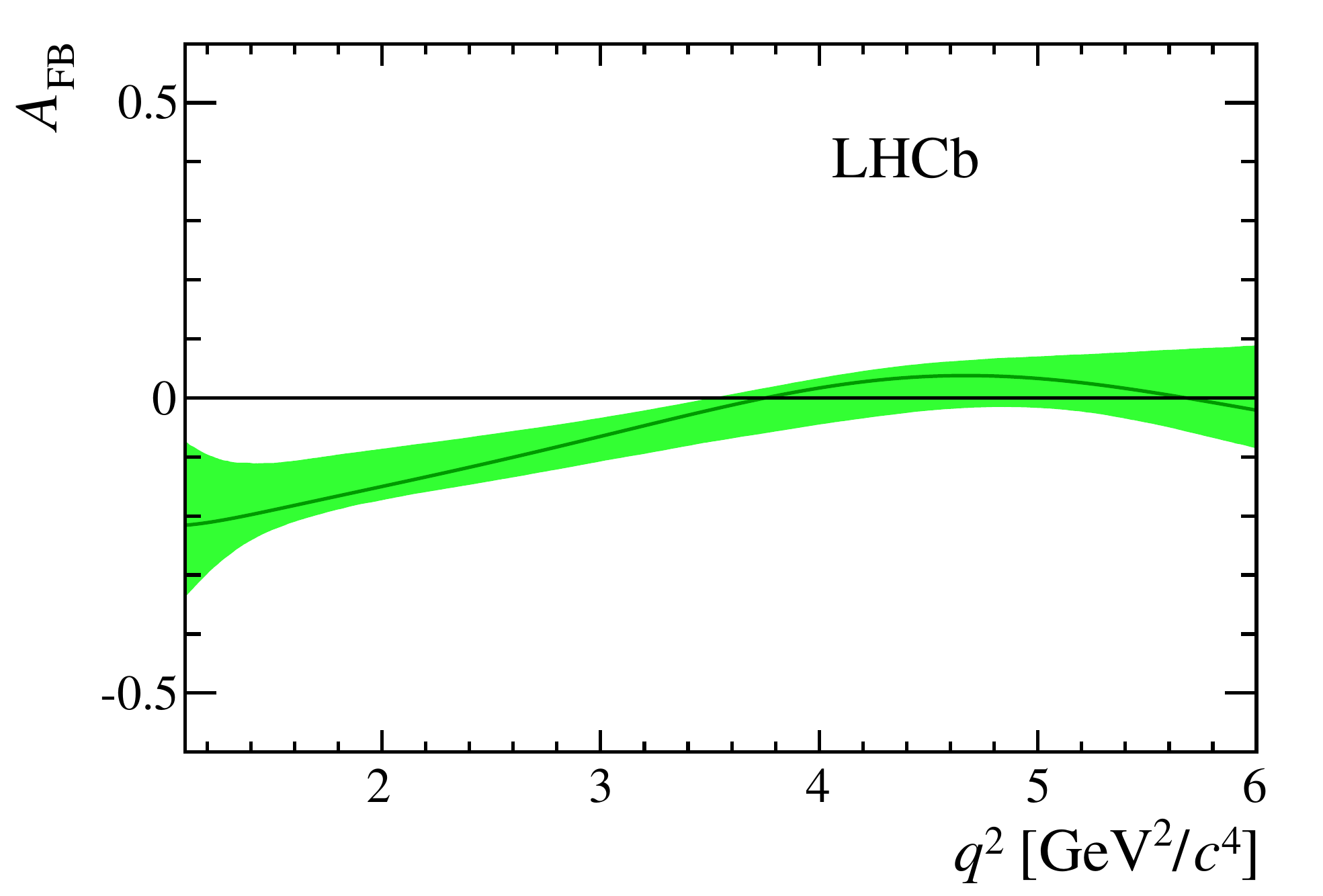} 
\end{center} 
\caption{
The observables $S_4$, $S_5$ and $A_{\rm FB}$ determined by fitting for the \qsq dependent decay amplitudes.  The line indicates the best-fit to the dataset. The band indicates the 68\% interval on the bootstraps at each point in \qsq. Note that, the correlation between points in the bands means it is not possible to extract the uncertainty on the zero-crossing points from these figures.
\label{fig:zcp:observables}}
\end{figure}

Figure~\ref{fig:zcp:observables} shows the observables $S_4$, $S_5$ and $A_{\rm FB}$ resulting from the fit to the \qsq-dependent decay amplitudes. 
The results are in agreement with those obtained from the likelihood fit of the angular observables and the moments analysis.
For $A_{\rm FB}$, the best-fit to the data (the line in the figure) has two zero-crossing points in the range $1.1 < \qsq < 6.0\gev^{2}/c^{4}$ with different slopes. 
As discussed in Sec.~\ref{sec:amplitudes}, only the solution consistent with the data in the range $\qsq > 6.0\gev^{2}/c^{4}$ is cited, \ie the solution with the positive slope (see Figs.~\ref{fig:results:Si} and \ref{fig:results:moments:Si}).
The zero-crossing points determined from the amplitude fit are 
\begin{align} 
\begin{split}
q_0^2(S_5) & \in [2.49,3.95]\gevgevcccc \text{~at~68\%~confidence~level~(C.L.)} \, , \\ 
q_0^2(A_{\rm FB}) & \in [3.40,4.87]\gevgevcccc \text{~at~68\%~C.L.}  \, . 
\end{split} 
\nonumber 
\end{align} 
It is not possible to determine if $S_{4}$ has a zero-crossing point at 68\% confidence level. If there is a zero-crossing then
\begin{align}
q_0^2(S_4) < 2.65\gevgevcccc \text{~at~95\%~C.L.}\, . \nonumber
\end{align}
The correlations between the measured values are less than 10\%.
The measured zero-crossing points are all consistent with their respective SM expectations.
Standard Model predictions for $q_0^2(A_{\rm FB})$  are typically in the range $3.9 - 4.4 \gev^{2}/c^{4}$~\cite{Bobeth:2011nj,Beneke:2004dp,Ali:2006ew} and have relative uncertainties below the 10\% level, for example, $q_{0}^{2}(A_{\rm FB}) = 4.36\,^{+0.33}_{-0.31} \gev^{2}/c^{4}$~\cite{Beneke:2004dp}.

\section{Compatibility with the Standard Model} 
\label{sec:discussion} 

The EOS software package~\cite{Bobeth:2010wg} is used to determine the level of compatibility of the data with the SM. 
It provides predictions for the observables integrated over the \qsq bins used in the analysis. 
A \chisq fit is performed to the \CP-averaged angular observables $F_{\rm L}$, $A_{\rm FB}$ and $S_{3}$--$S_9$ obtained from the likelihood fit to the data. 
The \chisq fit uses observables in the range $\qsq < 8.0
\gev^2/c^4$ and a wide \qsq bin covering the range $15.0 < \qsq < 19.0\gev^{2}/c^{4}$. 
Previous analyses~\cite{Descotes-Genon:2013wba,Altmannshofer:2013foa,Beaujean:2013soa,Hurth:2013ssa,Altmannshofer:2014rta,Mahmoudi:2014mja, Datta:2013kja} 
have shown that the existing measurements of decays involving a \decay{\bquark}{\squark} quark transition, including the previous LHCb $P'_5$ result from Ref.~\cite{LHCb-PAPER-2013-037}, can be accounted for  by modifying only the real part of the vector coupling strength of the decays, conventionally denoted ${\rm Re}({\cal C}_9)$. 
An analysis considering additional effective couplings would require a global fit to all of the measurements of $\bquark \to \squark$ quark transitions and is beyond the scope of this paper.
Note that modifying just the axial-vector coupling strength, ${\cal C}_{10}$, would lead to a branching fraction for the \decay{\Bs}{\mumu} decay that is excluded by existing measurements~\cite{LHCb-PAPER-2014-049}. 

In the \chisq fit, the correlations between the different observables are taken into account. The floating parameters are  ${\rm Re}({\cal C}_9)$ and a number of nuisance parameters associated with the form factors, CKM elements and possible sub-leading corrections to the amplitudes. 
The sub-leading corrections to the amplitudes are expected to be suppressed by the size of the \bquark-quark mass relative to the typical energy scale of QCD. 
The nuisance parameters are treated according to the prescription of Ref.~\cite{Beaujean:2013soa} and are included in the fit with Gaussian constraints.
In the \chisq minimisation procedure, the value of each observable (as derived from a particular choice of the theory parameters) is compared to the measured value. 
Depending on the sign of the difference between these values, either the lower or upper (asymmetric) uncertainty on the measurement is used to compute the \chisq.

\begin{figure}[!htb]
  \centering
  \includegraphics[width=0.75\linewidth]{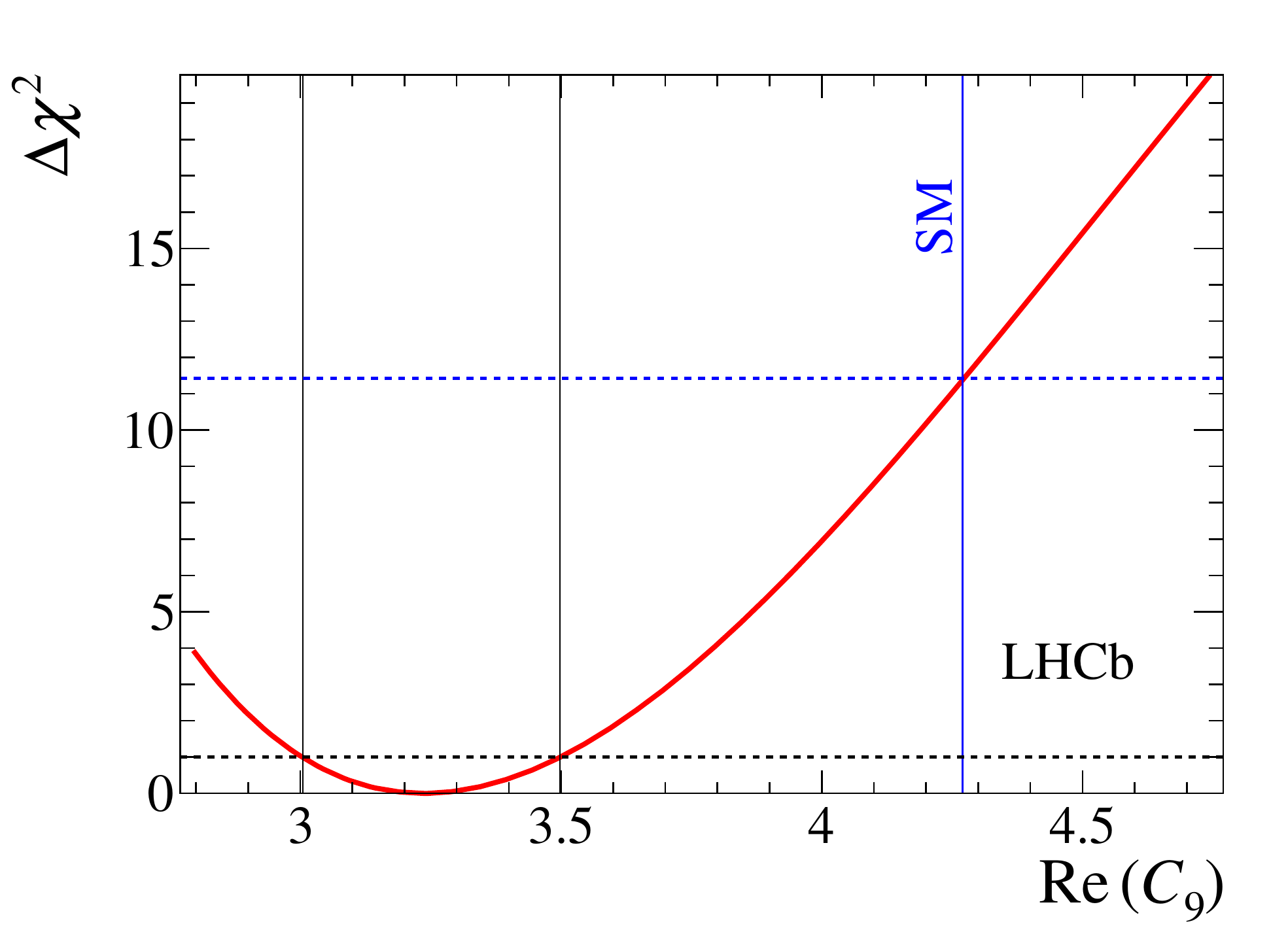}
  \caption{The $\Delta\chi^2$ distribution for the real part of the generalised vector-coupling strength, ${\cal C}_9$. 
 This is determined from a fit to the results of the maximum likelihood fit of the \CP-averaged observables.
    The SM central value is ${\rm Re}({\cal C}_9^{\rm SM}) = 4.27$~\cite{Beaujean:2013soa}. The best fit point is found to be at $\Delta{\rm Re}({\cal C}_9)= -1.04 \pm 0.25$.\label{fig:discussion:wcfit}}
\end{figure}

The minimum \chisq corresponds to a value of ${\rm Re}({\cal C}_9)$ shifted by $\Delta {\rm Re}({\cal C}_9) = -1.04 \pm 0.25$ from the SM central value of ${\rm Re}({\cal C}_9) = 4.27$~\cite{Beaujean:2013soa} (see Fig.~\ref{fig:discussion:wcfit}). 
From the difference in \chisq between the SM point and this best-fit point, the significance of this shift corresponds to 3.4 standard deviations.
As discussed in the literature~\cite{Descotes-Genon:2013wba,Altmannshofer:2013foa,Beaujean:2013soa,Hurth:2013ssa,Descotes-Genon:2014uoa,Lyon:2014hpa,Altmannshofer:2014cfa,Crivellin:2015mga,Gauld:2013qja,Altmannshofer:2014rta,Mahmoudi:2014mja, Datta:2013kja}, a shift in ${\cal C}_9$ could be caused by a contribution from a new vector particle or could result from an unexpectedly large hadronic effect.  

If a fit is instead performed to the \CP-averaged observables from the moment analysis in the same \qsq ranges, then $\Delta {\rm Re}({\cal C}_9) = -0.68\pm0.35$ is obtained. 
As expected, the uncertainty on $\Delta {\rm Re}({\cal C}_9)$  is larger than that from the likelihood fit. Taking into account the correlations between the two methods, the values of  $\Delta {\rm Re}({\cal C}_9)$ are statistically compatible.

\section{Conclusions}
\label{sec:conclusions}

This paper presents the first analysis of the full angular distribution of the \decay{\Bz}{\Kstarz\mumu} decay.
The analysis uses the complete LHCb Run 1 dataset and supersedes the results presented in Refs.~\cite{LHCb-PAPER-2013-019,LHCb-PAPER-2013-037}.
In addition to \CP-averaged observables, a complete set of \CP asymmetries of the angular distribution are measured for the first time. 
Correlations between the different observables are computed to allow the results to be included in global fits of $b \to s$ data. 

Three separate techniques are used to analyse the data.
An unbinned maximum likelihood fit to the full angular distribution is made in approximately $2\gev^{2}/c^{4}$ wide \qsq bins.
Observables are also determined by computing moments of the angular distribution in \qsq bins approximately $1\gev^{2}/c^{4}$ wide.
In addition, for the first time, a \qsq-dependent fit is performed to the angular distribution in order to determine the six complex decay amplitudes that describe the decay.  
The position in \qsq at which several observables cross zero is determined using these amplitudes. 

A global analysis of the \CP-averaged angular observables determined from the maximum likelihood fit indicates differences with the presently-available SM predictions at the level of 3.4 standard deviations.
These differences could be explained by an unexpectedly large hadronic effect that changes the SM predictions~\cite{Lyon:2014hpa,Altmannshofer:2014rta}.
The differences could also be explained by contributions to the decay from non-SM particles~\cite{Descotes-Genon:2013wba,Altmannshofer:2013foa,Beaujean:2013soa,Hurth:2013ssa,Descotes-Genon:2014uoa,Lyon:2014hpa,Altmannshofer:2014cfa,Crivellin:2015mga,Gauld:2013qja,Altmannshofer:2014rta,Mahmoudi:2014mja, Datta:2013kja}.

\section*{Acknowledgements}

\noindent We express our gratitude to our colleagues in the CERN
accelerator departments for the excellent performance of the LHC. We
thank the technical and administrative staff at the LHCb
institutes. We acknowledge support from CERN and from the national
agencies: CAPES, CNPq, FAPERJ and FINEP (Brazil); NSFC (China);
CNRS/IN2P3 (France); BMBF, DFG and MPG (Germany); INFN (Italy); 
FOM and NWO (The Netherlands); MNiSW and NCN (Poland); MEN/IFA (Romania); 
MinES and FANO (Russia); MinECo (Spain); SNSF and SER (Switzerland); 
NASU (Ukraine); STFC (United Kingdom); NSF (USA).
We acknowledge the computing resources that are provided by CERN, IN2P3 (France), KIT and DESY (Germany), INFN (Italy), SURF (The Netherlands), PIC (Spain), GridPP (United Kingdom), RRCKI (Russia), CSCS (Switzerland), IFIN-HH (Romania), CBPF (Brazil), PL-GRID (Poland) and OSC (USA). We are indebted to the communities behind the multiple open 
source software packages on which we depend. We are also thankful for the 
computing resources and the access to software R\&D tools provided by Yandex LLC (Russia).
Individual groups or members have received support from AvH Foundation (Germany),
EPLANET, Marie Sk\l{}odowska-Curie Actions and ERC (European Union), 
Conseil G\'{e}n\'{e}ral de Haute-Savoie, Labex ENIGMASS and OCEVU, 
R\'{e}gion Auvergne (France), RFBR (Russia), GVA, XuntaGal and GENCAT (Spain), The Royal Society 
and Royal Commission for the Exhibition of 1851 (United Kingdom).


\newpage 

\addcontentsline{toc}{section}{References}
\setboolean{inbibliography}{true}
\bibliographystyle{LHCb}
\bibliography{main,LHCb-PAPER,LHCb-CONF,LHCb-DP,LHCb-TDR}

\ifx\mcitethebibliography\mciteundefinedmacro
\PackageError{LHCb.bst}{mciteplus.sty has not been loaded}
{This bibstyle requires the use of the mciteplus package.}\fi
\providecommand{\href}[2]{#2}
\begin{mcitethebibliography}{10}
\mciteSetBstSublistMode{n}
\mciteSetBstMaxWidthForm{subitem}{\alph{mcitesubitemcount})}
\mciteSetBstSublistLabelBeginEnd{\mcitemaxwidthsubitemform\space}
{\relax}{\relax}

\bibitem{LHCb-PAPER-2013-019}
LHCb collaboration, R.~Aaij {\em et~al.},
  \ifthenelse{\boolean{articletitles}}{\emph{{Differential branching fraction
  and angular analysis of the decay $B^0 \to K^{*0} \mu^+\mu^-$}},
  }{}\href{http://dx.doi.org/10.1007/JHEP08(2013)131}{JHEP \textbf{08} (2013)
  131}, \href{http://arxiv.org/abs/1304.6325}{{\normalfont\ttfamily
  arXiv:1304.6325}}\relax
\mciteBstWouldAddEndPuncttrue
\mciteSetBstMidEndSepPunct{\mcitedefaultmidpunct}
{\mcitedefaultendpunct}{\mcitedefaultseppunct}\relax
\EndOfBibitem
\bibitem{Aubert:2006vb}
BaBar collaboration, B.~Aubert {\em et~al.},
  \ifthenelse{\boolean{articletitles}}{\emph{{Measurements of branching
  fractions, rate asymmetries, and angular distributions in the rare decays $B
  \to K \ell^{+} \ell^{-}$ and $B \to K^{*} \ell^{+} \ell^{-}$}},
  }{}\href{http://dx.doi.org/10.1103/PhysRevD.73.092001}{Phys.\ Rev.\
  \textbf{D73} (2006) 092001},
  \href{http://arxiv.org/abs/hep-ex/0604007}{{\normalfont\ttfamily
  arXiv:hep-ex/0604007}}\relax
\mciteBstWouldAddEndPuncttrue
\mciteSetBstMidEndSepPunct{\mcitedefaultmidpunct}
{\mcitedefaultendpunct}{\mcitedefaultseppunct}\relax
\EndOfBibitem
\bibitem{Lees:2015ymt}
BaBar collaboration, J.~P. Lees {\em et~al.},
  \ifthenelse{\boolean{articletitles}}{\emph{{Measurement of angular
  asymmetries in the decays $B \to K^{*}\ell^+\ell^-$}},
  }{}\href{http://arxiv.org/abs/1508.07960}{{\normalfont\ttfamily
  arXiv:1508.07960}}\relax
\mciteBstWouldAddEndPuncttrue
\mciteSetBstMidEndSepPunct{\mcitedefaultmidpunct}
{\mcitedefaultendpunct}{\mcitedefaultseppunct}\relax
\EndOfBibitem
\bibitem{:2009zv}
Belle collaboration, J.-T. Wei {\em et~al.},
  \ifthenelse{\boolean{articletitles}}{\emph{{Measurement of the differential
  branching fraction and forward-backward asymmetry for $B \to K^{(*)}
  \ell^+\ell^-$}},
  }{}\href{http://dx.doi.org/10.1103/PhysRevLett.103.171801}{Phys.\ Rev.\
  Lett.\  \textbf{103} (2009) 171801},
  \href{http://arxiv.org/abs/0904.0770}{{\normalfont\ttfamily
  arXiv:0904.0770}}\relax
\mciteBstWouldAddEndPuncttrue
\mciteSetBstMidEndSepPunct{\mcitedefaultmidpunct}
{\mcitedefaultendpunct}{\mcitedefaultseppunct}\relax
\EndOfBibitem
\bibitem{Aaltonen:2011ja}
CDF collaboration, T.~Aaltonen {\em et~al.},
  \ifthenelse{\boolean{articletitles}}{\emph{{Measurements of the angular
  distributions in the decays $B \to K^{(*)} \mu^+ \mu^-$ at CDF}},
  }{}\href{http://dx.doi.org/10.1103/PhysRevLett.108.081807}{Phys.\ Rev.\
  Lett.\  \textbf{108} (2012) 081807},
  \href{http://arxiv.org/abs/1108.0695}{{\normalfont\ttfamily
  arXiv:1108.0695}}\relax
\mciteBstWouldAddEndPuncttrue
\mciteSetBstMidEndSepPunct{\mcitedefaultmidpunct}
{\mcitedefaultendpunct}{\mcitedefaultseppunct}\relax
\EndOfBibitem
\bibitem{Chatrchyan:2013cda}
CMS collaboration, S.~Chatrchyan {\em et~al.},
  \ifthenelse{\boolean{articletitles}}{\emph{{Angular analysis and branching
  fraction measurement of the decay $B^0 \to K^{*0} \mu^+\mu^-$}},
  }{}\href{http://dx.doi.org/10.1016/j.physletb.2013.10.017}{Phys.\ Lett.\
  \textbf{B727} (2013) 77},
  \href{http://arxiv.org/abs/1308.3409}{{\normalfont\ttfamily
  arXiv:1308.3409}}\relax
\mciteBstWouldAddEndPuncttrue
\mciteSetBstMidEndSepPunct{\mcitedefaultmidpunct}
{\mcitedefaultendpunct}{\mcitedefaultseppunct}\relax
\EndOfBibitem
\bibitem{Khachatryan:2015isa}
CMS collaboration, V.~Khachatryan {\em et~al.},
  \ifthenelse{\boolean{articletitles}}{\emph{{Angular analysis of the decay
  $B^0 \to K^{*0} \mu^{+} \mu^{-} $ from pp collisions at $\sqrt{s}= $ 8 TeV}},
  }{}\href{http://arxiv.org/abs/1507.08126}{{\normalfont\ttfamily
  arXiv:1507.08126}}\relax
\mciteBstWouldAddEndPuncttrue
\mciteSetBstMidEndSepPunct{\mcitedefaultmidpunct}
{\mcitedefaultendpunct}{\mcitedefaultseppunct}\relax
\EndOfBibitem
\bibitem{LHCb-PAPER-2013-037}
LHCb collaboration, R.~Aaij {\em et~al.},
  \ifthenelse{\boolean{articletitles}}{\emph{{Measurement of
  form-factor-independent observables in the decay $B^0\to K^{*0}\mu^+\mu^-$}},
  }{}\href{http://dx.doi.org/10.1103/PhysRevLett.111.191801}{Phys.\ Rev.\
  Lett.\  \textbf{111} (2013) 191801},
  \href{http://arxiv.org/abs/1308.1707}{{\normalfont\ttfamily
  arXiv:1308.1707}}\relax
\mciteBstWouldAddEndPuncttrue
\mciteSetBstMidEndSepPunct{\mcitedefaultmidpunct}
{\mcitedefaultendpunct}{\mcitedefaultseppunct}\relax
\EndOfBibitem
\bibitem{Descotes-Genon:2013wba}
S.~Descotes-Genon, J.~Matias, and J.~Virto,
  \ifthenelse{\boolean{articletitles}}{\emph{{Understanding the $B \to
  K^*\mu^+\mu^-$ anomaly}},
  }{}\href{http://dx.doi.org/10.1103/PhysRevD.88.074002}{Phys.\ Rev.\
  \textbf{D88} (2013) 074002},
  \href{http://arxiv.org/abs/1307.5683}{{\normalfont\ttfamily
  arXiv:1307.5683}}\relax
\mciteBstWouldAddEndPuncttrue
\mciteSetBstMidEndSepPunct{\mcitedefaultmidpunct}
{\mcitedefaultendpunct}{\mcitedefaultseppunct}\relax
\EndOfBibitem
\bibitem{Altmannshofer:2013foa}
W.~Altmannshofer and D.~M. Straub,
  \ifthenelse{\boolean{articletitles}}{\emph{{New physics in $B \to
  K^*\mu\mu$?}},
  }{}\href{http://dx.doi.org/10.1140/epjc/s10052-013-2646-9}{Eur.\ Phys.\ J.\
  \textbf{C73} (2013) 2646},
  \href{http://arxiv.org/abs/1308.1501}{{\normalfont\ttfamily
  arXiv:1308.1501}}\relax
\mciteBstWouldAddEndPuncttrue
\mciteSetBstMidEndSepPunct{\mcitedefaultmidpunct}
{\mcitedefaultendpunct}{\mcitedefaultseppunct}\relax
\EndOfBibitem
\bibitem{Beaujean:2013soa}
F.~Beaujean, C.~Bobeth, and D.~van Dyk,
  \ifthenelse{\boolean{articletitles}}{\emph{{Comprehensive Bayesian analysis
  of rare (semi)leptonic and radiative $B$ decays}},
  }{}\href{http://dx.doi.org/10.1140/epjc/s10052-014-2897-0}{Eur.\ Phys.\ J.\
  \textbf{C74} (2014) 2897},
  \href{http://arxiv.org/abs/1310.2478}{{\normalfont\ttfamily
  arXiv:1310.2478}}\relax
\mciteBstWouldAddEndPuncttrue
\mciteSetBstMidEndSepPunct{\mcitedefaultmidpunct}
{\mcitedefaultendpunct}{\mcitedefaultseppunct}\relax
\EndOfBibitem
\bibitem{Hurth:2013ssa}
T.~Hurth and F.~Mahmoudi, \ifthenelse{\boolean{articletitles}}{\emph{{On the
  LHCb anomaly in B $\to K^*\ell^+\ell^-$}},
  }{}\href{http://dx.doi.org/10.1007/JHEP04(2014)097}{JHEP \textbf{04} (2014)
  097}, \href{http://arxiv.org/abs/1312.5267}{{\normalfont\ttfamily
  arXiv:1312.5267}}\relax
\mciteBstWouldAddEndPuncttrue
\mciteSetBstMidEndSepPunct{\mcitedefaultmidpunct}
{\mcitedefaultendpunct}{\mcitedefaultseppunct}\relax
\EndOfBibitem
\bibitem{Jager:2012uw}
S.~J\"{a}ger and J.~Martin~Camalich,
  \ifthenelse{\boolean{articletitles}}{\emph{{On $B\to Vll$ at small dilepton
  invariant mass, power corrections, and new physics}},
  }{}\href{http://dx.doi.org/10.1007/JHEP05(2013)043}{JHEP \textbf{05} (2013)
  043}, \href{http://arxiv.org/abs/1212.2263}{{\normalfont\ttfamily
  arXiv:1212.2263}}\relax
\mciteBstWouldAddEndPuncttrue
\mciteSetBstMidEndSepPunct{\mcitedefaultmidpunct}
{\mcitedefaultendpunct}{\mcitedefaultseppunct}\relax
\EndOfBibitem
\bibitem{Descotes-Genon:2014uoa}
S.~Descotes-Genon, L.~Hofer, J.~Matias, and J.~Virto,
  \ifthenelse{\boolean{articletitles}}{\emph{{On the impact of power
  corrections in the prediction of $B \to K^*\mu^+\mu^-$ observables}},
  }{}\href{http://dx.doi.org/10.1007/JHEP12(2014)125}{JHEP \textbf{12} (2014)
  125}, \href{http://arxiv.org/abs/1407.8526}{{\normalfont\ttfamily
  arXiv:1407.8526}}\relax
\mciteBstWouldAddEndPuncttrue
\mciteSetBstMidEndSepPunct{\mcitedefaultmidpunct}
{\mcitedefaultendpunct}{\mcitedefaultseppunct}\relax
\EndOfBibitem
\bibitem{Lyon:2014hpa}
J.~Lyon and R.~Zwicky, \ifthenelse{\boolean{articletitles}}{\emph{{Resonances
  gone topsy turvy - the charm of QCD or new physics in $b \to s \ell^+
  \ell^-$?}}, }{}\href{http://arxiv.org/abs/1406.0566}{{\normalfont\ttfamily
  arXiv:1406.0566}}\relax
\mciteBstWouldAddEndPuncttrue
\mciteSetBstMidEndSepPunct{\mcitedefaultmidpunct}
{\mcitedefaultendpunct}{\mcitedefaultseppunct}\relax
\EndOfBibitem
\bibitem{Altmannshofer:2014cfa}
W.~Altmannshofer, S.~Gori, M.~Pospelov, and I.~Yavin,
  \ifthenelse{\boolean{articletitles}}{\emph{{Quark flavor transitions in
  $L_\mu-L_\tau$ models}},
  }{}\href{http://dx.doi.org/10.1103/PhysRevD.89.095033}{Phys.\ Rev.\
  \textbf{D89} (2014) 095033},
  \href{http://arxiv.org/abs/1403.1269}{{\normalfont\ttfamily
  arXiv:1403.1269}}\relax
\mciteBstWouldAddEndPuncttrue
\mciteSetBstMidEndSepPunct{\mcitedefaultmidpunct}
{\mcitedefaultendpunct}{\mcitedefaultseppunct}\relax
\EndOfBibitem
\bibitem{Crivellin:2015mga}
A.~Crivellin, G.~D'Ambrosio, and J.~Heeck,
  \ifthenelse{\boolean{articletitles}}{\emph{{Explaining $h\to\mu^\pm\tau^\mp$,
  $B\to K^* \mu^+\mu^-$ and $B\to K \mu^+\mu^-/B\to K e^+e^-$ in a
  two-Higgs-doublet model with gauged $L_\mu-L_\tau$}},
  }{}\href{http://dx.doi.org/10.1103/PhysRevLett.114.151801}{Phys.\ Rev.\
  Lett.\  \textbf{114} (2015) 151801},
  \href{http://arxiv.org/abs/1501.00993}{{\normalfont\ttfamily
  arXiv:1501.00993}}\relax
\mciteBstWouldAddEndPuncttrue
\mciteSetBstMidEndSepPunct{\mcitedefaultmidpunct}
{\mcitedefaultendpunct}{\mcitedefaultseppunct}\relax
\EndOfBibitem
\bibitem{Gauld:2013qja}
R.~Gauld, F.~Goertz, and U.~Haisch,
  \ifthenelse{\boolean{articletitles}}{\emph{{An explicit $Z^{\prime}$-boson
  explanation of the $B \to K^* \mu^+ \mu^-$ anomaly}},
  }{}\href{http://dx.doi.org/10.1007/JHEP01(2014)069}{JHEP \textbf{01} (2014)
  069}, \href{http://arxiv.org/abs/1310.1082}{{\normalfont\ttfamily
  arXiv:1310.1082}}\relax
\mciteBstWouldAddEndPuncttrue
\mciteSetBstMidEndSepPunct{\mcitedefaultmidpunct}
{\mcitedefaultendpunct}{\mcitedefaultseppunct}\relax
\EndOfBibitem
\bibitem{Altmannshofer:2014rta}
W.~Altmannshofer and D.~M. Straub,
  \ifthenelse{\boolean{articletitles}}{\emph{{New physics in $b\rightarrow s$
  transitions after LHC Run 1}},
  }{}\href{http://dx.doi.org/10.1140/epjc/s10052-015-3602-7}{Eur.\ Phys.\ J.\
  \textbf{C75} (2015) 382},
  \href{http://arxiv.org/abs/1411.3161}{{\normalfont\ttfamily
  arXiv:1411.3161}}\relax
\mciteBstWouldAddEndPuncttrue
\mciteSetBstMidEndSepPunct{\mcitedefaultmidpunct}
{\mcitedefaultendpunct}{\mcitedefaultseppunct}\relax
\EndOfBibitem
\bibitem{Mahmoudi:2014mja}
F.~Mahmoudi, S.~Neshatpour, and J.~Virto,
  \ifthenelse{\boolean{articletitles}}{\emph{{$B \to K^{*} \mu^{+} \mu^{-}$
  optimised observables in the MSSM}},
  }{}\href{http://dx.doi.org/10.1140/epjc/s10052-014-2927-y}{Eur.\ Phys.\ J.\
  \textbf{C74} (2014) 2927},
  \href{http://arxiv.org/abs/1401.2145}{{\normalfont\ttfamily
  arXiv:1401.2145}}\relax
\mciteBstWouldAddEndPuncttrue
\mciteSetBstMidEndSepPunct{\mcitedefaultmidpunct}
{\mcitedefaultendpunct}{\mcitedefaultseppunct}\relax
\EndOfBibitem
\bibitem{Datta:2013kja}
A.~Datta, M.~Duraisamy, and D.~Ghosh,
  \ifthenelse{\boolean{articletitles}}{\emph{{Explaining the $B \to K^\ast
  \mu^+ \mu^-$ data with scalar interactions}},
  }{}\href{http://dx.doi.org/10.1103/PhysRevD.89.071501}{Phys.\ Rev.\
  \textbf{D89} (2014) 071501},
  \href{http://arxiv.org/abs/1310.1937}{{\normalfont\ttfamily
  arXiv:1310.1937}}\relax
\mciteBstWouldAddEndPuncttrue
\mciteSetBstMidEndSepPunct{\mcitedefaultmidpunct}
{\mcitedefaultendpunct}{\mcitedefaultseppunct}\relax
\EndOfBibitem
\bibitem{Altmannshofer:2008dz}
W.~Altmannshofer {\em et~al.},
  \ifthenelse{\boolean{articletitles}}{\emph{{Symmetries and asymmetries of $B
  \to K^{*} \mu^{+} \mu^{-}$ decays in the Standard Model and beyond}},
  }{}\href{http://dx.doi.org/10.1088/1126-6708/2009/01/019}{JHEP \textbf{01}
  (2009) 019}, \href{http://arxiv.org/abs/0811.1214}{{\normalfont\ttfamily
  arXiv:0811.1214}}\relax
\mciteBstWouldAddEndPuncttrue
\mciteSetBstMidEndSepPunct{\mcitedefaultmidpunct}
{\mcitedefaultendpunct}{\mcitedefaultseppunct}\relax
\EndOfBibitem
\bibitem{Kruger:2005ep}
F.~Kr{\"u}ger and J.~Matias,
  \ifthenelse{\boolean{articletitles}}{\emph{{Probing new physics via the
  transverse amplitudes of $\Bzb \to \Kstarb ( \to K^- \pi^+) \ell^+ \ell^-$ at
  large recoil}}, }{}\href{http://dx.doi.org/10.1103/PhysRevD.71.094009}{Phys.\
  Rev.\  \textbf{D71} (2005) 094009},
  \href{http://arxiv.org/abs/hep-ph/0502060}{{\normalfont\ttfamily
  arXiv:hep-ph/0502060}}\relax
\mciteBstWouldAddEndPuncttrue
\mciteSetBstMidEndSepPunct{\mcitedefaultmidpunct}
{\mcitedefaultendpunct}{\mcitedefaultseppunct}\relax
\EndOfBibitem
\bibitem{DescotesGenon:2012zf}
S.~Descotes-Genon, J.~Matias, M.~Ramon, and J.~Virto,
  \ifthenelse{\boolean{articletitles}}{\emph{{Implications from clean
  observables for the binned analysis of $B \to K^{*}\mu^+\mu^-$ at large
  recoil}}, }{}\href{http://dx.doi.org/10.1007/JHEP01(2013)048}{JHEP
  \textbf{01} (2013) 048},
  \href{http://arxiv.org/abs/1207.2753}{{\normalfont\ttfamily
  arXiv:1207.2753}}\relax
\mciteBstWouldAddEndPuncttrue
\mciteSetBstMidEndSepPunct{\mcitedefaultmidpunct}
{\mcitedefaultendpunct}{\mcitedefaultseppunct}\relax
\EndOfBibitem
\bibitem{Beaujean:2015xea}
F.~Beaujean, M.~Chrzaszcz, N.~Serra, and D.~van Dyk,
  \ifthenelse{\boolean{articletitles}}{\emph{{Extracting angular observables
  without a likelihood and applications to rare decays}},
  }{}\href{http://dx.doi.org/10.1103/PhysRevD.91.114012}{Phys.\ Rev.\
  \textbf{D91} (2015) 114012},
  \href{http://arxiv.org/abs/1503.04100}{{\normalfont\ttfamily
  arXiv:1503.04100}}\relax
\mciteBstWouldAddEndPuncttrue
\mciteSetBstMidEndSepPunct{\mcitedefaultmidpunct}
{\mcitedefaultendpunct}{\mcitedefaultseppunct}\relax
\EndOfBibitem
\bibitem{Ali:1999mm}
A.~Ali, P.~Ball, L.~T. Handoko, and G.~Hiller,
  \ifthenelse{\boolean{articletitles}}{\emph{{A Comparative study of the decays
  $B \to (K, K^{*}) \ell^+ \ell^-$ in standard model and supersymmetric
  theories}}, }{}\href{http://dx.doi.org/10.1103/PhysRevD.61.074024}{Phys.\
  Rev.\  \textbf{D61} (2000) 074024},
  \href{http://arxiv.org/abs/hep-ph/9910221}{{\normalfont\ttfamily
  arXiv:hep-ph/9910221}}\relax
\mciteBstWouldAddEndPuncttrue
\mciteSetBstMidEndSepPunct{\mcitedefaultmidpunct}
{\mcitedefaultendpunct}{\mcitedefaultseppunct}\relax
\EndOfBibitem
\bibitem{Kumar:2014bna}
G.~Kumar and N.~Mahajan,
  \ifthenelse{\boolean{articletitles}}{\emph{{$B\rightarrow K^{*}l^+ l^-$:
  Zeroes of angular observables as test of standard model}},
  }{}\href{http://arxiv.org/abs/1412.2955}{{\normalfont\ttfamily
  arXiv:1412.2955}}\relax
\mciteBstWouldAddEndPuncttrue
\mciteSetBstMidEndSepPunct{\mcitedefaultmidpunct}
{\mcitedefaultendpunct}{\mcitedefaultseppunct}\relax
\EndOfBibitem
\bibitem{Alves:2008zz}
LHCb collaboration, A.~A. Alves~Jr.\ {\em et~al.},
  \ifthenelse{\boolean{articletitles}}{\emph{{The \lhcb detector at the LHC}},
  }{}\href{http://dx.doi.org/10.1088/1748-0221/3/08/S08005}{JINST \textbf{3}
  (2008) S08005}\relax
\mciteBstWouldAddEndPuncttrue
\mciteSetBstMidEndSepPunct{\mcitedefaultmidpunct}
{\mcitedefaultendpunct}{\mcitedefaultseppunct}\relax
\EndOfBibitem
\bibitem{LHCb-DP-2014-002}
LHCb collaboration, R.~Aaij {\em et~al.},
  \ifthenelse{\boolean{articletitles}}{\emph{{LHCb detector performance}},
  }{}\href{http://dx.doi.org/10.1142/S0217751X15300227}{Int.\ J.\ Mod.\ Phys.\
  \textbf{A30} (2015) 1530022},
  \href{http://arxiv.org/abs/1412.6352}{{\normalfont\ttfamily
  arXiv:1412.6352}}\relax
\mciteBstWouldAddEndPuncttrue
\mciteSetBstMidEndSepPunct{\mcitedefaultmidpunct}
{\mcitedefaultendpunct}{\mcitedefaultseppunct}\relax
\EndOfBibitem
\bibitem{LHCb-DP-2012-004}
R.~Aaij {\em et~al.}, \ifthenelse{\boolean{articletitles}}{\emph{{The \lhcb
  trigger and its performance in 2011}},
  }{}\href{http://dx.doi.org/10.1088/1748-0221/8/04/P04022}{JINST \textbf{8}
  (2013) P04022}, \href{http://arxiv.org/abs/1211.3055}{{\normalfont\ttfamily
  arXiv:1211.3055}}\relax
\mciteBstWouldAddEndPuncttrue
\mciteSetBstMidEndSepPunct{\mcitedefaultmidpunct}
{\mcitedefaultendpunct}{\mcitedefaultseppunct}\relax
\EndOfBibitem
\bibitem{Sjostrand:2006za}
T.~Sj\"{o}strand, S.~Mrenna, and P.~Skands,
  \ifthenelse{\boolean{articletitles}}{\emph{{PYTHIA 6.4 physics and manual}},
  }{}\href{http://dx.doi.org/10.1088/1126-6708/2006/05/026}{JHEP \textbf{05}
  (2006) 026}, \href{http://arxiv.org/abs/hep-ph/0603175}{{\normalfont\ttfamily
  arXiv:hep-ph/0603175}}\relax
\mciteBstWouldAddEndPuncttrue
\mciteSetBstMidEndSepPunct{\mcitedefaultmidpunct}
{\mcitedefaultendpunct}{\mcitedefaultseppunct}\relax
\EndOfBibitem
\bibitem{Sjostrand:2007gs}
T.~Sj\"{o}strand, S.~Mrenna, and P.~Skands,
  \ifthenelse{\boolean{articletitles}}{\emph{{A brief introduction to PYTHIA
  8.1}}, }{}\href{http://dx.doi.org/10.1016/j.cpc.2008.01.036}{Comput.\ Phys.\
  Commun.\  \textbf{178} (2008) 852},
  \href{http://arxiv.org/abs/0710.3820}{{\normalfont\ttfamily
  arXiv:0710.3820}}\relax
\mciteBstWouldAddEndPuncttrue
\mciteSetBstMidEndSepPunct{\mcitedefaultmidpunct}
{\mcitedefaultendpunct}{\mcitedefaultseppunct}\relax
\EndOfBibitem
\bibitem{LHCb-PROC-2010-056}
I.~Belyaev {\em et~al.}, \ifthenelse{\boolean{articletitles}}{\emph{{Handling
  of the generation of primary events in Gauss, the LHCb simulation
  framework}}, }{}\href{http://dx.doi.org/10.1088/1742-6596/331/3/032047}{{J.\
  Phys.\ Conf.\ Ser.\ } \textbf{331} (2011) 032047}\relax
\mciteBstWouldAddEndPuncttrue
\mciteSetBstMidEndSepPunct{\mcitedefaultmidpunct}
{\mcitedefaultendpunct}{\mcitedefaultseppunct}\relax
\EndOfBibitem
\bibitem{Lange:2001uf}
D.~J. Lange, \ifthenelse{\boolean{articletitles}}{\emph{{The EvtGen particle
  decay simulation package}},
  }{}\href{http://dx.doi.org/10.1016/S0168-9002(01)00089-4}{Nucl.\ Instrum.\
  Meth.\  \textbf{A462} (2001) 152}\relax
\mciteBstWouldAddEndPuncttrue
\mciteSetBstMidEndSepPunct{\mcitedefaultmidpunct}
{\mcitedefaultendpunct}{\mcitedefaultseppunct}\relax
\EndOfBibitem
\bibitem{Golonka:2005pn}
P.~Golonka and Z.~Was, \ifthenelse{\boolean{articletitles}}{\emph{{PHOTOS Monte
  Carlo: A precision tool for QED corrections in $Z$ and $W$ decays}},
  }{}\href{http://dx.doi.org/10.1140/epjc/s2005-02396-4}{Eur.\ Phys.\ J.\
  \textbf{C45} (2006) 97},
  \href{http://arxiv.org/abs/hep-ph/0506026}{{\normalfont\ttfamily
  arXiv:hep-ph/0506026}}\relax
\mciteBstWouldAddEndPuncttrue
\mciteSetBstMidEndSepPunct{\mcitedefaultmidpunct}
{\mcitedefaultendpunct}{\mcitedefaultseppunct}\relax
\EndOfBibitem
\bibitem{Allison:2006ve}
Geant4 collaboration, J.~Allison {\em et~al.},
  \ifthenelse{\boolean{articletitles}}{\emph{{Geant4 developments and
  applications}}, }{}\href{http://dx.doi.org/10.1109/TNS.2006.869826}{IEEE
  Trans.\ Nucl.\ Sci.\  \textbf{53} (2006) 270}\relax
\mciteBstWouldAddEndPuncttrue
\mciteSetBstMidEndSepPunct{\mcitedefaultmidpunct}
{\mcitedefaultendpunct}{\mcitedefaultseppunct}\relax
\EndOfBibitem
\bibitem{Agostinelli:2002hh}
Geant4 collaboration, S.~Agostinelli {\em et~al.},
  \ifthenelse{\boolean{articletitles}}{\emph{{Geant4: A simulation toolkit}},
  }{}\href{http://dx.doi.org/10.1016/S0168-9002(03)01368-8}{Nucl.\ Instrum.\
  Meth.\  \textbf{A506} (2003) 250}\relax
\mciteBstWouldAddEndPuncttrue
\mciteSetBstMidEndSepPunct{\mcitedefaultmidpunct}
{\mcitedefaultendpunct}{\mcitedefaultseppunct}\relax
\EndOfBibitem
\bibitem{LHCb-PROC-2011-006}
M.~Clemencic {\em et~al.}, \ifthenelse{\boolean{articletitles}}{\emph{{The
  \lhcb simulation application, Gauss: Design, evolution and experience}},
  }{}\href{http://dx.doi.org/10.1088/1742-6596/331/3/032023}{{J.\ Phys.\ Conf.\
  Ser.\ } \textbf{331} (2011) 032023}\relax
\mciteBstWouldAddEndPuncttrue
\mciteSetBstMidEndSepPunct{\mcitedefaultmidpunct}
{\mcitedefaultendpunct}{\mcitedefaultseppunct}\relax
\EndOfBibitem
\bibitem{Breiman}
L.~Breiman, J.~H. Friedman, R.~A. Olshen, and C.~J. Stone, {\em Classification
  and regression trees}, Wadsworth international group, Belmont, California,
  USA, 1984\relax
\mciteBstWouldAddEndPuncttrue
\mciteSetBstMidEndSepPunct{\mcitedefaultmidpunct}
{\mcitedefaultendpunct}{\mcitedefaultseppunct}\relax
\EndOfBibitem
\bibitem{AdaBoost}
R.~E. Schapire and Y.~Freund, \ifthenelse{\boolean{articletitles}}{\emph{A
  decision-theoretic generalization of on-line learning and an application to
  boosting}, }{}\href{http://dx.doi.org/10.1006/jcss.1997.1504}{Jour.\ Comp.\
  and Syst.\ Sc.\  \textbf{55} (1997) 119}\relax
\mciteBstWouldAddEndPuncttrue
\mciteSetBstMidEndSepPunct{\mcitedefaultmidpunct}
{\mcitedefaultendpunct}{\mcitedefaultseppunct}\relax
\EndOfBibitem
\bibitem{LHCb-PAPER-2011-004}
LHCb collaboration, R.~Aaij {\em et~al.},
  \ifthenelse{\boolean{articletitles}}{\emph{{Search for the rare decays $B^0_s
  \to \mu^+\mu^-$ and $B^0 \to \mu^+\mu^-$}},
  }{}\href{http://dx.doi.org/10.1016/j.physletb.2011.04.031}{Phys.\ Lett.\
  \textbf{B699} (2011) 330},
  \href{http://arxiv.org/abs/1103.2465}{{\normalfont\ttfamily
  arXiv:1103.2465}}\relax
\mciteBstWouldAddEndPuncttrue
\mciteSetBstMidEndSepPunct{\mcitedefaultmidpunct}
{\mcitedefaultendpunct}{\mcitedefaultseppunct}\relax
\EndOfBibitem
\bibitem{Blum:1999:BHB:307400.307439}
A.~Blum, A.~Kalai, and J.~Langford,
  \ifthenelse{\boolean{articletitles}}{\emph{Beating the hold-out: Bounds for
  k-fold and progressive cross-validation}, }{} in {\em Proceedings of the
  Twelfth Annual Conference on Computational Learning Theory}, COLT '99, (New
  York, NY, USA), pp.~203--208, ACM, 1999\relax
\mciteBstWouldAddEndPuncttrue
\mciteSetBstMidEndSepPunct{\mcitedefaultmidpunct}
{\mcitedefaultendpunct}{\mcitedefaultseppunct}\relax
\EndOfBibitem
\bibitem{Aubert:2007hz}
BaBar collaboration, B.~Aubert {\em et~al.},
  \ifthenelse{\boolean{articletitles}}{\emph{{Measurement of decay amplitudes
  of $B \to J/\psi K^{*}, \psi(2S) K^{*}$, and $\chi_{c1} K^{*}$ with an
  angular analysis}},
  }{}\href{http://dx.doi.org/10.1103/PhysRevD.76.031102}{Phys.\ Rev.\
  \textbf{D76} (2007) 031102},
  \href{http://arxiv.org/abs/0704.0522}{{\normalfont\ttfamily
  arXiv:0704.0522}}\relax
\mciteBstWouldAddEndPuncttrue
\mciteSetBstMidEndSepPunct{\mcitedefaultmidpunct}
{\mcitedefaultendpunct}{\mcitedefaultseppunct}\relax
\EndOfBibitem
\bibitem{Chilikin:2014bkk}
Belle collaboration, K.~Chilikin {\em et~al.},
  \ifthenelse{\boolean{articletitles}}{\emph{{Observation of a new charged
  charmonium like state in $\bar{B}^0 \to J/\psi K^- \pi^+$ decays}},
  }{}\href{http://dx.doi.org/10.1103/PhysRevD.90.112009}{Phys.\ Rev.\
  \textbf{D90} (2014) 112009},
  \href{http://arxiv.org/abs/1408.6457}{{\normalfont\ttfamily
  arXiv:1408.6457}}\relax
\mciteBstWouldAddEndPuncttrue
\mciteSetBstMidEndSepPunct{\mcitedefaultmidpunct}
{\mcitedefaultendpunct}{\mcitedefaultseppunct}\relax
\EndOfBibitem
\bibitem{Aaij:2013cma}
LHCb collaboration, R.~Aaij {\em et~al.},
  \ifthenelse{\boolean{articletitles}}{\emph{{Measurement of the polarization
  amplitudes in $B^0 \to J/\psi K^{*}(892)^0$ decays}},
  }{}\href{http://dx.doi.org/10.1103/PhysRevD.88.052002}{Phys.\ Rev.\
  \textbf{D88} (2013) 052002},
  \href{http://arxiv.org/abs/1307.2782}{{\normalfont\ttfamily
  arXiv:1307.2782}}\relax
\mciteBstWouldAddEndPuncttrue
\mciteSetBstMidEndSepPunct{\mcitedefaultmidpunct}
{\mcitedefaultendpunct}{\mcitedefaultseppunct}\relax
\EndOfBibitem
\bibitem{LHCb-PAPER-2012-014}
LHCb collaboration, R.~Aaij {\em et~al.},
  \ifthenelse{\boolean{articletitles}}{\emph{{Measurement of the $B^0_s \to
  J/\psi \overline{K}^{*0}$ branching fraction and angular amplitudes}},
  }{}\href{http://dx.doi.org/10.1103/PhysRevD.86.071102}{Phys.\ Rev.\
  \textbf{D86} (2012) 071102(R)},
  \href{http://arxiv.org/abs/1208.0738}{{\normalfont\ttfamily
  arXiv:1208.0738}}\relax
\mciteBstWouldAddEndPuncttrue
\mciteSetBstMidEndSepPunct{\mcitedefaultmidpunct}
{\mcitedefaultendpunct}{\mcitedefaultseppunct}\relax
\EndOfBibitem
\bibitem{Aston1988493}
D.~Aston {\em et~al.}, \ifthenelse{\boolean{articletitles}}{\emph{{A study of
  $\Km\pip$ scattering in the reaction $\Km p\to \Km\pip n$ at $11\gevc$}},
  }{}\href{http://dx.doi.org/http://dx.doi.org/10.1016/0550-3213(88)90028-4}{Nuclear
  Physics B \textbf{296} (1988) 493 }\relax
\mciteBstWouldAddEndPuncttrue
\mciteSetBstMidEndSepPunct{\mcitedefaultmidpunct}
{\mcitedefaultendpunct}{\mcitedefaultseppunct}\relax
\EndOfBibitem
\bibitem{1998PhRvD..57.3873F}
G.~J. {Feldman} and R.~D. {Cousins},
  \ifthenelse{\boolean{articletitles}}{\emph{{Unified approach to the classical
  statistical analysis of small signals}},
  }{}\href{http://dx.doi.org/10.1103/PhysRevD.57.3873}{Phys.\ Rev.\ D
  \textbf{57} (1998) 3873},
  \href{http://arxiv.org/abs/physics/9711021}{{\normalfont\ttfamily
  arXiv:physics/9711021}}\relax
\mciteBstWouldAddEndPuncttrue
\mciteSetBstMidEndSepPunct{\mcitedefaultmidpunct}
{\mcitedefaultendpunct}{\mcitedefaultseppunct}\relax
\EndOfBibitem
\bibitem{plugin}
B.~Sen, M.~Walker, and M.~Woodroofe,
  \ifthenelse{\boolean{articletitles}}{\emph{{On the unified method with
  nuisance parameters}}, }{}Statistica Sinica \textbf{19} (2009) 301\relax
\mciteBstWouldAddEndPuncttrue
\mciteSetBstMidEndSepPunct{\mcitedefaultmidpunct}
{\mcitedefaultendpunct}{\mcitedefaultseppunct}\relax
\EndOfBibitem
\bibitem{Gratrex:2015hna}
J.~Gratrex, M.~Hopfer, and R.~Zwicky,
  \ifthenelse{\boolean{articletitles}}{\emph{{Generalised helicity formalism,
  higher moments and the $B \to K_{J_K}(\to K \pi) \bar{\ell}_1 \ell_2$ angular
  distributions}},
  }{}\href{http://arxiv.org/abs/1506.03970}{{\normalfont\ttfamily
  arXiv:1506.03970}}\relax
\mciteBstWouldAddEndPuncttrue
\mciteSetBstMidEndSepPunct{\mcitedefaultmidpunct}
{\mcitedefaultendpunct}{\mcitedefaultseppunct}\relax
\EndOfBibitem
\bibitem{Efron:1979}
B.~Efron, \ifthenelse{\boolean{articletitles}}{\emph{Bootstrap methods: Another
  look at the jackknife},
  }{}\href{http://dx.doi.org/10.1214/aos/1176344552}{Ann.\ Statist.\
  \textbf{7} (1979) 1}\relax
\mciteBstWouldAddEndPuncttrue
\mciteSetBstMidEndSepPunct{\mcitedefaultmidpunct}
{\mcitedefaultendpunct}{\mcitedefaultseppunct}\relax
\EndOfBibitem
\bibitem{Beaujean:2015gba}
F.~Beaujean, C.~Bobeth, and S.~Jahn,
  \ifthenelse{\boolean{articletitles}}{\emph{{Constraints on tensor and scalar
  couplings from $B\to K\bar\mu\mu$ and $B_s\to \bar\mu\mu$}},
  }{}\href{http://arxiv.org/abs/1508.01526}{{\normalfont\ttfamily
  arXiv:1508.01526}}\relax
\mciteBstWouldAddEndPuncttrue
\mciteSetBstMidEndSepPunct{\mcitedefaultmidpunct}
{\mcitedefaultendpunct}{\mcitedefaultseppunct}\relax
\EndOfBibitem
\bibitem{Egede:2010zc}
U.~Egede {\em et~al.}, \ifthenelse{\boolean{articletitles}}{\emph{{New physics
  reach of the decay mode $\bar{B} \to \bar{K}^{*0}\ell^+\ell^-$}},
  }{}\href{http://dx.doi.org/10.1007/JHEP10(2010)056}{JHEP \textbf{10} (2010)
  056}, \href{http://arxiv.org/abs/1005.0571}{{\normalfont\ttfamily
  arXiv:1005.0571}}\relax
\mciteBstWouldAddEndPuncttrue
\mciteSetBstMidEndSepPunct{\mcitedefaultmidpunct}
{\mcitedefaultendpunct}{\mcitedefaultseppunct}\relax
\EndOfBibitem
\bibitem{Egede:2015kha}
U.~Egede, M.~Patel, and K.~A. Petridis,
  \ifthenelse{\boolean{articletitles}}{\emph{{Method for an unbinned
  measurement of the $q^2$-dependent decay amplitudes of $\Bzb \to \Kstarzb
  \mu^+\mu^-$ decays}},
  }{}\href{http://dx.doi.org/10.1007/JHEP06(2015)084}{JHEP \textbf{06} (2015)
  084}, \href{http://arxiv.org/abs/1504.00574}{{\normalfont\ttfamily
  arXiv:1504.00574}}\relax
\mciteBstWouldAddEndPuncttrue
\mciteSetBstMidEndSepPunct{\mcitedefaultmidpunct}
{\mcitedefaultendpunct}{\mcitedefaultseppunct}\relax
\EndOfBibitem
\bibitem{Wang:2012swave}
C.-D. L\"u and W.~Wang, \ifthenelse{\boolean{articletitles}}{\emph{{Analysis of
  $B\rightarrow K^{*}_{J}(\rightarrow K\pi)\mu^{+}\mu^{-}$ in the higher kaon
  resonance\ region}},
  }{}\href{http://dx.doi.org/10.1103/PhysRevD.85.034014}{Phys.\ Rev.\ D
  \textbf{85} (2012) 034014},
  \href{http://arxiv.org/abs/1111.1513}{{\normalfont\ttfamily
  arXiv:1111.1513}}\relax
\mciteBstWouldAddEndPuncttrue
\mciteSetBstMidEndSepPunct{\mcitedefaultmidpunct}
{\mcitedefaultendpunct}{\mcitedefaultseppunct}\relax
\EndOfBibitem
\bibitem{LHCb-PAPER-2014-042}
LHCb collaboration, R.~Aaij {\em et~al.},
  \ifthenelse{\boolean{articletitles}}{\emph{{Measurement of the
  $\overline{B}^0$--$B^0$ and $\overline{B}^0_s$--$B^0_s$ production
  asymmetries in $pp$ collisions at $\sqrt{s}=7$ TeV}},
  }{}\href{http://dx.doi.org/10.1016/j.physletb.2014.10.005}{Phys.\ Lett.\
  \textbf{B739} (2014) 218},
  \href{http://arxiv.org/abs/1408.0275}{{\normalfont\ttfamily
  arXiv:1408.0275}}\relax
\mciteBstWouldAddEndPuncttrue
\mciteSetBstMidEndSepPunct{\mcitedefaultmidpunct}
{\mcitedefaultendpunct}{\mcitedefaultseppunct}\relax
\EndOfBibitem
\bibitem{LHCb-PAPER-2014-053}
LHCb collaboration, R.~Aaij {\em et~al.},
  \ifthenelse{\boolean{articletitles}}{\emph{{Measurement of the semileptonic
  $C\!P$ asymmetry in $B^0$--$\overline{B}^0$ mixing}},
  }{}\href{http://dx.doi.org/10.1103/PhysRevLett.114.041601}{Phys.\ Rev.\
  Lett.\  \textbf{114} (2015) 041601},
  \href{http://arxiv.org/abs/1409.8586}{{\normalfont\ttfamily
  arXiv:1409.8586}}\relax
\mciteBstWouldAddEndPuncttrue
\mciteSetBstMidEndSepPunct{\mcitedefaultmidpunct}
{\mcitedefaultendpunct}{\mcitedefaultseppunct}\relax
\EndOfBibitem
\bibitem{LHCb-PAPER-2014-013}
LHCb collaboration, R.~Aaij {\em et~al.},
  \ifthenelse{\boolean{articletitles}}{\emph{{Measurement of $C\!P$ asymmetry
  in $D^0 \to K^- K^+$ and $D^0 \to \pi^- \pi^+$ decays}},
  }{}\href{http://dx.doi.org/10.1007/JHEP07(2014)041}{JHEP \textbf{07} (2014)
  041}, \href{http://arxiv.org/abs/1405.2797}{{\normalfont\ttfamily
  arXiv:1405.2797}}\relax
\mciteBstWouldAddEndPuncttrue
\mciteSetBstMidEndSepPunct{\mcitedefaultmidpunct}
{\mcitedefaultendpunct}{\mcitedefaultseppunct}\relax
\EndOfBibitem
\bibitem{Bobeth:2010wg}
C.~Bobeth, G.~Hiller, and D.~van Dyk,
  \ifthenelse{\boolean{articletitles}}{\emph{{The benefits of $\Bb \to \Kstarb
  l^+ l^-$ decays at low recoil}},
  }{}\href{http://dx.doi.org/10.1007/JHEP07(2010)098}{JHEP \textbf{07} (2010)
  098}, \href{http://arxiv.org/abs/1006.5013}{{\normalfont\ttfamily
  arXiv:1006.5013}}\relax
\mciteBstWouldAddEndPuncttrue
\mciteSetBstMidEndSepPunct{\mcitedefaultmidpunct}
{\mcitedefaultendpunct}{\mcitedefaultseppunct}\relax
\EndOfBibitem
\bibitem{Jager:2014rwa}
S.~J{\"a}ger and J.~Martin~Camalich,
  \ifthenelse{\boolean{articletitles}}{\emph{{Reassessing the discovery
  potential of the $B \to K^{*} \ell^+\ell^-$ decays in the large-recoil
  region: SM challenges and BSM opportunities}},
  }{}\href{http://arxiv.org/abs/1412.3183}{{\normalfont\ttfamily
  arXiv:1412.3183}}\relax
\mciteBstWouldAddEndPuncttrue
\mciteSetBstMidEndSepPunct{\mcitedefaultmidpunct}
{\mcitedefaultendpunct}{\mcitedefaultseppunct}\relax
\EndOfBibitem
\bibitem{Das:2012kz}
D.~Das and R.~Sinha, \ifthenelse{\boolean{articletitles}}{\emph{{New physics
  effects and hadronic form factor uncertainties in $B\to K^* \ell^+ \ell^-$}},
  }{}\href{http://dx.doi.org/10.1103/PhysRevD.86.056006}{Phys.\ Rev.\
  \textbf{D86} (2012) 056006},
  \href{http://arxiv.org/abs/1205.1438}{{\normalfont\ttfamily
  arXiv:1205.1438}}\relax
\mciteBstWouldAddEndPuncttrue
\mciteSetBstMidEndSepPunct{\mcitedefaultmidpunct}
{\mcitedefaultendpunct}{\mcitedefaultseppunct}\relax
\EndOfBibitem
\bibitem{Horgan:2013pva}
R.~R. Horgan, Z.~Liu, S.~Meinel, and M.~Wingate,
  \ifthenelse{\boolean{articletitles}}{\emph{{Calculation of $B^0 \to K^{*0}
  \mu^+ \mu^-$ and $B_s^0 \to \phi \mu^+ \mu^-$ observables using form factors
  from lattice QCD}},
  }{}\href{http://dx.doi.org/10.1103/PhysRevLett.112.212003}{Phys.\ Rev.\
  Lett.\  \textbf{112} (2014) 212003},
  \href{http://arxiv.org/abs/1310.3887}{{\normalfont\ttfamily
  arXiv:1310.3887}}\relax
\mciteBstWouldAddEndPuncttrue
\mciteSetBstMidEndSepPunct{\mcitedefaultmidpunct}
{\mcitedefaultendpunct}{\mcitedefaultseppunct}\relax
\EndOfBibitem
\bibitem{Hambrock:2013zya}
C.~Hambrock, G.~Hiller, S.~Schacht, and R.~Zwicky,
  \ifthenelse{\boolean{articletitles}}{\emph{{$B \to K^*$ form factors from
  flavor data to QCD and back}},
  }{}\href{http://dx.doi.org/10.1103/PhysRevD.89.074014}{Phys.\ Rev.\
  \textbf{D89} (2014) 074014},
  \href{http://arxiv.org/abs/1308.4379}{{\normalfont\ttfamily
  arXiv:1308.4379}}\relax
\mciteBstWouldAddEndPuncttrue
\mciteSetBstMidEndSepPunct{\mcitedefaultmidpunct}
{\mcitedefaultendpunct}{\mcitedefaultseppunct}\relax
\EndOfBibitem
\bibitem{Hurth:2014vma}
T.~Hurth, F.~Mahmoudi, and S.~Neshatpour,
  \ifthenelse{\boolean{articletitles}}{\emph{{Global fits to b $\to s\ell\ell$
  data and signs for lepton non-universality}},
  }{}\href{http://dx.doi.org/10.1007/JHEP12(2014)053}{JHEP \textbf{12} (2014)
  053}, \href{http://arxiv.org/abs/1410.4545}{{\normalfont\ttfamily
  arXiv:1410.4545}}\relax
\mciteBstWouldAddEndPuncttrue
\mciteSetBstMidEndSepPunct{\mcitedefaultmidpunct}
{\mcitedefaultendpunct}{\mcitedefaultseppunct}\relax
\EndOfBibitem
\bibitem{Ball:2004rg}
P.~Ball and R.~Zwicky, \ifthenelse{\boolean{articletitles}}{\emph{{$B_{(d,s)}
  \to \rho, \omega, K^{*}, \phi$ decay form-factors from light-cone sum rules
  revisited}}, }{}\href{http://dx.doi.org/10.1103/PhysRevD.71.014029}{Phys.\
  Rev.\  \textbf{D71} (2005) 014029},
  \href{http://arxiv.org/abs/hep-ph/0412079}{{\normalfont\ttfamily
  arXiv:hep-ph/0412079}}\relax
\mciteBstWouldAddEndPuncttrue
\mciteSetBstMidEndSepPunct{\mcitedefaultmidpunct}
{\mcitedefaultendpunct}{\mcitedefaultseppunct}\relax
\EndOfBibitem
\bibitem{Straub:2015ica}
A.~Bharucha, D.~M. Straub, and R.~Zwicky,
  \ifthenelse{\boolean{articletitles}}{\emph{{$B\to V\ell^+\ell^-$ in the
  Standard Model from light-cone sum rules}},
  }{}\href{http://arxiv.org/abs/1503.05534}{{\normalfont\ttfamily
  arXiv:1503.05534}}\relax
\mciteBstWouldAddEndPuncttrue
\mciteSetBstMidEndSepPunct{\mcitedefaultmidpunct}
{\mcitedefaultendpunct}{\mcitedefaultseppunct}\relax
\EndOfBibitem
\bibitem{Horgan:2013hoa}
R.~R. Horgan, Z.~Liu, S.~Meinel, and M.~Wingate,
  \ifthenelse{\boolean{articletitles}}{\emph{{Lattice QCD calculation of form
  factors describing the rare decays $B \to K^* \ell^+ \ell^-$ and $B_s \to
  \phi \ell^+ \ell^-$}},
  }{}\href{http://dx.doi.org/10.1103/PhysRevD.89.094501}{Phys.\ Rev.\
  \textbf{D89} (2014) 094501},
  \href{http://arxiv.org/abs/1310.3722}{{\normalfont\ttfamily
  arXiv:1310.3722}}\relax
\mciteBstWouldAddEndPuncttrue
\mciteSetBstMidEndSepPunct{\mcitedefaultmidpunct}
{\mcitedefaultendpunct}{\mcitedefaultseppunct}\relax
\EndOfBibitem
\bibitem{Horgan:2015vla}
R.~R. Horgan, Z.~Liu, S.~Meinel, and M.~Wingate,
  \ifthenelse{\boolean{articletitles}}{\emph{{Rare $B$ decays using lattice QCD
  form factors}}, }{}PoS LATTICE2014 (2015) 372,
  \href{http://arxiv.org/abs/1501.00367}{{\normalfont\ttfamily
  arXiv:1501.00367}}\relax
\mciteBstWouldAddEndPuncttrue
\mciteSetBstMidEndSepPunct{\mcitedefaultmidpunct}
{\mcitedefaultendpunct}{\mcitedefaultseppunct}\relax
\EndOfBibitem
\bibitem{Khodjamirian:2010vf}
A.~Khodjamirian, T.~Mannel, A.~A. Pivovarov, and Y.-M. Wang,
  \ifthenelse{\boolean{articletitles}}{\emph{{Charm-loop effect in $B \to
  K^{(*)} \ell^{+} \ell^{-}$ and $B\to K^*\gamma$}},
  }{}\href{http://dx.doi.org/10.1007/JHEP09(2010)089}{JHEP \textbf{09} (2010)
  089}, \href{http://arxiv.org/abs/1006.4945}{{\normalfont\ttfamily
  arXiv:1006.4945}}\relax
\mciteBstWouldAddEndPuncttrue
\mciteSetBstMidEndSepPunct{\mcitedefaultmidpunct}
{\mcitedefaultendpunct}{\mcitedefaultseppunct}\relax
\EndOfBibitem
\bibitem{Descotes-Genon:2013vna}
S.~Descotes-Genon, T.~Hurth, J.~Matias, and J.~Virto,
  \ifthenelse{\boolean{articletitles}}{\emph{{Optimizing the basis of $B \to
  K^{*} \ell^+ \ell^-$ observables in the full kinematic range}},
  }{}\href{http://dx.doi.org/10.1007/JHEP05(2013)137}{JHEP \textbf{05} (2013)
  137}, \href{http://arxiv.org/abs/1303.5794}{{\normalfont\ttfamily
  arXiv:1303.5794}}\relax
\mciteBstWouldAddEndPuncttrue
\mciteSetBstMidEndSepPunct{\mcitedefaultmidpunct}
{\mcitedefaultendpunct}{\mcitedefaultseppunct}\relax
\EndOfBibitem
\bibitem{Bobeth:2011nj}
C.~Bobeth, G.~Hiller, D.~van Dyk, and C.~Wacker,
  \ifthenelse{\boolean{articletitles}}{\emph{{The decay $B \to K l^+ l^-$ at
  low hadronic recoil and model-independent $\Delta B = 1$ constraints}},
  }{}\href{http://dx.doi.org/10.1007/JHEP01(2012)107}{JHEP \textbf{01} (2012)
  107}, \href{http://arxiv.org/abs/1111.2558}{{\normalfont\ttfamily
  arXiv:1111.2558}}\relax
\mciteBstWouldAddEndPuncttrue
\mciteSetBstMidEndSepPunct{\mcitedefaultmidpunct}
{\mcitedefaultendpunct}{\mcitedefaultseppunct}\relax
\EndOfBibitem
\bibitem{Beneke:2004dp}
M.~Beneke, T.~Feldmann, and D.~Seidel,
  \ifthenelse{\boolean{articletitles}}{\emph{{Exclusive radiative and
  electroweak $b \to d$ and $b \to s$ penguin decays at NLO}},
  }{}\href{http://dx.doi.org/10.1140/epjc/s2005-02181-5}{Eur.\ Phys.\ J.\
  \textbf{C41} (2005) 173},
  \href{http://arxiv.org/abs/hep-ph/0412400}{{\normalfont\ttfamily
  arXiv:hep-ph/0412400}}\relax
\mciteBstWouldAddEndPuncttrue
\mciteSetBstMidEndSepPunct{\mcitedefaultmidpunct}
{\mcitedefaultendpunct}{\mcitedefaultseppunct}\relax
\EndOfBibitem
\bibitem{Ali:2006ew}
A.~Ali, G.~Kramer, and G.-h. Zhu,
  \ifthenelse{\boolean{articletitles}}{\emph{{$B \to K^{*} \ell^+ \ell^-$ decay
  in soft-collinear effective theory}},
  }{}\href{http://dx.doi.org/10.1140/epjc/s2006-02596-4}{Eur.\ Phys.\ J.\
  \textbf{C47} (2006) 625},
  \href{http://arxiv.org/abs/hep-ph/0601034}{{\normalfont\ttfamily
  arXiv:hep-ph/0601034}}\relax
\mciteBstWouldAddEndPuncttrue
\mciteSetBstMidEndSepPunct{\mcitedefaultmidpunct}
{\mcitedefaultendpunct}{\mcitedefaultseppunct}\relax
\EndOfBibitem
\bibitem{LHCb-PAPER-2014-049}
CMS and LHCb collaborations, V.~Khachatryan {\em et~al.},
  \ifthenelse{\boolean{articletitles}}{\emph{{Observation of the rare $B_s^0
  \to \mu^+\mu^-$ decay from the combined analysis of CMS and LHCb data}},
  }{}\href{http://dx.doi.org/10.1038/nature14474}{Nature \textbf{522} (2015)
  68}, \href{http://arxiv.org/abs/1411.4413}{{\normalfont\ttfamily
  arXiv:1411.4413}}\relax
\mciteBstWouldAddEndPuncttrue
\mciteSetBstMidEndSepPunct{\mcitedefaultmidpunct}
{\mcitedefaultendpunct}{\mcitedefaultseppunct}\relax
\EndOfBibitem
\end{mcitethebibliography}

\newpage

\clearpage

{\noindent\bf\Large Appendices}

\appendix

\section{Tables of results} 
\label{sec:appendix:results}

The results of the likelihood fits described in Sec.~\ref{sec:likelihood} are given in Tables~\ref{tab:results:likelihood:widebins}--\ref{tab:results:likelihood:optimised} below.  The results of the method of moments described in Sec.~\ref{sec:moments} are given in Tables~\ref{tab:results:moments:averages}--\ref{tab:results:moments:optimised} below.

\begin{table}[htb]
\caption{\CP-averaged angular observables evaluated by the unbinned maximum likelihood fit, in the range $1.1 < \qsq < 6.0 \gev^{2}/c^{4}$ and $15.0 < \qsq < 19.0 \gev^{2}/c^{4}$. The first uncertainties are statistical and the second systematic. \label{tab:results:likelihood:widebins}}
\begin{center}
\setlength\extrarowheight{2pt}
\begin{tabular}{c|cc}
& $1.1<q^2<6.0\gevgevcccc$ & $15.0<q^2<19.0\gevgevcccc$ \\ \hline
$F_{\rm L}$ &  $\phantom{-}0.690\,{}^{+0.035}_{-0.036} \pm 0.017$ &  $\phantom{-}0.344\,{}^{+0.028}_{-0.030} \pm 0.008$ \\
$S_{3}$ &  $\phantom{-}0.012\,{}^{+0.038}_{-0.038} \pm 0.004$ &  $-0.163\,{}^{+0.033}_{-0.033} \pm 0.009$ \\
$S_{4}$ &  $-0.155\,{}^{+0.057}_{-0.056} \pm 0.004$ &  $-0.284\,{}^{+0.038}_{-0.041} \pm 0.007$ \\
$S_{5}$ &  $-0.023\,{}^{+0.050}_{-0.049} \pm 0.005$ &  $-0.325\,{}^{+0.036}_{-0.037} \pm 0.009$ \\
$A_{\rm FB}$ &  $-0.075\,{}^{+0.032}_{-0.034} \pm 0.007$ &  $\phantom{-}0.355\,{}^{+0.027}_{-0.027} \pm 0.009$ \\
$S_{7}$ &  $-0.077\,{}^{+0.050}_{-0.049} \pm 0.006$ &  $\phantom{-}0.048\,{}^{+0.043}_{-0.043} \pm 0.006$ \\
$S_{8}$ &  $\phantom{-}0.028\,{}^{+0.058}_{-0.057} \pm 0.008$ &  $\phantom{-}0.028\,{}^{+0.044}_{-0.045} \pm 0.003$ \\
$S_{9}$ &  $-0.064\,{}^{+0.042}_{-0.041} \pm 0.004$ &  $-0.053\,{}^{+0.039}_{-0.039} \pm 0.002$ \\
\hline
$P_{1}$ &  $\phantom{-}0.080\,{}^{+0.248}_{-0.245} \pm 0.044$ &  $-0.497\,{}^{+0.102}_{-0.099} \pm 0.027$ \\
$P_{2}$ &  $-0.162\,{}^{+0.072}_{-0.073} \pm 0.010$ &  $\phantom{-}0.361\,{}^{+0.025}_{-0.026} \pm 0.010$ \\
$P_{3}$ &  $\phantom{-}0.205\,{}^{+0.135}_{-0.134} \pm 0.017$ &  $\phantom{-}0.081\,{}^{+0.060}_{-0.059} \pm 0.005$ \\
$P'_{4}$ &  $-0.336\,{}^{+0.124}_{-0.122} \pm 0.012$ &  $-0.597\,{}^{+0.080}_{-0.085} \pm 0.015$ \\
$P'_{5}$ &  $-0.049\,{}^{+0.107}_{-0.108} \pm 0.014$ &  $-0.684\,{}^{+0.078}_{-0.081} \pm 0.020$ \\
$P'_{6}$ &  $-0.166\,{}^{+0.108}_{-0.108} \pm 0.021$ &  $\phantom{-}0.101\,{}^{+0.090}_{-0.092} \pm 0.011$ \\
$P'_{8}$ &  $\phantom{-}0.060\,{}^{+0.122}_{-0.124} \pm 0.009$ &  $\phantom{-}0.059\,{}^{+0.094}_{-0.093} \pm 0.008$ \\
\hline
$A_{3}$ &  $-0.072\,{}^{+0.038}_{-0.038} \pm 0.004$ &  $-0.035\,{}^{+0.043}_{-0.042} \pm 0.010$ \\
$A_{4}$ &  $\phantom{-}0.012\,{}^{+0.057}_{-0.056} \pm 0.005$ &  $-0.079\,{}^{+0.047}_{-0.048} \pm 0.008$ \\
$A_{5}$ &  $-0.044\,{}^{+0.049}_{-0.047} \pm 0.005$ &  $-0.035\,{}^{+0.047}_{-0.047} \pm 0.010$ \\
$A_{6}$ &  $\phantom{-}0.020\,{}^{+0.061}_{-0.060} \pm 0.009$ &  $-0.110\,{}^{+0.052}_{-0.051} \pm 0.013$ \\
$A_{7}$ &  $-0.045\,{}^{+0.050}_{-0.050} \pm 0.006$ &  $-0.040\,{}^{+0.045}_{-0.044} \pm 0.006$ \\
$A_{8}$ &  $-0.047\,{}^{+0.058}_{-0.057} \pm 0.008$ &  $\phantom{-}0.025\,{}^{+0.048}_{-0.047} \pm 0.003$ \\
$A_{9}$ &  $-0.033\,{}^{+0.040}_{-0.042} \pm 0.004$ &  $\phantom{-}0.061\,{}^{+0.043}_{-0.044} \pm 0.002$ \\
\end{tabular}  
\end{center}
\end{table} 

\clearpage

\begin{table}[htb]
\caption{\CP-averaged angular observables evaluated by the unbinned maximum likelihood fit. The first uncertainties are statistical and the second systematic. \label{tab:results:likelihood:averages}}
\begin{center}
\setlength\extrarowheight{2pt}
\begin{tabular}{c|ccc}
 & $0.10<q^2<0.98\gevgevcccc$ & $1.1<q^2<2.5\gevgevcccc$ & $2.5<q^2<4.0\gevgevcccc$ \\ \hline 
$F_{\rm L}$ &  $\phantom{-}0.263\,{}^{+0.045}_{-0.044} \pm 0.017$ &  $\phantom{-}0.660\,{}^{+0.083}_{-0.077} \pm 0.022$ &  $\phantom{-}0.876\,{}^{+0.109}_{-0.097} \pm 0.017$ \\                                                                                                                                    
$S_{3}$ &  $-0.036\,{}^{+0.063}_{-0.063} \pm 0.005$ &  $-0.077\,{}^{+0.087}_{-0.105} \pm 0.005$ &  $\phantom{-}0.035\,{}^{+0.098}_{-0.089} \pm 0.007$ \\  
$S_{4}$ &  $\phantom{-}0.082\,{}^{+0.068}_{-0.069} \pm 0.009$ &  $-0.077\,{}^{+0.111}_{-0.113} \pm 0.005$ &  $-0.234\,{}^{+0.127}_{-0.144} \pm 0.006$ \\  
$S_{5}$ &  $\phantom{-}0.170\,{}^{+0.059}_{-0.058} \pm 0.018$ &  $\phantom{-}0.137\,{}^{+0.099}_{-0.094} \pm 0.009$ &  $-0.022\,{}^{+0.110}_{-0.103} \pm 0.008$ \\                                                                                                                                                  
$A_{\rm FB}$ &  $-0.003\,{}^{+0.058}_{-0.057} \pm 0.009$ &  $-0.191\,{}^{+0.068}_{-0.080} \pm 0.012$ &  $-0.118\,{}^{+0.082}_{-0.090} \pm 0.007$ \\       
$S_{7}$ &  $\phantom{-}0.015\,{}^{+0.059}_{-0.059} \pm 0.006$ &  $-0.219\,{}^{+0.094}_{-0.104} \pm 0.004$ &  $\phantom{-}0.068\,{}^{+0.120}_{-0.112} \pm 0.005$ \\                                                                                                                                                  
$S_{8}$ &  $\phantom{-}0.079\,{}^{+0.076}_{-0.075} \pm 0.007$ &  $-0.098\,{}^{+0.108}_{-0.123} \pm 0.005$ &  $\phantom{-}0.030\,{}^{+0.129}_{-0.131} \pm 0.006$ \\
$S_{9}$ &  $-0.083\,{}^{+0.058}_{-0.057} \pm 0.004$ &  $-0.119\,{}^{+0.087}_{-0.104} \pm 0.005$ &  $-0.092\,{}^{+0.105}_{-0.125} \pm 0.007$ \\
\multicolumn{1}{c}{} \\
 & $4.0<q^2<6.0\gevgevcccc$ & $6.0<q^2<8.0\gevgevcccc$ & $11.0<q^2<12.5\gevgevcccc$ \\ \hline
$F_{\rm L}$ &  $\phantom{-}0.611\,{}^{+0.052}_{-0.053} \pm 0.017$ &  $\phantom{-}0.579\,{}^{+0.046}_{-0.046} \pm 0.015$ &  $\phantom{-}0.493\,{}^{+0.049}_{-0.047} \pm 0.013$ \\
$S_{3}$ &  $\phantom{-}0.035\,{}^{+0.069}_{-0.068} \pm 0.007$ &  $-0.042\,{}^{+0.058}_{-0.059} \pm 0.011$ &  $-0.189\,{}^{+0.054}_{-0.058} \pm 0.005$ \\
$S_{4}$ &  $-0.219\,{}^{+0.086}_{-0.084} \pm 0.008$ &  $-0.296\,{}^{+0.063}_{-0.067} \pm 0.011$ &  $-0.283\,{}^{+0.084}_{-0.095} \pm 0.009$ \\
$S_{5}$ &  $-0.146\,{}^{+0.077}_{-0.078} \pm 0.011$ &  $-0.249\,{}^{+0.059}_{-0.060} \pm 0.012$ &  $-0.327\,{}^{+0.076}_{-0.079} \pm 0.009$ \\
$A_{\rm FB}$ &  $\phantom{-}0.025\,{}^{+0.051}_{-0.052} \pm 0.004$ &  $\phantom{-}0.152\,{}^{+0.041}_{-0.040} \pm 0.008$ &  $\phantom{-}0.318\,{}^{+0.044}_{-0.040} \pm 0.009$ \\
$S_{7}$ &  $-0.016\,{}^{+0.081}_{-0.080} \pm 0.004$ &  $-0.047\,{}^{+0.068}_{-0.066} \pm 0.003$ &  $-0.141\,{}^{+0.072}_{-0.074} \pm 0.005$ \\
$S_{8}$ &  $\phantom{-}0.167\,{}^{+0.094}_{-0.091} \pm 0.004$ &  $-0.085\,{}^{+0.072}_{-0.070} \pm 0.006$ &  $-0.007\,{}^{+0.070}_{-0.072} \pm 0.005$ \\
$S_{9}$ &  $-0.032\,{}^{+0.071}_{-0.071} \pm 0.004$ &  $-0.024\,{}^{+0.059}_{-0.060} \pm 0.005$ &  $-0.004\,{}^{+0.070}_{-0.073} \pm 0.006$ \\
\multicolumn{1}{c}{} \\
& $15.0<q^2<17.0\gevgevcccc$ & $17.0<q^2<19.0\gevgevcccc$ \\ \hline
$F_{\rm L}$ &  $\phantom{-}0.349\,{}^{+0.039}_{-0.039} \pm 0.009$ &  $\phantom{-}0.354\,{}^{+0.049}_{-0.048} \pm 0.025$ \\
$S_{3}$ &  $-0.142\,{}^{+0.044}_{-0.049} \pm 0.007$ &  $-0.188\,{}^{+0.074}_{-0.084} \pm 0.017$ \\
$S_{4}$ &  $-0.321\,{}^{+0.055}_{-0.074} \pm 0.007$ &  $-0.266\,{}^{+0.063}_{-0.072} \pm 0.010$ \\
$S_{5}$ &  $-0.316\,{}^{+0.051}_{-0.057} \pm 0.009$ &  $-0.323\,{}^{+0.063}_{-0.072} \pm 0.009$ \\
$A_{\rm FB}$ &  $\phantom{-}0.411\,{}^{+0.041}_{-0.037} \pm 0.008$ &  $\phantom{-}0.305\,{}^{+0.049}_{-0.048} \pm 0.013$ \\
$S_{7}$ &  $\phantom{-}0.061\,{}^{+0.058}_{-0.058} \pm 0.005$ &  $\phantom{-}0.044\,{}^{+0.073}_{-0.072} \pm 0.013$ \\
$S_{8}$ &  $\phantom{-}0.003\,{}^{+0.061}_{-0.061} \pm 0.003$ &  $\phantom{-}0.013\,{}^{+0.071}_{-0.070} \pm 0.005$ \\
$S_{9}$ &  $-0.019\,{}^{+0.054}_{-0.056} \pm 0.004$ &  $-0.094\,{}^{+0.065}_{-0.067} \pm 0.004$ \\
\end{tabular}  
\end{center}
\end{table}

\clearpage

\begin{table}[htb]
\caption{\CP-asymmetric angular observables evaluated by the unbinned maximum likelihood fit. The first uncertainties are statistical and the second systematic. \label{tab:results:likelihood:asymmetries}}
\begin{center}
\setlength\extrarowheight{2pt}
\begin{tabular}{c|ccc}
& $0.10<q^2<0.98\gevgevcccc$ & $1.1<q^2<2.5\gevgevcccc$ & $2.5<q^2<4.0\gevgevcccc$ \\ \hline
$A_{3}$ &  $\phantom{-}0.006\,{}^{+0.064}_{-0.065} \pm 0.005$ &  $\phantom{-}0.042\,{}^{+0.097}_{-0.087} \pm 0.005$ &  $-0.111\,{}^{+0.087}_{-0.109} \pm 0.006$ \\    
$A_{4}$ &  $-0.068\,{}^{+0.071}_{-0.073} \pm 0.009$ &  $\phantom{-}0.235\,{}^{+0.125}_{-0.109} \pm 0.005$ &  $-0.007\,{}^{+0.130}_{-0.135} \pm 0.007$ \\  
$A_{5}$ &  $\phantom{-}0.001\,{}^{+0.061}_{-0.059} \pm 0.018$ &  $-0.114\,{}^{+0.099}_{-0.105} \pm 0.009$ &  $-0.005\,{}^{+0.107}_{-0.106} \pm 0.008$ \\  
$A_{6s}$ &  $\phantom{-}0.122\,{}^{+0.076}_{-0.075} \pm 0.011$ &  $\phantom{-}0.037\,{}^{+0.102}_{-0.091} \pm 0.016$ &  $\phantom{-}0.022\,{}^{+0.115}_{-0.096} \pm 0.010$ \\
$A_{7}$ &  $\phantom{-}0.076\,{}^{+0.061}_{-0.060} \pm 0.006$ &  $-0.087\,{}^{+0.091}_{-0.093} \pm 0.004$ &  $-0.032\,{}^{+0.109}_{-0.115} \pm 0.005$ \\  
$A_{8}$ &  $-0.031\,{}^{+0.074}_{-0.074} \pm 0.007$ &  $-0.044\,{}^{+0.108}_{-0.113} \pm 0.005$ &  $-0.071\,{}^{+0.124}_{-0.131} \pm 0.006$ \\            
$A_{9}$ &  $\phantom{-}0.030\,{}^{+0.062}_{-0.061} \pm 0.004$ &  $-0.004\,{}^{+0.092}_{-0.098} \pm 0.005$ &  $-0.228\,{}^{+0.114}_{-0.152} \pm 0.007$ \\  
\multicolumn{1}{c}{} \\
& $4.0<q^2<6.0\gevgevcccc$ & $6.0<q^2<8.0\gevgevcccc$ & $11.0<q^2<12.5\gevgevcccc$ \\ \hline
$A_{3}$ &  $-0.173\,{}^{+0.070}_{-0.079} \pm 0.006$ &  $\phantom{-}0.064\,{}^{+0.067}_{-0.064} \pm 0.011$ &  $\phantom{-}0.132\,{}^{+0.075}_{-0.073} \pm 0.005$ \\  
$A_{4}$ &  $-0.168\,{}^{+0.086}_{-0.085} \pm 0.008$ &  $-0.037\,{}^{+0.073}_{-0.073} \pm 0.011$ &  $-0.100\,{}^{+0.082}_{-0.077} \pm 0.009$ \\    
$A_{5}$ &  $-0.059\,{}^{+0.071}_{-0.073} \pm 0.011$ &  $\phantom{-}0.129\,{}^{+0.067}_{-0.066} \pm 0.012$ &  $\phantom{-}0.027\,{}^{+0.077}_{-0.076} \pm 0.010$ \\  
$A_{6s}$ &  $-0.023\,{}^{+0.082}_{-0.075} \pm 0.005$ &  $\phantom{-}0.047\,{}^{+0.062}_{-0.060} \pm 0.011$ &  $\phantom{-}0.024\,{}^{+0.069}_{-0.067} \pm 0.013$ \\  
$A_{7}$ &  $\phantom{-}0.041\,{}^{+0.083}_{-0.082} \pm 0.004$ &  $\phantom{-}0.035\,{}^{+0.065}_{-0.067} \pm 0.003$ &  $-0.008\,{}^{+0.073}_{-0.073} \pm 0.005$ \\   
$A_{8}$ &  $\phantom{-}0.004\,{}^{+0.093}_{-0.095} \pm 0.005$ &  $-0.043\,{}^{+0.070}_{-0.069} \pm 0.006$ &  $\phantom{-}0.014\,{}^{+0.075}_{-0.073} \pm 0.005$ \\ 
$A_{9}$ &  $\phantom{-}0.062\,{}^{+0.078}_{-0.072} \pm 0.004$ &  $\phantom{-}0.110\,{}^{+0.061}_{-0.060} \pm 0.005$ &  $-0.057\,{}^{+0.057}_{-0.059} \pm 0.006$ \\   
\multicolumn{1}{c}{} \\
& $15.0<q^2<17.0\gevgevcccc$ & $17.0<q^2<19.0\gevgevcccc$ &  \\ \hline
$A_{3}$ &  $-0.034\,{}^{+0.056}_{-0.055} \pm 0.007$ &  $-0.056\,{}^{+0.075}_{-0.073} \pm 0.017$ & \\ 
$A_{4}$ &  $-0.071\,{}^{+0.064}_{-0.064} \pm 0.008$ &  $-0.071\,{}^{+0.073}_{-0.073} \pm 0.011$ &  \\ 
$A_{5}$ &  $-0.076\,{}^{+0.065}_{-0.063} \pm 0.010$ &  $\phantom{-}0.008\,{}^{+0.073}_{-0.075} \pm 0.010$ &  \\ 
$A_{6s}$ &  $-0.085\,{}^{+0.062}_{-0.060} \pm 0.012$ &  $-0.127\,{}^{+0.080}_{-0.076} \pm 0.018$ &  \\           
$A_{7}$ &  $-0.105\,{}^{+0.058}_{-0.059} \pm 0.005$ &  $\phantom{-}0.047\,{}^{+0.070}_{-0.069} \pm 0.013$ &  \\ 
$A_{8}$ &  $\phantom{-}0.048\,{}^{+0.063}_{-0.063} \pm 0.003$ &  $\phantom{-}0.022\,{}^{+0.072}_{-0.073} \pm 0.005$ &  \\ 
$A_{9}$ &  $\phantom{-}0.091\,{}^{+0.059}_{-0.059} \pm 0.004$ &  $\phantom{-}0.043\,{}^{+0.066}_{-0.067} \pm 0.005$ &  \\ 
\end{tabular}  
\end{center}
\end{table}

\clearpage

\begin{table}[htb]
\caption{Optimised angular observables evaluated by the unbinned maximum likelihood fit. The first uncertainties are statistical and the second systematic. \label{tab:results:likelihood:optimised}}
\begin{center}
\setlength\extrarowheight{2pt}
\begin{tabular}{c|ccc} 
& $0.10<q^2<0.98\gevgevcccc$ & $1.1<q^2<2.5\gevgevcccc$ & $2.5<q^2<4.0\gevgevcccc$ \\ \hline
$P_{1}$ &  $-0.099\,{}^{+0.168}_{-0.163} \pm 0.014$ &  $-0.451\,{}^{+0.519}_{-0.636} \pm 0.038$ &  $\phantom{-}0.571\,{}^{+2.404}_{-1.714} \pm 0.045$ \\
$P_{2}$ &  $-0.003\,{}^{+0.051}_{-0.052} \pm 0.007$ &  $-0.373\,{}^{+0.146}_{-0.199} \pm 0.027$ &  $-0.636\,{}^{+0.444}_{-1.735} \pm 0.015$ \\
$P_{3}$ &  $\phantom{-}0.113\,{}^{+0.079}_{-0.079} \pm 0.006$ &  $\phantom{-}0.350\,{}^{+0.330}_{-0.254} \pm 0.015$ &  $\phantom{-}0.745\,{}^{+2.587}_{-0.861} \pm 0.030$ \\
$P'_{4}$ &  $\phantom{-}0.185\,{}^{+0.158}_{-0.154} \pm 0.023$ &  $-0.163\,{}^{+0.232}_{-0.240} \pm 0.021$ &  $-0.713\,{}^{+0.410}_{-1.305} \pm 0.024$ \\
$P'_{5}$ &  $\phantom{-}0.387\,{}^{+0.132}_{-0.133} \pm 0.052$ &  $\phantom{-}0.289\,{}^{+0.220}_{-0.202} \pm 0.023$ &  $-0.066\,{}^{+0.343}_{-0.364} \pm 0.023$ \\
$P'_{6}$ &  $\phantom{-}0.034\,{}^{+0.134}_{-0.135} \pm 0.015$ &  $-0.463\,{}^{+0.202}_{-0.221} \pm 0.012$ &  $\phantom{-}0.205\,{}^{+0.962}_{-0.341} \pm 0.013$ \\
$P'_{8}$ &  $\phantom{-}0.180\,{}^{+0.174}_{-0.169} \pm 0.007$ &  $-0.208\,{}^{+0.224}_{-0.270} \pm 0.024$ &  $\phantom{-}0.091\,{}^{+0.650}_{-0.432} \pm 0.025$ \\
\multicolumn{1}{c}{} \\
 & $4.0<q^2<6.0\gevgevcccc$ & $6.0<q^2<8.0\gevgevcccc$ & $11.0<q^2<12.5\gevgevcccc$ \\ \hline
$P_{1}$ &  $\phantom{-}0.180\,{}^{+0.364}_{-0.348} \pm 0.027$ &  $-0.199\,{}^{+0.281}_{-0.275} \pm 0.025$ &  $-0.745\,{}^{+0.207}_{-0.230} \pm 0.015$ \\
$P_{2}$ &  $\phantom{-}0.042\,{}^{+0.088}_{-0.087} \pm 0.011$ &  $\phantom{-}0.241\,{}^{+0.061}_{-0.062} \pm 0.013$ &  $\phantom{-}0.418\,{}^{+0.053}_{-0.046} \pm 0.005$ \\
$P_{3}$ &  $\phantom{-}0.083\,{}^{+0.187}_{-0.184} \pm 0.023$ &  $\phantom{-}0.057\,{}^{+0.148}_{-0.139} \pm 0.013$ &  $\phantom{-}0.007\,{}^{+0.141}_{-0.138} \pm 0.010$ \\
$P'_{4}$ &  $-0.448\,{}^{+0.169}_{-0.172} \pm 0.020$ &  $-0.599\,{}^{+0.131}_{-0.135} \pm 0.010$ &  $-0.567\,{}^{+0.169}_{-0.187} \pm 0.014$ \\
$P'_{5}$ &  $-0.300\,{}^{+0.158}_{-0.159} \pm 0.023$ &  $-0.505\,{}^{+0.122}_{-0.122} \pm 0.024$ &  $-0.655\,{}^{+0.147}_{-0.160} \pm 0.015$ \\
$P'_{6}$ &  $-0.032\,{}^{+0.167}_{-0.166} \pm 0.007$ &  $-0.095\,{}^{+0.135}_{-0.135} \pm 0.011$ &  $-0.282\,{}^{+0.146}_{-0.151} \pm 0.007$ \\
$P'_{8}$ &  $\phantom{-}0.342\,{}^{+0.188}_{-0.185} \pm 0.009$ &  $-0.171\,{}^{+0.142}_{-0.143} \pm 0.006$ &  $-0.015\,{}^{+0.145}_{-0.142} \pm 0.005$ \\
\multicolumn{1}{c}{} \\
 & $15.0<q^2<17.0\gevgevcccc$ & $17.0<q^2<19.0\gevgevcccc$ \\ \hline
$P_{1}$ &  $-0.436\,{}^{+0.134}_{-0.147} \pm 0.018$ &  $-0.581\,{}^{+0.225}_{-0.263} \pm 0.037$ \\
$P_{2}$ &  $\phantom{-}0.421\,{}^{+0.042}_{-0.035} \pm 0.005$ &  $\phantom{-}0.314\,{}^{+0.046}_{-0.048} \pm 0.007$ \\
$P_{3}$ &  $\phantom{-}0.029\,{}^{+0.082}_{-0.084} \pm 0.006$ &  $\phantom{-}0.145\,{}^{+0.107}_{-0.102} \pm 0.008$ \\
$P'_{4}$ &  $-0.672\,{}^{+0.113}_{-0.151} \pm 0.016$ &  $-0.556\,{}^{+0.133}_{-0.156} \pm 0.016$ \\
$P'_{5}$ &  $-0.662\,{}^{+0.109}_{-0.127} \pm 0.017$ &  $-0.676\,{}^{+0.133}_{-0.152} \pm 0.017$ \\
$P'_{6}$ &  $\phantom{-}0.127\,{}^{+0.119}_{-0.122} \pm 0.006$ &  $\phantom{-}0.092\,{}^{+0.148}_{-0.152} \pm 0.025$ \\
$P'_{8}$ &  $\phantom{-}0.007\,{}^{+0.125}_{-0.129} \pm 0.005$ &  $\phantom{-}0.027\,{}^{+0.147}_{-0.147} \pm 0.009$ \\
\end{tabular}  
\end{center}
\end{table}

\clearpage

	\begin{table}[htb]
		\caption{\CP-averaged angular observables evaluated using the method of moments. The first uncertainties are statistical and the second systematic. \label{tab:results:moments:averages}}
		\begin{center}
			\setlength\extrarowheight{2pt}
			\scalebox{0.7}{
				\begin{tabular}{c|cccc}
					& $0.10<q^2<0.98\gev^{2}/c^4$ & $1.1<q^2<2.0\gev^{2}/c^4$ & $2.0<q^2<3.0\gev^{2}/c^4$ & $3.0<q^2<4.0\gev^{2}/c^4$ \\ \hline
					$F_{\rm L}$  & $\phantom{-}0.242\,{}^{+0.058}_{-0.056} \pm 0.026 $ & $\phantom{-}0.768\,{}^{+0.141}_{-0.130} \pm 0.025 $ & $\phantom{-}0.690\,{}^{+0.113}_{-0.082} \pm 0.023 $ & $\phantom{-}0.873\,{}^{+0.154}_{-0.105} \pm 0.023 $ \\
					$S_3$  & $-0.014\,{}^{+0.059}_{-0.060} \pm 0.008 $ & $\phantom{-}0.065\,{}^{+0.137}_{-0.127} \pm 0.007 $ & $\phantom{-}0.006\,{}^{+0.100}_{-0.100} \pm 0.007 $ & $\phantom{-}0.078\,{}^{+0.131}_{-0.122} \pm 0.008 $ \\
					$S_4$  & $\phantom{-}0.039\,{}^{+0.091}_{-0.090} \pm 0.015 $ & $\phantom{-}0.127\,{}^{+0.190}_{-0.180} \pm 0.027 $ & $-0.339\,{}^{+0.115}_{-0.140} \pm 0.041 $ & $-0.046\,{}^{+0.193}_{-0.196} \pm 0.046 $ \\
					$S_5$  & $\phantom{-}0.129\,{}^{+0.068}_{-0.066} \pm 0.011 $ & $\phantom{-}0.286\,{}^{+0.168}_{-0.172} \pm 0.009 $ & $\phantom{-}0.206\,{}^{+0.131}_{-0.115} \pm 0.009 $ & $-0.110\,{}^{+0.163}_{-0.169} \pm 0.004 $ \\
					$A_{\rm FB}$  & $-0.138\,{}^{+0.095}_{-0.092} \pm 0.072 $ & $-0.333\,{}^{+0.115}_{-0.130} \pm 0.012 $ & $-0.158\,{}^{+0.080}_{-0.090} \pm 0.008 $ & $-0.041\,{}^{+0.091}_{-0.091} \pm 0.002 $ \\
					$S_7$  & $\phantom{-}0.038\,{}^{+0.063}_{-0.062} \pm 0.009 $ & $-0.293\,{}^{+0.180}_{-0.176} \pm 0.005 $ & $-0.252\,{}^{+0.127}_{-0.151} \pm 0.002 $ & $\phantom{-}0.171\,{}^{+0.175}_{-0.158} \pm 0.002 $ \\
					$S_8$  & $\phantom{-}0.063\,{}^{+0.079}_{-0.080} \pm 0.009 $ & $-0.114\,{}^{+0.185}_{-0.196} \pm 0.006 $ & $-0.176\,{}^{+0.149}_{-0.165} \pm 0.006 $ & $\phantom{-}0.097\,{}^{+0.189}_{-0.184} \pm 0.002 $ \\
					$S_9$  & $-0.113\,{}^{+0.061}_{-0.063} \pm 0.004 $ & $-0.110\,{}^{+0.140}_{-0.138} \pm 0.001 $ & $-0.000\,{}^{+0.100}_{-0.102} \pm 0.003 $ & $-0.203\,{}^{+0.112}_{-0.132} \pm 0.002 $ \\
					\hline
					$S_{6c}$ & $-0.098^{+0.132}_{-0.134} \pm 0.005$ & $-0.010^{+0.223}_{-0.212} \pm 0.003$ & $-0.239^{+0.268}_{-0.263} \pm 0.001$ & $-0.031^{+0.359}_{-0.337} \pm 0.000$ \\
					\multicolumn{1}{c}{} \\
					& $4.0<q^2<5.0\gev^{2}/c^4$ & $5.0<q^2<6.0\gev^{2}/c^4$ & $6.0<q^2<7.0\gev^{2}/c^4$ & $7.0<q^2<8.0\gev^{2}/c^4$ \\ \hline
					$F_{\rm L}$  & $\phantom{-}0.899\,{}^{+0.106}_{-0.104} \pm 0.023 $ & $\phantom{-}0.644\,{}^{+0.130}_{-0.121} \pm 0.025 $ & $\phantom{-}0.644\,{}^{+0.089}_{-0.084} \pm 0.025 $ & $\phantom{-}0.609\,{}^{+0.103}_{-0.082} \pm 0.025 $ \\
					$S_3$  & $\phantom{-}0.200\,{}^{+0.101}_{-0.097} \pm 0.007 $ & $-0.122\,{}^{+0.119}_{-0.126} \pm 0.009 $ & $-0.069\,{}^{+0.089}_{-0.091} \pm 0.004 $ & $-0.054\,{}^{+0.097}_{-0.099} \pm 0.005 $ \\
					$S_4$  & $-0.148\,{}^{+0.154}_{-0.154} \pm 0.047 $ & $-0.273\,{}^{+0.174}_{-0.184} \pm 0.048 $ & $-0.311\,{}^{+0.111}_{-0.118} \pm 0.052 $ & $-0.236\,{}^{+0.116}_{-0.136} \pm 0.058 $ \\
					$S_5$  & $-0.306\,{}^{+0.138}_{-0.141} \pm 0.004 $ & $-0.095\,{}^{+0.137}_{-0.142} \pm 0.004 $ & $-0.339\,{}^{+0.108}_{-0.114} \pm 0.008 $ & $-0.386\,{}^{+0.105}_{-0.135} \pm 0.007 $ \\
					$A_{\rm FB}$  & $\phantom{-}0.052\,{}^{+0.080}_{-0.080} \pm 0.004 $ & $\phantom{-}0.057\,{}^{+0.094}_{-0.090} \pm 0.006 $ & $\phantom{-}0.058\,{}^{+0.064}_{-0.063} \pm 0.009 $ & $\phantom{-}0.241\,{}^{+0.080}_{-0.062} \pm 0.012 $ \\
					$S_7$  & $-0.082\,{}^{+0.129}_{-0.128} \pm 0.001 $ & $\phantom{-}0.038\,{}^{+0.135}_{-0.135} \pm 0.002 $ & $\phantom{-}0.009\,{}^{+0.123}_{-0.124} \pm 0.004 $ & $-0.094\,{}^{+0.123}_{-0.130} \pm 0.003 $ \\
					$S_8$  & $\phantom{-}0.107\,{}^{+0.144}_{-0.146} \pm 0.003 $ & $-0.037\,{}^{+0.160}_{-0.159} \pm 0.003 $ & $\phantom{-}0.080\,{}^{+0.131}_{-0.129} \pm 0.002 $ & $-0.295\,{}^{+0.119}_{-0.139} \pm 0.002 $ \\
					$S_9$  & $\phantom{-}0.181\,{}^{+0.105}_{-0.099} \pm 0.001 $ & $-0.080\,{}^{+0.117}_{-0.120} \pm 0.001 $ & $\phantom{-}0.061\,{}^{+0.091}_{-0.091} \pm 0.001 $ & $\phantom{-}0.030\,{}^{+0.100}_{-0.098} \pm 0.001 $ \\
					\hline
					$S_{6c}$ & $\phantom{-}0.485^{+0.309}_{-0.317} \pm 0.001$ &  $\phantom{-}0.447^{+0.328}_{-0.333} \pm 0.001$ & $\phantom{-}0.219^{+0.249}_{-0.250} \pm 0.002$ & $\phantom{-}0.249^{+0.209}_{-0.208} \pm 0.002$ \\
					\multicolumn{1}{c}{} \\
					& $11.0<q^2<11.75\gev^{2}/c^4$ & $11.75<q^2<12.5\gev^{2}/c^4$ & $15.0<q^2<16.0\gev^{2}/c^4$ & $16.0<q^2<17.0\gev^{2}/c^4$ \\ \hline
					$F_{\rm L}$  & $\phantom{-}0.502\,{}^{+0.090}_{-0.082} \pm 0.022 $ & $\phantom{-}0.734\,{}^{+0.107}_{-0.094} \pm 0.018 $ & $\phantom{-}0.385\,{}^{+0.067}_{-0.066} \pm 0.013 $ & $\phantom{-}0.295\,{}^{+0.058}_{-0.062} \pm 0.013 $ \\
					$S_3$  & $-0.217\,{}^{+0.077}_{-0.090} \pm 0.008 $ & $-0.157\,{}^{+0.090}_{-0.098} \pm 0.008 $ & $-0.060\,{}^{+0.085}_{-0.088} \pm 0.006 $ & $-0.250\,{}^{+0.079}_{-0.092} \pm 0.007 $ \\
					$S_4$  & $-0.252\,{}^{+0.095}_{-0.113} \pm 0.063 $ & $-0.309\,{}^{+0.099}_{-0.111} \pm 0.056 $ & $-0.321\,{}^{+0.082}_{-0.099} \pm 0.007 $ & $-0.246\,{}^{+0.083}_{-0.096} \pm 0.029 $ \\
					$S_5$  & $-0.235\,{}^{+0.095}_{-0.115} \pm 0.013 $ & $-0.366\,{}^{+0.096}_{-0.112} \pm 0.012 $ & $-0.360\,{}^{+0.074}_{-0.092} \pm 0.006 $ & $-0.254\,{}^{+0.069}_{-0.081} \pm 0.010 $ \\
					$A_{\rm FB}$  & $\phantom{-}0.370\,{}^{+0.076}_{-0.054} \pm 0.015 $ & $\phantom{-}0.293\,{}^{+0.064}_{-0.052} \pm 0.014 $ & $\phantom{-}0.396\,{}^{+0.068}_{-0.047} \pm 0.009 $ & $\phantom{-}0.451\,{}^{+0.071}_{-0.048} \pm 0.007 $ \\
					$S_7$  & $-0.110\,{}^{+0.108}_{-0.114} \pm 0.002 $ & $-0.212\,{}^{+0.110}_{-0.118} \pm 0.002 $ & $\phantom{-}0.040\,{}^{+0.092}_{-0.089} \pm 0.002 $ & $\phantom{-}0.144\,{}^{+0.091}_{-0.085} \pm 0.005 $ \\
					$S_8$  & $-0.079\,{}^{+0.120}_{-0.122} \pm 0.003 $ & $-0.090\,{}^{+0.108}_{-0.111} \pm 0.003 $ & $-0.057\,{}^{+0.093}_{-0.095} \pm 0.005 $ & $\phantom{-}0.055\,{}^{+0.090}_{-0.088} \pm 0.005 $ \\
					$S_9$  & $-0.084\,{}^{+0.097}_{-0.102} \pm 0.003 $ & $\phantom{-}0.030\,{}^{+0.093}_{-0.091} \pm 0.002 $ & $-0.054\,{}^{+0.083}_{-0.087} \pm 0.005 $ & $-0.014\,{}^{+0.084}_{-0.086} \pm 0.004 $ \\
					\hline
					$S_{6c}$ & $\phantom{-}0.082^{+0.220}_{-0.223} \pm 0.003$ & $\phantom{-}0.392^{+0.293}_{-0.294} \pm 0.004$ & $-0.273^{+0.164}_{-0.161} \pm 0.004$ & $-0.112^{+0.127}_{-0.129} \pm 0.003$  \\ 
					\multicolumn{1}{c}{} \\
					& $17.0<q^2<18.0\gev^{2}/c^4$ & $18.0<q^2<19.0\gev^{2}/c^4$ & $15.0<q^2<19.0\gev^{2}/c^4$  \\ \hline
					$F_{\rm L}$  & $\phantom{-}0.363\,{}^{+0.073}_{-0.072} \pm 0.017 $ & $\phantom{-}0.421\,{}^{+0.100}_{-0.100} \pm 0.013 $  & $\phantom{-}0.357\,{}^{+0.035}_{-0.035} \pm 0.011 $  \\
					$S_3$  & $-0.099\,{}^{+0.091}_{-0.092} \pm 0.011 $ & $-0.131\,{}^{+0.128}_{-0.130} \pm 0.012 $   & $-0.135\,{}^{+0.046}_{-0.050} \pm 0.012 $\\
					$S_4$  & $-0.229\,{}^{+0.090}_{-0.096} \pm 0.045 $ & $-0.607\,{}^{+0.153}_{-0.170} \pm 0.059 $  & $-0.314\,{}^{+0.046}_{-0.054} \pm 0.027 $ \\
					$S_5$  & $-0.305\,{}^{+0.081}_{-0.088} \pm 0.015 $ & $-0.534\,{}^{+0.131}_{-0.150} \pm 0.015 $  & $-0.335\,{}^{+0.041}_{-0.047} \pm 0.007 $ \\
					$A_{\rm FB}$  & $\phantom{-}0.274\,{}^{+0.069}_{-0.061} \pm 0.008 $ & $\phantom{-}0.354\,{}^{+0.111}_{-0.099} \pm 0.012 $   & $\phantom{-}0.367\,{}^{+0.037}_{-0.029} \pm 0.007 $\\
					$S_7$  & $\phantom{-}0.022\,{}^{+0.094}_{-0.093} \pm 0.011 $ & $\phantom{-}0.058\,{}^{+0.123}_{-0.124} \pm 0.006 $   & $\phantom{-}0.066\,{}^{+0.049}_{-0.046} \pm 0.014 $\\
					$S_8$  & $-0.007\,{}^{+0.098}_{-0.098} \pm 0.001 $ & $\phantom{-}0.149\,{}^{+0.139}_{-0.138} \pm 0.010 $ & $\phantom{-}0.024\,{}^{+0.040}_{-0.048} \pm 0.009 $\\
					$S_9$  & $-0.090\,{}^{+0.092}_{-0.095} \pm 0.002 $ & $-0.079\,{}^{+0.122}_{-0.121} \pm 0.007 $  & $-0.056\,{}^{+0.046}_{-0.047} \pm 0.014 $\\
					\hline
					$S_{6c}$ & $-0.195^{+0.169}_{-0.170}\pm 0.003$ &  $\phantom{-}0.187^{+0.201}_{-0.207}\pm 0.001$  & $-0.125\,{}^{+0.082}_{-0.084} \pm 0.032 $ \\					
				\end{tabular}
			}
		\end{center}
	\end{table}
	
	\clearpage

	\begin{table}[htb]
		\caption{\CP-asymmetries  evaluated using the method of moments. The first uncertainties are  statistical and the second systematic. \label{tab:results:moments:asymmetries}}
		\begin{center}
			\setlength\extrarowheight{2pt}
			\scalebox{0.7}{
				\begin{tabular}{c|cccc}
					& $0.10<q^2<0.98\gev^{2}/c^4$ & $1.1<q^2<2.0\gev^{2}/c^4$ & $2.0<q^2<3.0\gev^{2}/c^4$ & $3.0<q^2<4.0\gev^{2}/c^4$ \\ \hline
					$A_3$  & $-0.040\,{}^{+0.059}_{-0.061} \pm 0.007 $ & $-0.134\,{}^{+0.126}_{-0.136} \pm 0.003 $ & $-0.018\,{}^{+0.101}_{-0.100} \pm 0.001 $ & $-0.118\,{}^{+0.120}_{-0.132} \pm 0.007 $ \\
					$A_4$  & $-0.047\,{}^{+0.090}_{-0.092} \pm 0.013 $ & $\phantom{-}0.283\,{}^{+0.191}_{-0.181} \pm 0.028 $ & $\phantom{-}0.261\,{}^{+0.146}_{-0.123} \pm 0.042 $ & $\phantom{-}0.002\,{}^{+0.194}_{-0.196} \pm 0.045 $ \\
					$A_5$  & $-0.008\,{}^{+0.066}_{-0.066} \pm 0.011 $ & $-0.110\,{}^{+0.166}_{-0.176} \pm 0.008 $ & $\phantom{-}0.028\,{}^{+0.124}_{-0.120} \pm 0.008 $ & $\phantom{-}0.015\,{}^{+0.167}_{-0.168} \pm 0.005 $ \\
					$A_{6s}$  & $\phantom{-}0.167\,{}^{+0.128}_{-0.127} \pm 0.016 $ & $\phantom{-}0.213\,{}^{+0.161}_{-0.155} \pm 0.005 $ & $\phantom{-}0.077\,{}^{+0.121}_{-0.111} \pm 0.004 $ & $-0.047\,{}^{+0.121}_{-0.125} \pm 0.001 $ \\
					$A_7$  & $\phantom{-}0.112\,{}^{+0.064}_{-0.062} \pm 0.010 $ & $-0.193\,{}^{+0.167}_{-0.200} \pm 0.006 $ & $-0.162\,{}^{+0.130}_{-0.144} \pm 0.003 $ & $-0.004\,{}^{+0.165}_{-0.162} \pm 0.003 $ \\
					$A_8$  & $\phantom{-}0.021\,{}^{+0.080}_{-0.080} \pm 0.012 $ & $\phantom{-}0.130\,{}^{+0.203}_{-0.180} \pm 0.008 $ & $-0.060\,{}^{+0.152}_{-0.161} \pm 0.006 $ & $\phantom{-}0.005\,{}^{+0.188}_{-0.185} \pm 0.003 $ \\
					$A_9$  & $\phantom{-}0.043\,{}^{+0.062}_{-0.062} \pm 0.009 $ & $-0.126\,{}^{+0.136}_{-0.153} \pm 0.010 $ & $\phantom{-}0.013\,{}^{+0.102}_{-0.101} \pm 0.007 $ & $-0.129\,{}^{+0.115}_{-0.125} \pm 0.003 $ \\
					\multicolumn{1}{c}{} \\
					& $4.0<q^2<5.0\gev^{2}/c^4$ & $5.0<q^2<6.0\gev^{2}/c^4$ & $6.0<q^2<7.0\gev^{2}/c^4$ & $7.0<q^2<8.0\gev^{2}/c^4$ \\ \hline
					$A_3$  & $-0.064\,{}^{+0.098}_{-0.098} \pm 0.005 $ & $-0.076\,{}^{+0.119}_{-0.122} \pm 0.004 $ & $-0.073\,{}^{+0.089}_{-0.091} \pm 0.007 $ & $\phantom{-}0.168\,{}^{+0.104}_{-0.093} \pm 0.005 $ \\
					$A_4$  & $\phantom{-}0.076\,{}^{+0.155}_{-0.154} \pm 0.047 $ & $-0.457\,{}^{+0.174}_{-0.187} \pm 0.048 $ & $-0.104\,{}^{+0.121}_{-0.120} \pm 0.052 $ & $\phantom{-}0.058\,{}^{+0.132}_{-0.127} \pm 0.058 $ \\
					$A_5$  & $\phantom{-}0.051\,{}^{+0.143}_{-0.142} \pm 0.005 $ & $-0.011\,{}^{+0.139}_{-0.139} \pm 0.006 $ & $\phantom{-}0.040\,{}^{+0.117}_{-0.116} \pm 0.006 $ & $\phantom{-}0.084\,{}^{+0.122}_{-0.117} \pm 0.007 $ \\
					$A_{6s}$  & $-0.085\,{}^{+0.107}_{-0.107} \pm 0.009 $ & $\phantom{-}0.116\,{}^{+0.124}_{-0.121} \pm 0.003 $ & $\phantom{-}0.063\,{}^{+0.087}_{-0.083} \pm 0.005 $ & $\phantom{-}0.100\,{}^{+0.096}_{-0.088} \pm 0.006 $ \\
					$A_7$  & $-0.146\,{}^{+0.13}_{-0.13} \pm 0.003 $ & $\phantom{-}0.058\,{}^{+0.135}_{-0.135} \pm 0.003 $ & $\phantom{-}0.181\,{}^{+0.125}_{-0.122} \pm 0.005 $ & $\phantom{-}0.064\,{}^{+0.129}_{-0.125} \pm 0.004 $ \\
					$A_8$  & $\phantom{-}0.183\,{}^{+0.150}_{-0.146} \pm 0.001 $ & $-0.195\,{}^{+0.156}_{-0.167} \pm 0.007 $ & $\phantom{-}0.004\,{}^{+0.131}_{-0.130} \pm 0.003 $ & $\phantom{-}0.078\,{}^{+0.131}_{-0.127} \pm 0.002 $ \\
					$A_9$  & $\phantom{-}0.160\,{}^{+0.103}_{-0.100} \pm 0.008 $ & $-0.001\,{}^{+0.118}_{-0.120} \pm 0.002 $ & $\phantom{-}0.125\,{}^{+0.092}_{-0.090} \pm 0.005 $ & $\phantom{-}0.195\,{}^{+0.108}_{-0.093} \pm 0.003 $ \\
					\multicolumn{1}{c}{} \\
					& $11.0<q^2<11.75\gev^{2}/c^4$ & $11.75<q^2<12.5\gev^{2}/c^4$ & $15.0<q^2<16.0\gev^{2}/c^4$ & $16.0<q^2<17.0\gev^{2}/c^4$ \\ \hline
					$A_3$  & $\phantom{-}0.124\,{}^{+0.090}_{-0.083} \pm 0.008 $ & $\phantom{-}0.124\,{}^{+0.096}_{-0.090} \pm 0.008 $ & $-0.108\,{}^{+0.085}_{-0.091} \pm 0.005 $ & $\phantom{-}0.016\,{}^{+0.087}_{-0.087} \pm 0.006 $ \\
					$A_4$  & $-0.058\,{}^{+0.101}_{-0.105} \pm 0.063 $ & $-0.242\,{}^{+0.102}_{-0.112} \pm 0.056 $ & $\phantom{-}0.059\,{}^{+0.094}_{-0.091} \pm 0.007 $ & $-0.110\,{}^{+0.087}_{-0.093} \pm 0.028 $ \\
					$A_5$  & $-0.042\,{}^{+0.102}_{-0.106} \pm 0.013 $ & $\phantom{-}0.097\,{}^{+0.105}_{-0.102} \pm 0.012 $ & $\phantom{-}0.039\,{}^{+0.087}_{-0.085} \pm 0.007 $ & $-0.138\,{}^{+0.073}_{-0.079} \pm 0.008 $ \\
					$A_{6s}$  & $\phantom{-}0.016\,{}^{+0.091}_{-0.089} \pm 0.003 $ & $-0.099\,{}^{+0.076}_{-0.081} \pm 0.003 $ & $-0.117\,{}^{+0.076}_{-0.081} \pm 0.010 $ & $-0.035\,{}^{+0.079}_{-0.081} \pm 0.010 $ \\
					$A_7$  & $-0.064\,{}^{+0.110}_{-0.113} \pm 0.002 $ & $\phantom{-}0.144\,{}^{+0.116}_{-0.110} \pm 0.002 $ & $-0.124\,{}^{+0.087}_{-0.094} \pm 0.004 $ & $-0.081\,{}^{+0.087}_{-0.092} \pm 0.006 $ \\
					$A_8$  & $\phantom{-}0.072\,{}^{+0.123}_{-0.121} \pm 0.003 $ & $-0.017\,{}^{+0.112}_{-0.107} \pm 0.003 $ & $\phantom{-}0.087\,{}^{+0.095}_{-0.092} \pm 0.005 $ & $\phantom{-}0.013\,{}^{+0.089}_{-0.089} \pm 0.005 $ \\
					$A_9$  & $-0.082\,{}^{+0.097}_{-0.102} \pm 0.003 $ & $-0.014\,{}^{+0.092}_{-0.092} \pm 0.002 $ & $\phantom{-}0.145\,{}^{+0.089}_{-0.082} \pm 0.007 $ & $\phantom{-}0.058\,{}^{+0.086}_{-0.084} \pm 0.004 $ \\
					\multicolumn{1}{c}{} \\
					& $17.0<q^2<18.0\gev^{2}/c^4$ & $18.0<q^2<19.0\gev^{2}/c^4$ & $15.0<q^2<19.0\gev^{2}/c^4$\\ \hline
					$A_3$  & $-0.145\,{}^{+0.090}_{-0.094} \pm 0.010 $ & $\phantom{-}0.050\,{}^{+0.133}_{-0.129} \pm 0.011 $ & $-0.053\,{}^{+0.047}_{-0.048} \pm 0.011 $\\
					$A_4$  & $-0.071\,{}^{+0.093}_{-0.095} \pm 0.045 $ & $-0.120\,{}^{+0.155}_{-0.162} \pm 0.059 $  & $-0.049\,{}^{+0.049}_{-0.052} \pm 0.027 $\\
					$A_5$  & $-0.020\,{}^{+0.085}_{-0.086} \pm 0.015 $ & $\phantom{-}0.186\,{}^{+0.134}_{-0.131} \pm 0.015 $ & $-0.010\,{}^{+0.044}_{-0.044} \pm 0.008 $\\
					$A_{6s}$  & $-0.109\,{}^{+0.087}_{-0.091} \pm 0.009 $ & $-0.167\,{}^{+0.132}_{-0.139} \pm 0.007 $ & $-0.096\,{}^{+0.045}_{-0.044} \pm 0.008 $ \\
					$A_7$  & $-0.025\,{}^{+0.093}_{-0.094} \pm 0.011 $ & $\phantom{-}0.022\,{}^{+0.125}_{-0.122} \pm 0.006 $ &  $-0.062\,{}^{+0.048}_{-0.049} \pm 0.014 $\\
					$A_8$  & $\phantom{-}0.131\,{}^{+0.100}_{-0.097} \pm 0.004 $ & $-0.029\,{}^{+0.140}_{-0.138} \pm 0.010 $ &  $\phantom{-}0.059\,{}^{+0.050}_{-0.049} \pm 0.010 $\\
					$A_9$  & $\phantom{-}0.116\,{}^{+0.095}_{-0.092} \pm 0.006 $ & $-0.147\,{}^{+0.121}_{-0.128} \pm 0.007 $ &  $\phantom{-}0.065\,{}^{+0.048}_{-0.045} \pm 0.016 $\\
				\end{tabular}
			}
		\end{center}
	\end{table} 
	
	\clearpage

	\begin{table}[htb]
		\caption{Optimised observables evaluated using the method of moments. The first uncertainties are statistical and the second systematic. \label{tab:results:moments:optimised}}
		\begin{center}
			\setlength\extrarowheight{2pt}
			\scalebox{0.7}{
				\begin{tabular}{c|cccc}
					& $0.10<q^2<0.98\gev^{2}/c^4$ & $1.1<q^2<2.0\gev^{2}/c^4$ & $2.0<q^2<3.0\gev^{2}/c^4$ & $3.0<q^2<4.0\gev^{2}/c^4$ \\ \hline                              
					$P_1$  & $-0.038\,{}^{+0.157}_{-0.158} \pm 0.020 $ & $\phantom{-}0.439\,{}^{+1.916}_{-1.013} \pm 0.012 $ & $\phantom{-}0.055\,{}^{+0.677}_{-0.756} \pm 0.007 $ & $\phantom{-}0.421\,{}^{+18.35}_{-1.190} \pm 0.018 $ \\                                                                                                                                     
					$P_2$  & $-0.119\,{}^{+0.080}_{-0.081} \pm 0.063 $ & $-0.667\,{}^{+0.149}_{-1.939} \pm 0.017 $ & $-0.323\,{}^{+0.147}_{-0.316} \pm 0.033 $ & $-0.117\,{}^{+0.485}_{-4.435} \pm 0.015 $ \\
					$P_3$  & $\phantom{-}0.147\,{}^{+0.086}_{-0.080} \pm 0.005 $ & $\phantom{-}0.363\,{}^{+1.088}_{-0.506} \pm 0.001 $ & $\phantom{-}0.005\,{}^{+0.362}_{-0.364} \pm 0.012 $ & $\phantom{-}0.905\,{}^{+17.51}_{-0.258} \pm 0.009 $ \\
					$P'_4$  & $\phantom{-}0.086\,{}^{+0.221}_{-0.209} \pm 0.026 $ & $\phantom{-}0.266\,{}^{+0.648}_{-0.406} \pm 0.057 $ & $-0.765\,{}^{+0.271}_{-0.359} \pm 0.099 $ & $-0.134\,{}^{+0.810}_{-1.343} \pm 0.108 $ \\
					$P'_5$  & $\phantom{-}0.300\,{}^{+0.171}_{-0.152} \pm 0.023 $ & $\phantom{-}0.606\,{}^{+0.769}_{-0.326} \pm 0.017 $ & $\phantom{-}0.461\,{}^{+0.313}_{-0.256} \pm 0.019 $ & $-0.295\,{}^{+0.508}_{-7.112} \pm 0.023 $ \\
					$P'_6$  & $\phantom{-}0.086\,{}^{+0.152}_{-0.145} \pm 0.024 $ & $-0.632\,{}^{+0.347}_{-0.753} \pm 0.009 $ & $-0.549\,{}^{+0.276}_{-0.393} \pm 0.005 $ & $\phantom{-}0.449\,{}^{+19.04}_{-0.397} \pm 0.007 $ \\
					$P'_8$  & $\phantom{-}0.143\,{}^{+0.195}_{-0.184} \pm 0.022 $ & $-0.244\,{}^{+0.433}_{-0.645} \pm 0.012 $ & $-0.393\,{}^{+0.332}_{-0.388} \pm 0.002 $ & $\phantom{-}0.303\,{}^{+1.394}_{-0.719} \pm 0.006 $ \\
					\multicolumn{1}{c}{} \\  
					& $4.0<q^2<5.0\gev^{2}/c^4$ & $5.0<q^2<6.0\gev^{2}/c^4$ & $6.0<q^2<7.0\gev^{2}/c^4$ & $7.0<q^2<8.0\gev^{2}/c^4$ \\ \hline
					$P_1$  & $\phantom{-}2.296\,{}^{+17.71}_{-0.694} \pm 0.024 $ & $-0.540\,{}^{+0.521}_{-1.100} \pm 0.025 $ & $-0.353\,{}^{+0.469}_{-0.602} \pm 0.026 $ & $-0.284\,{}^{+0.513}_{-0.548} \pm 0.025 $ \\
					$P_2$  & $\phantom{-}0.174\,{}^{+3.034}_{-0.376} \pm 0.010 $ & $\phantom{-}0.089\,{}^{+0.227}_{-0.155} \pm 0.012 $ & $\phantom{-}0.104\,{}^{+0.136}_{-0.115} \pm 0.013 $ & $\phantom{-}0.393\,{}^{+0.231}_{-0.093} \pm 0.013 $ \\
					$P_3$  & $-0.801\,{}^{+0.221}_{-17.42} \pm 0.007 $ & $\phantom{-}0.178\,{}^{+0.465}_{-0.286} \pm 0.007 $ & $-0.161\,{}^{+0.246}_{-0.291} \pm 0.001 $ & $-0.063\,{}^{+0.244}_{-0.298} \pm 0.002 $ \\
					$P'_4$  & $-0.415\,{}^{+0.438}_{-1.911} \pm 0.104 $ & $-0.561\,{}^{+0.345}_{-0.465} \pm 0.101 $ & $-0.641\,{}^{+0.222}_{-0.294} \pm 0.106 $ & $-0.503\,{}^{+0.253}_{-0.288} \pm 0.118 $ \\
					$P'_5$  & $-0.799\,{}^{+0.266}_{-18.19} \pm 0.022 $ & $-0.197\,{}^{+0.287}_{-0.334} \pm 0.018 $ & $-0.713\,{}^{+0.228}_{-0.268} \pm 0.015 $ & $-0.808\,{}^{+0.226}_{-0.303} \pm 0.010 $ \\
					$P'_6$  & $-0.215\,{}^{+0.397}_{-1.243} \pm 0.006 $ & $\phantom{-}0.074\,{}^{+0.309}_{-0.288} \pm 0.005 $ & $\phantom{-}0.017\,{}^{+0.267}_{-0.261} \pm 0.007 $ & $-0.201\,{}^{+0.261}_{-0.274} \pm 0.007 $ \\
					$P'_8$  & $\phantom{-}0.293\,{}^{+1.522}_{-0.441} \pm 0.006 $ & $-0.068\,{}^{+0.338}_{-0.372} \pm 0.006 $ & $\phantom{-}0.162\,{}^{+0.289}_{-0.267} \pm 0.005 $ & $-0.623\,{}^{+0.255}_{-0.295} \pm 0.005 $ \\
					\multicolumn{1}{c}{} \\  
					& $11.0<q^2<11.75\gev^{2}/c^4$ & $11.75<q^2<12.5\gev^{2}/c^4$ & $15.0<q^2<16.0\gev^{2}/c^4$ & $16.0<q^2<17.0\gev^{2}/c^4$ \\ \hline
					$P_1$  & $-0.869\,{}^{+0.304}_{-0.408} \pm 0.030 $ & $-1.002\,{}^{+0.502}_{-1.360} \pm 0.030 $ & $-0.199\,{}^{+0.280}_{-0.285} \pm 0.014 $ & $-0.726\,{}^{+0.239}_{-0.241} \pm 0.014 $ \\
					$P_2$  & $\phantom{-}0.494\,{}^{+0.134}_{-0.071} \pm 0.013 $ & $\phantom{-}0.637\,{}^{+0.599}_{-0.100} \pm 0.008 $ & $\phantom{-}0.433\,{}^{+0.074}_{-0.054} \pm 0.005 $ & $\phantom{-}0.430\,{}^{+0.063}_{-0.049} \pm 0.007 $ \\
					$P_3$  & $\phantom{-}0.166\,{}^{+0.221}_{-0.192} \pm 0.005 $ & $-0.105\,{}^{+0.349}_{-0.42} \pm 0.004 $ & $\phantom{-}0.087\,{}^{+0.144}_{-0.135} \pm 0.007 $ & $\phantom{-}0.019\,{}^{+0.122}_{-0.119} \pm 0.006 $ \\
					$P'_4$  & $-0.522\,{}^{+0.203}_{-0.222} \pm 0.128 $ & $-0.701\,{}^{+0.215}_{-0.342} \pm 0.114 $ & $-0.673\,{}^{+0.178}_{-0.199} \pm 0.013 $ & $-0.552\,{}^{+0.191}_{-0.213} \pm 0.055 $ \\
					$P'_5$  & $-0.485\,{}^{+0.203}_{-0.224} \pm 0.028 $ & $-0.827\,{}^{+0.205}_{-0.357} \pm 0.026 $ & $-0.758\,{}^{+0.165}_{-0.179} \pm 0.013 $ & $-0.567\,{}^{+0.157}_{-0.186} \pm 0.014 $ \\
					$P'_6$  & $-0.233\,{}^{+0.227}_{-0.224} \pm 0.004 $ & $-0.473\,{}^{+0.233}_{-0.344} \pm 0.004 $ & $\phantom{-}0.083\,{}^{+0.189}_{-0.187} \pm 0.004 $ & $\phantom{-}0.328\,{}^{+0.192}_{-0.195} \pm 0.012 $ \\
					$P'_8$  & $-0.159\,{}^{+0.241}_{-0.250} \pm 0.007 $ & $-0.211\,{}^{+0.255}_{-0.274} \pm 0.007 $ & $-0.120\,{}^{+0.192}_{-0.198} \pm 0.010 $ & $\phantom{-}0.122\,{}^{+0.199}_{-0.196} \pm 0.010 $ \\
					\multicolumn{1}{c}{} \\  
					& $17.0<q^2<18.0\gev^{2}/c^4$ & $18.0<q^2<19.0\gev^{2}/c^4$ & $15.0 < \qsq < 19.0\gev^2/c^{4}$ \\ \hline
					$P_1$  & $-0.313\,{}^{+0.286}_{-0.293} \pm 0.019 $ & $-0.450\,{}^{+0.440}_{-0.447} \pm 0.022 $ & $-0.424\,{}^{+0.139}_{-0.150} \pm 0.028 $ \\
					$P_2$  & $\phantom{-}0.288\,{}^{+0.075}_{-0.064} \pm 0.006 $ & $\phantom{-}0.393\,{}^{+0.159}_{-0.100} \pm 0.011 $ & $\phantom{-}0.385\,{}^{+0.035}_{-0.036} \pm 0.010 $\\
					$P_3$  & $\phantom{-}0.144\,{}^{+0.149}_{-0.147} \pm 0.002 $ & $\phantom{-}0.134\,{}^{+0.219}_{-0.208} \pm 0.010 $ & $\phantom{-}0.089\,{}^{+0.071}_{-0.072} \pm 0.019 $\\
					$P'_4$  & $-0.486\,{}^{+0.190}_{-0.200} \pm 0.092 $ & $-1.221\,{}^{+0.280}_{-0.388} \pm 0.119 $ & $-0.663\,{}^{+0.102}_{-0.105} \pm 0.055 $ \\
					$P'_5$  & $-0.646\,{}^{+0.176}_{-0.190} \pm 0.027 $ & $-1.070\,{}^{+0.237}_{-0.349} \pm 0.029 $ & $-0.709\,{}^{+0.093}_{-0.090} \pm 0.016 $\\
					$P'_6$  & $\phantom{-}0.047\,{}^{+0.198}_{-0.197} \pm 0.023 $ & $\phantom{-}0.128\,{}^{+0.246}_{-0.265} \pm 0.012 $  & $\phantom{-}0.140\,{}^{+0.101}_{-0.100} \pm 0.032 $\\
					$P'_8$  & $-0.006\,{}^{+0.199}_{-0.215} \pm 0.001 $ & $\phantom{-}0.300\,{}^{+0.297}_{-0.276} \pm 0.022 $ & $\phantom{-}0.049\,{}^{+0.106}_{-0.102} \pm 0.021 $\\
				\end{tabular}
			}
			
		\end{center}
	\end{table}

\clearpage

\section{Likelihood fit projections}
\label{sec:projections}

\begin{figure}[htb]
\begin{center}
\includegraphics[width=0.49\linewidth]{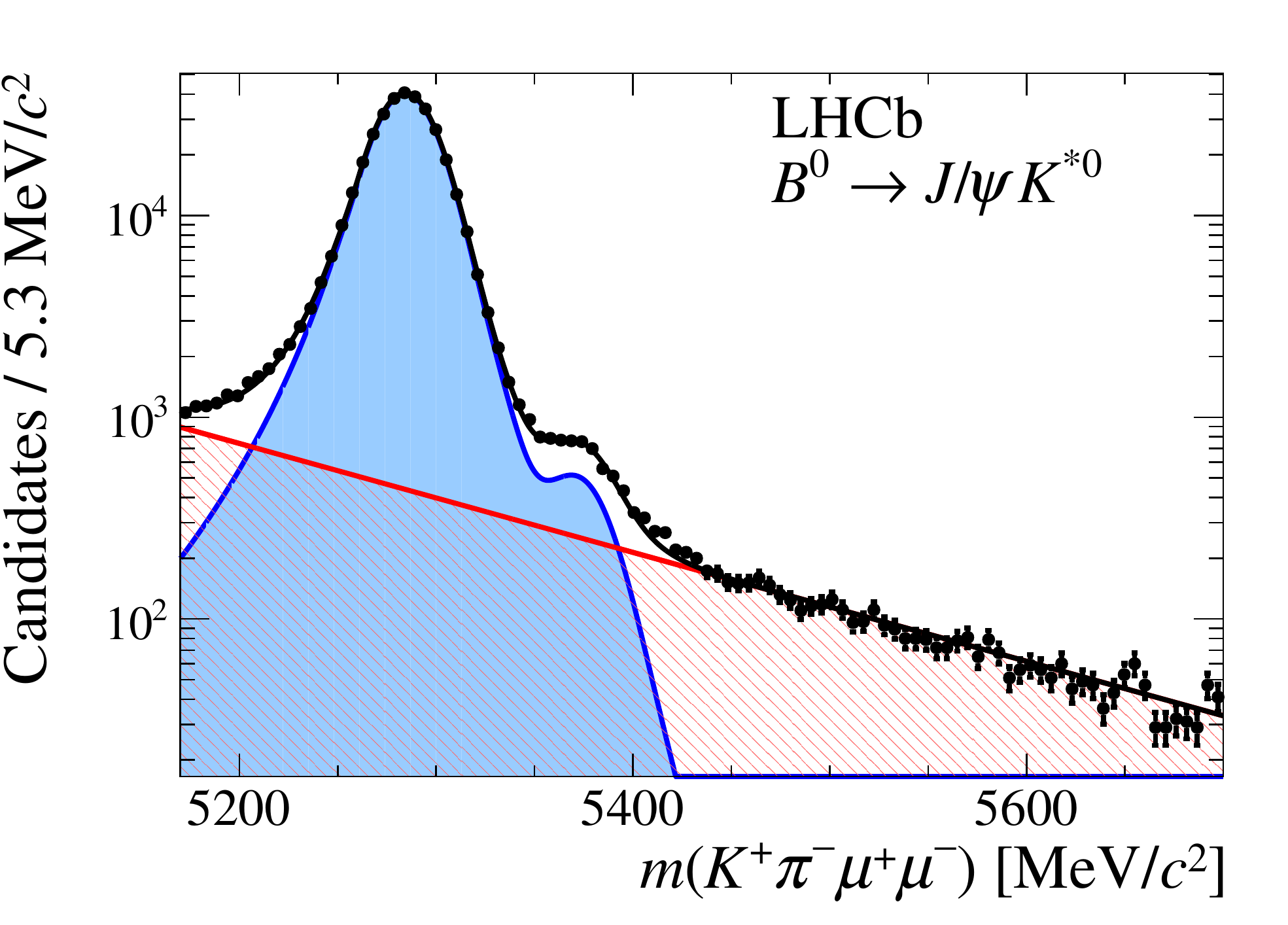} 
\includegraphics[width=0.48\linewidth]{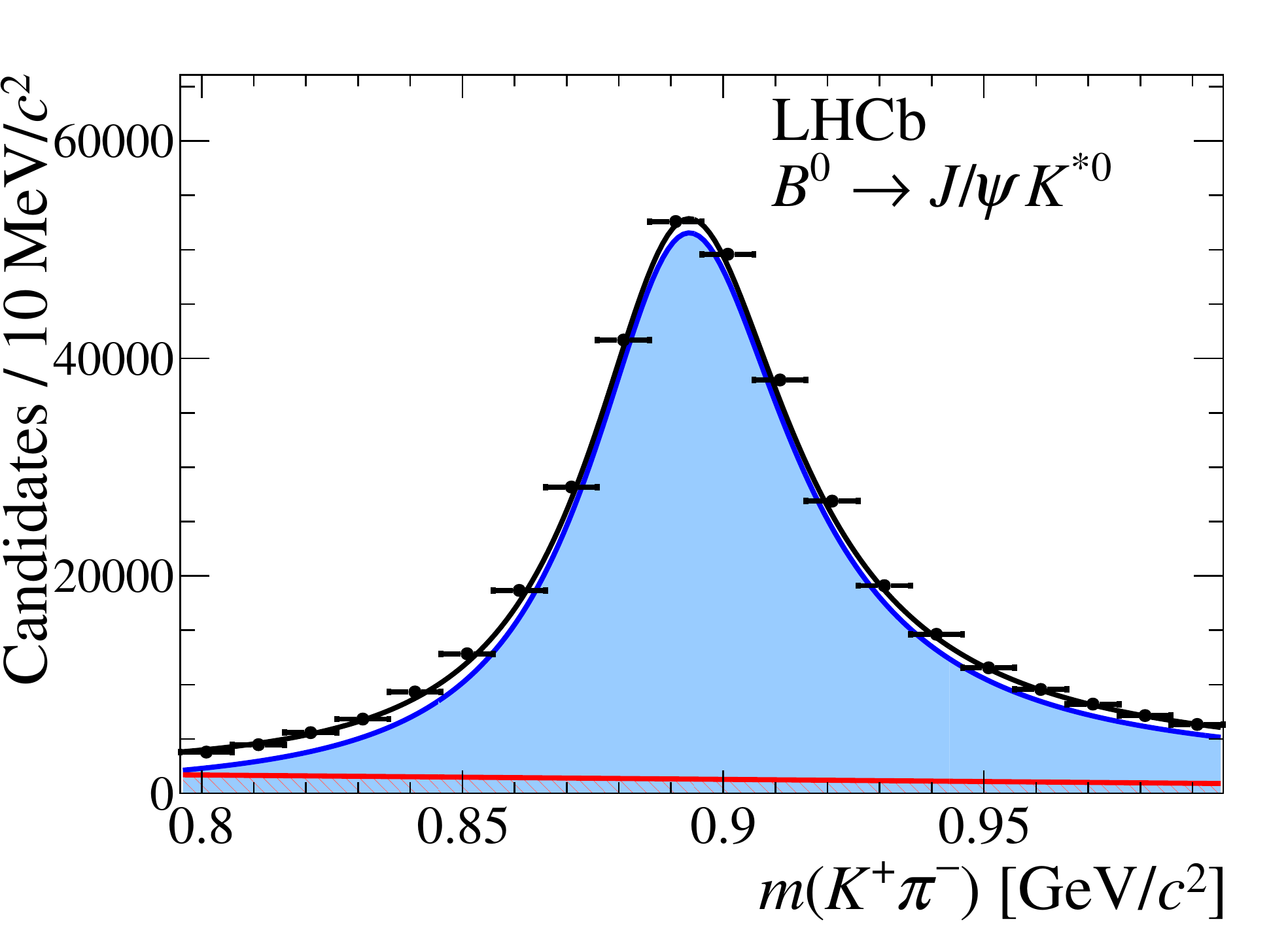}  \\
\includegraphics[width=0.48\linewidth]{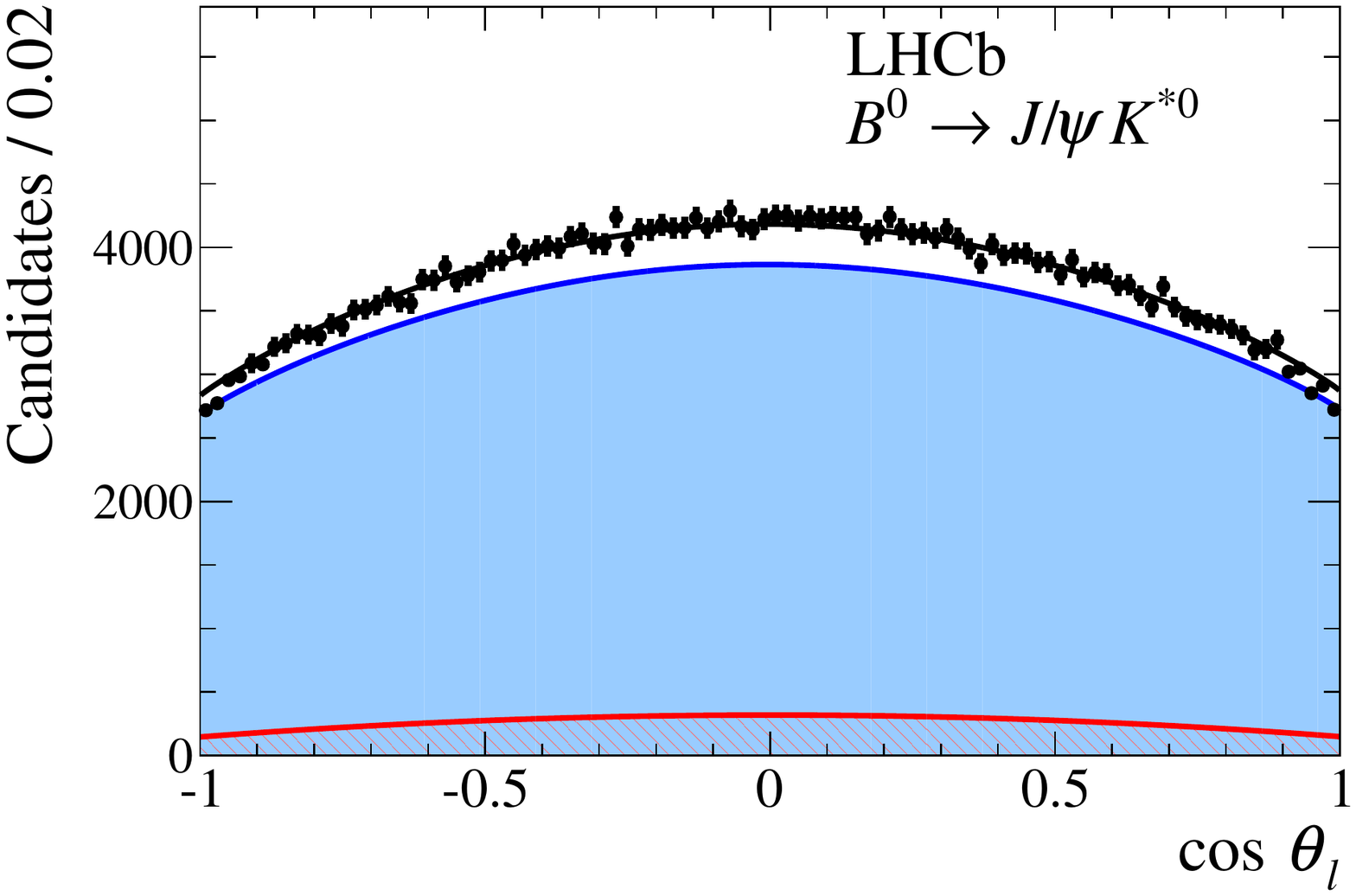} 
\includegraphics[width=0.48\linewidth]{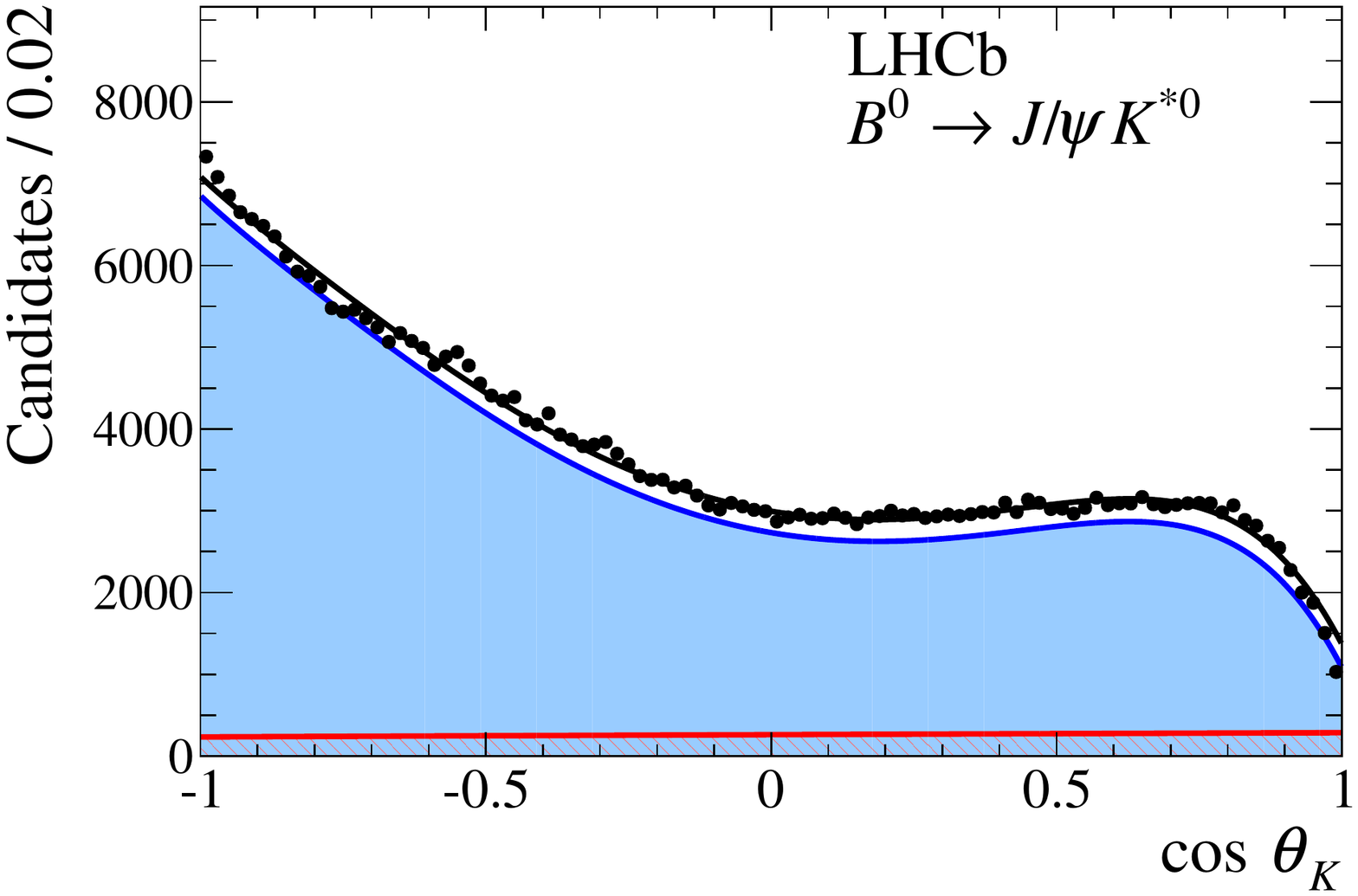} \\
\includegraphics[width=0.48\linewidth]{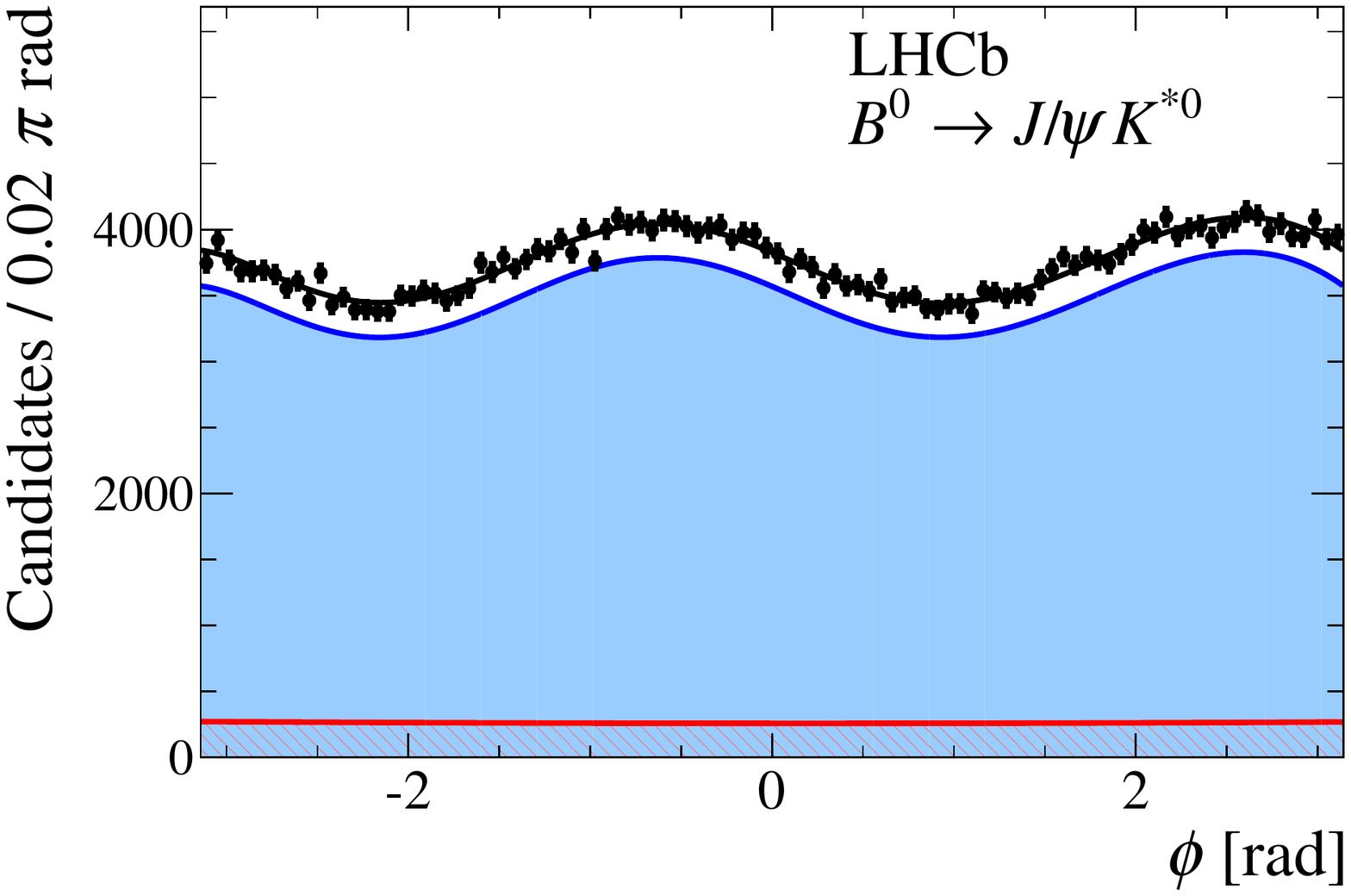} 
\end{center}
\vspace*{-0.7cm}
\caption{Angular and mass distribution of \decay{\Bz}{\jpsi\Kstarz} candidates in data. 
  A small signal component is also included in the fit to account for \decay{\Bsb}{\jpsi\Kstarz} decays. 
  Overlaid are the projections of the total fitted distribution (black line) and its different components.
  The signal is shown by the solid blue component and the background by the red hatched component. 
  \label{fig:supp:jpsikstar}}
\end{figure}

\begin{figure}[htb]
\begin{center}
\includegraphics[width=0.48\linewidth]{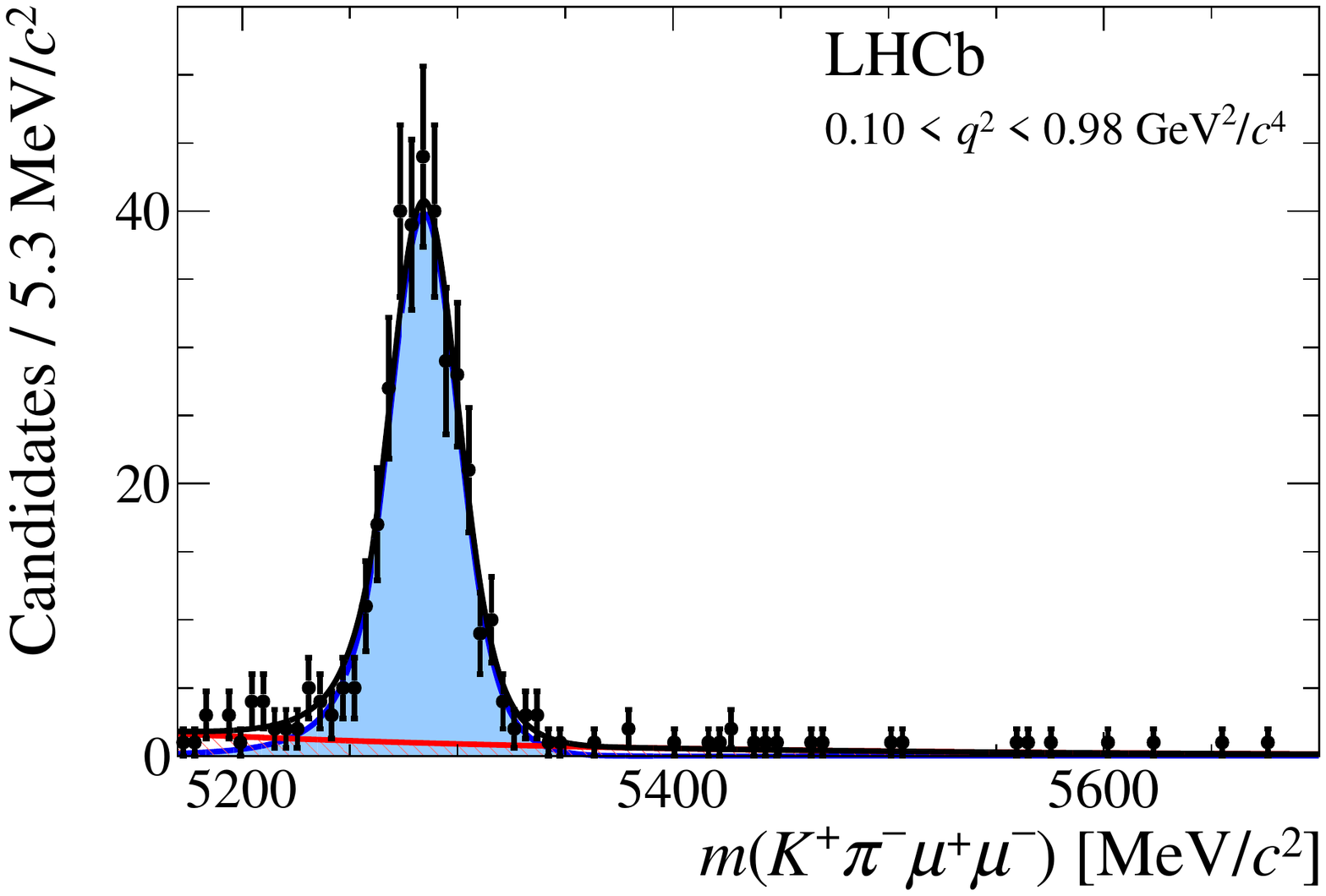} 
\includegraphics[width=0.48\linewidth]{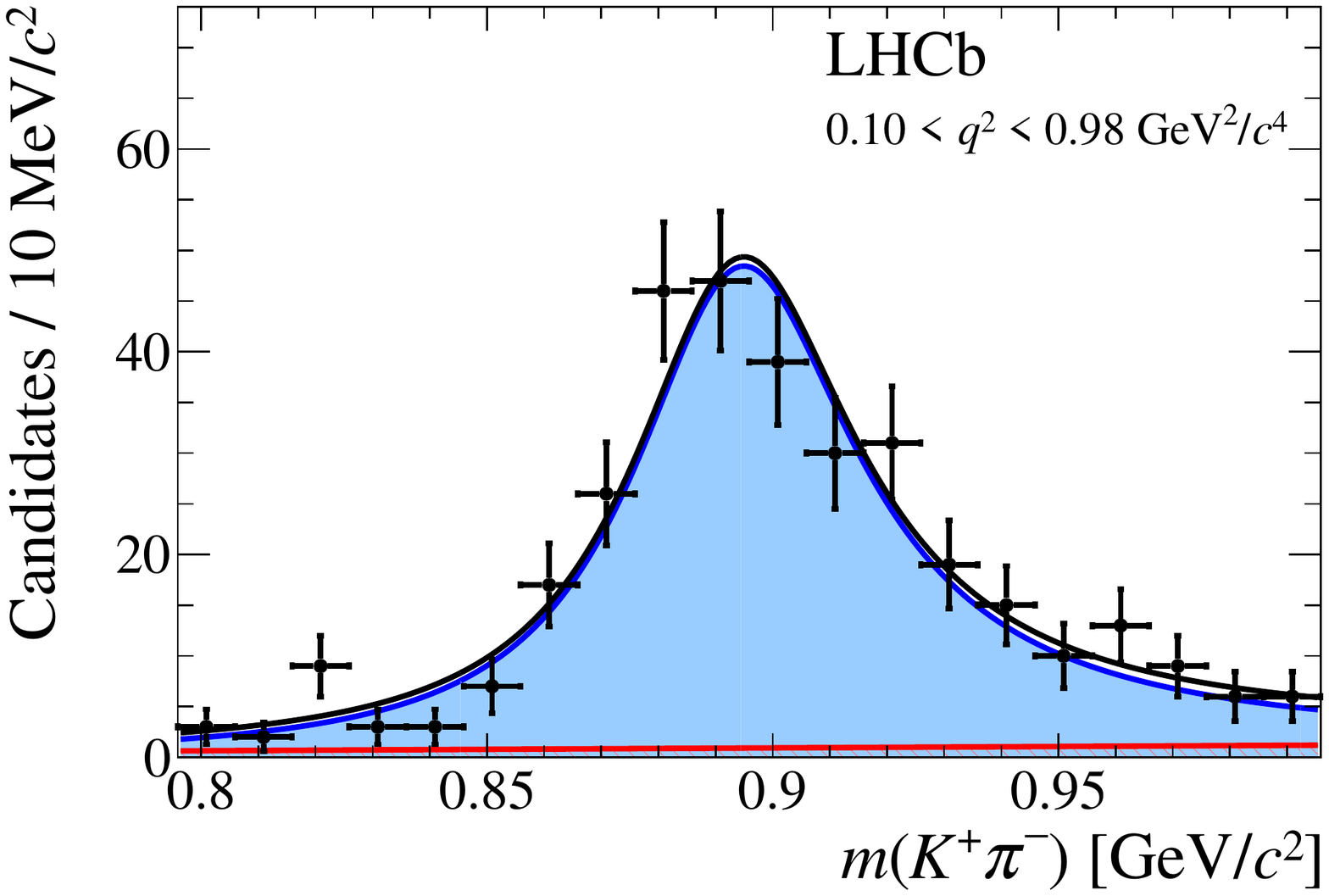}  \\ 
\includegraphics[width=0.48\linewidth]{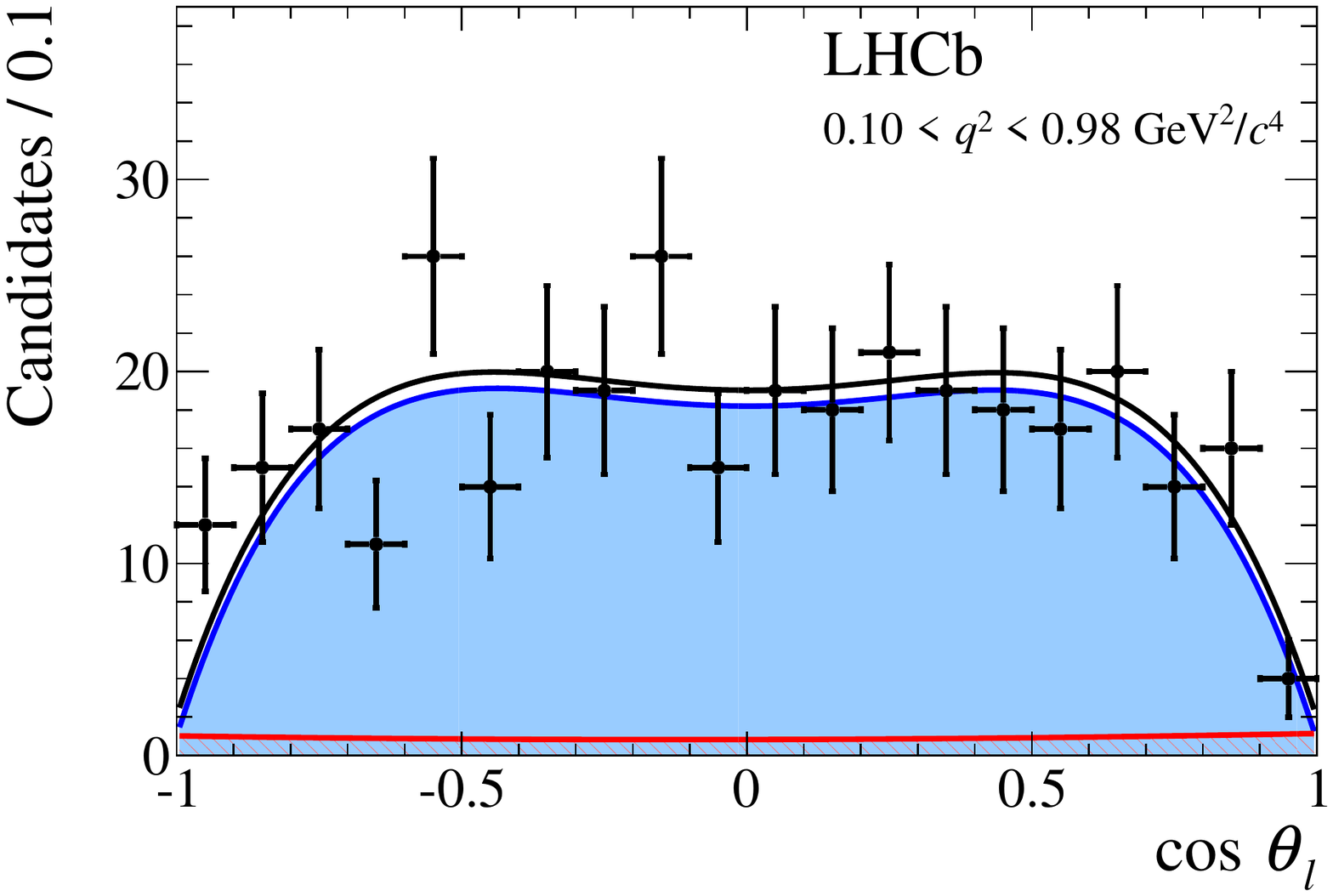} 
\includegraphics[width=0.48\linewidth]{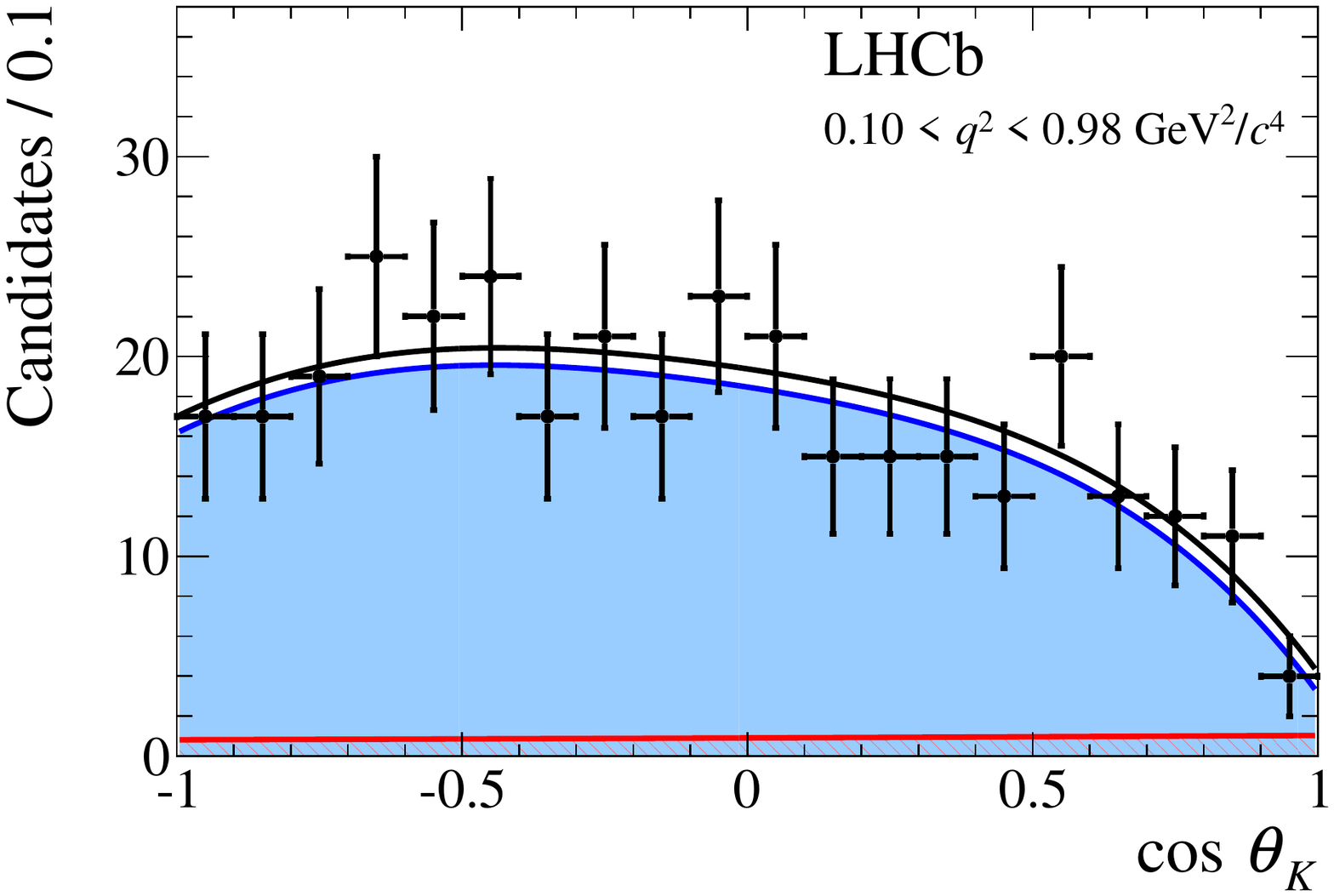} \\
\includegraphics[width=0.48\linewidth]{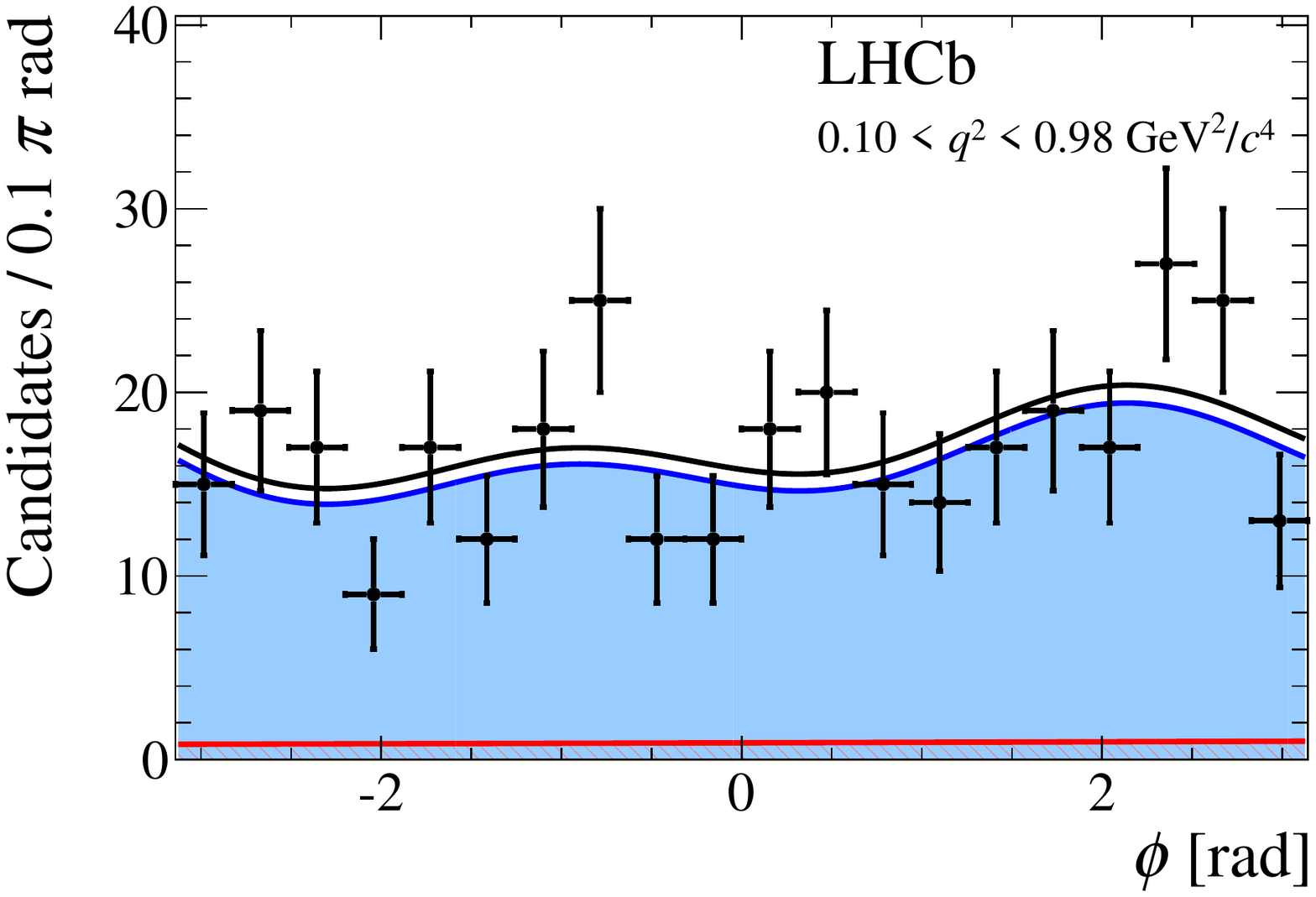}
\end{center}
\vspace*{-0.7cm}
\caption{Angular and mass distributions for $0.10<q^2<0.98\gevgevcccc$.
    The distributions of $m(\Kp\pim)$ and the three decay angles are given for candidates in the signal mass window $\pm50\mevcc$ around the known $\Bd$ mass. 
    Overlaid are the projections of the total fitted distribution (black line) and its different components.
    The signal is shown by the solid blue component and the background by the red hatched component. 
    \label{fig:supp:kstarmumu:bin0}}
\end{figure}

\begin{figure}[htb]
\begin{center} 
\includegraphics[width=0.48\linewidth]{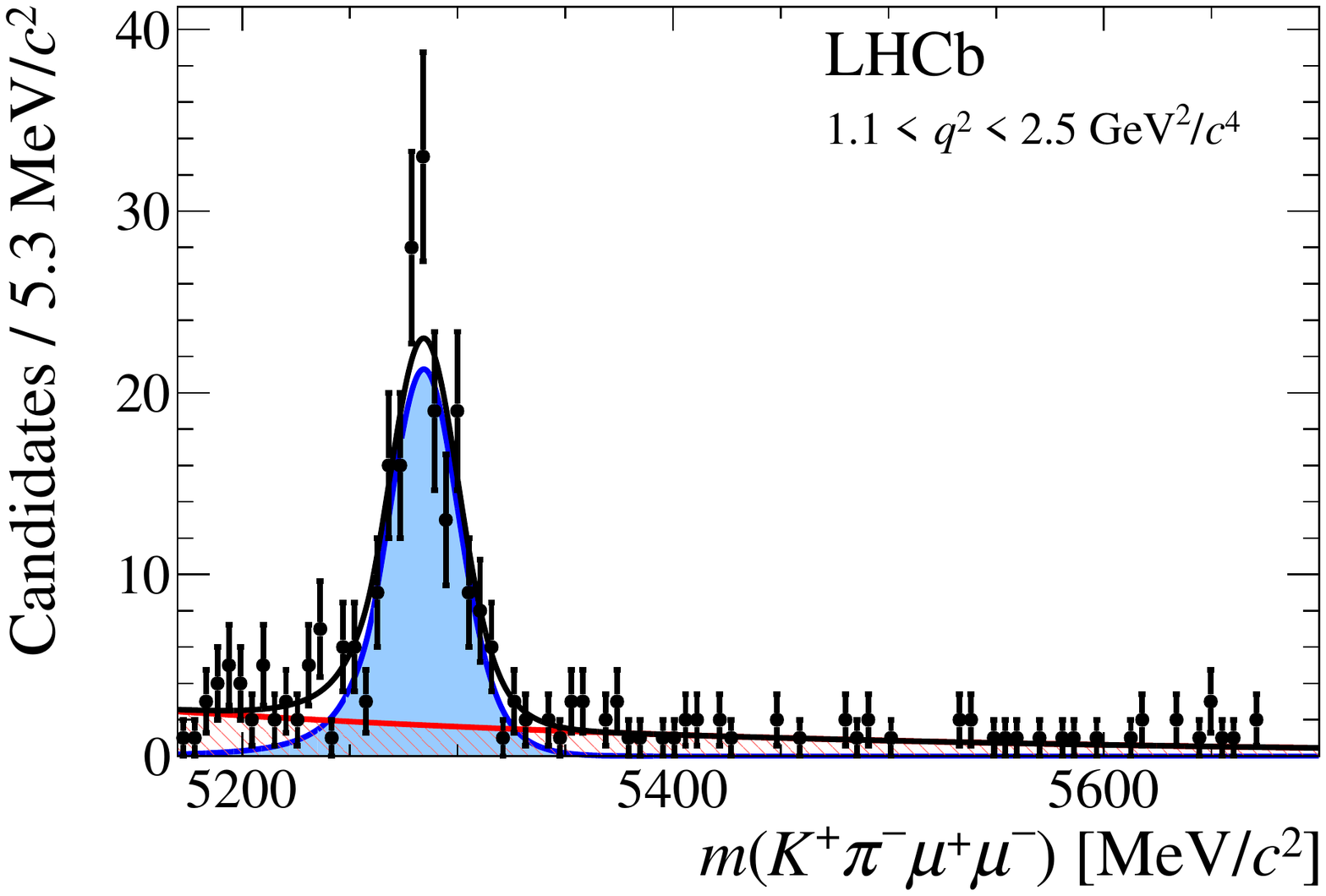} 
\includegraphics[width=0.48\linewidth]{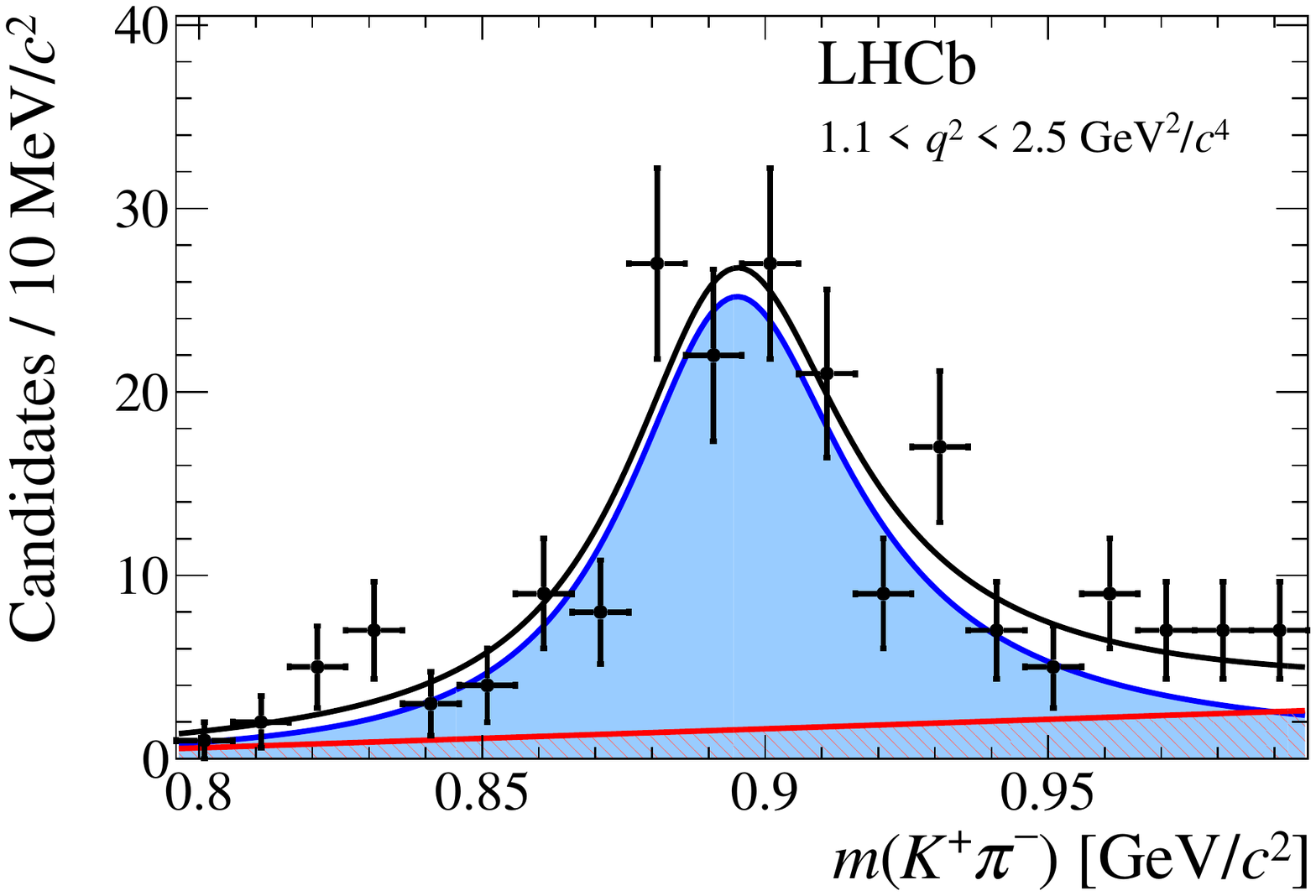}  \\ 
\includegraphics[width=0.48\linewidth]{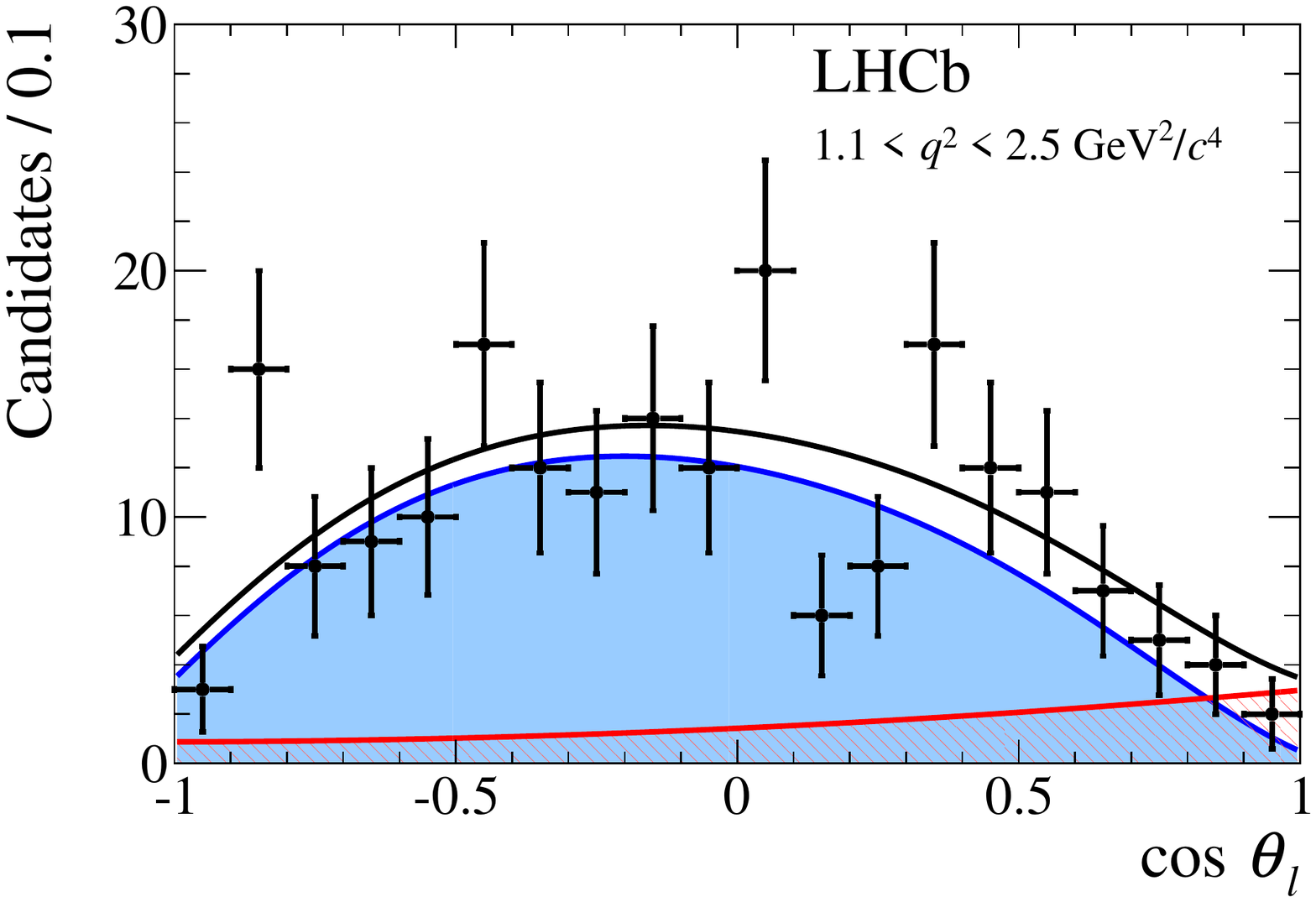} 
\includegraphics[width=0.48\linewidth]{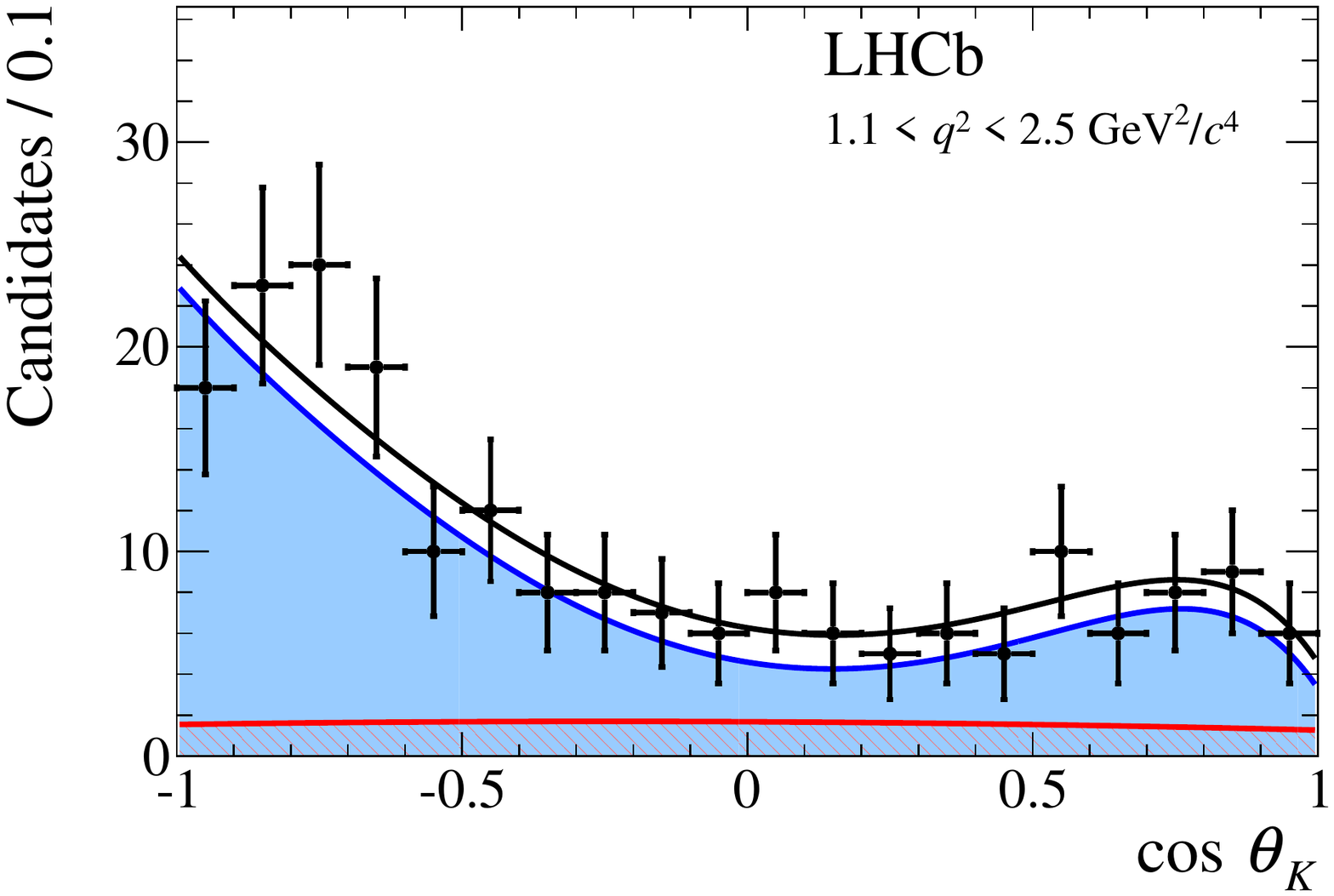} \\
\includegraphics[width=0.48\linewidth]{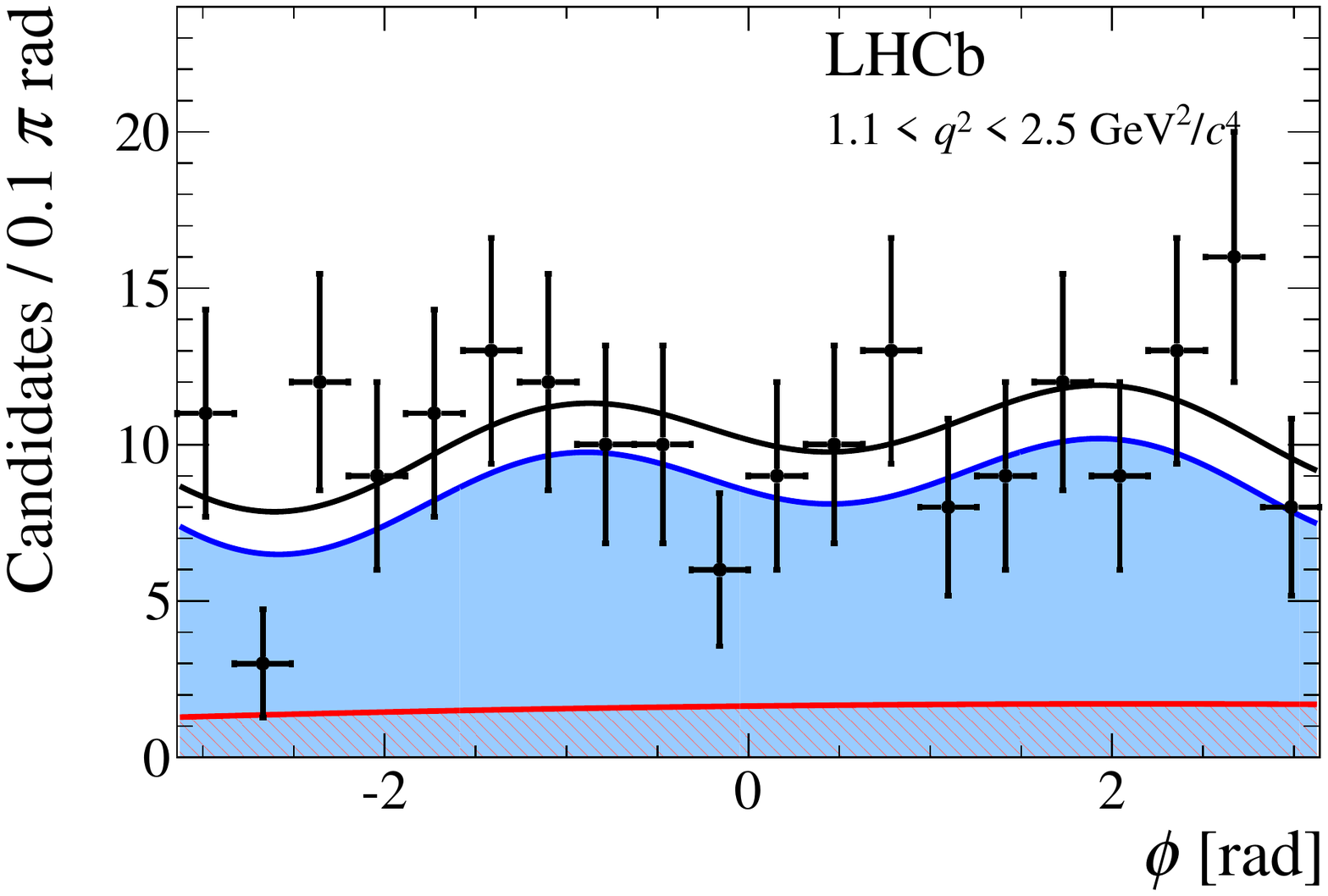}
\end{center}
\vspace*{-0.7cm} 
\caption{Angular and mass distributions for $1.1<q^2<2.5\gevgevcccc$.
    The distributions of $m(\Kp\pim)$ and the three decay angles are given for candidates in the signal mass window $\pm50\mevcc$ around the known $\Bd$ mass. 
    Overlaid are the projections of the total fitted distribution (black line) and its different components.
    The signal is shown by the solid blue component and the background by the red hatched component. 
    \label{fig:supp:kstarmumu:bin1}}
\end{figure}

\begin{figure}[htb]
\begin{center}
\includegraphics[width=0.48\linewidth]{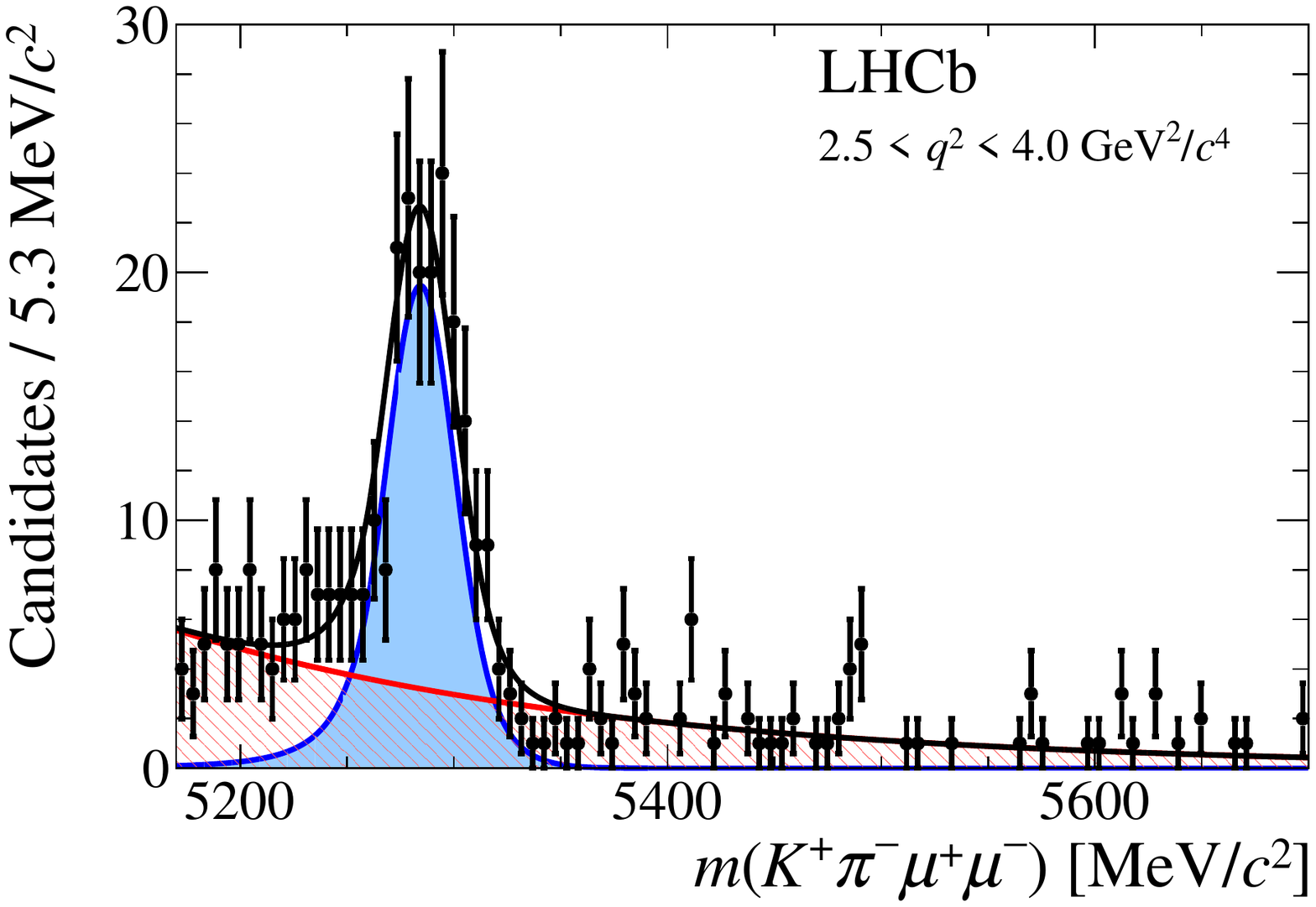} 
\includegraphics[width=0.48\linewidth]{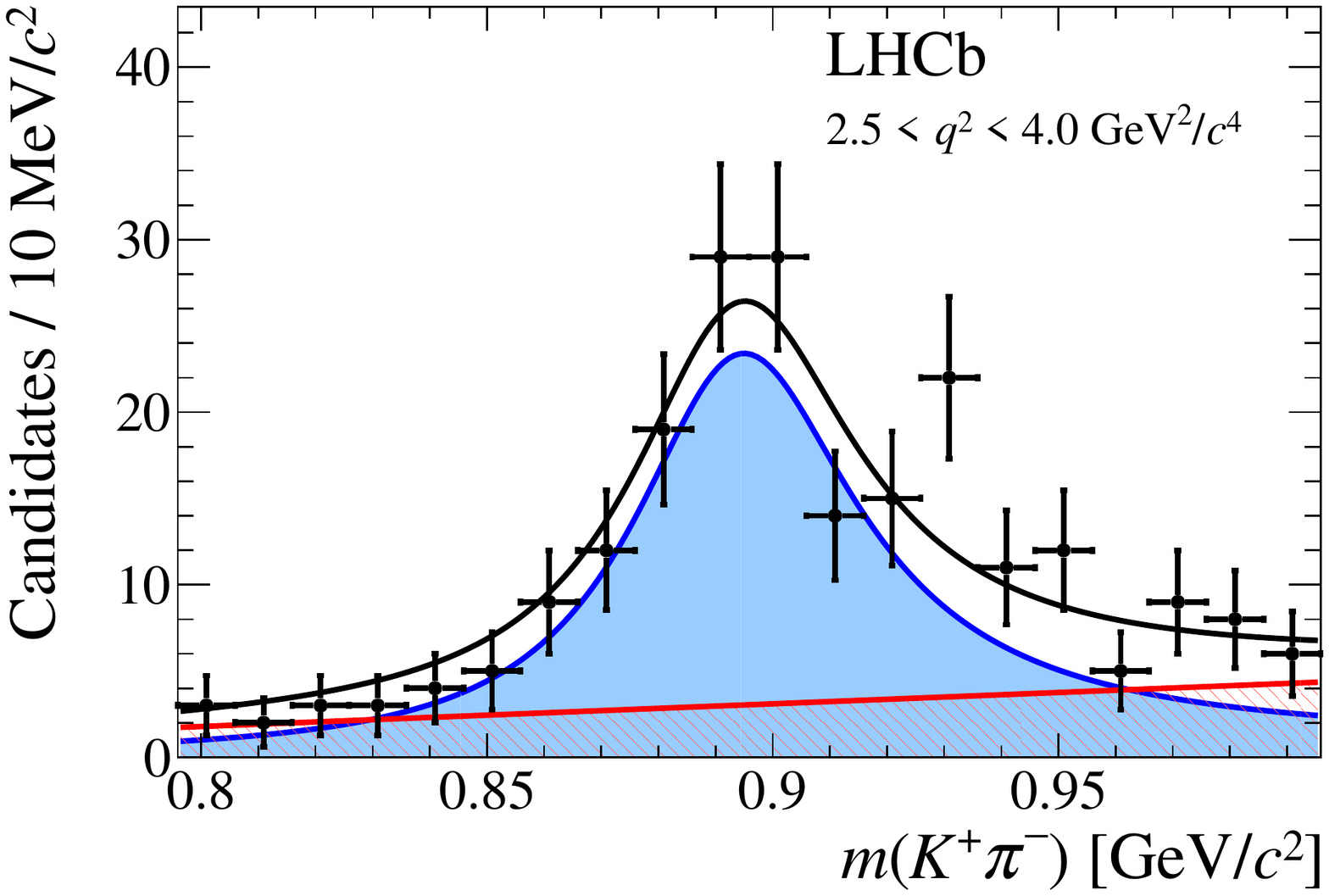}  \\ 
\includegraphics[width=0.48\linewidth]{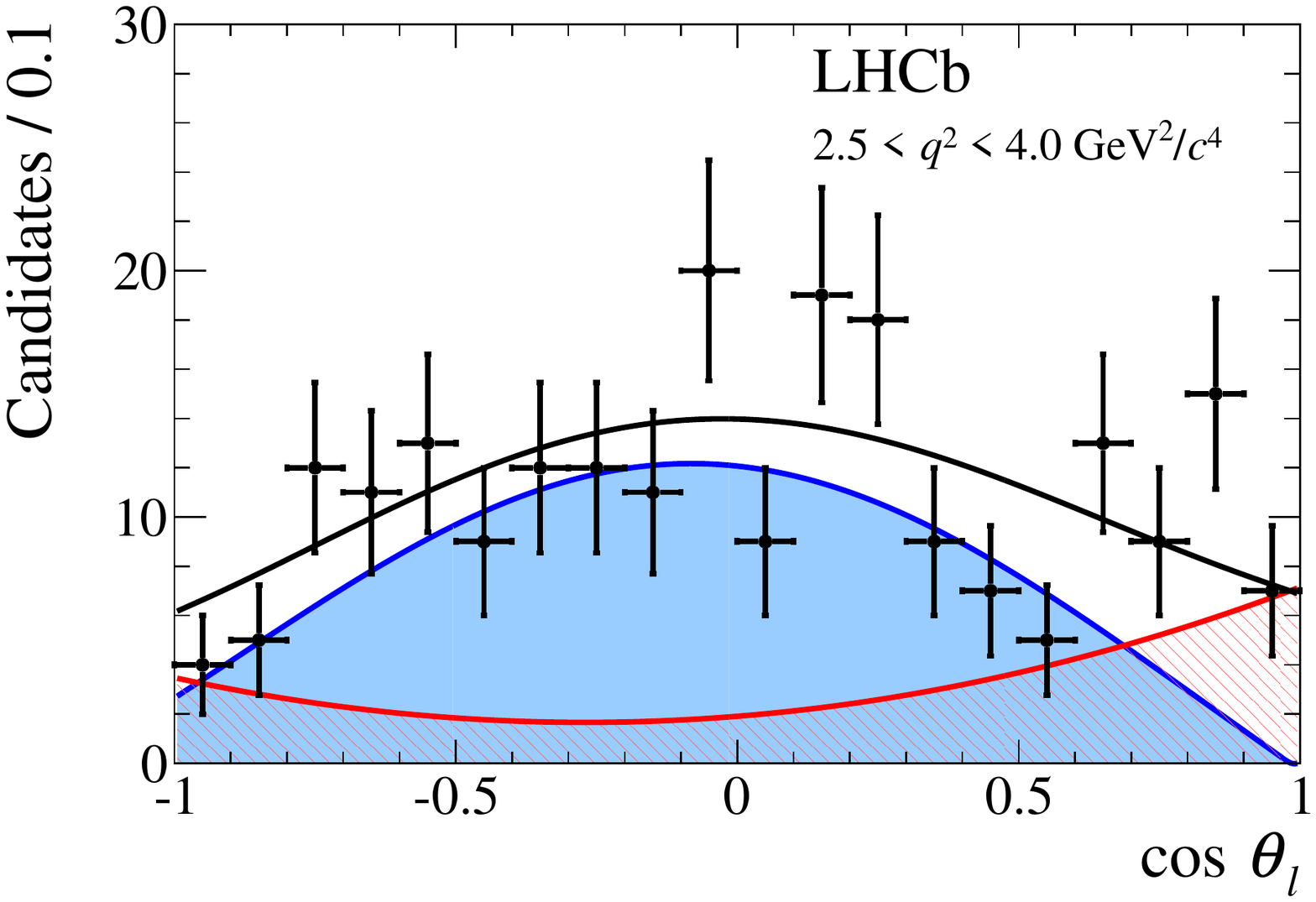} 
\includegraphics[width=0.48\linewidth]{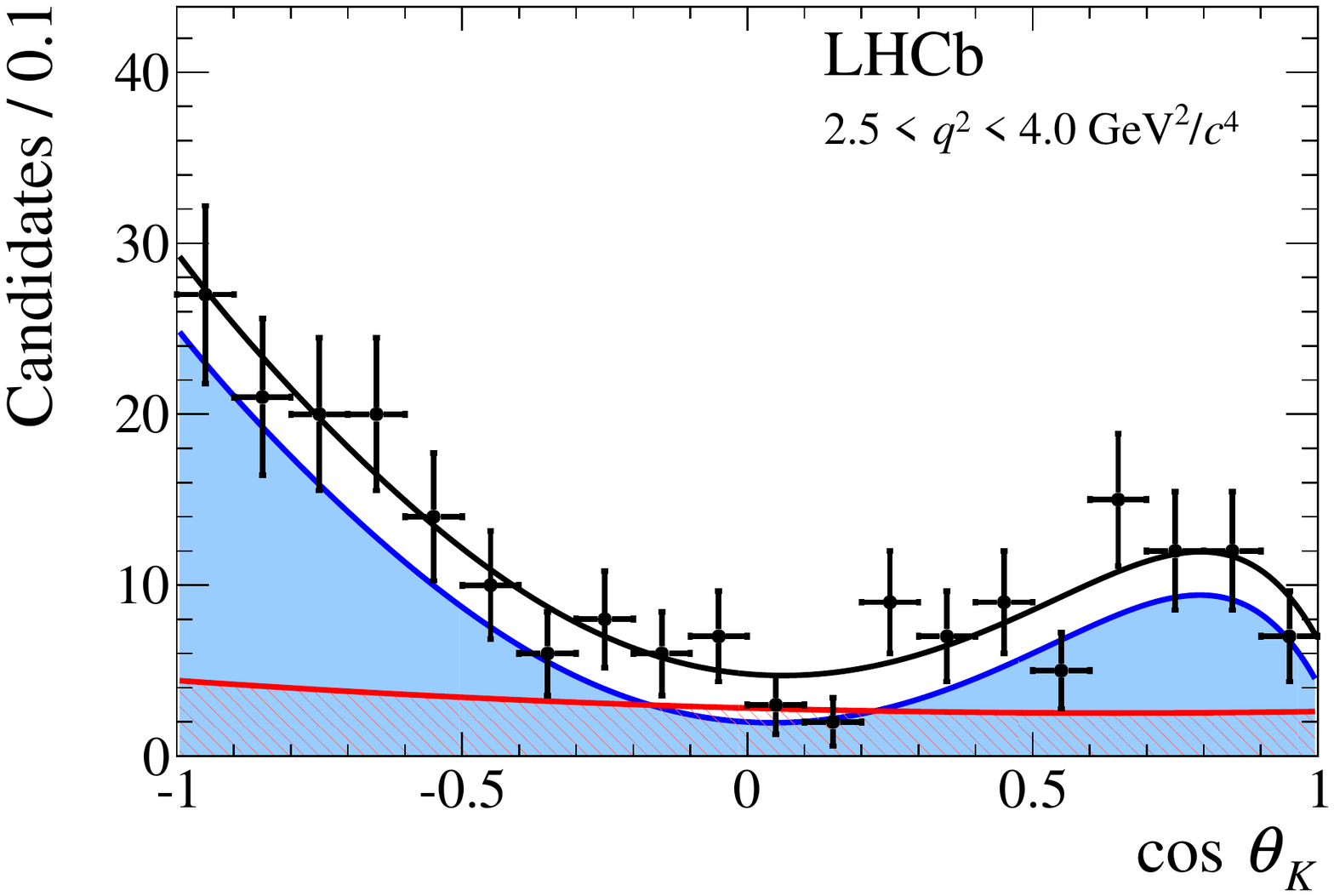} \\
\includegraphics[width=0.48\linewidth]{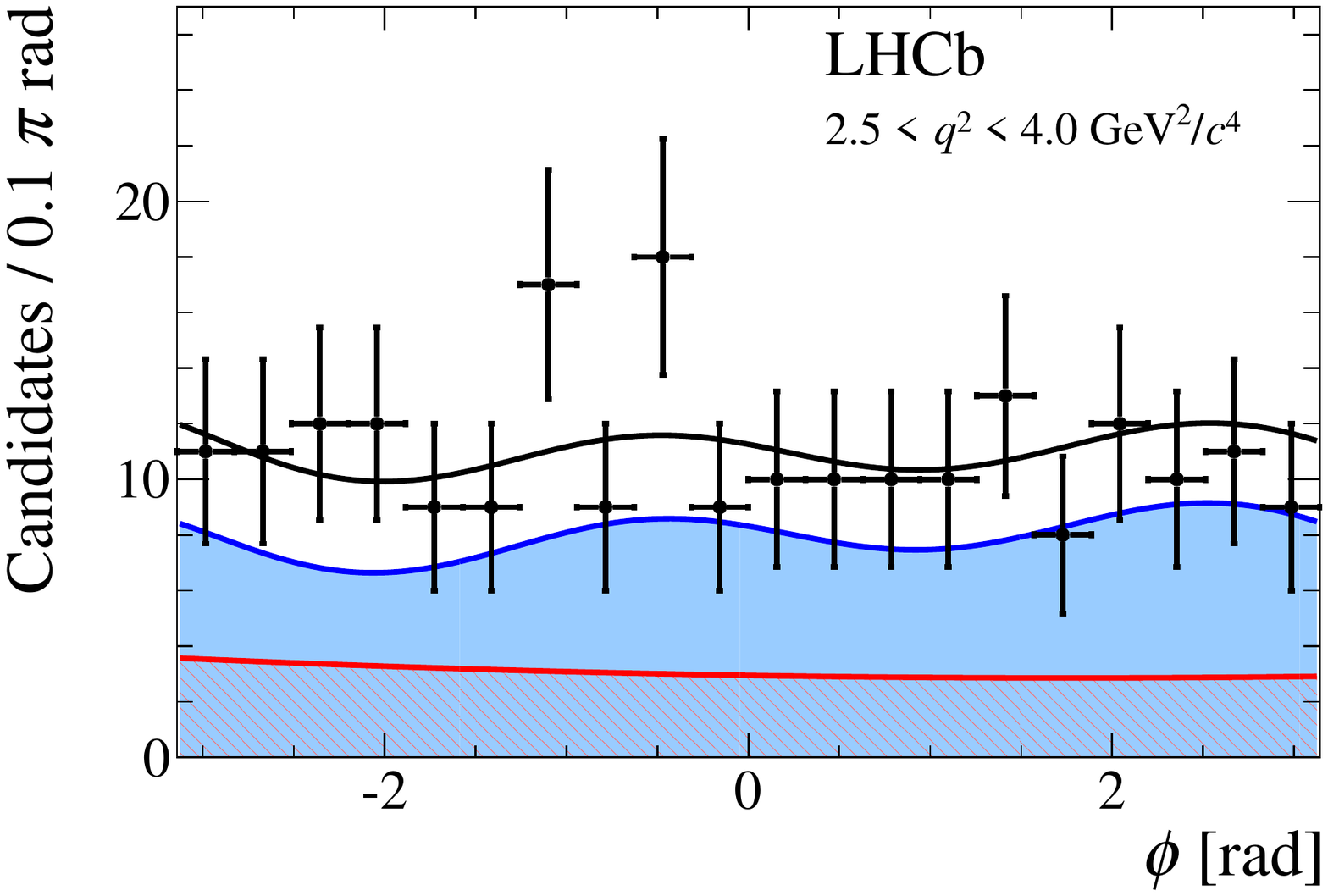}
\end{center}
\vspace*{-0.7cm}
\caption{Angular and mass distributions for $2.5<q^2<4.0\gevgevcccc$.
    The distributions of $m(\Kp\pim)$ and the three decay angles are given for candidates in the signal mass window $\pm50\mevcc$ around the known $\Bd$ mass. 
    Overlaid are the projections of the total fitted distribution (black line) and its different components.
    The signal is shown by the solid blue component and the background by the red hatched component. 
     \label{fig:supp:kstarmumu:bin2}}
\end{figure}

\begin{figure}[htb]
\begin{center}
\includegraphics[width=0.48\linewidth]{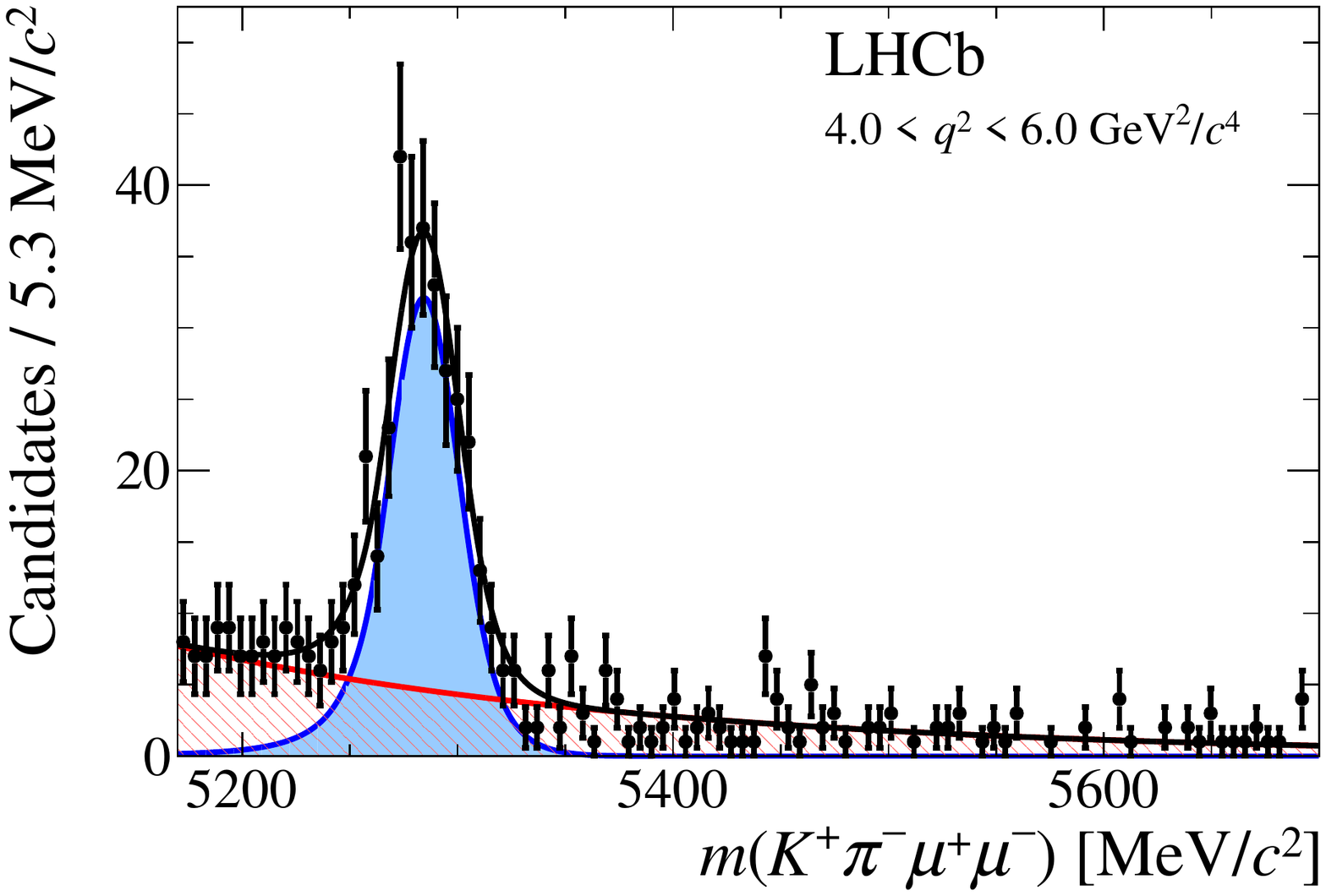} 
\includegraphics[width=0.48\linewidth]{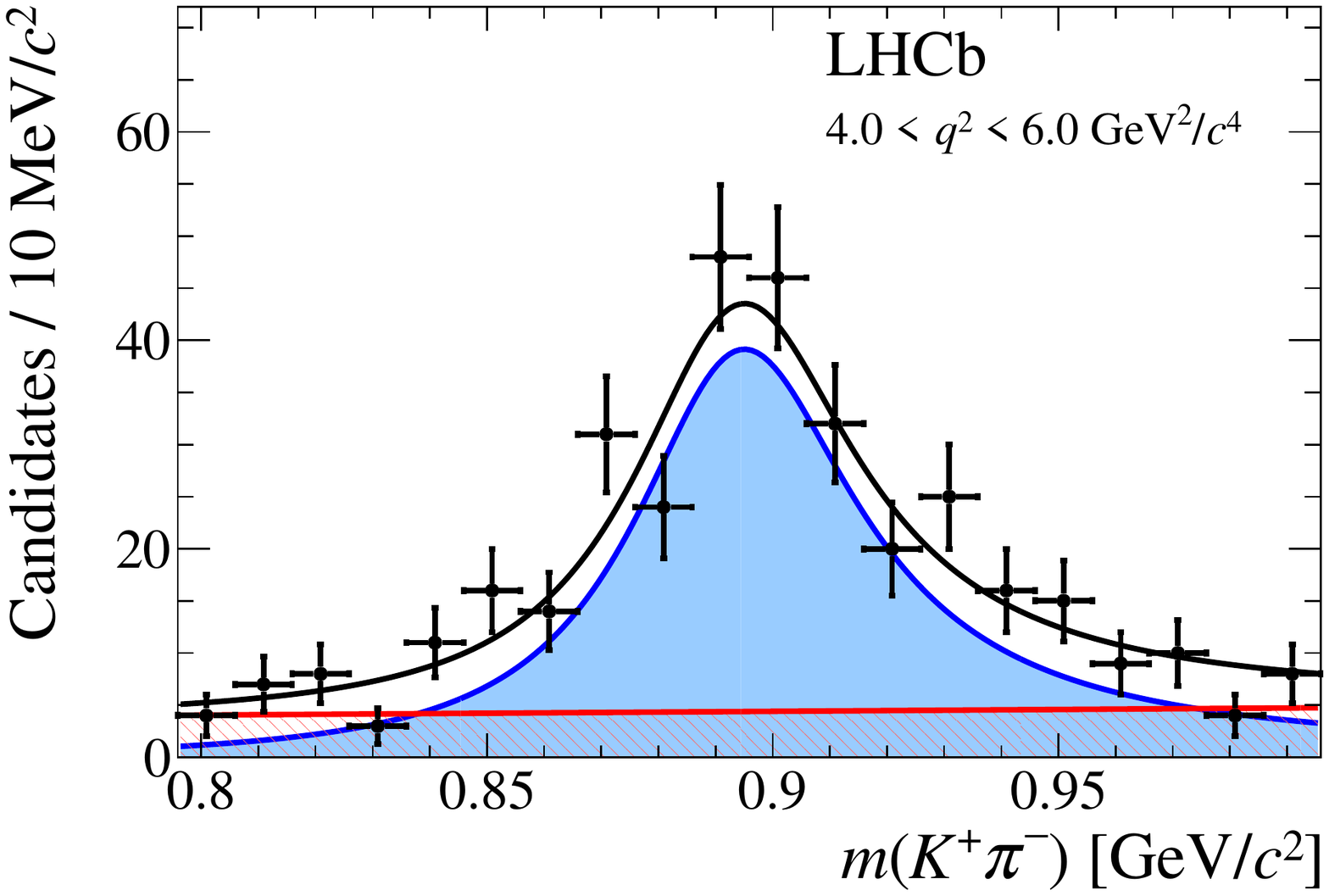}  \\
\includegraphics[width=0.48\linewidth]{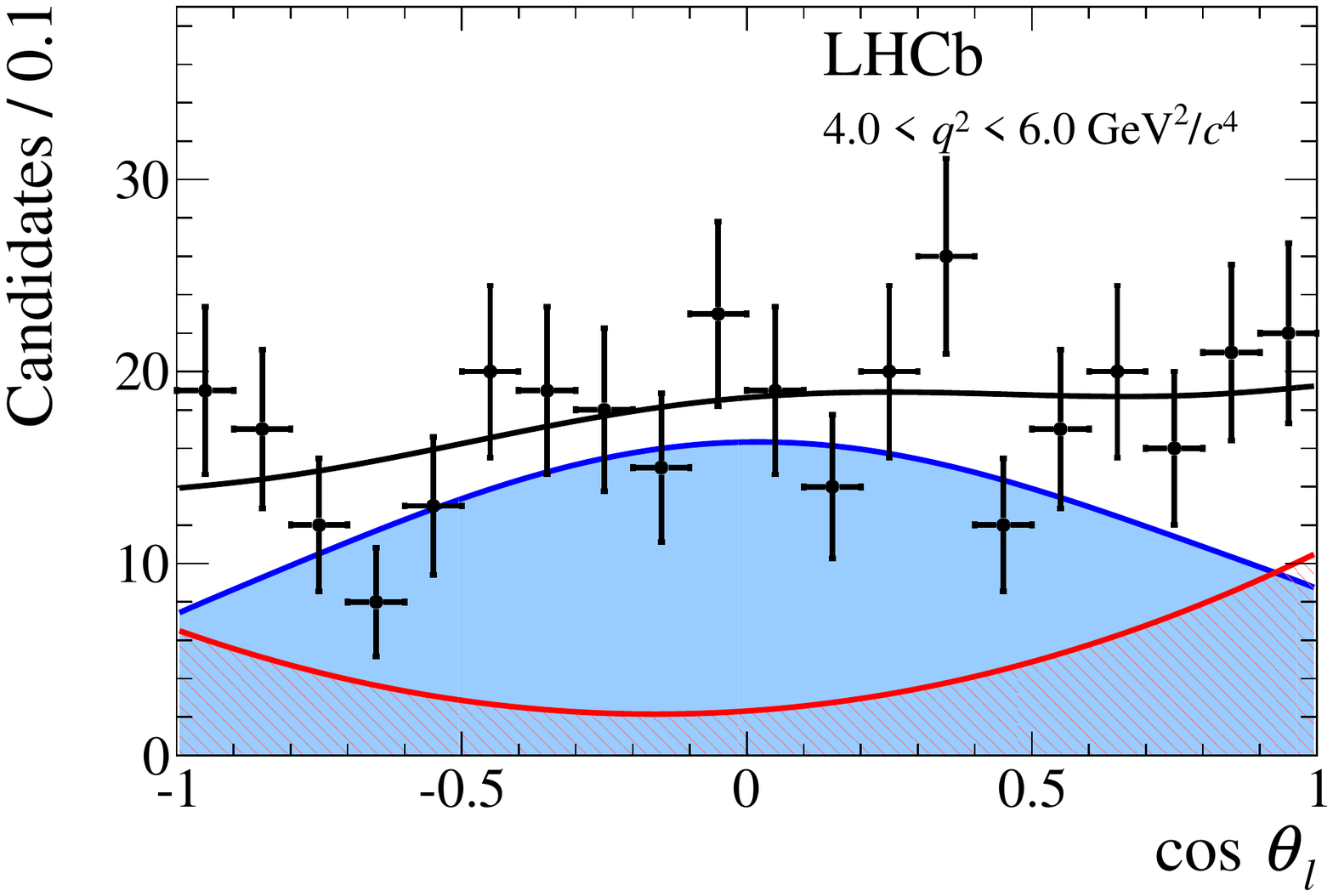} 
\includegraphics[width=0.48\linewidth]{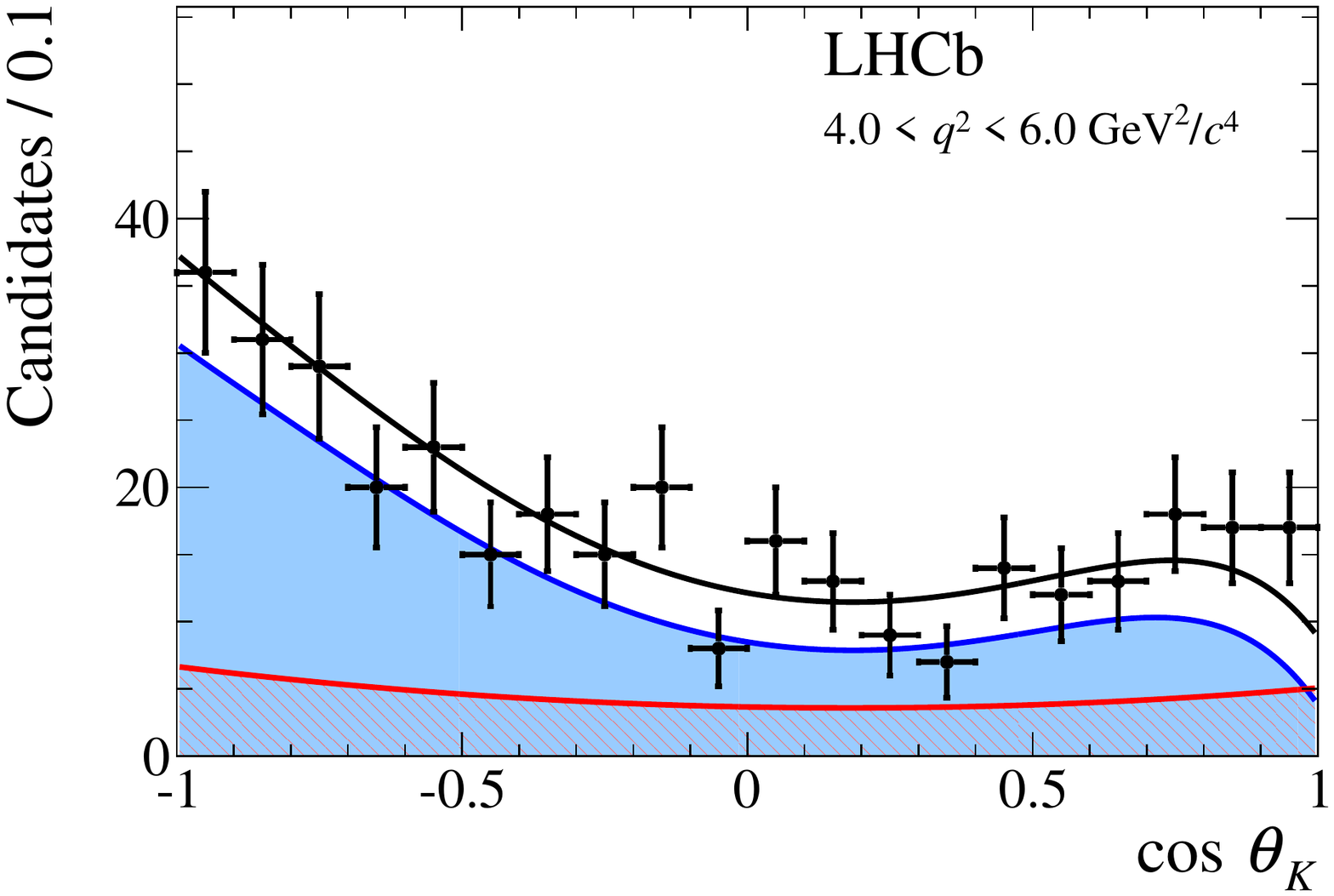} \\
\includegraphics[width=0.48\linewidth]{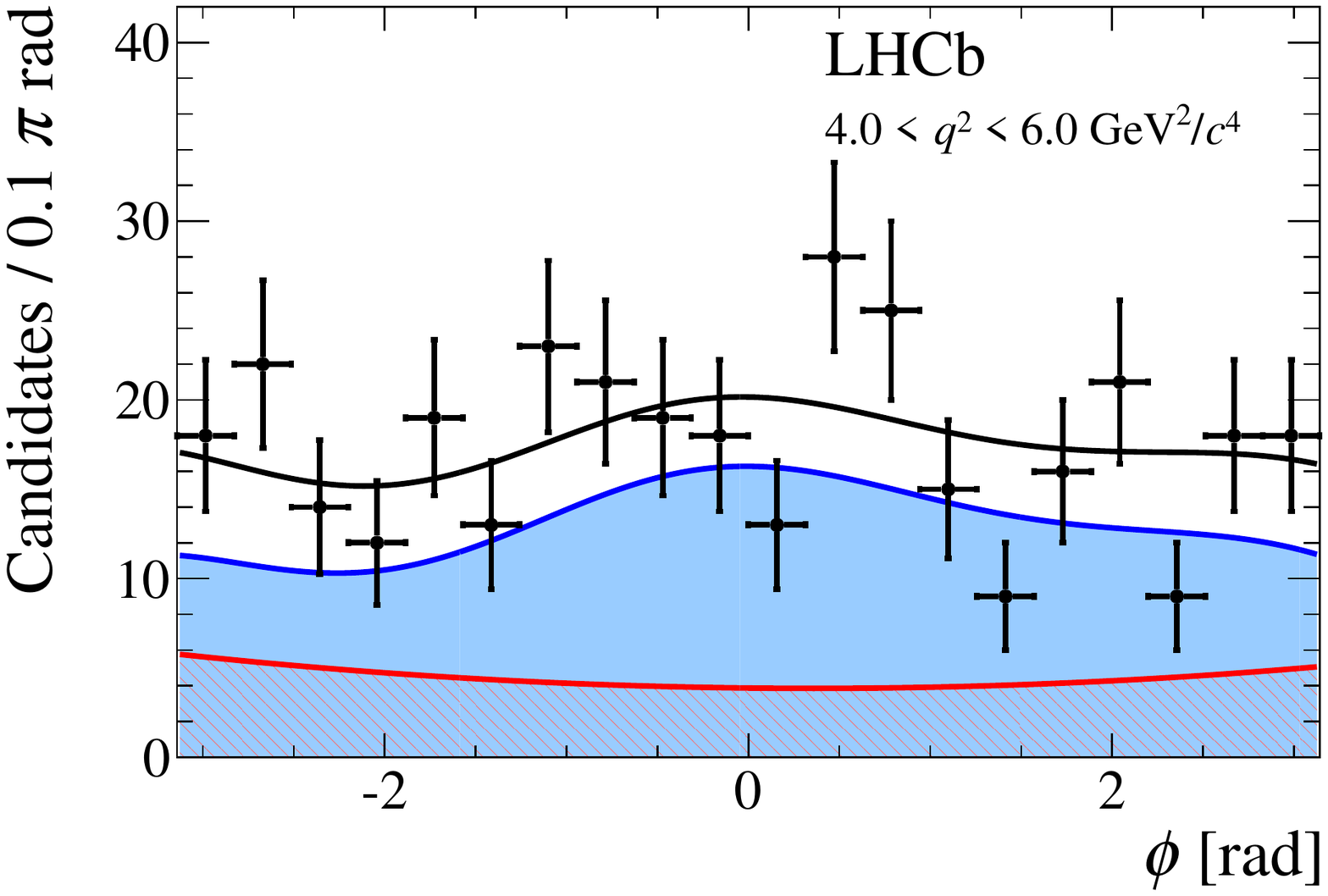}
\end{center}
\vspace*{-0.7cm}
\caption{Angular and mass distributions for $4.0<q^2<6.0\gevgevcccc$.
   The distributions of $m(\Kp\pim)$ and the three decay angles are given for candidates in the signal mass window $\pm50\mevcc$ around the known $\Bd$ mass. 
    Overlaid are the projections of the total fitted distribution (black line) and its different components.
    The signal is shown by the solid blue component and the background by the red hatched component. 
    \label{fig:supp:kstarmumu:bin3}}
\end{figure}

\begin{figure}[htb]
\begin{center}
\includegraphics[width=0.48\linewidth]{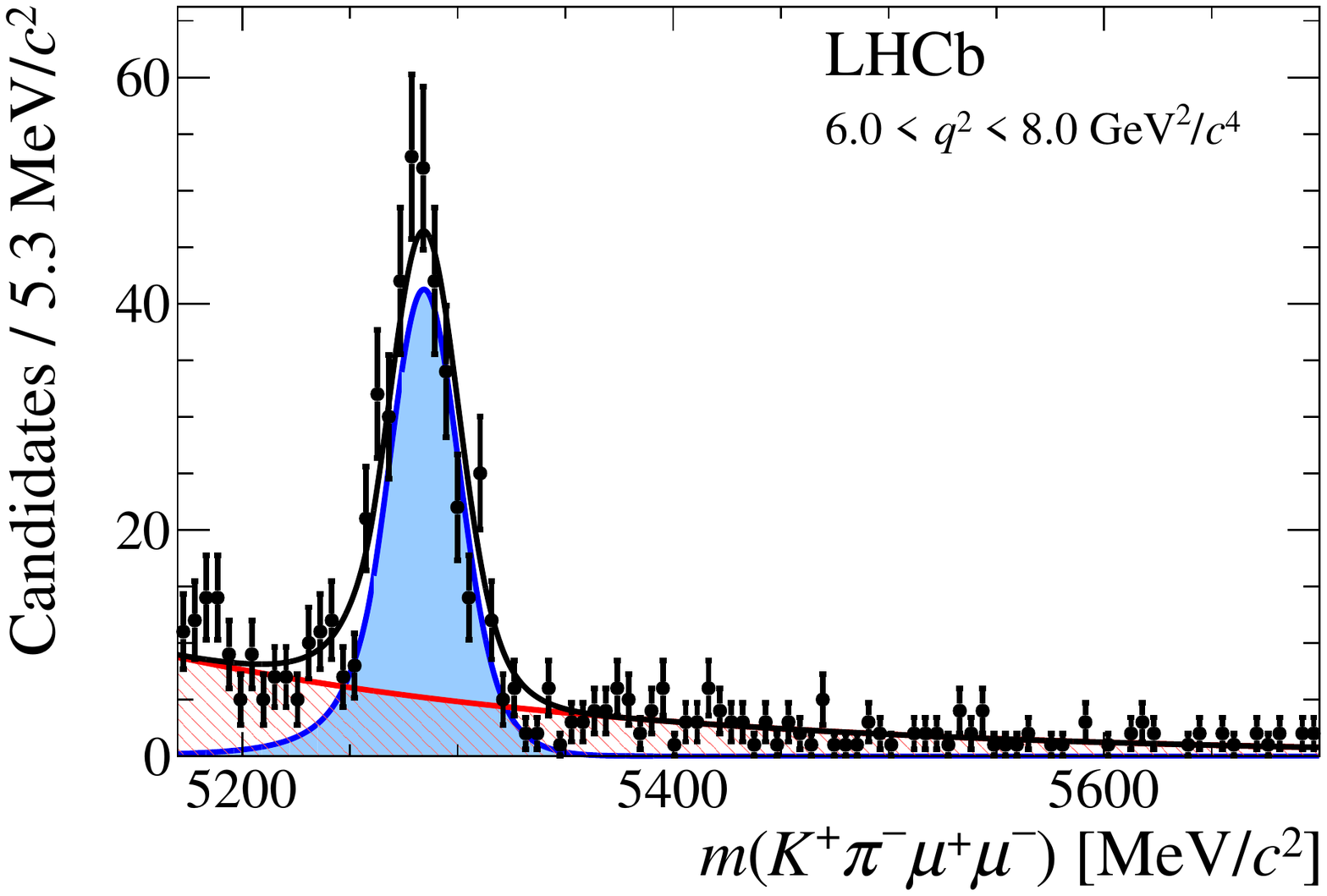} 
\includegraphics[width=0.48\linewidth]{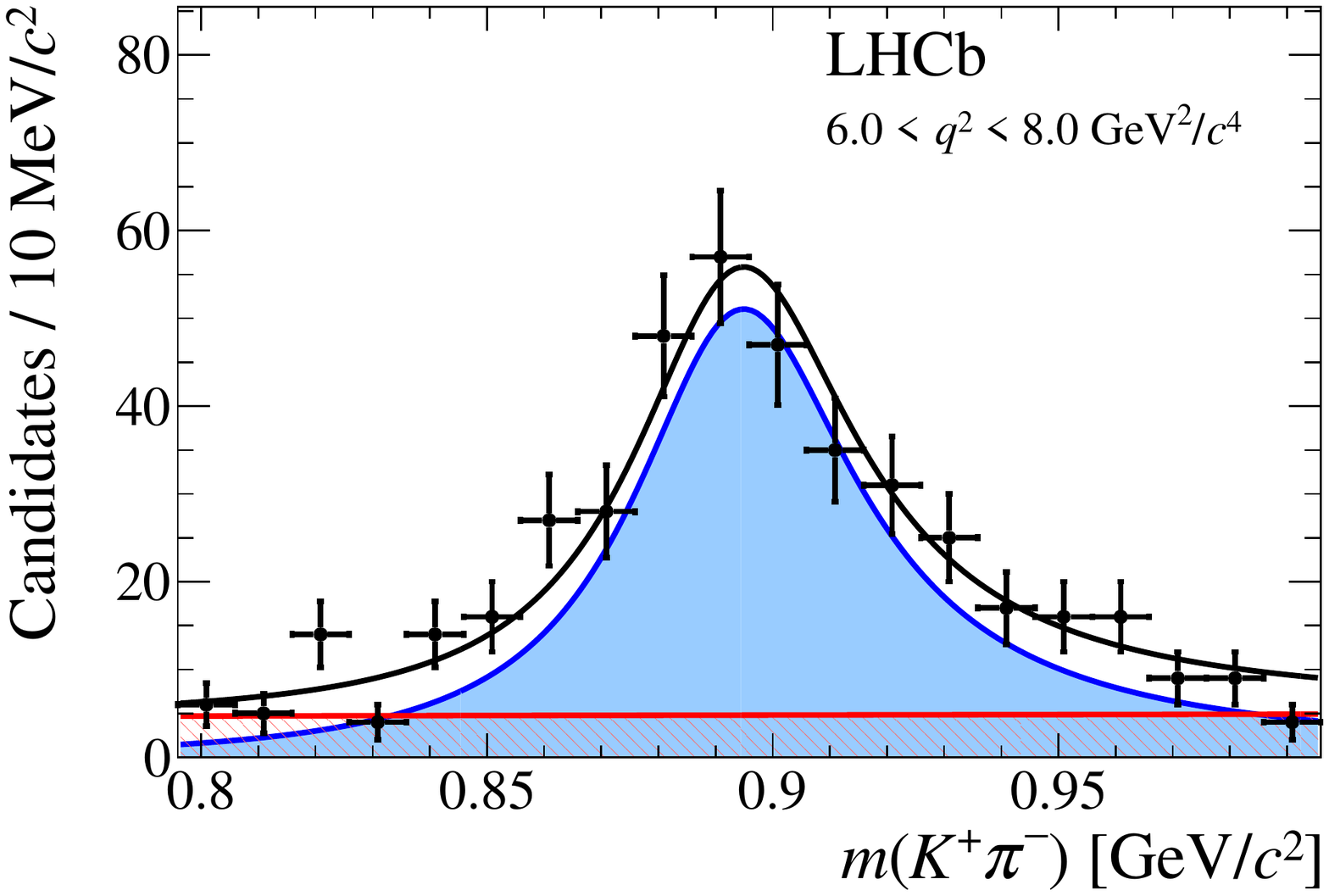} \\
\includegraphics[width=0.48\linewidth]{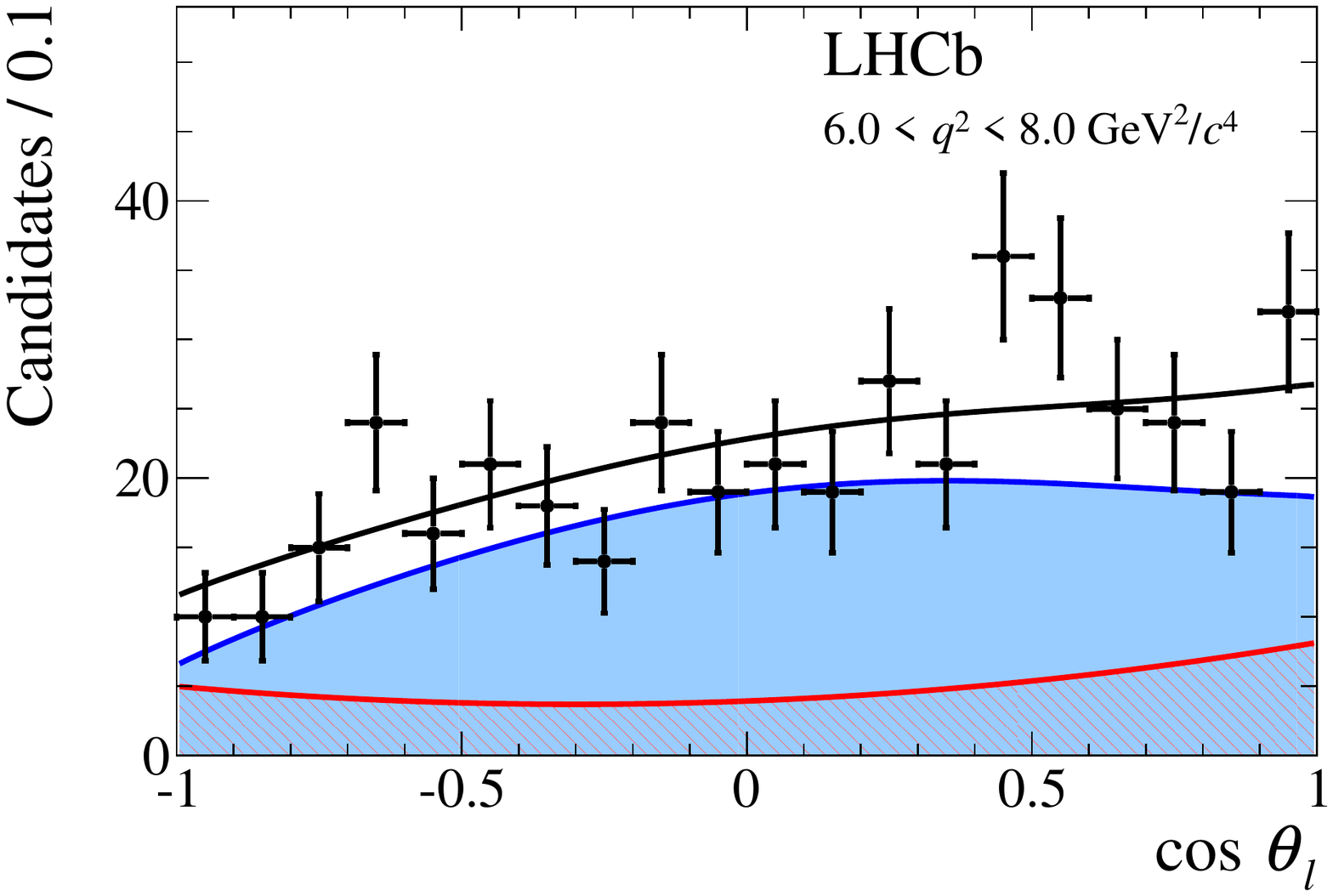} 
\includegraphics[width=0.48\linewidth]{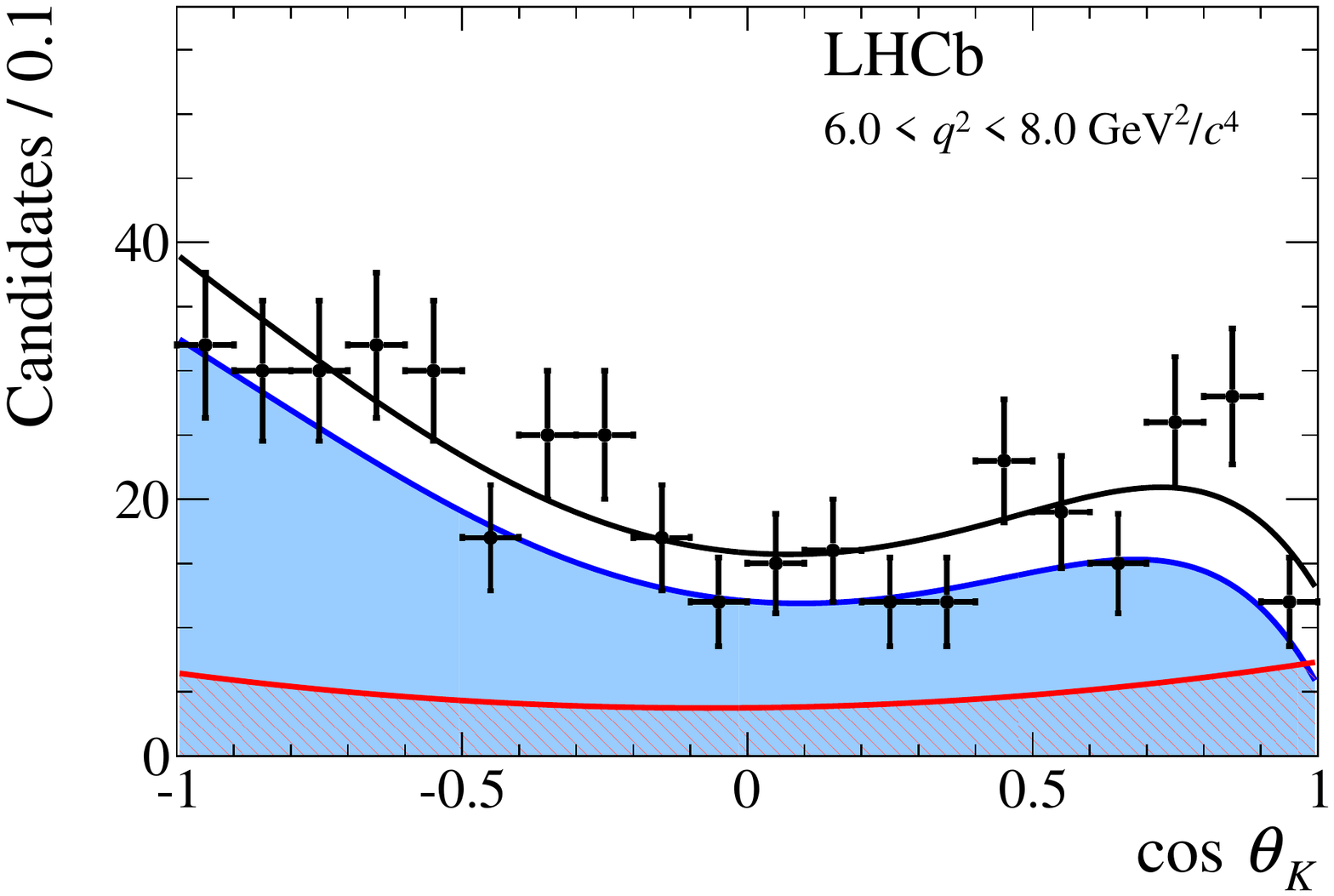} \\
\includegraphics[width=0.48\linewidth]{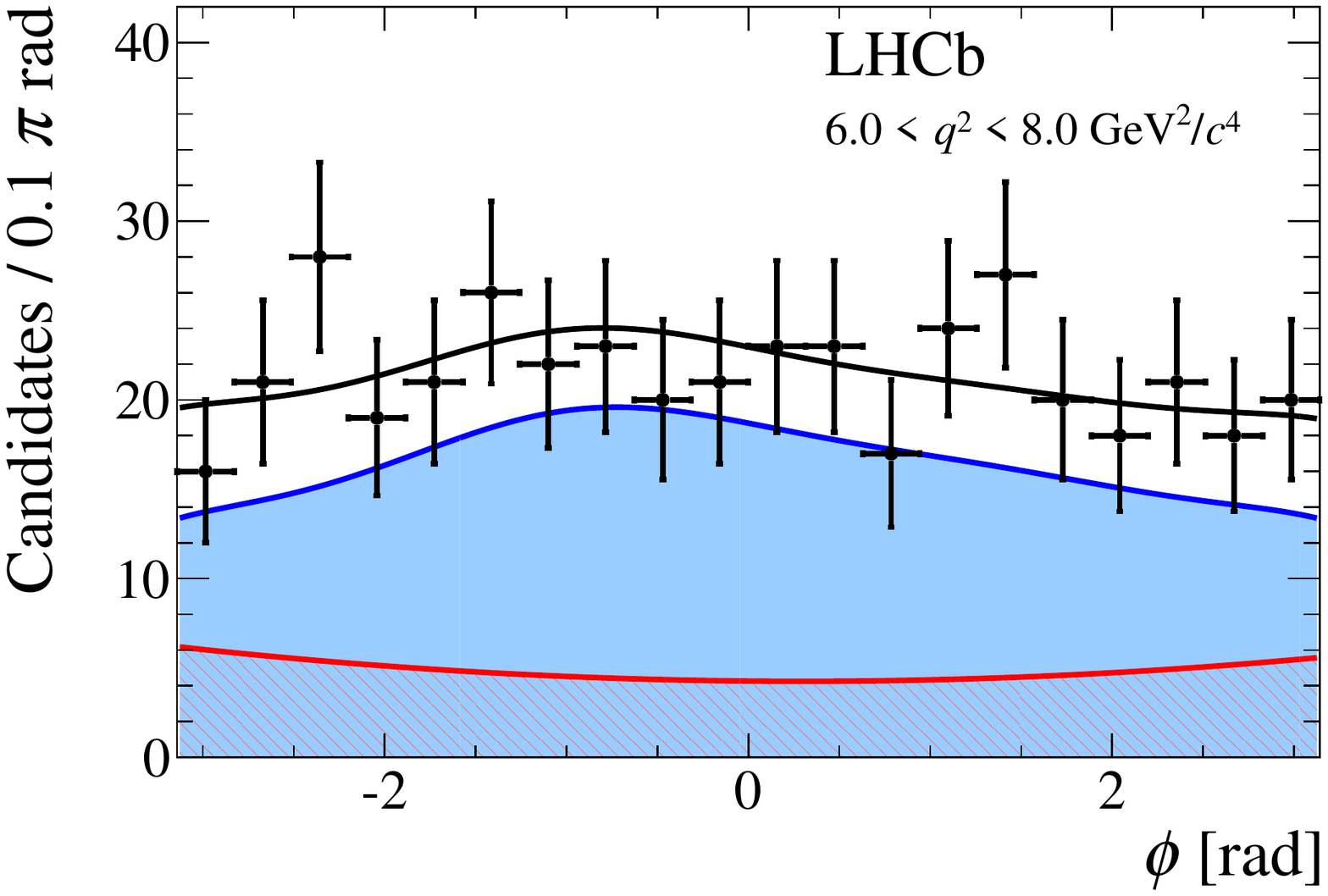}
\end{center}
\vspace*{-0.7cm}
\caption{Angular and mass distributions for $6.0<q^2<8.0\gevgevcccc$.
     The distributions of $m(\Kp\pim)$ and the three decay angles are given for candidates in the signal mass window $\pm50\mevcc$ around the known $\Bd$ mass. 
    Overlaid are the projections of the total fitted distribution (black line) and its different components.
    The signal is shown by the blue shaded area and the background by the red hatched area.
    \label{fig:supp:kstarmumu:bin4}}
\end{figure}

\begin{figure}[htb]
\begin{center}
\includegraphics[width=0.48\linewidth]{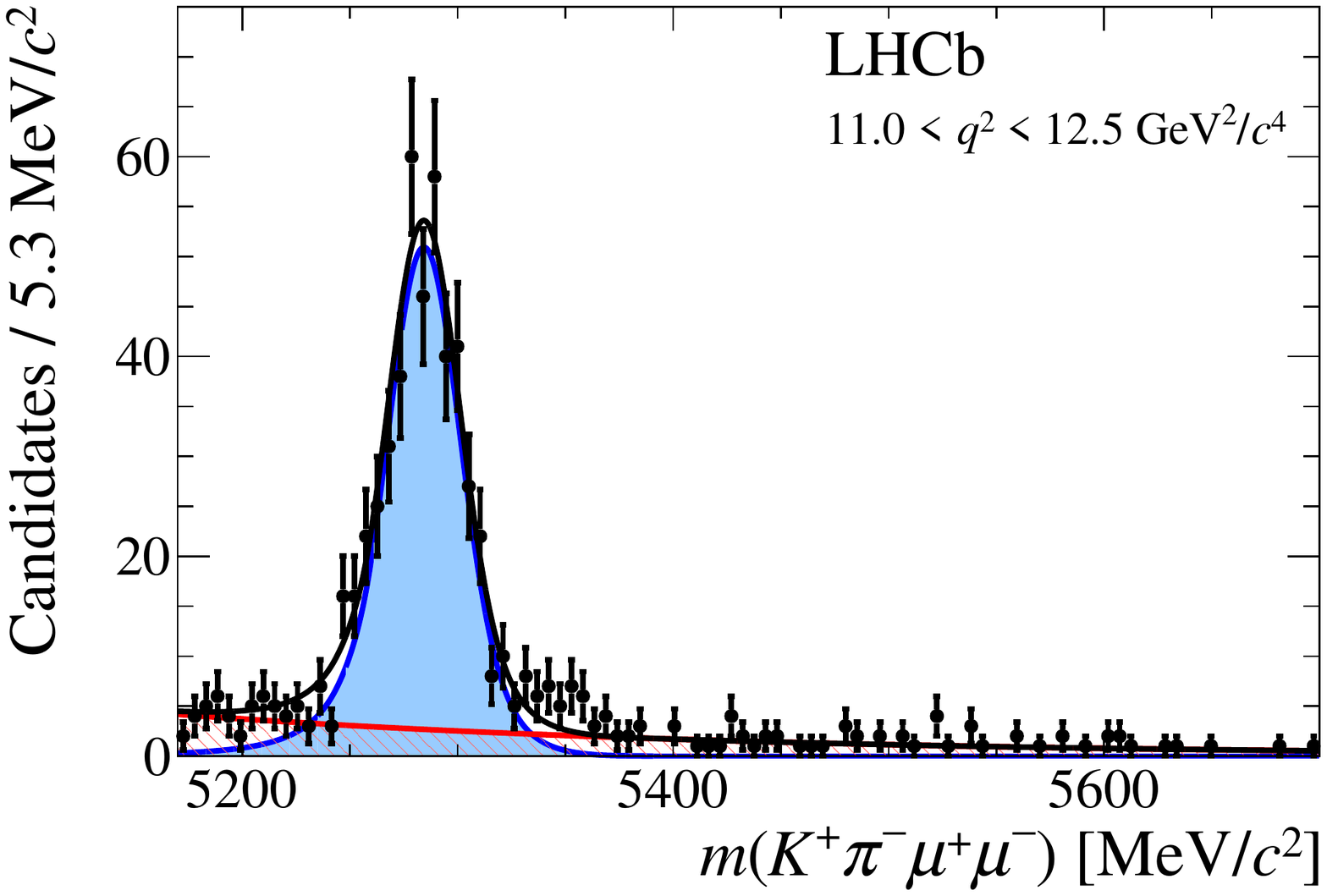} 
\includegraphics[width=0.48\linewidth]{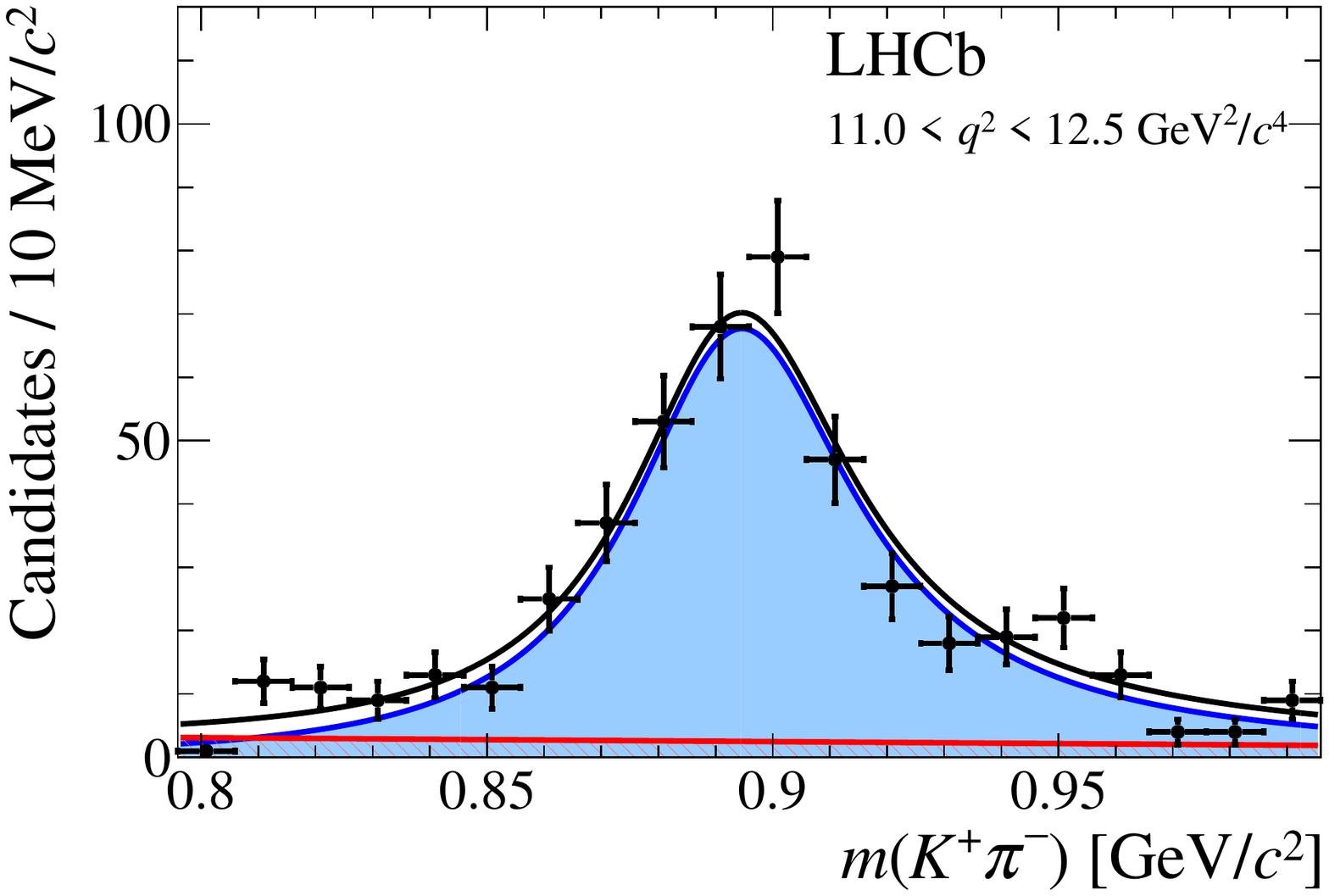} \\
\includegraphics[width=0.48\linewidth]{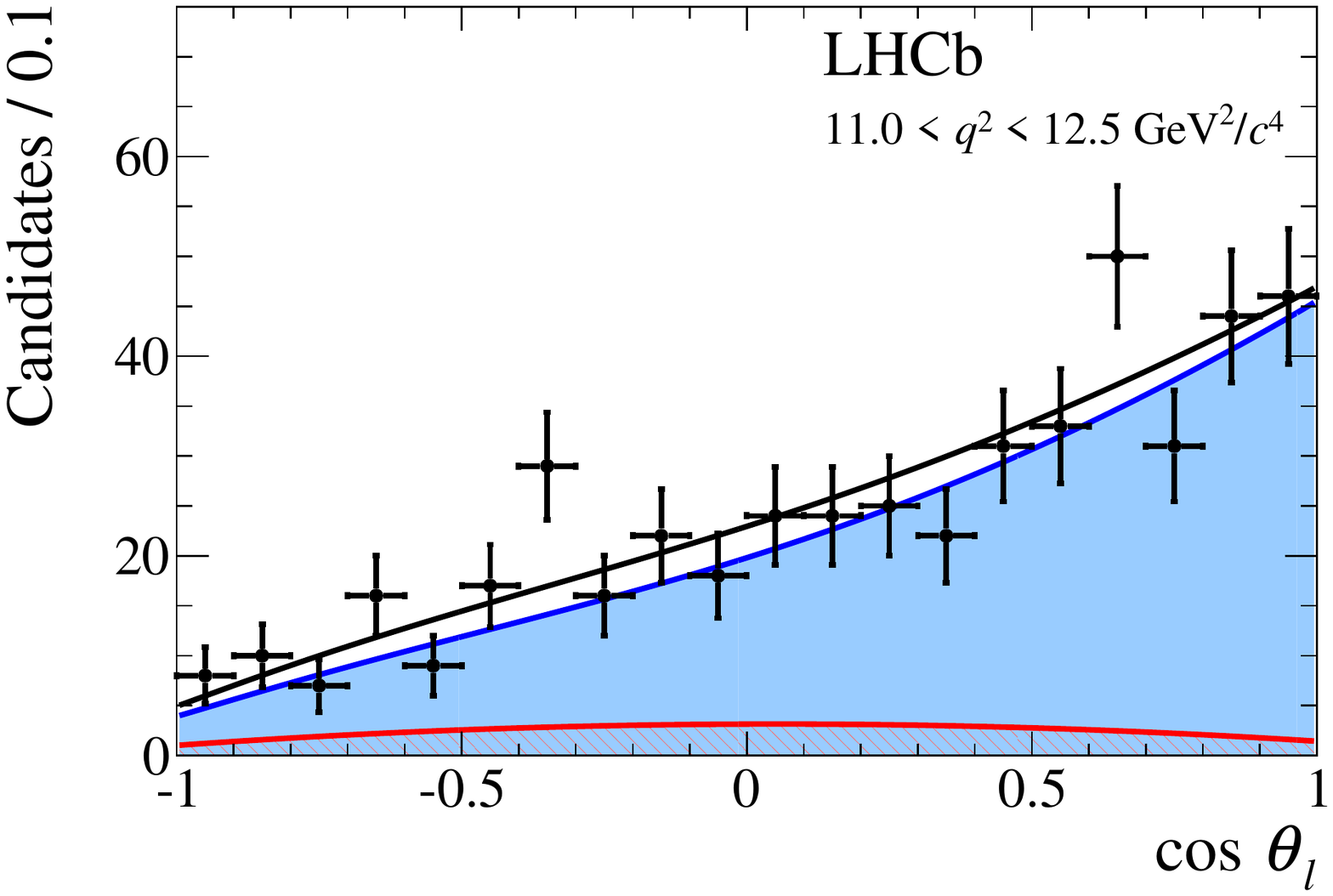}
\includegraphics[width=0.48\linewidth]{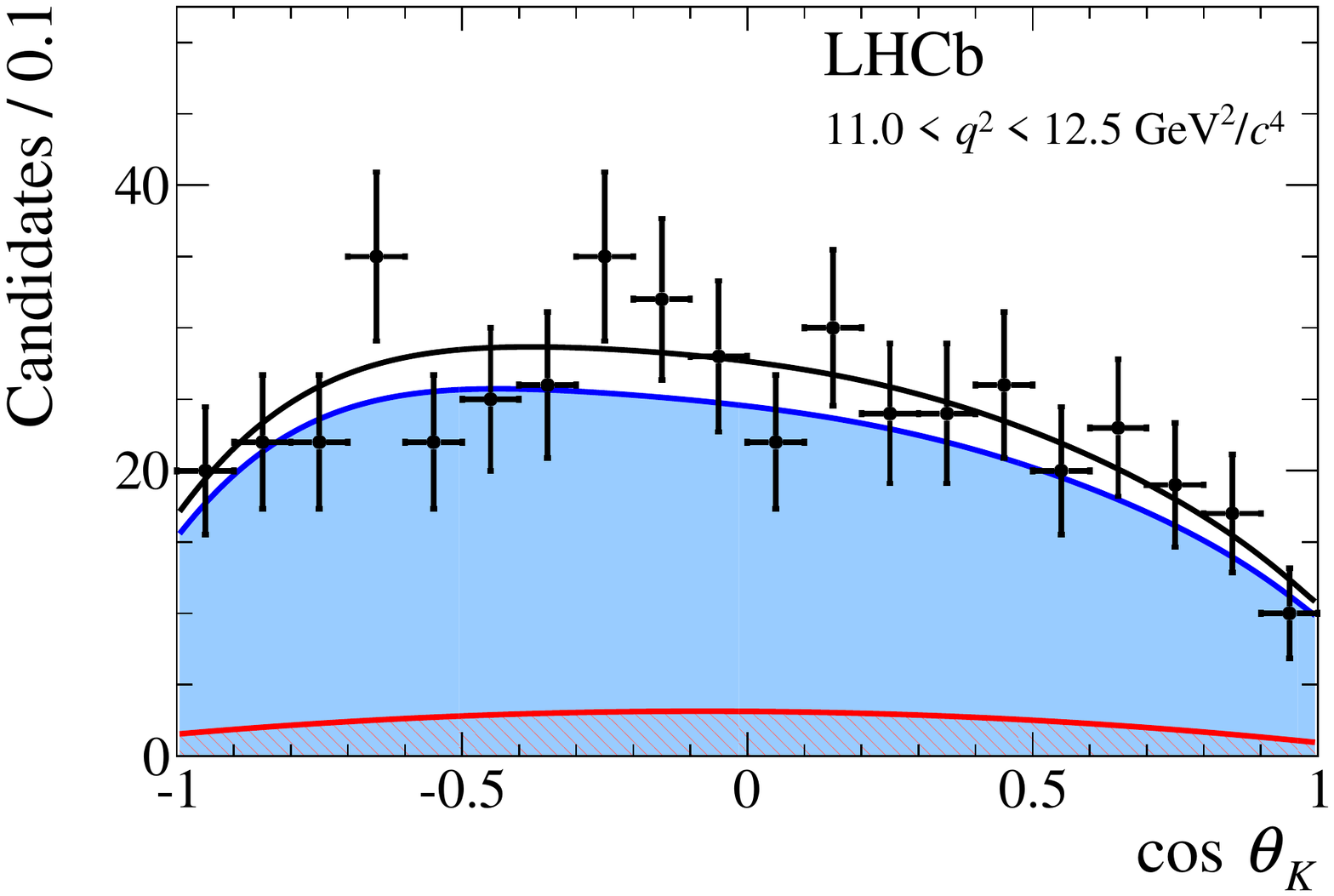} \\  
\includegraphics[width=0.48\linewidth]{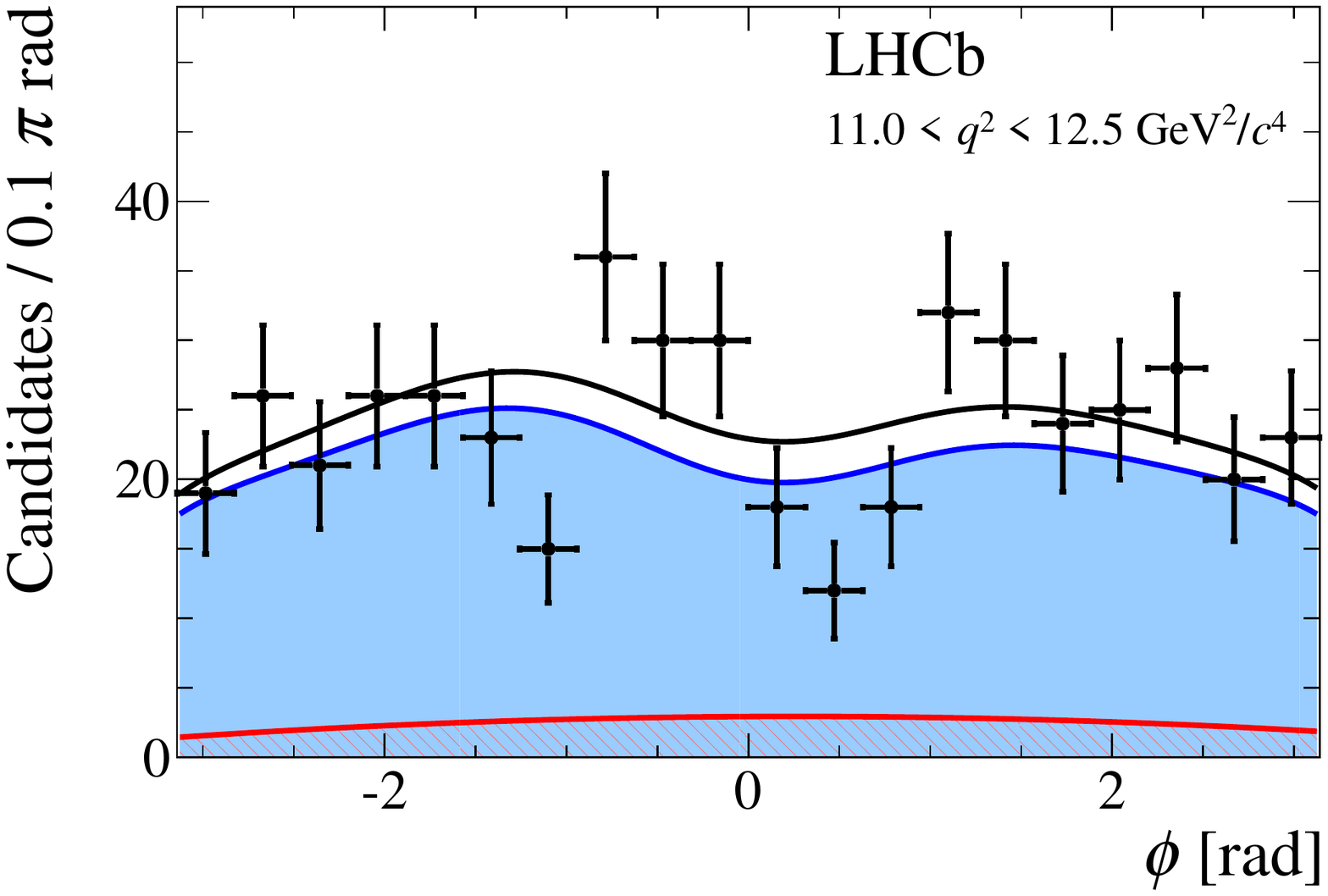}
\end{center}
\vspace*{-0.7cm}
\caption{Angular and mass distributions for $11.0<q^2<12.5\gevgevcccc$.
    The distributions of $m(\Kp\pim)$ and the three decay angles are given for candidates in the signal mass window $\pm50\mevcc$ around the known $\Bd$ mass. 
    Overlaid are the projections of the total fitted distribution (black line) and its different components.
    The signal is shown by the blue shaded area and the background by the red hatched area.
    \label{fig:supp:kstarmumu:bin5}}
\end{figure}

\begin{figure}[htb]
\begin{center}
\includegraphics[width=0.48\linewidth]{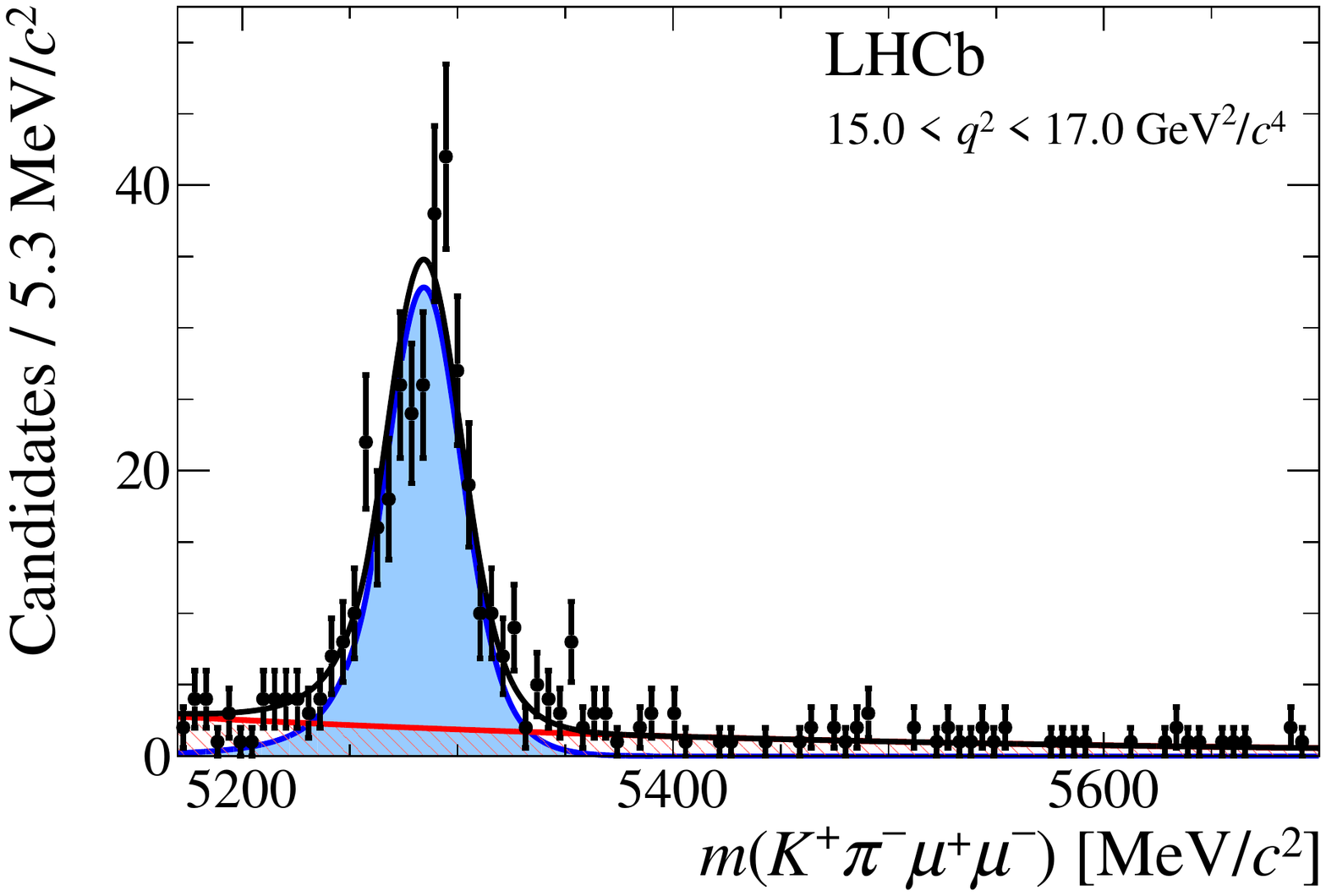} 
\includegraphics[width=0.48\linewidth]{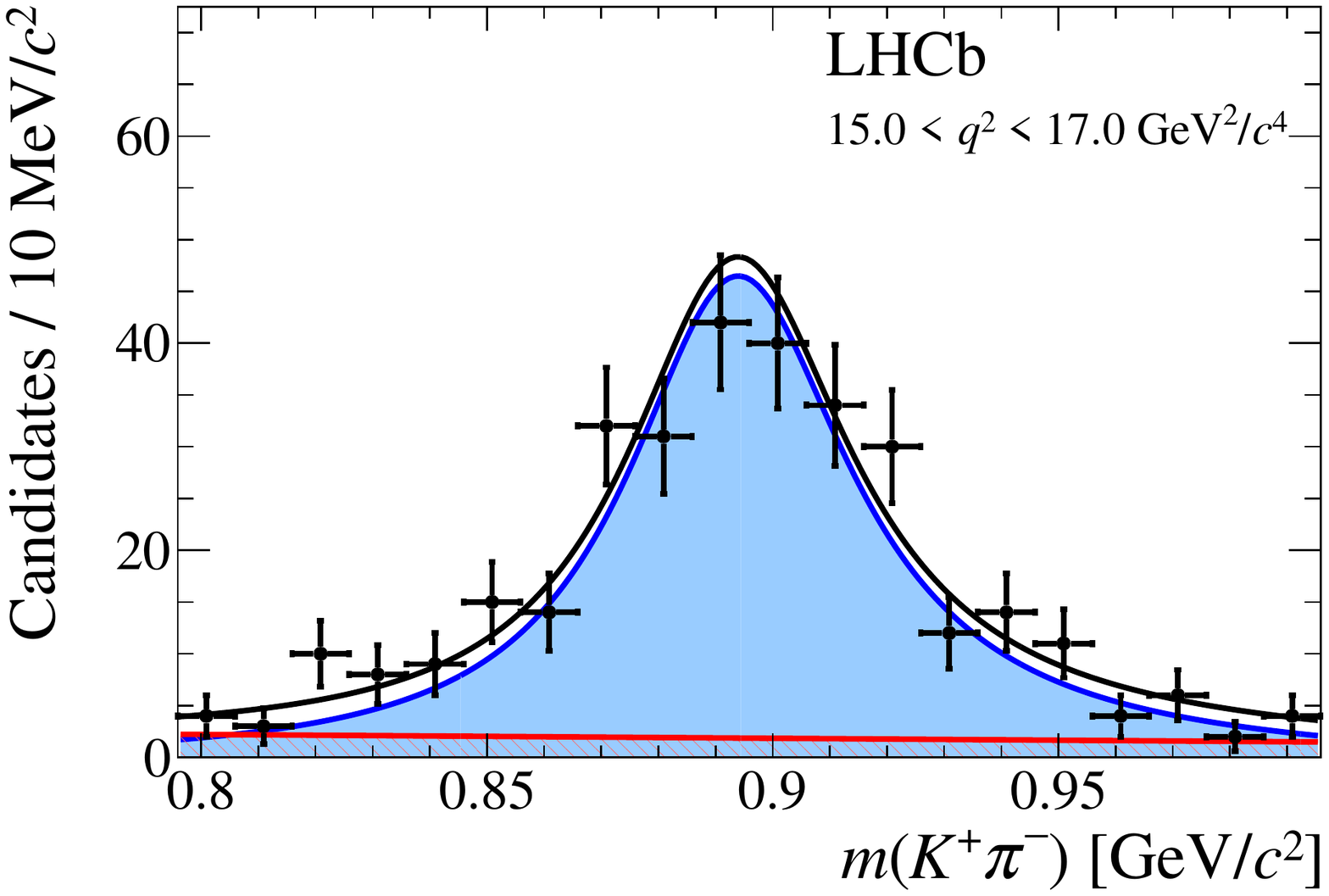} \\
\includegraphics[width=0.48\linewidth]{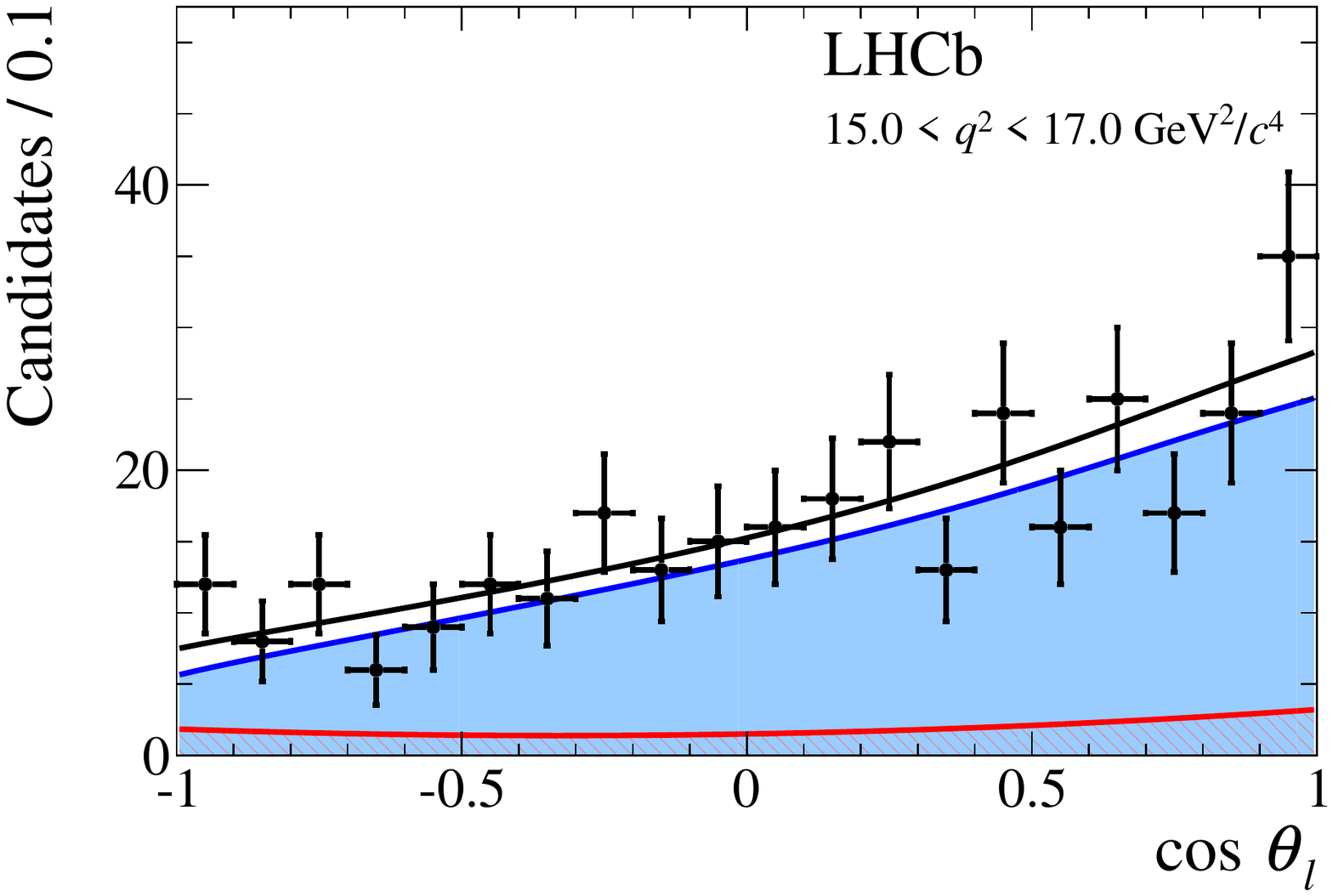} 
\includegraphics[width=0.48\linewidth]{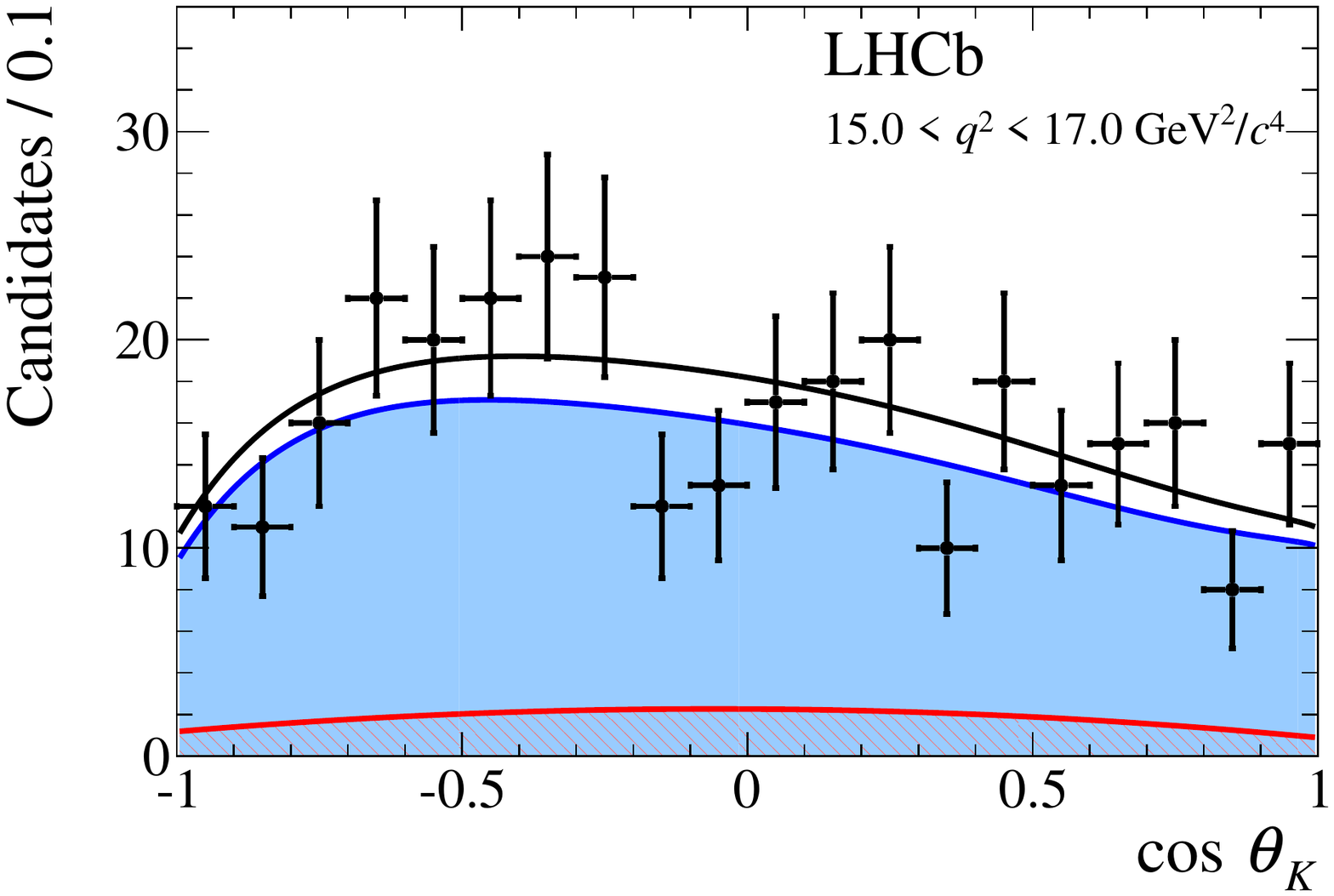} \\
\includegraphics[width=0.48\linewidth]{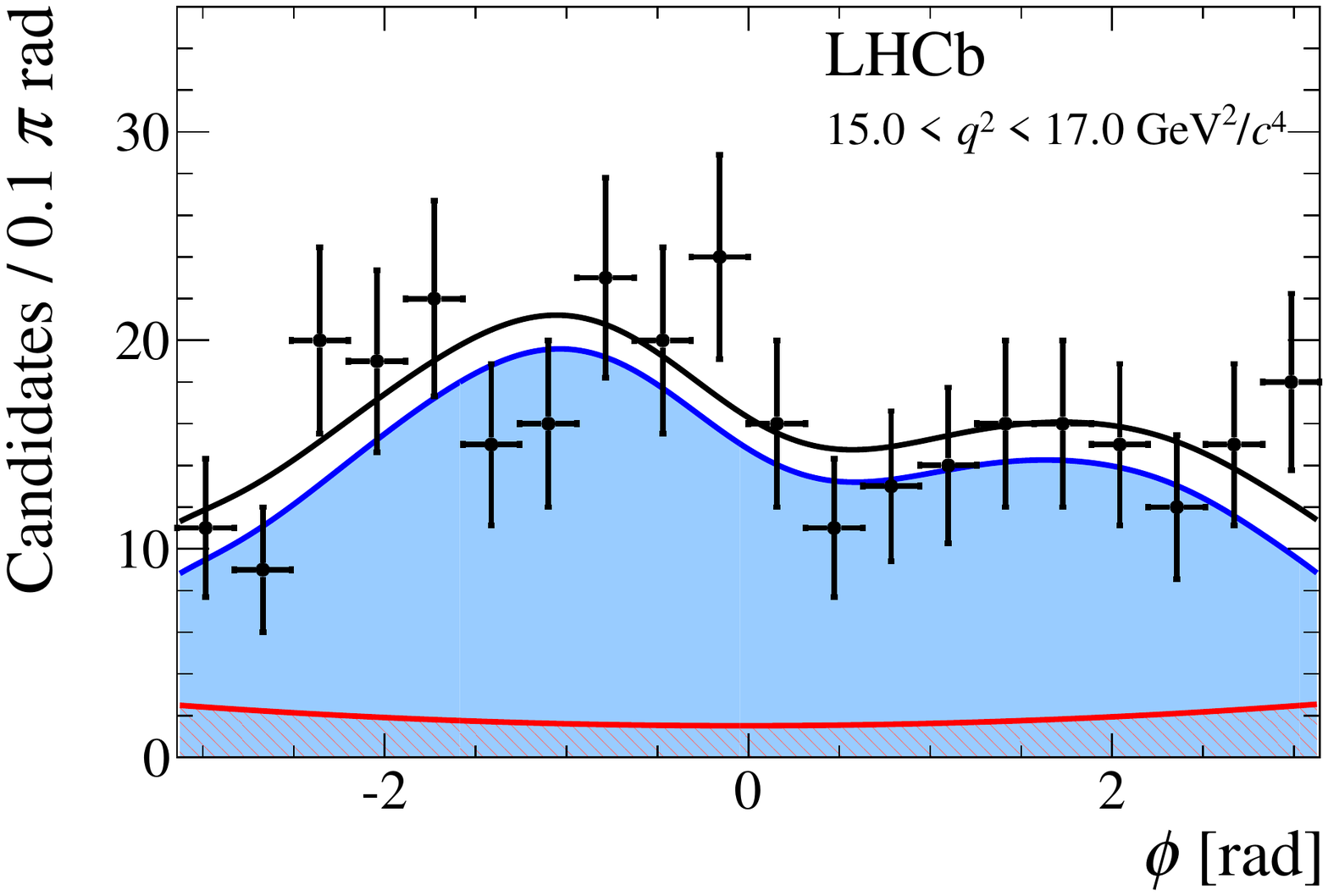}
\end{center}
\vspace*{-0.7cm}
\caption{Angular and mass distributions for $15.0<q^2<17.0\gevgevcccc$.
    The distributions of $m(\Kp\pim)$ and the three decay angles are given for candidates in the signal mass window $\pm50\mevcc$ around the known $\Bd$ mass. 
    Overlaid are the projections of the total fitted distribution (black line) and its different components.
    The signal is shown by the blue shaded area and the background by the red hatched area.
    \label{fig:supp:kstarmumu:bin6}}
\end{figure}

\begin{figure}[htb]
\begin{center}
\includegraphics[width=0.48\linewidth]{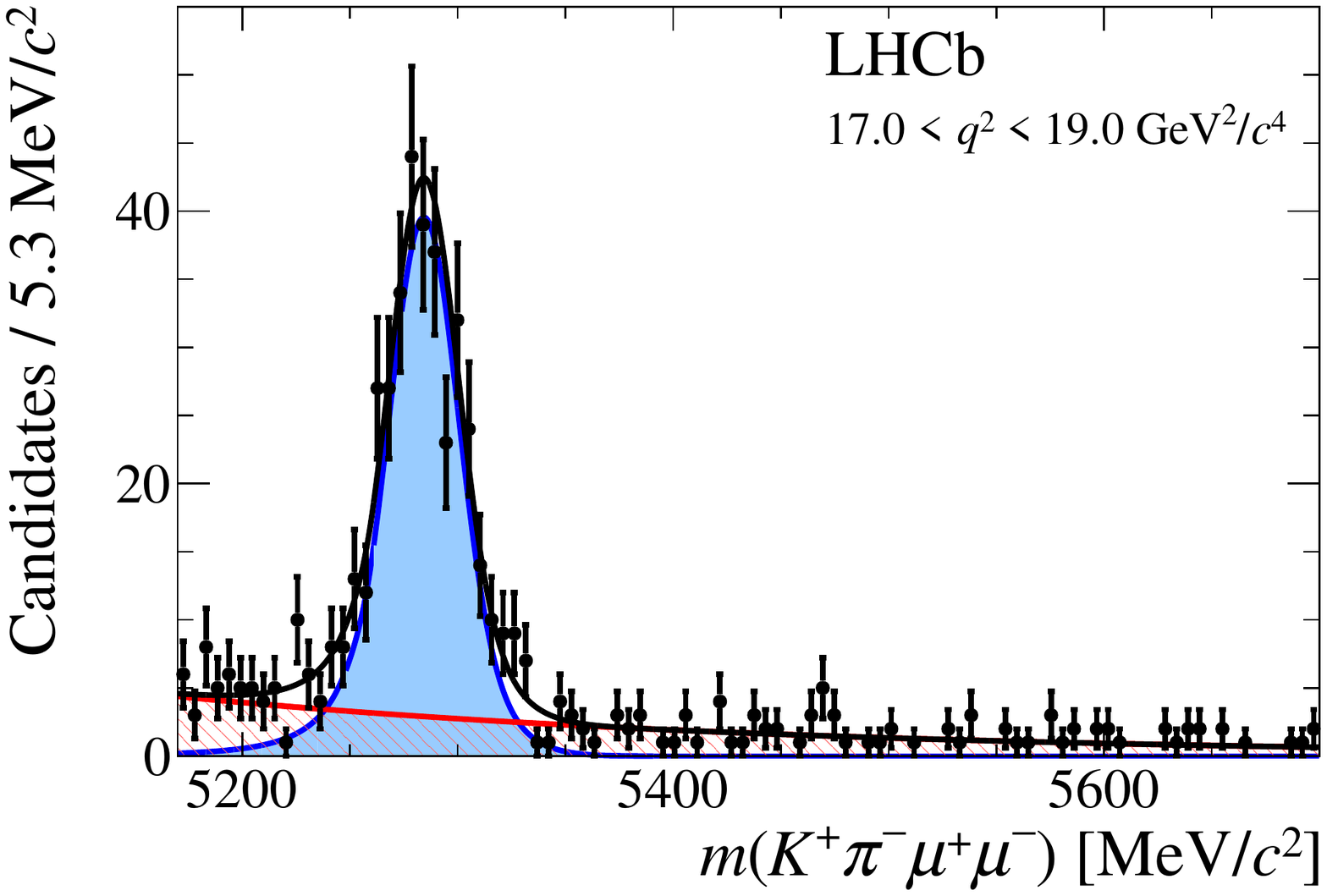} 
\includegraphics[width=0.48\linewidth]{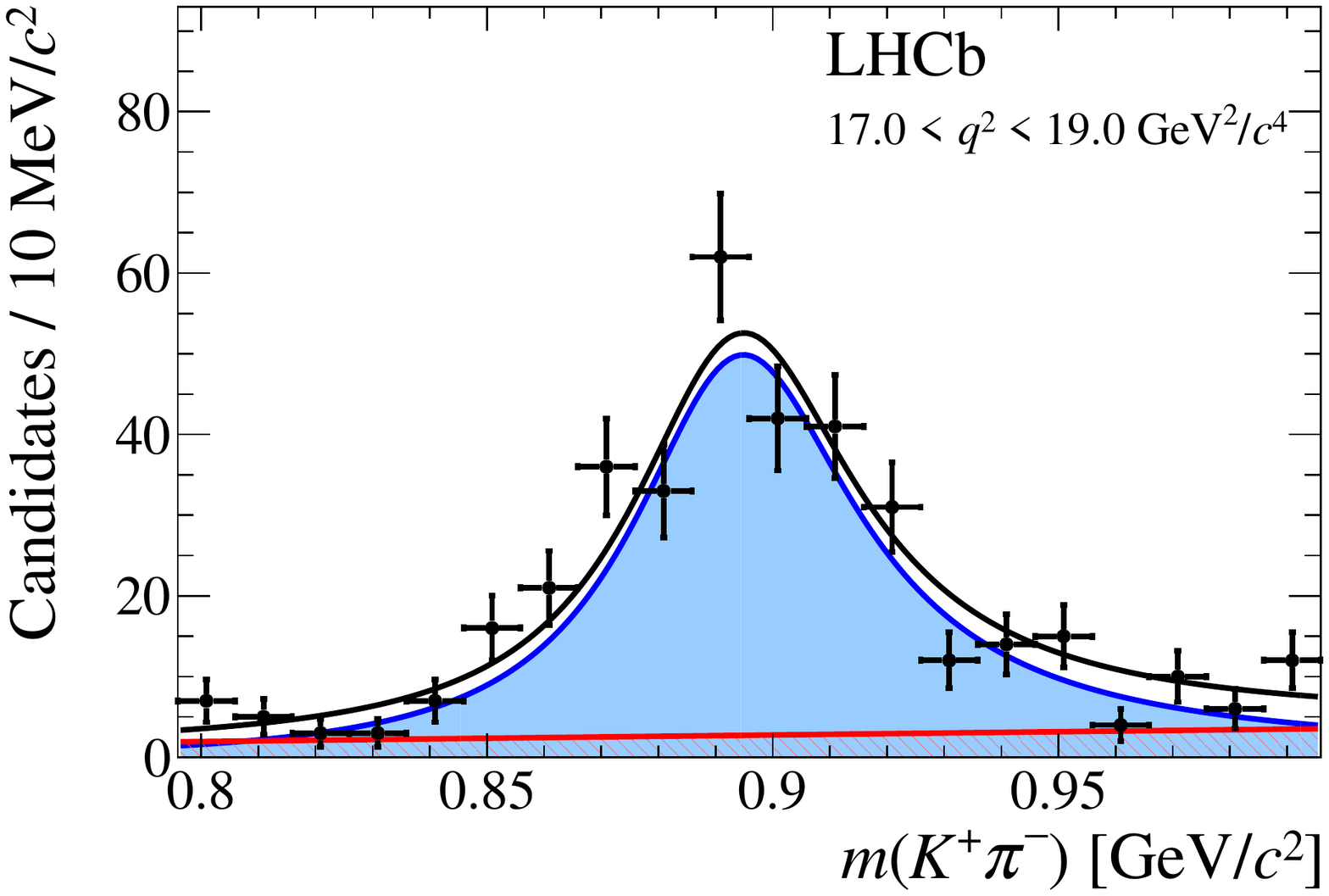} \\
\includegraphics[width=0.48\linewidth]{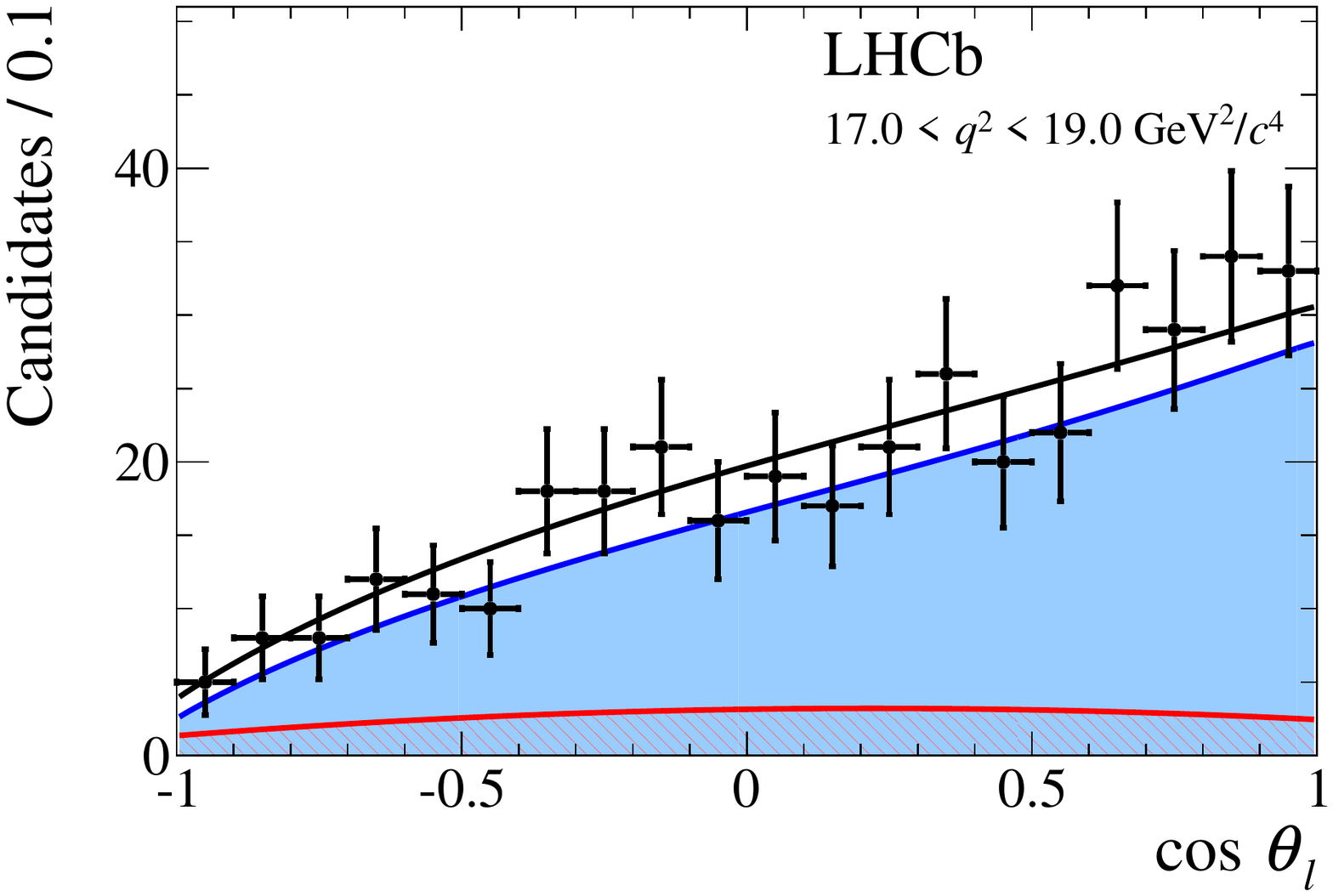} 
\includegraphics[width=0.48\linewidth]{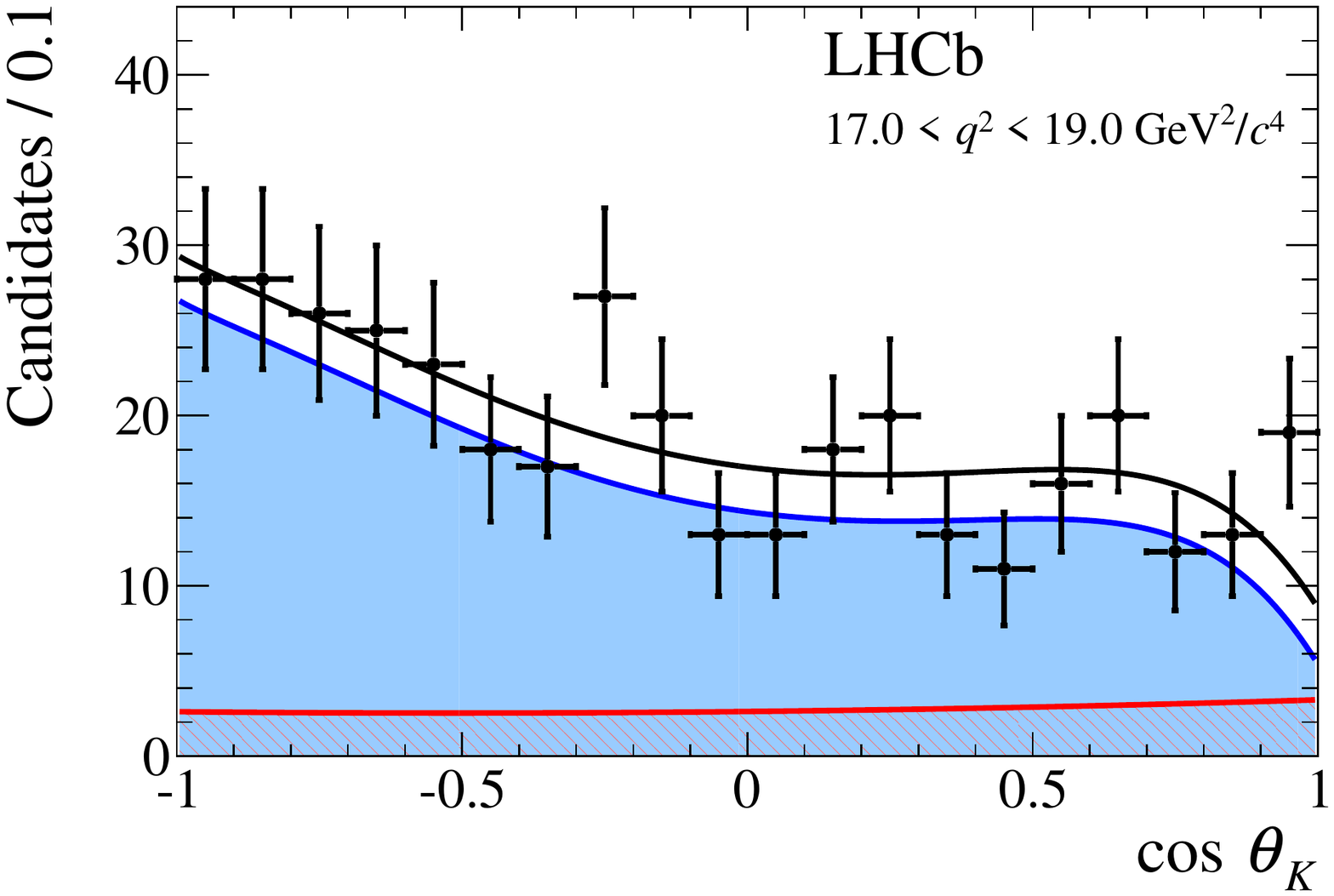} \\
\includegraphics[width=0.48\linewidth]{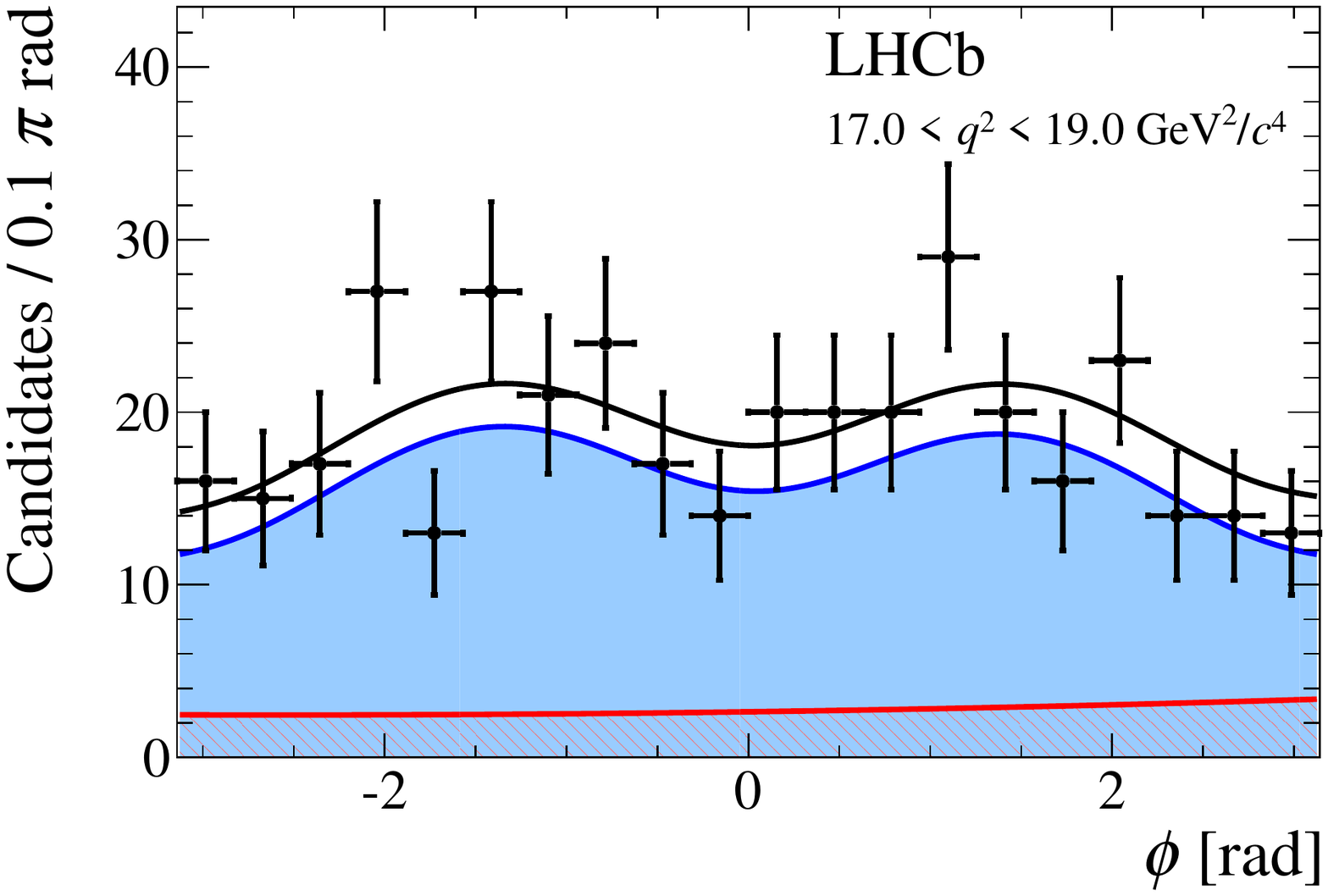}
\end{center}
\vspace*{-0.7cm}
\caption{Angular and mass distributions for $17.0<q^2<19.0\gevgevcccc$.
    The distributions of $m(\Kp\pim)$ and the three decay angles are given for candidates in the signal mass window $\pm50\mevcc$ around the known $\Bd$ mass. 
    Overlaid are the projections of the total fitted distribution (black line) and its different components.
    The signal is shown by the blue shaded area and the background by the red hatched area.    \label{fig:supp:kstarmumu:bin7}}
\end{figure}

\begin{figure}[htb]
\begin{center} 
\includegraphics[width=0.48\linewidth]{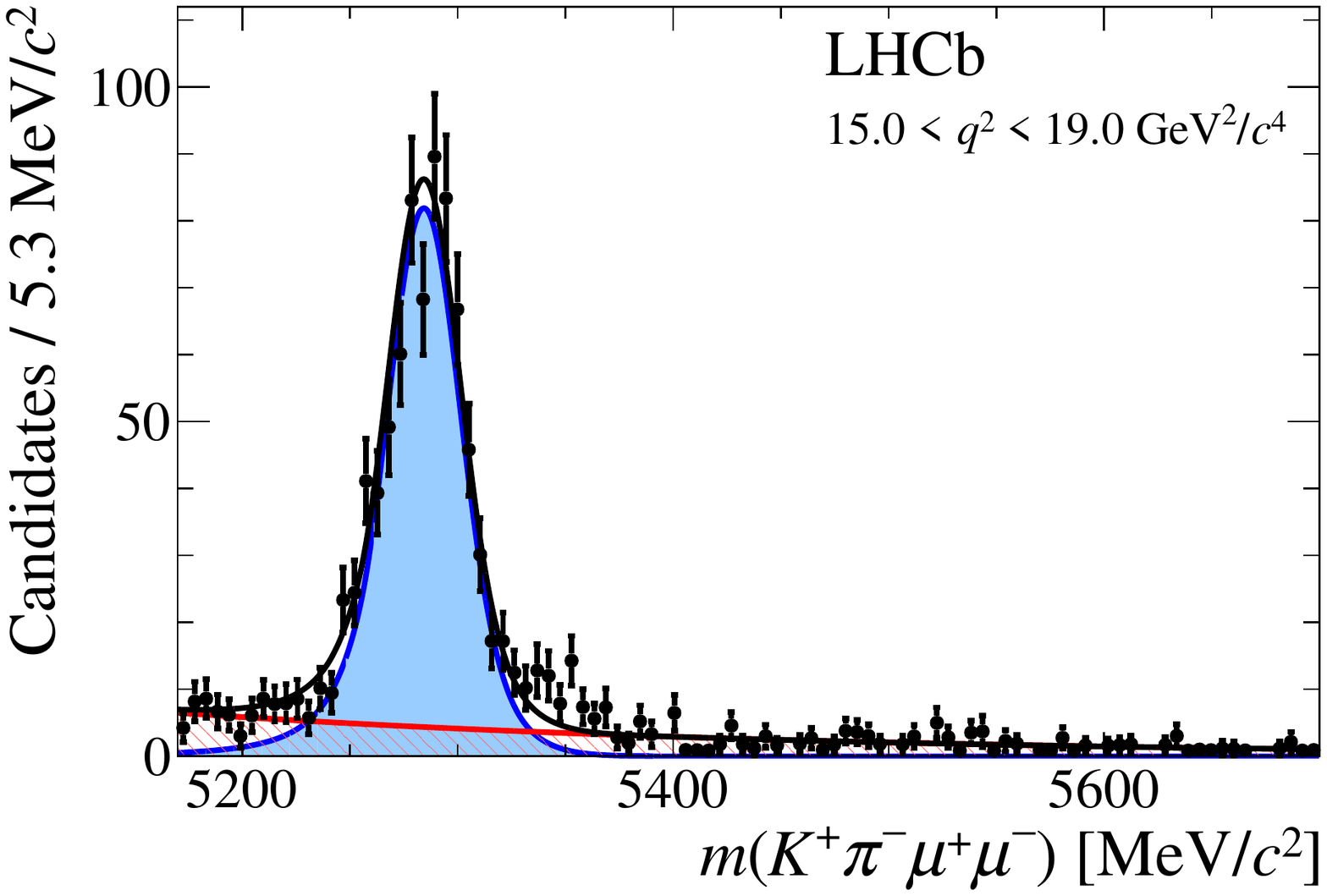} 
\includegraphics[width=0.48\linewidth]{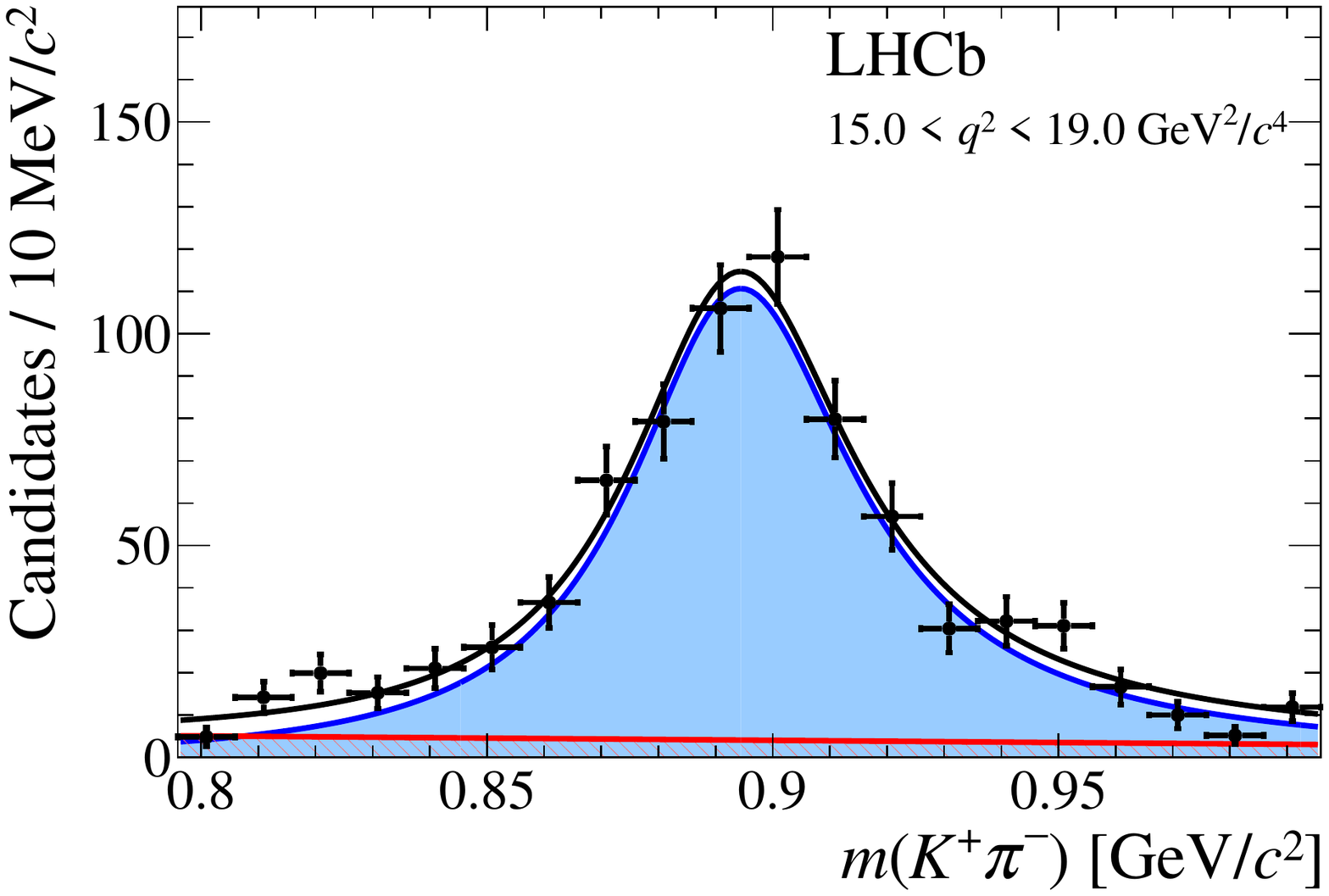} \\
\includegraphics[width=0.48\linewidth]{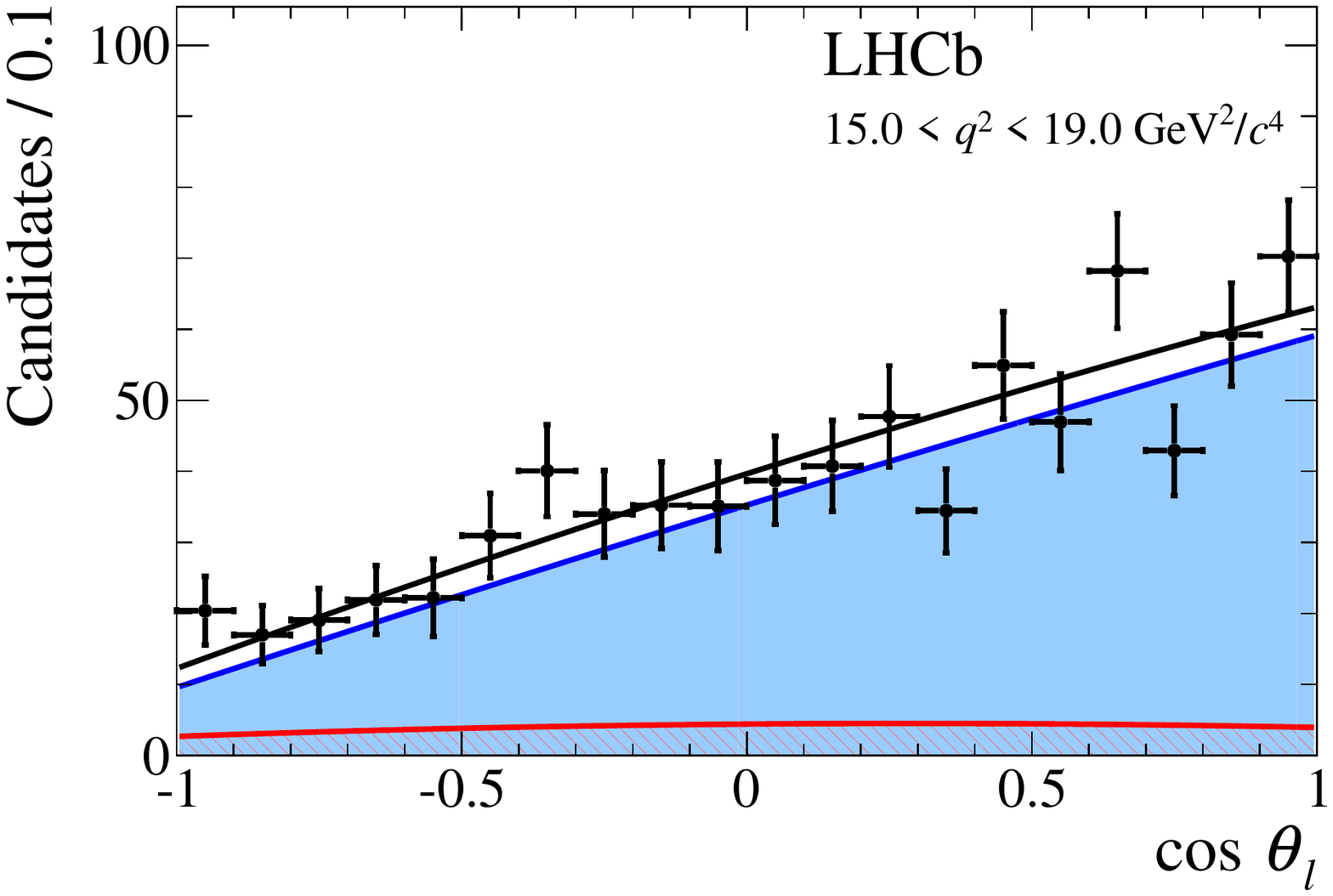} 
\includegraphics[width=0.48\linewidth]{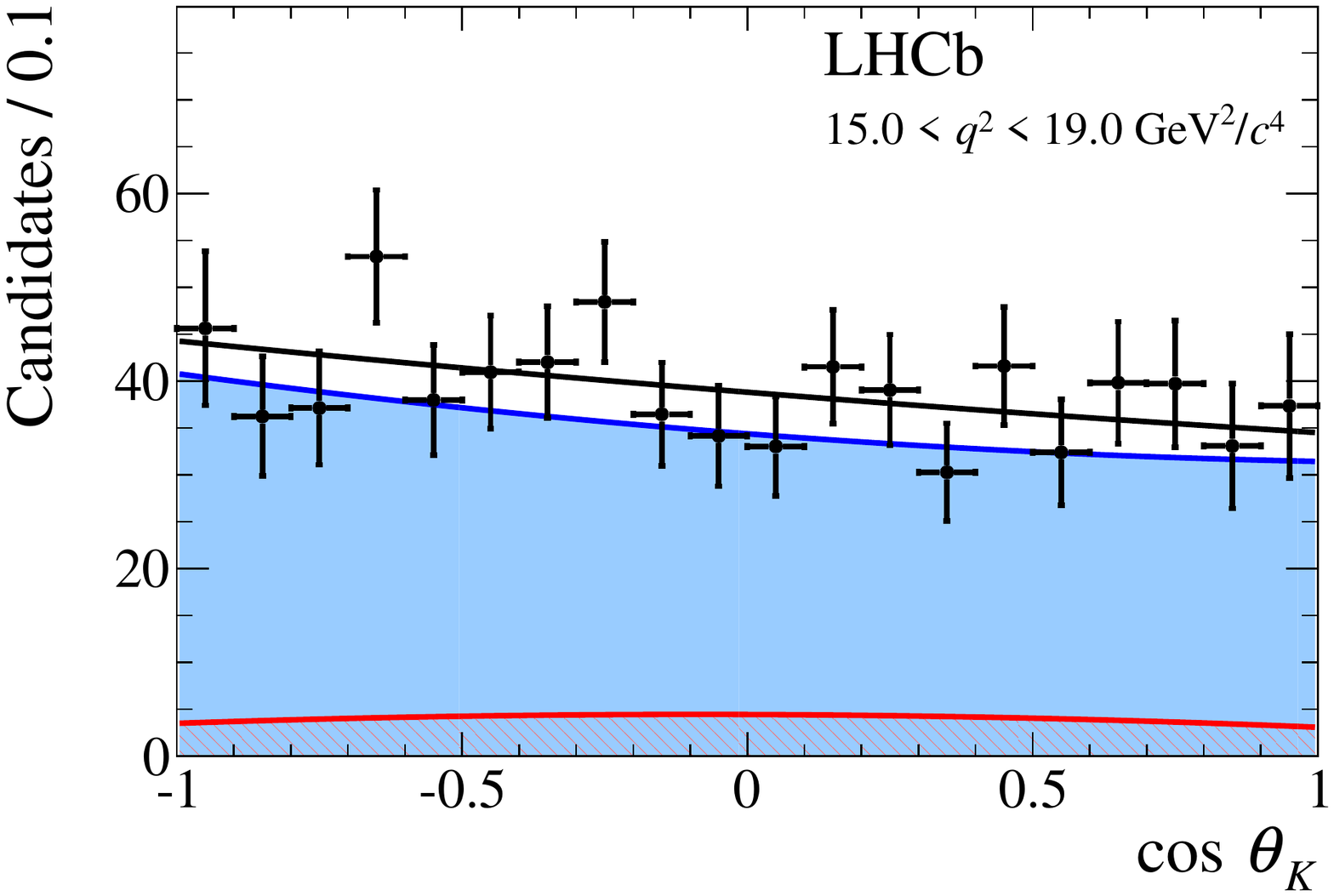} \\
\includegraphics[width=0.48\linewidth]{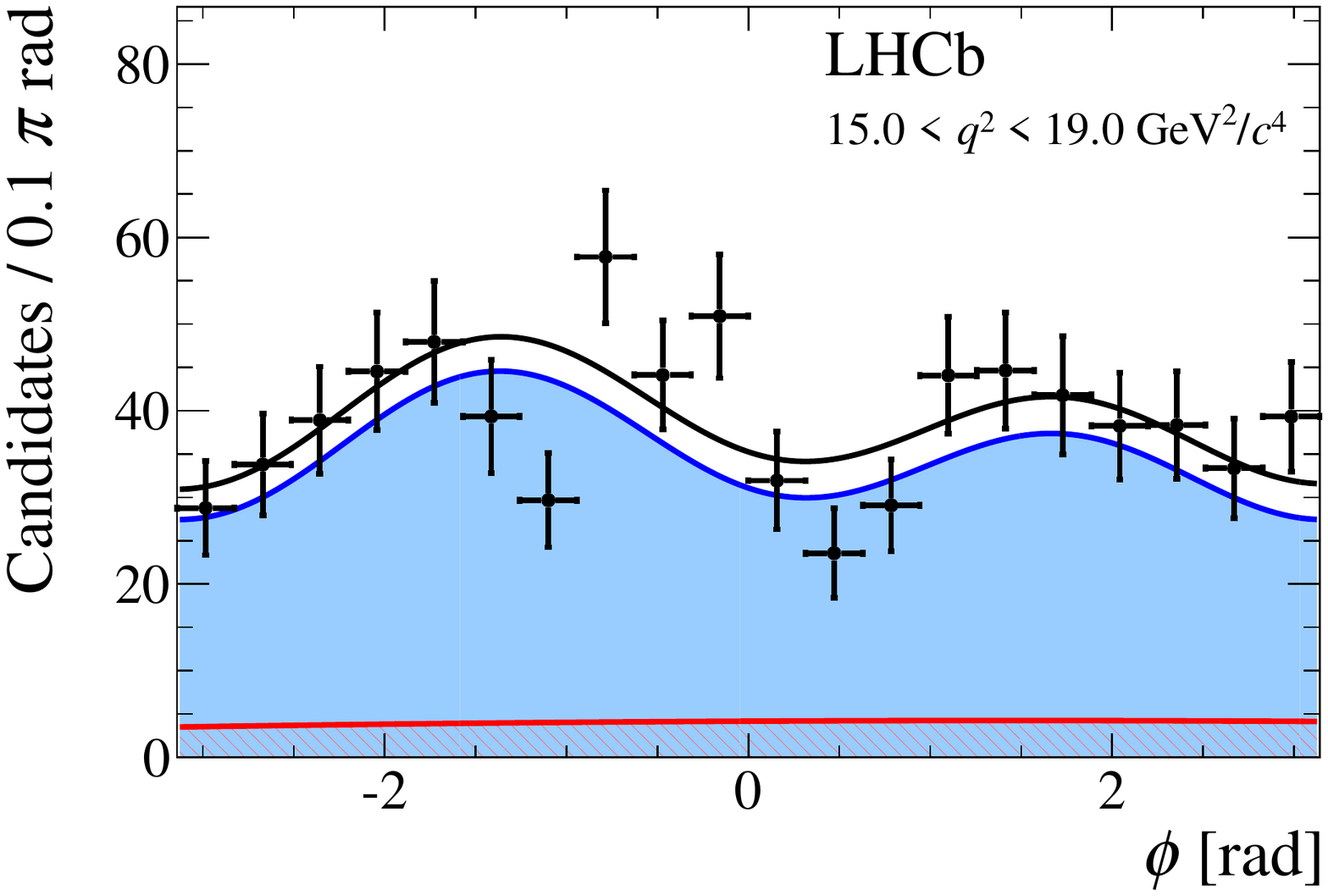}
\end{center}
\vspace*{-0.7cm}
\caption{Angular and mass distributions for $15.0 <q^2< 19.0\gevgevcccc$.
    The distributions of $m(\Kp\pim)$ and the three decay angles are given for candidates in the signal mass window $\pm50\mevcc$ around the known $\Bd$ mass. 
    Overlaid are the projections of the total fitted distribution (black line) and its different components.
  The signal is shown by the blue shaded area and the background by the red hatched area.
    \label{fig:supp:kstarmumu:bin9}}
\end{figure}

Figure~\ref{fig:supp:jpsikstar} shows the projection of the fitted probability density function on the angular and mass distributions for the \decay{\Bz}{\jpsi\Kstarz} control sample. Similarly, Figs.~\ref{fig:supp:kstarmumu:bin0}--\ref{fig:supp:kstarmumu:bin9} show the projections for the $\decay{\Bd}{\Kstarz\mumu}$ data in the different \qsq bins. 
The candidates in Fig.~\ref{fig:supp:kstarmumu:bin9} have been weighted to correct for the detector acceptance according to Sec.~\ref{sec:acceptance}.
In the other figures, the acceptance is included in the shape of the signal distribution.

\clearpage 

\section{Correlation matrices for the \boldmath{\CP}-averaged observables from the maximum likelihood fit} 
\label{sec:appendix:likelihood:correlation}

Correlation matrices between the \CP-averaged observables in the different \qsq bins are provided in Tables~\ref{appendix:likelihood:correlation:average:1}--\ref{appendix:likelihood:correlation:average:10} for the likelihood fit.

\begin{table}[!htb]
\caption{
Correlation matrix for the \CP-averaged observables from the maximum likelihood fit in the bin $0.10<q^2<0.98\gevgevcccc$ . 
\label{appendix:likelihood:correlation:average:1}
}
\centering 
\begin{tabular}{l|rrrrrrrr}
& $F_{\rm L}$ & $S_3$ & $S_4$ & $S_5$ & $A_{\rm FB}$ & $S_7$ & $S_8$ & $S_9$ \\ 
\hline
 $F_{\rm L}$ & $  1.00$ & $  0.06$ & $  0.00$ & $  0.03$ & $  0.04$ & $ -0.02$ & $  0.07$ & $  0.08$ \\
       $S_3$ & $      $ & $  1.00$ & $  0.01$ & $  0.10$ & $ -0.00$ & $ -0.07$ & $ -0.01$ & $ -0.03$ \\
       $S_4$ & $      $ & $      $ & $  1.00$ & $  0.08$ & $  0.11$ & $ -0.00$ & $  0.07$ & $  0.02$ \\
       $S_5$ & $      $ & $      $ & $      $ & $  1.00$ & $  0.05$ & $ -0.01$ & $  0.00$ & $  0.04$ \\
$A_{\rm FB}$ & $      $ & $      $ & $      $ & $      $ & $  1.00$ & $  0.03$ & $ -0.07$ & $  0.02$ \\
       $S_7$ & $      $ & $      $ & $      $ & $      $ & $      $ & $  1.00$ & $  0.01$ & $  0.11$ \\
       $S_8$ & $      $ & $      $ & $      $ & $      $ & $      $ & $      $ & $  1.00$ & $  0.02$ \\
       $S_9$ & $      $ & $      $ & $      $ & $      $ & $      $ & $      $ & $      $ & $  1.00$ \\
\end{tabular}
\end{table}

\begin{table}[!htb]
\caption{
Correlation matrix for the \CP-averaged observables from the maximum likelihood fit in the bin $1.1<q^2<2.5\gevgevcccc$. 
\label{appendix:likelihood:correlation:average:2}
}
\centering 
\begin{tabular}{l|rrrrrrrr}
& $F_{\rm L}$ & $S_3$ & $S_4$ & $S_5$ & $A_{\rm FB}$ & $S_7$ & $S_8$ & $S_9$ \\ 
\hline
 $F_{\rm L}$ & $  1.00$ & $  0.09$ & $  0.07$ & $  0.07$ & $  0.09$ & $ -0.05$ & $ -0.04$ & $  0.08$ \\
       $S_3$ & $      $ & $  1.00$ & $ -0.04$ & $  0.04$ & $  0.01$ & $  0.13$ & $  0.09$ & $  0.12$ \\
       $S_4$ & $      $ & $      $ & $  1.00$ & $ -0.22$ & $ -0.01$ & $ -0.00$ & $ -0.05$ & $  0.03$ \\
       $S_5$ & $      $ & $      $ & $      $ & $  1.00$ & $ -0.14$ & $ -0.11$ & $ -0.03$ & $ -0.21$ \\
$A_{\rm FB}$ & $      $ & $      $ & $      $ & $      $ & $  1.00$ & $ -0.03$ & $ -0.10$ & $ -0.11$ \\
       $S_7$ & $      $ & $      $ & $      $ & $      $ & $      $ & $  1.00$ & $ -0.11$ & $  0.23$ \\
       $S_8$ & $      $ & $      $ & $      $ & $      $ & $      $ & $      $ & $  1.00$ & $ -0.04$ \\
       $S_9$ & $      $ & $      $ & $      $ & $      $ & $      $ & $      $ & $      $ & $  1.00$ \\
\end{tabular}
\end{table}

\begin{table}[!htb]
\caption{
Correlation matrix for the \CP-averaged observables from the maximum likelihood fit in the bin $2.5<q^2<4.0\gevgevcccc$.
\label{appendix:likelihood:correlation:average:3}
}
\centering 
\begin{tabular}{l|rrrrrrrr}
& $F_{\rm L}$ & $S_3$ & $S_4$ & $S_5$ & $A_{\rm FB}$ & $S_7$ & $S_8$ & $S_9$ \\ \hline
 $F_{\rm L}$ & $  1.00$ & $ -0.13$ & $ -0.14$ & $  0.01$ & $ -0.03$ & $  0.10$ & $ -0.03$ & $ -0.01$ \\
       $S_3$ & $      $ & $  1.00$ & $ -0.06$ & $  0.09$ & $  0.07$ & $ -0.02$ & $  0.01$ & $ -0.07$ \\
       $S_4$ & $      $ & $      $ & $  1.00$ & $ -0.19$ & $ -0.09$ & $ -0.05$ & $  0.12$ & $  0.07$ \\
       $S_5$ & $      $ & $      $ & $      $ & $  1.00$ & $ -0.01$ & $  0.05$ & $ -0.02$ & $  0.10$ \\
$A_{\rm FB}$ & $      $ & $      $ & $      $ & $      $ & $  1.00$ & $ -0.01$ & $ -0.10$ & $  0.10$ \\
       $S_7$ & $      $ & $      $ & $      $ & $      $ & $      $ & $  1.00$ & $  0.07$ & $ -0.05$ \\
       $S_8$ & $      $ & $      $ & $      $ & $      $ & $      $ & $      $ & $  1.00$ & $ -0.01$ \\
       $S_9$ & $      $ & $      $ & $      $ & $      $ & $      $ & $      $ & $      $ & $  1.00$ \\
\end{tabular}
\end{table}

\begin{table}[!htb]
\caption{
Correlation matrix for the \CP-averaged observables from the maximum likelihood fit in the bin $4.0 <q^2< 6.0\gevgevcccc$.
\label{appendix:likelihood:correlation:average:4}
}
\centering 
\begin{tabular}{l|rrrrrrrr}
& $F_{\rm L}$ & $S_3$ & $S_4$ & $S_5$ & $A_{\rm FB}$ & $S_7$ & $S_8$ & $S_9$ \\ \hline
 $F_{\rm L}$ & $  1.00$ & $ -0.03$ & $  0.09$ & $  0.10$ & $ -0.05$ & $ -0.10$ & $  0.04$ & $  0.00$ \\
       $S_3$ & $      $ & $  1.00$ & $ -0.04$ & $ -0.03$ & $  0.09$ & $ -0.10$ & $ -0.00$ & $ -0.12$ \\
       $S_4$ & $      $ & $      $ & $  1.00$ & $  0.10$ & $ -0.10$ & $ -0.02$ & $ -0.04$ & $  0.04$ \\
       $S_5$ & $      $ & $      $ & $      $ & $  1.00$ & $ -0.06$ & $ -0.03$ & $ -0.01$ & $ -0.04$ \\
$A_{\rm FB}$ & $      $ & $      $ & $      $ & $      $ & $  1.00$ & $  0.03$ & $  0.07$ & $ -0.03$ \\
       $S_7$ & $      $ & $      $ & $      $ & $      $ & $      $ & $  1.00$ & $  0.06$ & $ -0.15$ \\
       $S_8$ & $      $ & $      $ & $      $ & $      $ & $      $ & $      $ & $  1.00$ & $  0.03$ \\
       $S_9$ & $      $ & $      $ & $      $ & $      $ & $      $ & $      $ & $      $ & $  1.00$ \\
\end{tabular}
\end{table}

\begin{table}[!htb]
\caption{
Correlation matrix for the \CP-averaged observables from the maximum likelihood fit in the bin $6.0<q^2<8.0\gevgevcccc$.
\label{appendix:likelihood:correlation:average:5}
}
\centering 
\begin{tabular}{l|rrrrrrrr}
& $F_{\rm L}$ & $S_3$ & $S_4$ & $S_5$ & $A_{\rm FB}$ & $S_7$ & $S_8$ & $S_9$ \\ \hline
 $F_{\rm L}$ & $  1.00$ & $  0.03$ & $  0.06$ & $  0.03$ & $ -0.31$ & $ -0.08$ & $ -0.01$ & $ -0.06$ \\
       $S_3$ & $      $ & $  1.00$ & $ -0.16$ & $ -0.23$ & $  0.01$ & $  0.02$ & $  0.02$ & $ -0.07$ \\
       $S_4$ & $      $ & $      $ & $  1.00$ & $ -0.13$ & $ -0.12$ & $ -0.01$ & $ -0.11$ & $  0.01$ \\
       $S_5$ & $      $ & $      $ & $      $ & $  1.00$ & $ -0.16$ & $ -0.14$ & $ -0.01$ & $ -0.04$ \\
$A_{\rm FB}$ & $      $ & $      $ & $      $ & $      $ & $  1.00$ & $ -0.01$ & $  0.04$ & $  0.02$ \\
       $S_7$ & $      $ & $      $ & $      $ & $      $ & $      $ & $  1.00$ & $  0.10$ & $ -0.05$ \\
       $S_8$ & $      $ & $      $ & $      $ & $      $ & $      $ & $      $ & $  1.00$ & $ -0.10$ \\
       $S_9$ & $      $ & $      $ & $      $ & $      $ & $      $ & $      $ & $      $ & $  1.00$ \\
\end{tabular}
\end{table}

\begin{table}[!htb]
\caption{
Correlation matrix for the \CP-averaged observables from the maximum likelihood fit in the bin $11.0 <q^2< 12.5 \gevgevcccc$.
\label{appendix:likelihood:correlation:average:6}
}
\centering 
\begin{tabular}{l|rrrrrrrr}
& $F_{\rm L}$ & $S_3$ & $S_4$ & $S_5$ & $A_{\rm FB}$ & $S_7$ & $S_8$ & $S_9$ \\ \hline
 $F_{\rm L}$ & $  1.00$ & $  0.25$ & $  0.02$ & $ -0.02$ & $ -0.62$ & $  0.03$ & $  0.05$ & $  0.02$ \\
       $S_3$ & $      $ & $  1.00$ & $  0.05$ & $ -0.35$ & $ -0.24$ & $ -0.04$ & $  0.06$ & $ -0.02$ \\
       $S_4$ & $      $ & $      $ & $  1.00$ & $ -0.02$ & $  0.06$ & $ -0.05$ & $ -0.12$ & $ -0.08$ \\
       $S_5$ & $      $ & $      $ & $      $ & $  1.00$ & $  0.01$ & $ -0.04$ & $ -0.09$ & $ -0.24$ \\
$A_{\rm FB}$ & $      $ & $      $ & $      $ & $      $ & $  1.00$ & $ -0.01$ & $ -0.06$ & $  0.07$ \\
       $S_7$ & $      $ & $      $ & $      $ & $      $ & $      $ & $  1.00$ & $  0.27$ & $ -0.19$ \\
       $S_8$ & $      $ & $      $ & $      $ & $      $ & $      $ & $      $ & $  1.00$ & $ -0.09$ \\
       $S_9$ & $      $ & $      $ & $      $ & $      $ & $      $ & $      $ & $      $ & $  1.00$ \\
\end{tabular}
\end{table}

\begin{table}[!htb]
\caption{
Correlation matrix for the \CP-averaged observables from the maximum likelihood fit in the bin $15.0 <q^2< 17.0 \gevgevcccc$.
\label{appendix:likelihood:correlation:average:7}
}
\centering 
\begin{tabular}{l|rrrrrrrr}
& $F_{\rm L}$ & $S_3$ & $S_4$ & $S_5$ & $A_{\rm FB}$ & $S_7$ & $S_8$ & $S_9$ \\ \hline
 $F_{\rm L}$ & $  1.00$ & $  0.26$ & $ -0.10$ & $  0.09$ & $ -0.50$ & $ -0.02$ & $ -0.06$ & $  0.14$ \\
       $S_3$ & $      $ & $  1.00$ & $ -0.08$ & $ -0.03$ & $ -0.00$ & $ -0.04$ & $ -0.05$ & $  0.10$ \\
       $S_4$ & $      $ & $      $ & $  1.00$ & $  0.26$ & $ -0.16$ & $ -0.05$ & $  0.19$ & $  0.05$ \\
       $S_5$ & $      $ & $      $ & $      $ & $  1.00$ & $ -0.20$ & $  0.12$ & $ -0.01$ & $  0.05$ \\
$A_{\rm FB}$ & $      $ & $      $ & $      $ & $      $ & $  1.00$ & $  0.05$ & $ -0.02$ & $ -0.08$ \\
       $S_7$ & $      $ & $      $ & $      $ & $      $ & $      $ & $  1.00$ & $  0.25$ & $ -0.23$ \\
       $S_8$ & $      $ & $      $ & $      $ & $      $ & $      $ & $      $ & $  1.00$ & $ -0.11$ \\
       $S_9$ & $      $ & $      $ & $      $ & $      $ & $      $ & $      $ & $      $ & $  1.00$ \\
\end{tabular}
\end{table}

\begin{table}[!htb]
\caption{
Correlation matrix for the \CP-averaged observables from the maximum likelihood fit in the bin $17.0 <q^2< 19.0\gevgevcccc$.
\label{appendix:likelihood:correlation:average:8}
}
\centering 
\begin{tabular}{l|rrrrrrrr}
& $F_{\rm L}$ & $S_3$ & $S_4$ & $S_5$ & $A_{\rm FB}$ & $S_7$ & $S_8$ & $S_9$ \\ \hline
 $F_{\rm L}$ & $  1.00$ & $  0.07$ & $  0.06$ & $  0.04$ & $ -0.35$ & $  0.07$ & $  0.07$ & $  0.08$ \\
       $S_3$ & $      $ & $  1.00$ & $ -0.15$ & $ -0.39$ & $ -0.05$ & $ -0.06$ & $ -0.04$ & $ -0.07$ \\
       $S_4$ & $      $ & $      $ & $  1.00$ & $  0.10$ & $ -0.17$ & $  0.03$ & $  0.18$ & $ -0.04$ \\
       $S_5$ & $      $ & $      $ & $      $ & $  1.00$ & $ -0.11$ & $  0.04$ & $  0.01$ & $ -0.00$ \\
$A_{\rm FB}$ & $      $ & $      $ & $      $ & $      $ & $  1.00$ & $ -0.02$ & $ -0.09$ & $ -0.03$ \\
       $S_7$ & $      $ & $      $ & $      $ & $      $ & $      $ & $  1.00$ & $  0.34$ & $ -0.15$ \\
       $S_8$ & $      $ & $      $ & $      $ & $      $ & $      $ & $      $ & $  1.00$ & $ -0.11$ \\
       $S_9$ & $      $ & $      $ & $      $ & $      $ & $      $ & $      $ & $      $ & $  1.00$ \\
\end{tabular}
\end{table}

\begin{table}[!htb]
\caption{
Correlation matrix for the \CP-averaged observables from the maximum likelihood fit in the bin $1.1 <q^2< 6.0\gevgevcccc$.
\label{appendix:likelihood:correlation:average:9}
}
\centering 
\begin{tabular}{l|rrrrrrrr}
& $F_{\rm L}$ & $S_3$ & $S_4$ & $S_5$ & $A_{\rm FB}$ & $S_7$ & $S_8$ & $S_9$ \\ \hline
 $F_{\rm L}$ & $  1.00$ & $ -0.04$ & $  0.05$ & $  0.03$ & $  0.05$ & $ -0.04$ & $ -0.01$ & $  0.08$ \\
       $S_3$ & $      $ & $  1.00$ & $ -0.05$ & $ -0.00$ & $  0.05$ & $  0.01$ & $  0.01$ & $ -0.01$ \\
       $S_4$ & $      $ & $      $ & $  1.00$ & $ -0.05$ & $ -0.11$ & $ -0.02$ & $ -0.01$ & $  0.05$ \\
       $S_5$ & $      $ & $      $ & $      $ & $  1.00$ & $ -0.07$ & $ -0.01$ & $ -0.02$ & $ -0.04$ \\
$A_{\rm FB}$ & $      $ & $      $ & $      $ & $      $ & $  1.00$ & $  0.02$ & $ -0.02$ & $ -0.04$ \\
       $S_7$ & $      $ & $      $ & $      $ & $      $ & $      $ & $  1.00$ & $  0.04$ & $ -0.01$ \\
       $S_8$ & $      $ & $      $ & $      $ & $      $ & $      $ & $      $ & $  1.00$ & $ -0.03$ \\
       $S_9$ & $      $ & $      $ & $      $ & $      $ & $      $ & $      $ & $      $ & $  1.00$ \\
\end{tabular}
\end{table}

\begin{table}[!htb]
\caption{
Correlation matrix for the \CP-averaged observables from the maximum likelihood fit in the bin $15.0 <q^2< 19.0\gevgevcccc$.
\label{appendix:likelihood:correlation:average:10}
}
\centering 
\begin{tabular}{l|rrrrrrrr}
& $F_{\rm L}$ & $S_3$ & $S_4$ & $S_5$ & $A_{\rm FB}$ & $S_7$ & $S_8$ & $S_9$ \\ \hline
 $F_{\rm L}$ & $  1.00$ & $  0.17$ & $ -0.03$ & $ -0.02$ & $ -0.39$ & $  0.01$ & $ -0.00$ & $  0.11$ \\
       $S_3$ & $      $ & $  1.00$ & $ -0.15$ & $ -0.19$ & $  0.05$ & $ -0.02$ & $ -0.04$ & $ -0.02$ \\
       $S_4$ & $      $ & $      $ & $  1.00$ & $  0.06$ & $ -0.12$ & $  0.03$ & $  0.14$ & $  0.01$ \\
       $S_5$ & $      $ & $      $ & $      $ & $  1.00$ & $ -0.12$ & $  0.12$ & $  0.04$ & $  0.02$ \\
$A_{\rm FB}$ & $      $ & $      $ & $      $ & $      $ & $  1.00$ & $  0.00$ & $ -0.02$ & $ -0.01$ \\
       $S_7$ & $      $ & $      $ & $      $ & $      $ & $      $ & $  1.00$ & $  0.24$ & $ -0.19$ \\
       $S_8$ & $      $ & $      $ & $      $ & $      $ & $      $ & $      $ & $  1.00$ & $ -0.13$ \\
       $S_9$ & $      $ & $      $ & $      $ & $      $ & $      $ & $      $ & $      $ & $  1.00$ \\
\end{tabular}
\end{table}

\clearpage

\section{Correlation matrices for the \boldmath{\CP}-asymmetric observables from the maximum likelihood fit} 
\label{sec:appendix:correlation:asymmetries}

Correlation matrices between $F_{\rm L}$ and the \CP-asymmetric observables in the different \qsq bins are provided in Tables~\ref{appendix:likelihood:correlation:asymmetry:1}--\ref{appendix:likelihood:correlation:asymmetry:10} for the likelihood fit.

\begin{table}[!htb]
\caption{
Correlation matrix for the \CP-asymmetric observables from the maximum likelihood fit in the bin $0.10<q^2<0.98\gevgevcccc$. 
\label{appendix:likelihood:correlation:asymmetry:1}
}
\centering 
\begin{tabular}{l|rrrrrrrr}
& $F_{\rm L}$ & $A_3$ & $A_4$ & $A_5$ & $A_{6s}$ & $A_7$ & $A_8$ & $A_9$ \\ \hline
 $F_{\rm L}$ & $  1.00$ & $ -0.00$ & $  0.02$ & $  0.01$ & $ -0.07$ & $ -0.01$ & $ -0.01$ & $ -0.03$ \\
       $A_3$ & $      $ & $  1.00$ & $ -0.04$ & $ -0.07$ & $  0.00$ & $ -0.03$ & $  0.02$ & $ -0.05$ \\
       $A_4$ & $      $ & $      $ & $  1.00$ & $  0.05$ & $ -0.08$ & $  0.02$ & $  0.09$ & $ -0.03$ \\
       $A_5$ & $      $ & $      $ & $      $ & $  1.00$ & $ -0.04$ & $  0.08$ & $  0.03$ & $  0.02$ \\
       $A_{6s}$ & $      $ & $      $ & $      $ & $      $ & $  1.00$ & $  -0.04$ & $ -0.07$ & $  0.01$ \\
       $A_7$ & $      $ & $      $ & $      $ & $      $ & $      $ & $  1.00$ & $  0.00$ & $ -0.14$ \\
       $A_8$ & $      $ & $      $ & $      $ & $      $ & $      $ & $      $ & $  1.00$ & $ -0.01$ \\
       $A_9$ & $      $ & $      $ & $      $ & $      $ & $      $ & $      $ & $      $ & $  1.00$ \\
\end{tabular} 
\end{table}

\begin{table}[!htb]
\caption{
Correlation matrix for the \CP-asymmetric observables from the maximum likelihood fit in the bin $1.1<q^2<2.5\gevgevcccc$. 
\label{appendix:likelihood:correlation:asymmetry:2}
}
\centering 
\begin{tabular}{l|rrrrrrrr}
& $F_{\rm L}$ & $A_3$ & $A_4$ & $A_5$ & $A_{6s}$ & $A_7$ & $A_8$ & $A_9$ \\ \hline
 $F_{\rm L}$ & $  1.00$ & $  0.07$ & $ -0.14$ & $ -0.06$ & $ -0.04$ & $  0.06$ & $ -0.04$ & $ -0.10$ \\
       $A_3$ & $      $ & $  1.00$ & $ -0.05$ & $ -0.11$ & $  0.01$ & $ -0.04$ & $  0.05$ & $ -0.05$ \\
       $A_4$ & $      $ & $      $ & $  1.00$ & $  0.09$ & $ -0.26$ & $  0.03$ & $ -0.15$ & $  0.10$ \\
       $A_5$ & $      $ & $      $ & $      $ & $  1.00$ & $  0.03$ & $  0.03$ & $  0.06$ & $  0.01$ \\
       $A_{6s}$ & $      $ & $      $ & $      $ & $      $ & $  1.00$ & $  0.10$ & $  0.11$ & $  0.01$ \\
       $A_7$ & $      $ & $      $ & $      $ & $      $ & $      $ & $  1.00$ & $  0.19$ & $  0.12$ \\
       $A_8$ & $      $ & $      $ & $      $ & $      $ & $      $ & $      $ & $  1.00$ & $  0.03$ \\
       $A_9$ & $      $ & $      $ & $      $ & $      $ & $      $ & $      $ & $      $ & $  1.00$ \\
\end{tabular}
\end{table}

\begin{table}[!htb]
\caption{
Correlation matrix for the \CP-asymmetric observables from the maximum likelihood fit in the bin $2.5<q^2<4.0\gevgevcccc$. 
\label{appendix:likelihood:correlation:asymmetry:3}
}
\centering 
\begin{tabular}{l|rrrrrrrr}
& $F_{\rm L}$ & $A_3$ & $A_4$ & $A_5$ & $A_{6s}$ & $A_7$ & $A_8$ & $A_9$ \\ \hline
 $F_{\rm L}$ & $  1.00$ & $  0.13$ & $ -0.04$ & $  0.07$ & $  0.10$ & $ -0.08$ & $  0.08$ & $  0.11$ \\
       $A_3$ & $      $ & $  1.00$ & $  0.19$ & $ -0.00$ & $ -0.07$ & $ -0.03$ & $  0.09$ & $  0.11$ \\
       $A_4$ & $      $ & $      $ & $  1.00$ & $  0.21$ & $ -0.12$ & $  0.02$ & $  0.13$ & $  0.09$ \\
       $A_5$ & $      $ & $      $ & $      $ & $  1.00$ & $  0.11$ & $  0.08$ & $  0.01$ & $  0.06$ \\
       $A_{6s}$ & $      $ & $      $ & $      $ & $      $ & $  1.00$ & $  -0.05$ & $  -0.28$ & $  -0.05$ \\
       $A_7$ & $      $ & $      $ & $      $ & $      $ & $      $ & $  1.00$ & $  0.31$ & $ -0.03$ \\
       $A_8$ & $      $ & $      $ & $      $ & $      $ & $      $ & $      $ & $  1.00$ & $  0.10$ \\
       $A_9$ & $      $ & $      $ & $      $ & $      $ & $      $ & $      $ & $      $ & $  1.00$ \\
\end{tabular} 
\end{table}

\begin{table}[!htb]
\caption{
Correlation matrix for the \CP-asymmetric observables from the maximum likelihood fit in the bin $4.0 <q^2< 6.0\gevgevcccc$. 
\label{appendix:likelihood:correlation:asymmetry:4}
}
\centering
\begin{tabular}{l|rrrrrrrr}
& $F_{\rm L}$ & $A_3$ & $A_4$ & $A_5$ & $A_{6s}$ & $A_7$ & $A_8$ & $A_9$ \\ \hline
 $F_{\rm L}$ & $  1.00$ & $  0.03$ & $  0.02$ & $  0.03$ & $  0.01$ & $  0.05$ & $  0.08$ & $  0.01$ \\
       $A_3$ & $      $ & $  1.00$ & $  0.08$ & $  0.19$ & $  0.10$ & $ -0.16$ & $  0.06$ & $ -0.08$ \\
       $A_4$ & $      $ & $      $ & $  1.00$ & $  0.06$ & $ -0.01$ & $ -0.01$ & $ -0.01$ & $ -0.03$ \\
       $A_5$ & $      $ & $      $ & $      $ & $  1.00$ & $  0.08$ & $  0.00$ & $  0.01$ & $ -0.12$ \\
       $A_{6s}$ & $      $ & $      $ & $      $ & $      $ & $  1.00$ & $  -0.12$ & $  0.06$ & $  -0.05$ \\
       $A_7$ & $      $ & $      $ & $      $ & $      $ & $      $ & $  1.00$ & $  0.05$ & $ -0.01$ \\
       $A_8$ & $      $ & $      $ & $      $ & $      $ & $      $ & $      $ & $  1.00$ & $  0.13$ \\
       $A_9$ & $      $ & $      $ & $      $ & $      $ & $      $ & $      $ & $      $ & $  1.00$ \\
\end{tabular} 
\end{table}

\begin{table}[!htb]
\caption{
Correlation matrix for the \CP-asymmetric observables from the maximum likelihood fit in the bin $6.0 <q^2< 8.0\gevgevcccc$. 
\label{appendix:likelihood:correlation:asymmetry:5}
}
\centering
\begin{tabular}{l|rrrrrrrr}
& $F_{\rm L}$ & $A_3$ & $A_4$ & $A_5$ & $A_{6s}$ & $A_7$ & $A_8$ & $A_9$ \\ \hline
 $F_{\rm L}$ & $  1.00$ & $ -0.07$ & $ -0.06$ & $  0.03$ & $ -0.03$ & $ -0.01$ & $  0.02$ & $ -0.15$ \\
       $A_3$ & $      $ & $  1.00$ & $  0.08$ & $  0.12$ & $ -0.07$ & $ -0.01$ & $  0.01$ & $ -0.06$ \\
       $A_4$ & $      $ & $      $ & $  1.00$ & $ -0.07$ & $  0.08$ & $  0.01$ & $ -0.00$ & $  0.05$ \\
       $A_5$ & $      $ & $      $ & $      $ & $  1.00$ & $  0.13$ & $ -0.03$ & $  0.01$ & $  0.01$ \\
       $A_{6s}$ & $      $ & $      $ & $      $ & $      $ & $  1.00$ & $  0.02$ & $  0.05$ & $  -0.06$ \\
       $A_7$ & $      $ & $      $ & $      $ & $      $ & $      $ & $  1.00$ & $ -0.11$ & $  0.12$ \\
       $A_8$ & $      $ & $      $ & $      $ & $      $ & $      $ & $      $ & $  1.00$ & $  0.06$ \\
       $A_9$ & $      $ & $      $ & $      $ & $      $ & $      $ & $      $ & $      $ & $  1.00$ \\
\end{tabular} 
\end{table}

\begin{table}[!htb]
\caption{
Correlation matrix for the \CP-asymmetric observables from the maximum likelihood fit in the bin $11.0 <q^2< 12.5 \gevgevcccc$. 
\label{appendix:likelihood:correlation:asymmetry:6}
}
\centering
\begin{tabular}{l|rrrrrrrr}
& $F_{\rm L}$ & $A_3$ & $A_4$ & $A_5$ & $A_{6s}$ & $A_7$ & $A_8$ & $A_9$ \\ \hline
 $F_{\rm L}$ & $  1.00$ & $ -0.09$ & $ -0.01$ & $  0.00$ & $ -0.01$ & $ -0.02$ & $  0.01$ & $  0.06$ \\
       $A_3$ & $      $ & $  1.00$ & $  0.05$ & $  0.19$ & $  0.02$ & $ -0.16$ & $  0.08$ & $  0.11$ \\
       $A_4$ & $      $ & $      $ & $  1.00$ & $ -0.26$ & $  0.20$ & $ -0.10$ & $  0.02$ & $ -0.01$ \\
       $A_5$ & $      $ & $      $ & $      $ & $  1.00$ & $  0.11$ & $ -0.03$ & $ -0.06$ & $  0.04$ \\
       $A_{6s}$ & $      $ & $      $ & $      $ & $      $ & $  1.00$ & $  -0.06$ & $  0.11$ & $  -0.02$ \\
       $A_7$ & $      $ & $      $ & $      $ & $      $ & $      $ & $  1.00$ & $ -0.22$ & $  0.19$ \\
       $A_8$ & $      $ & $      $ & $      $ & $      $ & $      $ & $      $ & $  1.00$ & $  0.04$ \\
       $A_9$ & $      $ & $      $ & $      $ & $      $ & $      $ & $      $ & $      $ & $  1.00$ \\
\end{tabular} 
\end{table}

\begin{table}[!htb]
\caption{
Correlation matrix for the \CP-asymmetric observables from the maximum likelihood fit in the bin $15.0 <q^2< 17.0 \gevgevcccc$. 
\label{appendix:likelihood:correlation:asymmetry:7}
}
\centering
\begin{tabular}{l|rrrrrrrr}
& $F_{\rm L}$ & $A_3$ & $A_4$ & $A_5$ & $A_{6s}$ & $A_7$ & $A_8$ & $A_9$ \\ \hline
 $F_{\rm L}$ & $  1.00$ & $  0.04$ & $  0.02$ & $  0.02$ & $  0.06$ & $ -0.07$ & $  0.04$ & $ -0.07$ \\
       $A_3$ & $      $ & $  1.00$ & $  0.05$ & $  0.16$ & $  0.02$ & $ -0.00$ & $ -0.02$ & $  0.04$ \\
       $A_4$ & $      $ & $      $ & $  1.00$ & $ -0.25$ & $  0.16$ & $ -0.07$ & $  0.10$ & $ -0.01$ \\
       $A_5$ & $      $ & $      $ & $      $ & $  1.00$ & $  0.12$ & $  0.09$ & $ -0.07$ & $ -0.07$ \\
       $A_{6s}$ & $      $ & $      $ & $      $ & $      $ & $  1.00$ & $  -0.01$ & $  0.04$ & $  0.03$ \\
       $A_7$ & $      $ & $      $ & $      $ & $      $ & $      $ & $  1.00$ & $ -0.17$ & $  0.11$ \\
       $A_8$ & $      $ & $      $ & $      $ & $      $ & $      $ & $      $ & $  1.00$ & $  0.08$ \\
       $A_9$ & $      $ & $      $ & $      $ & $      $ & $      $ & $      $ & $      $ & $  1.00$ \\
\end{tabular} 
\end{table}

\begin{table}[!htb]
\caption{
Correlation matrix for the \CP-asymmetric observables from the maximum likelihood fit in the bin $17.0 <q^2< 19.0 \gevgevcccc$. 
\label{appendix:likelihood:correlation:asymmetry:8}
}
\centering
\begin{tabular}{l|rrrrrrrr}
& $F_{\rm L}$ & $A_3$ & $A_4$ & $A_5$ & $A_{6s}$ & $A_7$ & $A_8$ & $A_9$ \\ \hline
 $F_{\rm L}$ & $  1.00$ & $ -0.00$ & $  0.03$ & $ -0.00$ & $  0.09$ & $ -0.04$ & $  0.07$ & $ -0.03$ \\
       $A_3$ & $      $ & $  1.00$ & $  0.14$ & $  0.18$ & $ -0.01$ & $ -0.07$ & $ -0.05$ & $ -0.06$ \\
       $A_4$ & $      $ & $      $ & $  1.00$ & $ -0.09$ & $  0.10$ & $ -0.11$ & $  0.04$ & $ -0.09$ \\
       $A_5$ & $      $ & $      $ & $      $ & $  1.00$ & $  0.11$ & $  0.04$ & $ -0.11$ & $ -0.04$ \\
       $A_{6s}$ & $      $ & $      $ & $      $ & $      $ & $  1.00$ & $  -0.01$ & $  -0.08$ & $  -0.08$ \\
       $A_7$ & $      $ & $      $ & $      $ & $      $ & $      $ & $  1.00$ & $ -0.03$ & $  0.14$ \\
       $A_8$ & $      $ & $      $ & $      $ & $      $ & $      $ & $      $ & $  1.00$ & $  0.01$ \\
       $A_9$ & $      $ & $      $ & $      $ & $      $ & $      $ & $      $ & $      $ & $  1.00$ \\
\end{tabular} 
\end{table}

\begin{table}[!htb]
\caption{
Correlation matrix for the \CP-asymmetric observables from the maximum likelihood fit in the bin $1.1 <q^2< 6.0 \gevgevcccc$. 
\label{appendix:likelihood:correlation:asymmetry:9}
}
\centering
\begin{tabular}{l|rrrrrrrr}
& $F_{\rm L}$ & $A_3$ & $A_4$ & $A_5$ & $A_{6s}$ & $A_7$ & $A_8$ & $A_9$ \\ \hline
 $F_{\rm L}$ & $  1.00$ & $  0.04$ & $  0.00$ & $ -0.00$ & $ -0.01$ & $  0.01$ & $  0.04$ & $ -0.01$ \\
       $A_3$ & $      $ & $  1.00$ & $  0.05$ & $  0.04$ & $  0.04$ & $ -0.08$ & $  0.05$ & $  0.01$ \\
       $A_4$ & $      $ & $      $ & $  1.00$ & $  0.12$ & $ -0.09$ & $ -0.02$ & $  0.03$ & $  0.03$ \\
       $A_5$ & $      $ & $      $ & $      $ & $  1.00$ & $  0.05$ & $  0.05$ & $ -0.01$ & $  0.01$ \\
       $A_{6s}$ & $      $ & $      $ & $      $ & $      $ & $  1.00$ & $  -0.02$ & $  0.05$ & $  -0.03$ \\
       $A_7$ & $      $ & $      $ & $      $ & $      $ & $      $ & $  1.00$ & $  0.18$ & $  0.00$ \\
       $A_8$ & $      $ & $      $ & $      $ & $      $ & $      $ & $      $ & $  1.00$ & $  0.06$ \\
       $A_9$ & $      $ & $      $ & $      $ & $      $ & $      $ & $      $ & $      $ & $  1.00$ \\
\end{tabular} 
\end{table}

\begin{table}[!htb]
\caption{
Correlation matrix for the \CP-asymmetric observables from the maximum likelihood fit in the bin $15.0 <q^2< 19.0 \gevgevcccc$. 
\label{appendix:likelihood:correlation:asymmetry:10}
}
\centering
\begin{tabular}{l|rrrrrrrr}
& $F_{\rm L}$ & $A_3$ & $A_4$ & $A_5$ & $A_{6s}$ & $A_7$ & $A_8$ & $A_9$ \\ \hline
 $F_{\rm L}$ & $  1.00$ & $  0.02$ & $  0.01$ & $ -0.01$ & $  0.06$ & $ -0.04$ & $  0.04$ & $ -0.04$ \\
       $A_3$ & $      $ & $  1.00$ & $  0.07$ & $  0.17$ & $  0.03$ & $ -0.02$ & $ -0.02$ & $  0.02$ \\
       $A_4$ & $      $ & $      $ & $  1.00$ & $ -0.19$ & $  0.13$ & $ -0.06$ & $  0.06$ & $ -0.03$ \\
       $A_5$ & $      $ & $      $ & $      $ & $  1.00$ & $  0.11$ & $  0.06$ & $ -0.07$ & $ -0.05$ \\
       $A_{6s}$ & $      $ & $      $ & $      $ & $      $ & $  1.00$ & $  -0.00$ & $  -0.01$ & $  -0.01$ \\
       $A_7$ & $      $ & $      $ & $      $ & $      $ & $      $ & $  1.00$ & $ -0.11$ & $  0.13$ \\
       $A_8$ & $      $ & $      $ & $      $ & $      $ & $      $ & $      $ & $  1.00$ & $  0.05$ \\
       $A_9$ & $      $ & $      $ & $      $ & $      $ & $      $ & $      $ & $      $ & $  1.00$ \\
\end{tabular} 
\end{table}

\clearpage

\section{Correlation matrices for the optimised angular observables from the maximum likelihood fit} 
\label{sec:appendix:correlation:optimised}

Correlation matrices between $F_{\rm L}$ and the optimised $P^{(\prime)}_{i}$ basis of observables in the different \qsq bins are provided in Tables~\ref{appendix:likelihood:correlation:optimised:1}--\ref{appendix:likelihood:correlation:optimised:10} for the likelihood fit.

\begin{table}[!htb]
\caption{
Correlation matrix for the optimised angular observables from the maximum likelihood fit in the bin $0.10<q^2<0.98 \gevgevcccc$. 
\label{appendix:likelihood:correlation:optimised:1}
}
\centering
\begin{tabular}{l|rrrrrrrr}
& $F_{\rm L}$ & $P_1$ & $P_2$ & $P_3$ & $P_4^\prime$ & $P_5^\prime$ & $P_6^\prime$ & $P_8^\prime$ \\ \hline
 $F_{\rm L}$ & $  1.00$ & $  0.02$ & $  0.03$ & $  0.00$ & $ -0.07$ & $ -0.12$ & $ -0.03$ & $  0.01$ \\
       $P_1$ & $      $ & $  1.00$ & $ -0.00$ & $  0.04$ & $  0.01$ & $  0.09$ & $ -0.07$ & $ -0.02$ \\
       $P_2$ & $      $ & $      $ & $  1.00$ & $ -0.02$ & $  0.11$ & $  0.04$ & $  0.03$ & $ -0.08$ \\
       $P_3$ & $      $ & $      $ & $      $ & $  1.00$ & $ -0.02$ & $ -0.04$ & $ -0.11$ & $ -0.01$ \\
$P_4^\prime$ & $      $ & $      $ & $      $ & $      $ & $  1.00$ & $  0.09$ & $  0.00$ & $  0.07$ \\
$P_5^\prime$ & $      $ & $      $ & $      $ & $      $ & $      $ & $  1.00$ & $ -0.00$ & $ -0.00$ \\
$P_6^\prime$ & $      $ & $      $ & $      $ & $      $ & $      $ & $      $ & $  1.00$ & $  0.01$ \\
$P_8^\prime$ & $      $ & $      $ & $      $ & $      $ & $      $ & $      $ & $      $ & $  1.00$ \\
\end{tabular} 
\end{table}

\begin{table}[!htb]
\caption{
Correlation matrix for the optimised angular observables from the maximum likelihood fit in the bin $1.1<q^2<2.5 \gevgevcccc$. 
\label{appendix:likelihood:correlation:optimised:2}
}
\centering
\begin{tabular}{l|rrrrrrrr}
& $F_{\rm L}$ & $P_1$ & $P_2$ & $P_3$ & $P_4^\prime$ & $P_5^\prime$ & $P_6^\prime$ & $P_8^\prime$ \\ \hline
 $F_{\rm L}$ & $  1.00$ & $ -0.11$ & $ -0.48$ & $  0.23$ & $  0.03$ & $  0.15$ & $ -0.17$ & $ -0.09$ \\
       $P_1$ & $      $ & $  1.00$ & $  0.06$ & $ -0.13$ & $ -0.05$ & $  0.01$ & $  0.15$ & $  0.11$ \\
       $P_2$ & $      $ & $      $ & $  1.00$ & $ -0.02$ & $ -0.03$ & $ -0.19$ & $  0.06$ & $ -0.04$ \\
       $P_3$ & $      $ & $      $ & $      $ & $  1.00$ & $ -0.01$ & $  0.24$ & $ -0.26$ & $  0.01$ \\
$P_4^\prime$ & $      $ & $      $ & $      $ & $      $ & $  1.00$ & $ -0.22$ & $ -0.00$ & $ -0.05$ \\
$P_5^\prime$ & $      $ & $      $ & $      $ & $      $ & $      $ & $  1.00$ & $ -0.13$ & $ -0.04$ \\
$P_6^\prime$ & $      $ & $      $ & $      $ & $      $ & $      $ & $      $ & $  1.00$ & $ -0.10$ \\
$P_8^\prime$ & $      $ & $      $ & $      $ & $      $ & $      $ & $      $ & $      $ & $  1.00$ \\
\end{tabular} 
\end{table}

\begin{table}[!htb]
\caption{
Correlation matrix for the optimised angular observables from the maximum likelihood fit in the bin $2.5<q^2<4.0 \gevgevcccc$. 
\label{appendix:likelihood:correlation:optimised:3}
}
\centering
\begin{tabular}{l|rrrrrrrr}
& $F_{\rm L}$ & $P_1$ & $P_2$ & $P_3$ & $P_4^\prime$ & $P_5^\prime$ & $P_6^\prime$ & $P_8^\prime$ \\ \hline
 $F_{\rm L}$ & $  1.00$ & $  0.23$ & $ -0.79$ & $  0.61$ & $ -0.60$ & $ -0.05$ & $  0.29$ & $  0.06$ \\
       $P_1$ & $      $ & $  1.00$ & $ -0.14$ & $  0.19$ & $ -0.20$ & $  0.08$ & $  0.06$ & $  0.02$ \\
       $P_2$ & $      $ & $      $ & $  1.00$ & $ -0.53$ & $  0.43$ & $  0.04$ & $ -0.23$ & $ -0.11$ \\
       $P_3$ & $      $ & $      $ & $      $ & $  1.00$ & $ -0.41$ & $ -0.11$ & $  0.21$ & $  0.04$ \\
$P_4^\prime$ & $      $ & $      $ & $      $ & $      $ & $  1.00$ & $ -0.12$ & $ -0.21$ & $  0.06$ \\
$P_5^\prime$ & $      $ & $      $ & $      $ & $      $ & $      $ & $  1.00$ & $  0.03$ & $ -0.03$ \\
$P_6^\prime$ & $      $ & $      $ & $      $ & $      $ & $      $ & $      $ & $  1.00$ & $  0.08$ \\
$P_8^\prime$ & $      $ & $      $ & $      $ & $      $ & $      $ & $      $ & $      $ & $  1.00$ \\
\end{tabular} 
\end{table}

\begin{table}[!htb]
\caption{
Correlation matrix for the optimised angular observables from the maximum likelihood fit in the bin $4.0 <q^2< 6.0 \gevgevcccc$. 
\label{appendix:likelihood:correlation:optimised:4}
}
\centering
\begin{tabular}{l|rrrrrrrr}
& $F_{\rm L}$ & $P_1$ & $P_2$ & $P_3$ & $P_4^\prime$ & $P_5^\prime$ & $P_6^\prime$ & $P_8^\prime$ \\ \hline
 $F_{\rm L}$ & $  1.00$ & $  0.04$ & $  0.02$ & $  0.06$ & $  0.02$ & $  0.06$ & $ -0.10$ & $  0.09$ \\
       $P_1$ & $      $ & $  1.00$ & $  0.09$ & $  0.12$ & $ -0.04$ & $ -0.03$ & $ -0.11$ & $  0.00$ \\
       $P_2$ & $      $ & $      $ & $  1.00$ & $  0.03$ & $ -0.09$ & $ -0.05$ & $  0.02$ & $  0.07$ \\
       $P_3$ & $      $ & $      $ & $      $ & $  1.00$ & $ -0.04$ & $  0.04$ & $  0.14$ & $ -0.02$ \\
$P_4^\prime$ & $      $ & $      $ & $      $ & $      $ & $  1.00$ & $  0.10$ & $ -0.02$ & $ -0.04$ \\
$P_5^\prime$ & $      $ & $      $ & $      $ & $      $ & $      $ & $  1.00$ & $ -0.03$ & $ -0.01$ \\
$P_6^\prime$ & $      $ & $      $ & $      $ & $      $ & $      $ & $      $ & $  1.00$ & $  0.06$ \\
$P_8^\prime$ & $      $ & $      $ & $      $ & $      $ & $      $ & $      $ & $      $ & $  1.00$ \\
\end{tabular} 
\end{table}

\begin{table}[!htb]
\caption{
Correlation matrix for the optimised angular observables from the maximum likelihood fit in the bin $6.0 <q^2< 8.0 \gevgevcccc$. 
\label{appendix:likelihood:correlation:optimised:5}
}
\centering
\begin{tabular}{l|rrrrrrrr}
& $F_{\rm L}$ & $P_1$ & $P_2$ & $P_3$ & $P_4^\prime$ & $P_5^\prime$ & $P_6^\prime$ & $P_8^\prime$ \\ \hline
 $F_{\rm L}$ & $  1.00$ & $ -0.05$ & $  0.11$ & $  0.11$ & $ -0.01$ & $ -0.03$ & $ -0.09$ & $ -0.03$ \\
       $P_1$ & $      $ & $  1.00$ & $  0.02$ & $  0.06$ & $ -0.16$ & $ -0.23$ & $  0.03$ & $  0.02$ \\
       $P_2$ & $      $ & $      $ & $  1.00$ & $  0.01$ & $ -0.11$ & $ -0.16$ & $ -0.05$ & $  0.04$ \\
       $P_3$ & $      $ & $      $ & $      $ & $  1.00$ & $ -0.01$ & $  0.03$ & $  0.05$ & $  0.10$ \\
$P_4^\prime$ & $      $ & $      $ & $      $ & $      $ & $  1.00$ & $ -0.13$ & $ -0.01$ & $ -0.11$ \\
$P_5^\prime$ & $      $ & $      $ & $      $ & $      $ & $      $ & $  1.00$ & $ -0.13$ & $ -0.01$ \\
$P_6^\prime$ & $      $ & $      $ & $      $ & $      $ & $      $ & $      $ & $  1.00$ & $  0.10$ \\
$P_8^\prime$ & $      $ & $      $ & $      $ & $      $ & $      $ & $      $ & $      $ & $  1.00$ \\
\end{tabular} 
\end{table}

\begin{table}[!htb]
\caption{
Correlation matrix for the optimised angular observables from the maximum likelihood fit in the bin $11.0 <q^2< 12.5 \gevgevcccc$. 
\label{appendix:likelihood:correlation:optimised:6}
}
\centering
\begin{tabular}{l|rrrrrrrr}
& $F_{\rm L}$ & $P_1$ & $P_2$ & $P_3$ & $P_4^\prime$ & $P_5^\prime$ & $P_6^\prime$ & $P_8^\prime$ \\ \hline
 $F_{\rm L}$ & $  1.00$ & $ -0.12$ & $  0.13$ & $ -0.02$ & $  0.03$ & $ -0.02$ & $  0.03$ & $  0.05$ \\
       $P_1$ & $      $ & $  1.00$ & $ -0.13$ & $  0.03$ & $  0.04$ & $ -0.35$ & $ -0.05$ & $  0.05$ \\
       $P_2$ & $      $ & $      $ & $  1.00$ & $ -0.12$ & $  0.10$ & $ -0.01$ & $  0.01$ & $ -0.04$ \\
       $P_3$ & $      $ & $      $ & $      $ & $  1.00$ & $  0.08$ & $  0.24$ & $  0.19$ & $  0.09$ \\
$P_4^\prime$ & $      $ & $      $ & $      $ & $      $ & $  1.00$ & $ -0.02$ & $ -0.05$ & $ -0.12$ \\
$P_5^\prime$ & $      $ & $      $ & $      $ & $      $ & $      $ & $  1.00$ & $ -0.04$ & $ -0.09$ \\
$P_6^\prime$ & $      $ & $      $ & $      $ & $      $ & $      $ & $      $ & $  1.00$ & $  0.27$ \\
$P_8^\prime$ & $      $ & $      $ & $      $ & $      $ & $      $ & $      $ & $      $ & $  1.00$ \\
\end{tabular} 
\end{table}

\begin{table}[!htb]
\caption{
Correlation matrix for the optimised angular observables from the maximum likelihood fit in the bin $15.0 <q^2< 17.0 \gevgevcccc$. 
\label{appendix:likelihood:correlation:optimised:7}
}
\centering
\begin{tabular}{l|rrrrrrrr}
& $F_{\rm L}$ & $P_1$ & $P_2$ & $P_3$ & $P_4^\prime$ & $P_5^\prime$ & $P_6^\prime$ & $P_8^\prime$ \\ \hline
 $F_{\rm L}$ & $  1.00$ & $  0.06$ & $  0.19$ & $ -0.12$ & $  0.07$ & $  0.25$ & $ -0.05$ & $ -0.07$ \\
       $P_1$ & $      $ & $  1.00$ & $  0.16$ & $ -0.07$ & $ -0.06$ & $ -0.04$ & $ -0.04$ & $ -0.04$ \\
       $P_2$ & $      $ & $      $ & $  1.00$ & $ -0.01$ & $ -0.22$ & $ -0.12$ & $  0.04$ & $ -0.07$ \\
       $P_3$ & $      $ & $      $ & $      $ & $  1.00$ & $ -0.07$ & $ -0.07$ & $  0.23$ & $  0.11$ \\
$P_4^\prime$ & $      $ & $      $ & $      $ & $      $ & $  1.00$ & $  0.28$ & $ -0.06$ & $  0.18$ \\
$P_5^\prime$ & $      $ & $      $ & $      $ & $      $ & $      $ & $  1.00$ & $  0.10$ & $ -0.02$ \\
$P_6^\prime$ & $      $ & $      $ & $      $ & $      $ & $      $ & $      $ & $  1.00$ & $  0.25$ \\
$P_8^\prime$ & $      $ & $      $ & $      $ & $      $ & $      $ & $      $ & $      $ & $  1.00$ \\
\end{tabular} 
\end{table}

\begin{table}[!htb]
\caption{
Correlation matrix for the optimised angular observables from the maximum likelihood fit in the bin $17.0 <q^2< 19.0 \gevgevcccc$. 
\label{appendix:likelihood:correlation:optimised:8}
}
\centering
\begin{tabular}{l|rrrrrrrr}
& $F_{\rm L}$ & $P_1$ & $P_2$ & $P_3$ & $P_4^\prime$ & $P_5^\prime$ & $P_6^\prime$ & $P_8^\prime$ \\ \hline
 $F_{\rm L}$ & $  1.00$ & $ -0.14$ & $  0.14$ & $  0.03$ & $  0.20$ & $  0.21$ & $  0.05$ & $  0.07$ \\
       $P_1$ & $      $ & $  1.00$ & $ -0.05$ & $  0.07$ & $ -0.18$ & $ -0.41$ & $ -0.07$ & $ -0.06$ \\
       $P_2$ & $      $ & $      $ & $  1.00$ & $  0.00$ & $ -0.13$ & $ -0.06$ & $  0.01$ & $ -0.05$ \\
       $P_3$ & $      $ & $      $ & $      $ & $  1.00$ & $  0.05$ & $  0.01$ & $  0.16$ & $  0.12$ \\
$P_4^\prime$ & $      $ & $      $ & $      $ & $      $ & $  1.00$ & $  0.14$ & $  0.03$ & $  0.19$ \\
$P_5^\prime$ & $      $ & $      $ & $      $ & $      $ & $      $ & $  1.00$ & $  0.05$ & $  0.02$ \\
$P_6^\prime$ & $      $ & $      $ & $      $ & $      $ & $      $ & $      $ & $  1.00$ & $  0.34$ \\
$P_8^\prime$ & $      $ & $      $ & $      $ & $      $ & $      $ & $      $ & $      $ & $  1.00$ \\
\end{tabular} 
\end{table}

\begin{table}[!htb]
\caption{
Correlation matrix for the optimised angular observables from the maximum likelihood fit in the bin $1.1 <q^2< 6.0 \gevgevcccc$. 
\label{appendix:likelihood:correlation:optimised:9}
}
\centering
\begin{tabular}{l|rrrrrrrr}
& $F_{\rm L}$ & $P_1$ & $P_2$ & $P_3$ & $P_4^\prime$ & $P_5^\prime$ & $P_6^\prime$ & $P_8^\prime$ \\ \hline
 $F_{\rm L}$ & $  1.00$ & $ -0.01$ & $ -0.20$ & $  0.07$ & $ -0.03$ & $  0.01$ & $ -0.08$ & $  0.00$ \\
       $P_1$ & $      $ & $  1.00$ & $  0.05$ & $  0.00$ & $ -0.05$ & $ -0.00$ & $  0.01$ & $  0.01$ \\
       $P_2$ & $      $ & $      $ & $  1.00$ & $  0.03$ & $ -0.10$ & $ -0.07$ & $  0.04$ & $ -0.02$ \\
       $P_3$ & $      $ & $      $ & $      $ & $  1.00$ & $ -0.05$ & $  0.04$ & $  0.00$ & $  0.03$ \\
$P_4^\prime$ & $      $ & $      $ & $      $ & $      $ & $  1.00$ & $ -0.05$ & $ -0.02$ & $ -0.01$ \\
$P_5^\prime$ & $      $ & $      $ & $      $ & $      $ & $      $ & $  1.00$ & $ -0.01$ & $ -0.02$ \\
$P_6^\prime$ & $      $ & $      $ & $      $ & $      $ & $      $ & $      $ & $  1.00$ & $  0.04$ \\
$P_8^\prime$ & $      $ & $      $ & $      $ & $      $ & $      $ & $      $ & $      $ & $  1.00$ \\
\end{tabular} 
\end{table}

\begin{table}[!htb]
\caption{
Correlation matrix for the optimised angular observables from the maximum likelihood fit in the bin $15.0 <q^2< 19.0 \gevgevcccc$. 
\label{appendix:likelihood:correlation:optimised:10}
}
\centering
\begin{tabular}{l|rrrrrrrr}
& $F_{\rm L}$ & $P_1$ & $P_2$ & $P_3$ & $P_4^\prime$ & $P_5^\prime$ & $P_6^\prime$ & $P_8^\prime$ \\ \hline
 $F_{\rm L}$ & $  1.00$ & $ -0.03$ & $  0.14$ & $ -0.05$ & $  0.11$ & $  0.15$ & $ -0.01$ & $ -0.01$ \\
       $P_1$ & $      $ & $  1.00$ & $  0.13$ & $  0.04$ & $ -0.14$ & $ -0.19$ & $ -0.02$ & $ -0.04$ \\
       $P_2$ & $      $ & $      $ & $  1.00$ & $ -0.05$ & $ -0.13$ & $ -0.11$ & $  0.01$ & $ -0.03$ \\
       $P_3$ & $      $ & $      $ & $      $ & $  1.00$ & $ -0.02$ & $ -0.03$ & $  0.19$ & $  0.13$ \\
$P_4^\prime$ & $      $ & $      $ & $      $ & $      $ & $  1.00$ & $  0.08$ & $  0.03$ & $  0.14$ \\
$P_5^\prime$ & $      $ & $      $ & $      $ & $      $ & $      $ & $  1.00$ & $  0.11$ & $  0.04$ \\
$P_6^\prime$ & $      $ & $      $ & $      $ & $      $ & $      $ & $      $ & $  1.00$ & $  0.24$ \\
$P_8^\prime$ & $      $ & $      $ & $      $ & $      $ & $      $ & $      $ & $      $ & $  1.00$ \\
\end{tabular} 
\end{table}

\clearpage

\section{Correlation matrices for the \boldmath{\CP}-averaged observables from the method of moments}
\label{sec:appendix:bootstrap:correlation}

Correlation matrices between the \CP-averaged observables in the different \qsq bins are in Tables~\ref{appendix:moments:correlation:average:1}--\ref{appendix:moments:correlation:average:15} for the moment analysis. 
The correlations are determined by a bootstrapping technique.

\begin{table}[!htb]
\caption{
Correlation matrix for the \CP-averaged observables obtained for the method of moments in the bin $0.10<q^2<0.98\gevgevcccc$. 
\label{appendix:moments:correlation:average:1}
}
\centering 
\begin{tabular}{l|rrrrrrrr}
 & $F_{\rm L}$ & $S_{3}$ & $S_{4}$ & $S_{5}$ & $A_{\rm FB}$ & $S_{7}$ & $S_{8}$ & $S_{9}$ \\ 
\hline
$F_{\rm L}$ & $1.00$ & $0.02 $ & $0.00$ & $-0.05 $ &  $0.23 $ & $0.02 $ & $-0.01 $ & $0.05$ \\
$S_{3}$ & & $1.00$ &  $0.04 $ &         $0.09 $ &      $-0.01 $ &    $0.01 $ &    $-0.04 $ &    $0.05 $ \\
$S_{4}$ & & & $1.00$ & $-0.24 $ &     $-0.05 $ &    $-0.07 $ &          $0.06 $ &    $0.03 $  \\
$S_{5}$ & & & & $1.00$ &  $0.12 $ &     $0.00$ &           $-0.09 $ &         $-0.02 $ \\
$A_{\rm FB}$ & & & & & $1.00$ &  $0.09 $ &           $-0.07 $ &         $-0.04 $  \\ 
$S_{7}$ & & & & & & $1.00$ &  $-0.09$ &    $0.10$  \\ 
$S_{8}$ & & & & & & & $1.00$ & $0.03$  \\
$S_{9}$ & & & & & & & & $1.00$  \\
\end{tabular}
\end{table}

\begin{table}[!htb]
\caption{
Correlation matrix for the \CP-averaged observables obtained for the method of moments in the bin $1.1<q^2<2.0\gevgevcccc$. 
\label{appendix:moments:correlation:average:2}
}
\centering 
\begin{tabular}{l|rrrrrrrr}
& $F_{\rm L}$ & $S_{3}$ & $S_{4}$ & $S_{5}$ & $A_{\rm FB}$ & $S_{7}$ & $S_{8}$ & $S_{9}$ \\ 
\hline
$F_{\rm L}$ & $1.00$ &       $-0.02 $ &      $0.06 $ &         $0.16 $ &    $-0.05 $ &      $-0.03 $ &            $-0.04 $ &           $0.09 $\\
$S_{3}$ & & $1.00$ &   $-0.01 $ &        $0.03 $ &    $0.08 $ &       $0.10$ &              $-0.03 $ &           $0.08 $\\
$S_{4}$ & & & $1.00$ &   $0.00$ &    $-0.03 $ &      $-0.13 $ &            $0.00 $ &            $-0.06 $ \\
$S_{5}$ & & & & $1.00$ &  $-0.07 $ &      $-0.05 $ &            $-0.11 $ &           $-0.06 $ \\
$A_{\rm FB}$ & & & & & $1.00$ &   $0.04 $ &      $-0.06 $ &           $-0.10$ \\
$S_{7}$ & & & & & & $1.00$ & $-0.05 $ &           $0.01 $\\
$S_{8}$ & & & & & & & $1.00$ & $-0.01$  \\
$S_{9}$ & & & & & & & & $1.00$  \\
\end{tabular}
\end{table}

\begin{table}[!htb]
\caption{
Correlation matrix for the \CP-averaged observables obtained for the method of moments in the bin $2.0<q^2<3.0\gevgevcccc$. 
\label{appendix:moments:correlation:average:3}
}
\centering 
\begin{tabular}{l|rrrrrrrr}
 & $F_{\rm L}$ & $S_{3}$ & $S_{4}$ & $S_{5}$ & $A_{\rm FB}$ & $S_{7}$ & $S_{8}$ & $S_{9}$ \\ 
\hline
$F_{\rm L}$ & $1.00$ & $-0.12 $ &      $-0.14 $ &        $-0.04 $ &     $0.11 $ &     $-0.18 $ &            $0.04 $ &    $0.01 $\\
$S_{3}$ & & $1.00$ & $-0.09 $ &        $0.06 $ &      $0.04 $ &     $0.08 $ &             $-0.04 $ &         $-0.01 $ \\
$S_{4}$ & & & $1.00$ & $-0.08 $ &     $0.00$ &      $0.03 $ &             $-0.05 $ &         $-0.06 $\\
$S_{5}$ & & & & $1.00$ &  $-0.10$ &     $-0.08 $ &            $0.04 $ &          $-0.08 $\\
$A_{\rm FB}$ & & & & & $1.00$ & $-0.01 $ &            $-0.10$ &          $0.04 $\\
$S_{7}$ & & & & & & $1.00$ & $-0.12 $ &         $0.01 $\\
$S_{8}$ & & & & & & & $1.00$ &  $-0.07 $\\
$S_{9}$ & & & & & & & & $1.00$  \\
\end{tabular}
\end{table}

\begin{table}[!htb]
\caption{
Correlation matrix for the \CP-averaged observables obtained for the method of moments in the bin $3.0<q^2<4.0\gevgevcccc$. 
\label{appendix:moments:correlation:average:4}
}
\centering 
\begin{tabular}{l|rrrrrrrr}
& $F_{\rm L}$ & $S_{3}$ & $S_{4}$ & $S_{5}$ & $A_{\rm FB}$ & $S_{7}$ & $S_{8}$ & $S_{9}$ \\ 
\hline
$F_{\rm L}$ & $1.00$ &  $0.10 $ &        $-0.01 $ &        $-0.03 $ &     $0.01 $ &     $0.01 $ &    $-0.16 $ &    $0.05 $\\
$S_{3}$ & & $1.00$ &  $-0.01 $ &        $-0.04 $ &     $0.03 $ &     $-0.08 $ &          $-0.04 $ &     $-0.11 $ \\
$S_{4}$ & & & $1.00$ &  $0.18 $ &      $0.05 $ &     $0.01 $ &           $-0.05 $ &     $-0.02 $\\
$S_{5}$ & & & & $1.00$ &  $0.00$ &      $-0.01 $ &          $0.02 $ &      $-0.03 $ \\
$A_{\rm FB}$ & & & & & $1.00$ & $0.04 $ &           $-0.03 $ &     $-0.01 $\\
$S_{7}$ & & & & & & $1.00$ & $0.18 $ &      $-0.08 $\\
$S_{8}$ & & & & & & & $1.00$ & $-0.03$  \\
$S_{9}$ & & & & & & & & $1.00$  \\
\end{tabular}
\end{table}

\begin{table}[!htb]
\caption{
Correlation matrix for the \CP-averaged observables obtained for the method of moments in the bin $4.0<q^2<5.0\gevgevcccc$. 
\label{appendix:moments:correlation:average:5}
}
\centering 
\begin{tabular}{l|rrrrrrrr}
 & $F_{\rm L}$ & $S_{3}$ & $S_{4}$ & $S_{5}$ & $A_{\rm FB}$ & $S_{7}$ & $S_{8}$ & $S_{9}$ \\ 
\hline
$F_{\rm L}$ & $1.00$ & $-0.01 $ &      $0.03 $ &         $-0.12 $ &     $-0.02 $ &    $-0.07 $ &            $0.04 $ &    $0.10$ \\
$S_{3}$ & & $1.00$ & $-0.10$ &         $-0.11 $ &     $0.03 $ &     $0.08 $ &             $-0.12 $ &        $0.07 $\\
$S_{4}$ & & & $1.00$ & $0.15 $ &      $-0.03 $ &    $-0.07 $ &            $0.21 $ &         $0.04 $\\
$S_{5}$ & & & & $1.00$ & $-0.03 $ &    $0.10$ &              $-0.02 $ &        $-0.09 $\\
$A_{\rm FB}$ & & & & & $1.00$ & $0.11 $ &             $-0.15 $ &        $0.00$ \\
$S_{7}$ & & & & & & $1.00$ & $0.07 $ &         $-0.07 $\\
$S_{8}$ & & & & & & & $1.00$ & $0.0$  \\
$S_{9}$ & & & & & & & & $1.00$  \\
\end{tabular}
\end{table}

\begin{table}[!htb]
\caption{
Correlation matrix for the \CP-averaged observables obtained for the method of moments in the bin $5.0<q^2<6.0\gevgevcccc$. 
\label{appendix:moments:correlation:average:6}
}
\centering 
\begin{tabular}{l|rrrrrrrr}
& $F_{\rm L}$ & $S_{3}$ & $S_{4}$ & $S_{5}$ & $A_{\rm FB}$ & $S_{7}$ & $S_{8}$ & $S_{9}$ \\ 
\hline
$F_{\rm L}$ & $1.00$ &  $-0.03 $ &      $-0.06 $ &        $-0.04 $ &     $-0.01 $ &    $-0.03 $ &            $-0.09 $ &           $-0.02 $ \\
$S_{3}$ &  & $1.00$ &   $-0.01 $ &        $-0.06 $ &     $-0.11 $ &    $-0.05 $ &            $0.02 $ &            $0.11 $\\
$S_{4}$ & & & $1.00$ &  $0.10$ &       $-0.03 $ &    $0.08 $ &             $0.02 $ &            $0.01 $\\
$S_{5}$ & & & & $1.00$ & $-0.08 $ &    $-0.03 $ &            $0.06 $ &            $0.07 $\\
$A_{\rm FB}$ & & & & & $1.00$ &  $0.01$ &             $0.00$ &             $0.00 $\\
$S_{7}$ & & & & & & $1.00$ & $0.07$ &            $-0.09 $\\
$S_{8}$ & & & & & & & $1.00$ & $-0.13$  \\
$S_{9}$ & & & & & & & & $1.00$  \\
\end{tabular}
\end{table}

\begin{table}[!htb]
\caption{
Correlation matrix for the \CP-averaged observables obtained for the method of moments in the bin $6.0<q^2<7.0\gevgevcccc$. 
\label{appendix:moments:correlation:average:7}
}
\centering 
\begin{tabular}{l|rrrrrrrr}
& $F_{\rm L}$ & $S_{3}$ & $S_{4}$ & $S_{5}$ & $A_{\rm FB}$ & $S_{7}$ & $S_{8}$ & $S_{9}$ \\ 
\hline
$F_{\rm L}$ & $1.00$ &  $0.00$ &       $-0.24 $ &        $-0.14 $ &     $-0.04 $ &    $0.08 $ &    $0.07 $ &     $-0.03 $ \\
$S_{3}$ & & $1.00$ & $-0.09 $ &        $-0.17 $ &     $-0.08 $ &    $0.02 $ &    $-0.04 $ &    $-0.02 $\\
$S_{4}$ & & & $1.00$ & $0.13 $ &      $-0.12 $ &    $-0.03 $ &          $-0.01 $ &    $-0.04 $\\
$S_{5}$ & & & & $1.00$ &  $-0.07 $ &    $-0.01 $ &          $-0.01 $ &    $-0.04 $\\
$A_{\rm FB}$ & & & & & $1.00$ & $0.02 $ &           $-0.01 $ &    $-0.05 $\\
$S_{7}$ & & & & & & $1.00$ & $0.21$ & $-0.11$  \\
$S_{8}$ & & & & & & & $1.00$ & $-0.06$  \\
$S_{9}$ & & & & & & & & $1.00$  \\
\end{tabular}
\end{table}

\begin{table}[!htb]
\caption{
Correlation matrix for the \CP-averaged observables obtained for the method of moments in the bin $7.0<q^2<8.0\gevgevcccc$. 
\label{appendix:moments:correlation:average:8}
}
\centering 
\begin{tabular}{l|rrrrrrrr}
& $F_{\rm L}$ & $S_{3}$ & $S_{4}$ & $S_{5}$ & $A_{\rm FB}$ & $S_{7}$ & $S_{8}$ & $S_{9}$ \\ 
\hline
$F_{\rm L}$ & $1.00$ & $0.07 $ &       $-0.13 $ &      $-0.22 $ &      $-0.08 $ &      $-0.07 $ &      $-0.01 $ &      $0.09 $ \\
$S_{3}$ & & $1.00$ & $-0.12 $ &      $-0.15 $ &      $0.07 $ &       $0.05 $ &       $0.02 $ &       $-0.01 $  \\
$S_{4}$ & & & $1.00$ &  $0.15 $ &       $-0.09 $ &      $-0.05 $ &      $0.06 $ &       $0.00$  \\
$S_{5}$ & & & & $1.00$ & $-0.15 $ &      $0.13 $ &       $0.00$ &        $0.03 $ \\
$A_{\rm FB}$ & & & & & $1.00$ &  $-0.02 $ &      $-0.16 $ &      $0.04 $ \\
$S_{7}$ & & & & & & $1.00$ & $0.07 $ &       $-0.11 $\\
$S_{8}$ & & & & & & & $1.00$ & $-0.07$  \\
$S_{9}$ & & & & & & & & $1.00$  \\
\end{tabular}
\end{table}

\begin{table}[!htb]
\caption{
Correlation matrix for the \CP-averaged observables obtained for the method of moments in the bin $11.00 <q^2<11.75\gevgevcccc$. 
\label{appendix:moments:correlation:average:9}
}
\centering 
\begin{tabular}{l|rrrrrrrr}
 & $F_{\rm L}$ & $S_{3}$ & $S_{4}$ & $S_{5}$ & $A_{\rm FB}$ & $S_{7}$ & $S_{8}$ & $S_{9}$ \\ 
\hline
$F_{\rm L}$ & $1.00$ &   $0.15 $ &       $0.16 $ &       $0.03 $ &       $-0.34 $ &      $-0.05 $ &      $-0.12 $ &      $-0.01 $ \\
$S_{3}$ & & $1.00$ &   $-0.06 $ &      $-0.21 $ &      $-0.06 $ &      $0.04 $ &       $0.00$ &       $-0.02 $ \\
$S_{4}$ & & & $1.00$ & $0.19 $ &       $-0.19 $ &      $-0.11 $ &      $-0.15 $ &      $-0.04 $\\
$S_{5}$ & & & & $1.00$ & $-0.11 $ &      $-0.13 $ &      $-0.10$ &       $-0.09 $\\
$A_{\rm FB}$ & & & & & $1.00$ & $0.03 $ &       $-0.03 $ &      $-0.04 $ \\
$S_{7}$ & & & & & & $1.00$ &   $0.24 $ &       $-0.03 $\\
$S_{8}$ & & & & & & & $1.00$ & $-0.10$  \\
$S_{9}$ & & & & & & & & $1.00$  \\
\end{tabular}
\end{table}

\begin{table}[!htb]
\caption{
Correlation matrix for the \CP-averaged observables obtained for the method of moments in the bin $11.75 <q^2<12.50\gevgevcccc$. 
\label{appendix:moments:correlation:average:10}
}
\centering 
\begin{tabular}{l|rrrrrrrr}
& $F_{\rm L}$ & $S_{3}$ & $S_{4}$ & $S_{5}$ & $A_{\rm FB}$ & $S_{7}$ & $S_{8}$ & $S_{9}$ \\ 
\hline
$F_{\rm L}$ & $1.00$ &  $0.04 $ &       $-0.05 $ &      $-0.01 $ &      $-0.17 $ &      $-0.08 $ &      $0.05 $ &       $0.00$\\
$S_{3}$ & & $1.00$ &  $-0.13 $ &      $-0.14 $ &      $0.00$ &        $0.02 $ &       $0.00$ &       $0.05 $  \\
$S_{4}$ & & & $1.00$ &  $0.16 $ &       $-0.22 $ &      $0.10$ &        $0.18 $ &       $-0.02 $\\
$S_{5}$ & & & & $1.00$ &  $-0.17 $ &      $0.16 $ &       $0.08 $ &       $-0.10$ \\
$A_{\rm FB}$ & & & & & $1.00$ &  $-0.08 $ &      $-0.12 $ &      $0.07 $ \\
$S_{7}$ & & & & & & $1.00$ & $0.16 $ &       $-0.16 $\\
$S_{8}$ & & & & & & & $1.00$ & $-0.08$  \\
$S_{9}$ & & & & & & & & $1.00$  \\
\end{tabular}
\end{table}

\begin{table}[!htb]
\caption{
Correlation matrix for the \CP-averaged observables obtained for the method of moments in the bin $15.0 <q^2<16.0\gevgevcccc$. 
\label{appendix:moments:correlation:average:11}
}
\centering 
\begin{tabular}{l|rrrrrrrr}
& $F_{\rm L}$ & $S_{3}$ & $S_{4}$ & $S_{5}$ & $A_{\rm FB}$ & $S_{7}$ & $S_{8}$ & $S_{9}$ \\ 
\hline
$F_{\rm L}$ & $1.00$ & $0.05 $ &       $-0.01 $ &      $-0.09 $ &      $-0.34 $ &      $0.01 $ &       $0.03 $ &       $-0.01 $\\
$S_{3}$ & & $1.00$ & $-0.15 $ &      $-0.29 $ &      $0.06 $ &       $-0.03 $ &      $0.02 $ &       $-0.09 $\\
$S_{4}$ & & & $1.00$ & $0.33 $ &       $-0.06 $ &      $-0.02 $ &      $-0.17 $ &      $-0.01 $ \\
$S_{5}$ & & & & $1.00$ & $-0.10$ &       $-0.13 $ &      $-0.02 $ &      $-0.05 $\\
$A_{\rm FB}$ & & & & & $1.00$ & $-0.01 $ &      $-0.03 $ &      $-0.04 $ \\
$S_{7}$ & & & & & & $1.00$ & $0.12 $ &       $-0.10$ \\
$S_{8}$ & & & & & & & $1.00$ & $-0.12$  \\
$S_{9}$ & & & & & & & & $1.00$  \\
\end{tabular}
\end{table}

\begin{table}[!htb]
\caption{
Correlation matrix for the \CP-averaged observables obtained for the method of moments in the bin $16.0 <q^2<17.0\gevgevcccc$. 
\label{appendix:moments:correlation:average:12}
}
\centering 
\begin{tabular}{l|rrrrrrrr}
 & $F_{\rm L}$ & $S_{3}$ & $S_{4}$ & $S_{5}$ & $A_{\rm FB}$ & $S_{7}$ & $S_{8}$ & $S_{9}$ \\ 
\hline
$F_{\rm L}$ & $1.00$ &  $0.16 $ &       $-0.02 $ &      $0.01 $ &       $-0.33 $ &      $0.16 $ &       $0.03 $ &       $-0.01 $\\
$S_{3}$ & & $1.00$ & $-0.12 $ &      $-0.13 $ &      $0.04 $ &       $0.05 $ &       $-0.01 $ &      $-0.03 $\\
$S_{4}$ & & & $1.00$ &  $0.21 $ &       $-0.20 $ &       $0.08 $ &       $-0.02 $ &      $0.06 $ \\
$S_{5}$ & & & & $1.00$ &  $-0.14 $ &      $0.02 $ &       $0.07 $ &       $0.20 $ \\
$A_{\rm FB}$ & & & & & $1.00$ &  $-0.05 $ &      $0.01 $ &       $-0.02 $ \\
$S_{7}$ & & & & & & $1.00$ & $0.15$ & $-0.13$  \\
$S_{8}$ & & & & & & & $1.00$ & $-0.08$  \\
$S_{9}$ & & & & & & & & $1.00$  \\
\end{tabular}
\end{table}

\begin{table}[!htb]
\caption{
Correlation matrix for the \CP-averaged observables obtained for the method of moments in the bin $17.0 <q^2<18.0\gevgevcccc$. 
\label{appendix:moments:correlation:average:13}
}
\centering 
\begin{tabular}{l|rrrrrrrr}
 & $F_{\rm L}$ & $S_{3}$ & $S_{4}$ & $S_{5}$ & $A_{\rm FB}$ & $S_{7}$ & $S_{8}$ & $S_{9}$ \\ 
\hline
$F_{\rm L}$ & $1.00$ &  $0.06 $ &       $-0.08 $ &      $0.05 $ &       $-0.21 $ &      $-0.05 $ &      $0.06 $ &       $0.00$\\
$S_{3}$ & & $1.00$ &  $-0.12 $ &      $-0.19 $ &      $0.03 $ &       $0.09 $ &       $0.01 $ &       $-0.08 $\\
$S_{4}$ & & & $1.00$ &  $0.14 $ &       $-0.07 $ &      $0.05 $ &       $-0.12 $ &      $0.00$\\
$S_{5}$ & & & & $1.00$ &   $-0.06 $ &      $-0.17 $ &      $0.07 $ &       $0.06 $\\
$A_{\rm FB}$ & & & & & $1.00$ & $0.01 $ &       $0.03 $ &       $0.03 $\\
$S_{7}$ & & & & & & $1.00$ &  $0.11 $ &       $-0.20$\\
$S_{8}$ & & & & & & & $1.00$ & $-0.05$  \\
$S_{9}$ & & & & & & & & $1.00$  \\
\end{tabular}
\end{table}

\begin{table}[!htb]
\caption{
Correlation matrix for the \CP-averaged observables obtained for the method of moments in the bin $18.0 <q^2<19.0\gevgevcccc$. 
\label{appendix:moments:correlation:average:14}
}
\centering 
\begin{tabular}{l|rrrrrrrr}
 & $F_{\rm L}$ & $S_{3}$ & $S_{4}$ & $S_{5}$ & $A_{\rm FB}$ & $S_{7}$ & $S_{8}$ & $S_{9}$ \\ 
\hline
$F_{\rm L}$ & $1.00$ &  $0.20 $ &        $-0.21 $ &      $-0.16 $ &      $-0.21 $ &      $0.01 $ &       $0.10$ &        $0.02 $\\
$S_{3}$ & & $1.00$ & $-0.18 $ &      $-0.21 $ &      $-0.03 $ &      $0.00$ &       $0.05 $ &       $-0.02 $\\
$S_{4}$ & & & $1.00$ & $0.36 $ &       $-0.18 $ &      $0.03 $ &       $0.00$ &        $0.00$\\
$S_{5}$ & & & & $1.00$ & $-0.24 $ &      $-0.01 $ &      $-0.03 $ &      $0.02 $\\
$A_{\rm FB}$ & & & & & $1.00$ &  $-0.04 $ &      $0.02 $ &       $0.05 $ \\
$S_{7}$ & & & & & & $1.00$ & $0.19 $ &       $-0.17 $\\
$S_{8}$ & & & & & & & $1.00$ & $-0.01$  \\
$S_{9}$ & & & & & & & & $1.00$  \\
\end{tabular}
\end{table}

\begin{table}[!htb]
\caption{
Correlation matrix for the \CP-averaged observables obtained for the method of moments in the bin $15.0 <q^2<19.0\gevgevcccc$. 
\label{appendix:moments:correlation:average:15}
}
\centering                                                                                                                              
\begin{tabular}{l|rrrrrrrr}                                                                                                                     
 & $F_{\rm L}$ & $S_{3}$ & $S_{4}$ & $S_{5}$ & $A_{\rm FB}$ & $S_{7}$ & $S_{8}$ & $S_{9}$ \\                                                    
\hline                                                                                                                                          
$F_{\rm L}$ & $1.00$ &  $0.11 $ &   $-0.07 $ &   $-0.04 $ &   $-0.28 $ &   $0.04 $ &   $0.04 $ &   $-0.01 $ \\
$S_{3}$ & & $1.00$ &   $-0.15 $ &   $-0.21 $ &   $0.04 $ &   $0.02 $ &   $0.01 $ &   $-0.06 $ \\
$S_{4}$ & & & $1.00$ &$0.24 $ &   $-0.11 $ &   $0.04 $ &   $-0.07 $ &   $0.02 $\\
$S_{5}$ & & & & $1.00$ & $-0.11 $ &   $-0.08 $ &   $0.03 $ &   $0.07 $\\
$A_{\rm FB}$ & & & & & $1.00$ &  $-0.02 $ &   $0.00$ &   $0.00$ \\
$S_{7}$ & & & & & & $1.00$ &  $0.14 $ &   $-0.15 $\\
$S_{8}$ & & & & & & & $1.00$ & $-0.07$  \\                                                                                                      
$S_{9}$ & & & & & & & & $1.00$  \\                                                                                                              
\end{tabular} 
\end{table}                                                                     
                                                                                  
\clearpage

\section{Correlation matrices for the \boldmath{\CP}-asymmetric observables from the method of moments}
\label{sec:appendix:bootstrap:correlation:asymmetries}

Correlation matrices between the \CP asymmetries  in the different \qsq bins are provided in Tables~\ref{appendix:moments:correlation:asymmetry:1}--\ref{appendix:moments:correlation:asymmetry:15}  for the moment analysis. The correlations are determined by a bootstrapping technique.

\begin{table}[!htb]
\caption{
Correlation matrix for the \CP-asymmetric observables obtained for the method of moments in the bin $0.10<q^2<0.98\gevgevcccc$. 
\label{appendix:moments:correlation:asymmetry:1}
}
\centering   
\begin{tabular}{l|rrrrrrr}
 & $A_{3}$ & $A_{4}$ & $A_{5}$ & $A_{6s}$ & $A_{7}$ & $A_{8}$ & $A_{9}$ \\
\hline
$A_{3}$ & $1.00 $ & $0.04 $ & $0.09 $ & $-0.02 $ &  $0.01 $ & $-0.04 $ & $0.05 $  \\
$A_{4}$ & & $1.00 $ &  $-0.24 $ & $-0.07 $ & $-0.08 $ & $0.07 $ & $0.02 $  \\
$A_{5}$ & & & $1.00 $ & $0.07 $ & $0.00$ &  $-0.07 $ & $-0.01 $  \\
$A_{6s}$ & & & & $1.00 $ & $0.08 $ & $-0.11 $ & $0.00 $   \\
$A_{7}$ & & &  & & $1.00 $ & $-0.09 $ & $0.12 $  \\
$A_{8}$ & & & & & &  $1.00 $ & $0.01 $  \\
$A_{9}$ & & & & & & &  $1.00 $   \\
\end{tabular}
\end{table}

\begin{table}[!htb]
\caption{
Correlation matrix for the \CP-asymmetric observables obtained for the method of moments in the bin $1.1<q^2<2.0\gevgevcccc$. 
\label{appendix:moments:correlation:asymmetry:2}
}
\centering   
\begin{tabular}{l|rrrrrrr}
& $A_{3}$ & $A_{4}$ & $A_{5}$ & $A_{6s}$ & $A_{7}$ & $A_{8}$ & $A_{9}$ \\
\hline
$A_{3}$ & $1.00 $ &  $-0.01 $ &  $0.04 $ &  $0.06 $ &   $0.12 $ &  $-0.05 $ & $0.08 $  \\
$A_{4}$ & & $1.00 $ &   $-0.06 $ &   $0.04 $ &   $-0.16 $ &  $0.04 $ & $-0.10$  \\
$A_{5}$ & &  &  $1.00 $ &  $-0.05 $ &  $0.01 $ &   $-0.11 $ &  $-0.07 $  \\
$A_{6s}$ & &  &  & $1.00$ &  $-0.06 $ &   $-0.07 $ & $-0.09 $ \\
$A_{7}$ & &  &  &  &  $1.00$ &  $-0.12 $ & $0.10 $  \\
$A_{8}$ & &  &  &  &  &  $1.00$ &  $-0.04 $  \\
$A_{9}$ & &  &  &  &  &  & $1.00$ \\
\end{tabular}
\end{table}

\begin{table}[!htb]
\caption{
Correlation matrix for the \CP-asymmetric observables obtained for the method of moments in the bin $2.0<q^2<3.0\gevgevcccc$. 
\label{appendix:moments:correlation:asymmetry:3}
}
\centering
\begin{tabular}{l|rrrrrrr}
 & $A_{3}$ & $A_{4}$ & $A_{5}$ & $A_{6s}$ & $A_{7}$ & $A_{8}$ & $A_{9}$ \\
\hline
$A_{3}$ & $1.00 $ &  $-0.10$ &  $0.06 $ &  $0.03$ &  $0.07 $ &  $-0.04$ &  $-0.02$  \\
$A_{4}$ & &  $1.00 $ &  $-0.07 $ &  $0.07$ &  $0.06 $ & $-0.06$ & $-0.05$  \\
$A_{5}$ & &  & $1.00 $ &  $-0.10$ &  $-0.07 $ & $0.04$ & $-0.07$  \\
$A_{6s}$ & &  &  &   $1.00 $ &  $-0.03 $ & $-0.11$ & $0.04$  \\
$A_{7}$ & &  &  &  & $1.00$ &  $-0.15$ &   $0.02$  \\
$A_{8}$ & &  &  &  &  &  $1.00$ & $-0.07$  \\
$A_{9}$ & &  &  &  &  &   &  $1.00$ \\
\end{tabular}
\end{table}

\begin{table}[!htb]
\caption{
Correlation matrix for the \CP-asymmetric observables obtained for the method of moments in the bin $3.0<q^2<4.0\gevgevcccc$. 
\label{appendix:moments:correlation:asymmetry:4}
}
\centering
\begin{tabular}{l|rrrrrrr}
&  $A_{3}$ & $A_{4}$ & $A_{5}$ & $A_{6s}$ & $A_{7}$ & $A_{8}$ & $A_{9}$ \\
\hline
$A_{3}$ & $1.00$ &  $0.00$ & $-0.04 $ & $0.03 $ & $-0.12 $ & $-0.05 $ & $-0.06 $  \\
$A_{4}$ & &  $1.00$ & $0.18 $ & $0.06 $ & $0.01 $ & $-0.05 $ & $-0.01 $  \\
$A_{5}$ & &  &  $1.00$ & $0.01 $ & $-0.01 $ & $0.01 $ & $-0.01 $  \\
$A_{6s}$ & &  &  &  $1.00$ &  $0.03 $ &  $-0.05 $ &  $0.00 $  \\
$A_{7}$ & &  &  &  &  $1.00$ &  $0.18 $ &  $-0.05 $ \\
$A_{8}$ & &  &  &  &  &  $1.00$ &  $-0.03 $  \\
$A_{9}$ & &  &  &  &  &   &  $1.00$  \\
\end{tabular}
\end{table}

\begin{table}[!htb]
\caption{
Correlation matrix for the \CP-asymmetric observables obtained for the method of moments in the bin $4.0<q^2<5.0\gevgevcccc$. 
\label{appendix:moments:correlation:asymmetry:5}
}
\centering
\begin{tabular}{l|rrrrrrr}
 & $A_{3}$ & $A_{4}$ & $A_{5}$ & $A_{6s}$ & $A_{7}$ & $A_{8}$ & $A_{9}$ \\
\hline
$A_{3}$ & $1.00$ & $-0.12 $ &   $-0.11 $ &   $0.02 $ & $0.06 $ & $-0.12 $ & $0.06 $ \\
$A_{4}$ & &  $1.00$ &  $0.17 $ &  $-0.03 $ &  $-0.06 $ & $0.19 $ & $0.03 $  \\
$A_{5}$ & &  & $1.00$ &  $-0.04 $ & $0.14 $ &   $-0.06 $ &  $-0.09 $ \\
$A_{6s}$ & &  &  &  $1.00$ &  $0.10$ & $-0.14 $ &   $0.00 $  \\
$A_{7}$ & &  &  &  &  $1.00$ &   $0.04 $ &  $-0.08 $  \\
$A_{8}$ & &  &  &  &   &  $1.00$ &  $0.02 $  \\
$A_{9}$ & &  &  &  &   &   &  $1.00$ \\
\end{tabular}
\end{table}

\begin{table}[!htb]
\caption{
Correlation matrix for the \CP-asymmetric observables obtained for the method of moments in the bin $5.0<q^2<6.0\gevgevcccc$. 
\label{appendix:moments:correlation:asymmetry:6}
}
\centering
\begin{tabular}{l|rrrrrrr}
& $A_{3}$ & $A_{4}$ & $A_{5}$ & $A_{6s}$ & $A_{7}$ & $A_{8}$ & $A_{9}$ \\
\hline
$A_{3}$ & $1.00 $ &         $-0.03 $ &      $-0.07 $ &      $-0.09 $ &      $-0.04 $ &      $0.03 $ &       $0.11 $  \\
$A_{4}$ & &  $1.00$ &        $0.10$ &        $-0.03 $ &      $0.08 $ &       $0.07 $ &       $0.03 $ \\
$A_{5}$ & &  &  $1.00$ &        $-0.08 $ &      $-0.04 $ &      $0.07 $ &       $0.07 $  \\
$A_{6s}$ & &  &  &  $1.00$ &        $0.01 $ &       $-0.01 $ &      $-0.01 $  \\
$A_{7}$ & &  &  &  &  $1.00$ &        $0.07 $ &       $-0.09 $  \\
$A_{8}$ & &  &  &  &  &  $1.00$ &   $-0.12 $ \\
$A_{9}$ & &  &  &  &  &  &  $1.00$ \\
\end{tabular}
\end{table}

\begin{table}[!htb]
\caption{
Correlation matrix for the \CP-asymmetric observables obtained for the method of moments in the bin $6.0<q^2<7.0\gevgevcccc$. 
\label{appendix:moments:correlation:asymmetry:7}
}
\centering
\begin{tabular}{l|rrrrrrr}
& $A_{3}$ & $A_{4}$ & $A_{5}$ & $A_{6s}$ & $A_{7}$ & $A_{8}$ & $A_{9}$ \\
\hline
$A_{3}$ & $1.00$ &         $-0.08 $ &      $-0.15 $ &      $-0.09 $ &      $0.02 $ &       $-0.05 $ &      $-0.02 $  \\
$A_{4}$ &  & $1.00$ &        $0.21 $ &       $-0.15 $ &      $-0.03 $ &      $-0.04 $ &      $-0.04 $  \\
$A_{5}$ &  & &   $1.00$ &        $-0.10$ &       $-0.02 $ &      $-0.03 $ &      $-0.05 $  \\
$A_{6s}$ &  &  &  &  $1.00 $ &        $0.03 $ &       $0.00$ &        $-0.05 $  \\
$A_{7}$ &  &  &  &  &  $1.00 $ &        $0.22 $ &       $-0.11 $  \\
$A_{8}$ &  &  &  &  &  &  $1.00$ &        $-0.05 $  \\
$A_{9}$ &  &  &  &   &  &  &  $1.00$  \\
\end{tabular}
\end{table}

\begin{table}[!htb]
\caption{
Correlation matrix for the \CP-asymmetric observables obtained for the method of moments in the bin $7.0<q^2<8.0\gevgevcccc$. 
\label{appendix:moments:correlation:asymmetry:8}
}
\centering
\begin{tabular}{l|rrrrrrr}
& $A_{3}$ & $A_{4}$ & $A_{5}$ & $A_{6s} $ & $A_{7}$ & $A_{8}$ & $A_{9}$ \\
\hline
$A_{3}$ & $1.00$ &         $-0.07 $ &      $-0.11 $ &      $0.04 $ &       $0.06 $ &       $0.04 $ &       $-0.01 $  \\
$A_{4}$ &  &       $1.00$ &        $0.18 $ &       $-0.07 $ &      $-0.02 $ &      $0.05 $ &       $0.01 $ \\
$A_{5}$ &  &  &       $1.00$ &        $-0.11 $ &      $0.14 $ &       $-0.02 $ &      $0.02 $  \\
$A_{6s}$ &  &  &  &      $1.00$ &        $-0.03 $ &      $-0.14 $ &      $0.07 $ \\
$A_{7}$ &  &  &  &  &      $1.00$ &        $0.07 $ &       $-0.11 $ \\
$A_{8}$ &  &  &  &  &   &       $1.00$ &        $-0.08 $  \\
$A_{9}$ &  &  &   &  &  &  &    $1.00 $  \\
\end{tabular}
\end{table}

\begin{table}[!htb]
\caption{
Correlation matrix for the \CP-asymmetric observables obtained for the method of moments in the bin $11.00 <q^2<11.75\gevgevcccc$. 
\label{appendix:moments:correlation:asymmetry:9}
}
\centering
\begin{tabular}{l|rrrrrrr}
 & $A_{3}$ & $A_{4}$ & $A_{5}$ & $A_{6s}$ & $A_{7}$ & $A_{8}$ & $A_{9}$ \\
\hline
$A_{3}$ & $1.00$ &         $-0.08 $ &      $-0.20$ &       $-0.10$ &       $0.06 $ &       $0.03 $ &       $-0.02 $  \\
$A_{4}$ & &       $1.00$ &        $0.16 $ &       $-0.14 $ &      $-0.10$ &       $-0.15 $ &      $-0.04 $  \\
$A_{5}$ & &  &    $1.00$ &        $-0.09 $ &      $-0.11 $ &      $-0.09 $ &      $-0.10$  \\
$A_{6s}$ & &  &   &    $1.00$ &        $-0.02 $ &      $-0.07 $ &      $-0.05 $  \\
$A_{7}$ & &  &   &   &      $1.00$ &        $0.25 $ &       $-0.02 $  \\
$A_{8}$ & &  &   &   &   &   $1.00$ &        $-0.09 $  \\
$A_{9}$ & &  &   &   &   &   &   $1.00$ \\
\end{tabular}
\end{table}

\begin{table}[!htb]
\caption{
Correlation matrix for the \CP-asymmetric observables obtained for the method of moments in the bin $11.75 <q^2<12.50\gevgevcccc$. 
\label{appendix:moments:correlation:asymmetry:10}
}
\centering
\begin{tabular}{l|rrrrrrr}
& $A_{3}$ & $A_{4}$ & $A_{5}$ & $A_{6s}$ & $A_{7}$ & $A_{8}$ & $A_{9}$ \\
\hline
$A_{3}$ & $1.00$ &   $-0.12 $ &      $-0.16 $ &      $0.01 $ &       $0.01 $ &       $0.03 $ &       $0.06 $ \\
$A_{4}$ &  &   $1.00$ & $0.17 $ &       $-0.21 $ &      $0.08 $ &       $0.15 $ &       $-0.05 $ \\
$A_{5}$ &  &  &  $1.00$ &  $-0.17 $ &      $0.14 $ &       $0.12 $ &       $-0.09 $ \\
$A_{6s}$ &  &  &  &    $1.00$ &  $-0.07 $ &      $-0.17 $ &      $0.05 $ \\
$A_{7}$ &  &  &  &   & $1.00 $ &  $0.19 $ & $-0.15 $ \\
$A_{8}$ &  &  &  &   & & $1.00 $ & $-0.08 $ \\
$A_{9}$ &  &  &  &   &   & &  $1.00 $ \\
\end{tabular}
\end{table}

\begin{table}[!htb]
\caption{
Correlation matrix for the \CP-asymmetric observables obtained for the method of moments in the bin $15.0 <q^2<16.0\gevgevcccc$. 
\label{appendix:moments:correlation:asymmetry:11}
}
\centering
\begin{tabular}{l|rrrrrrr}
& $A_{3}$ & $A_{4}$ & $A_{5}$ & $A_{6s}$ & $A_{7}$ & $A_{8}$ & $A_{9}$ \\
\hline
$A_{3}$ & $1.00$ &         $-0.14 $ &      $-0.26 $ &      $0.05 $ &       $-0.02 $ &      $0.02 $ &       $-0.10 $  \\
$A_{4}$ &  &     $1.00$ &        $0.36 $ &       $-0.12 $ &      $-0.02 $ &      $-0.17 $ &      $0.00$  \\
$A_{5}$ & &  &       $1.00$ &        $-0.16 $ &      $-0.12 $ &      $-0.02 $ &      $-0.04 $  \\
$A_{6s}$ & &  &   &      $1.00 $ &        $-0.02 $ &      $-0.03 $ &      $-0.05 $  \\
$A_{7}$ & &  &   &    &      $1.00$ &        $0.13 $ &       $-0.09 $ \\
$A_{8}$ & &  &    &   &   &  $1.00$ &        $-0.12 $ \\
$A_{9}$ & &  &    &   &   &    &  $1.00$  \\
\end{tabular}
\end{table}

\begin{table}[!htb]
\caption{
Correlation matrix for the \CP-asymmetric observables obtained for the method of moments in the bin $16.0 <q^2<17.0\gevgevcccc$. 
\label{appendix:moments:correlation:asymmetry:12}
}
\centering
\begin{tabular}{l|rrrrrrr}
 & $A_{3}$ & $A_{4}$ & $A_{5}$ & $A_{6s}$ & $A_{7}$ & $A_{8}$ & $A_{9}$ \\
\hline
$A_{3}$ & $1.00 $ &         $-0.08 $ &      $-0.09 $ &      $0.00$ &       $0.01 $ &       $-0.03 $ &      $-0.04 $  \\
$A_{4}$ &  &     $1.00 $ &        $0.21 $ &       $-0.22 $ &      $0.05 $ &       $-0.02 $ &      $0.06 $  \\
$A_{5}$ &  &    &       $1.00 $ &        $-0.14 $ &      $-0.01 $ &      $0.05 $ &       $0.19 $ \\
$A_{6s}$ &  &    &    &      $1.00 $ &        $0.02 $ &       $0.02 $ &       $-0.01 $  \\
$A_{7}$ &  &    &    &    &   $1.00 $ &        $0.15 $ &       $-0.13 $ \\
$A_{8}$ &  &    &    &    &   &   $1.00 $ &        $-0.08 $ \\
$A_{9}$ &  &    &    &    &   &   &   $1.00 $  \\
\end{tabular}
\end{table}

\begin{table}[!htb]
\caption{
Correlation matrix for the \CP-asymmetric observables obtained for the method of moments in the bin $17.0 <q^2<18.0\gevgevcccc$. 
\label{appendix:moments:correlation:asymmetry:13}
}
\centering
\begin{tabular}{l|rrrrrrr}
 & $A_{3}$ & $A_{4}$ & $A_{5}$ & $A_{6s}$ & $A_{7}$ & $A_{8}$ & $A_{9}$ \\
\hline
$A_{3}$ & $1.00 $ &         $-0.10$ &       $-0.16 $ &      $-0.01 $ &      $0.00$ &        $0.00 $ &        $-0.06 $ \\
$A_{4}$ & &        $1.00 $ &        $0.18 $ &       $-0.10$ &       $0.07 $ &       $-0.14 $ &      $0.03 $ \\
$A_{5}$ & &  &       $1.00$ &        $-0.10$ &       $-0.16 $ &      $0.05 $ &       $0.09 $  \\
$A_{6s}$ & &  &  &       $1.00 $ &        $0.00$ &        $0.05 $ &       $0.01 $ \\
$A_{7}$ & &  &  &  &   $1.00$ &        $0.09 $ &       $-0.20$  \\
$A_{8}$ & &  &  &  &  &  $1.00$ &        $-0.06 $  \\
$A_{9}$ & &  &  &  &  &  &  $1.00 $ \\
\end{tabular}
\end{table}

\begin{table}[!htb]
\caption{
Correlation matrix for the \CP-asymmetric observables obtained for the method of moments in the bin $18.0 <q^2<19.0\gevgevcccc$. 
\label{appendix:moments:correlation:asymmetry:14}
}
\centering
\begin{tabular}{l|rrrrrrr}
 &  $A_{3}$ & $A_{4}$ & $A_{5}$ & $A_{6s}$ & $A_{7}$ & $A_{8}$ & $A_{9}$ \\
\hline
$A_{3}$ & $1.00 $ &         $-0.18 $ &      $-0.20$ &       $-0.06 $ &      $-0.01 $ &      $0.04 $ &       $-0.03 $  \\
$A_{4}$ &  &       $1.00 $ &        $0.28 $ &       $-0.10$ &       $-0.02 $ &      $0.01 $ &       $0.07 $  \\
$A_{5}$ &  &  &       $1.00$ &        $-0.15 $ &      $-0.05 $ &      $0.00$ &       $0.04 $  \\
$A_{6s}$ &  &  &   &   $1.00 $ &        $-0.01 $ &      $-0.01 $ &      $0.03 $  \\
$A_{7}$ &  &  &  &  &   $1.00 $ &        $0.21 $ &       $-0.19 $  \\
$A_{8}$ &  &  &  &  &   &  $1.00 $ & $-0.03 $  \\
$A_{9}$ &  &  &  &  &  &  & $1.00$  \\
\end{tabular}
\end{table}

\begin{table}[!htb]
\caption{
Correlation matrix for the \CP-asymmetric observables obtained for the method of moments in the bin $15.0 <q^2<19.0\gevgevcccc$. 
\label{appendix:moments:correlation:asymmetry:15}
}
\centering
\begin{tabular}{l|rrrrrrr}
 &  $A_{3}$ & $A_{4}$ & $A_{5}$ & $A_{6s}$ & $A_{7}$ & $A_{8}$ & $A_{9}$ \\
\hline
$A_{3}$ &  $1.00 $ &   $-0.12 $ &   $-0.18 $ &   $0.00 $ &   $0.01 $ &   $0.01 $ &   $-0.05 $  \\
$A_{4}$ &  &   $1.00 $ &   $0.26 $ &   $-0.14 $ &   $0.02 $ &   $-0.08 $ &   $0.03 $ \\
$A_{5}$ &  &  &  $1.00 $ &   $-0.13 $ &   $-0.09 $ &   $0.02 $ &   $0.07 $ \\
$A_{6s}$ &  &  &   &  $1.00 $ &   $0.0 $ &   $0.01 $ &   $-0.01 $ \\
$A_{7}$ &  &  &  &  &   $1.00 $ &   $0.14 $ &   $-0.15 $  \\
$A_{8}$ &  &  &  &  &   &  $1.00 $ &   $-0.07 $  \\
$A_{9}$ &  &  &  &  &  &  &  $1.00 $ \\
\end{tabular}
\end{table}

\clearpage

\section{Correlation matrices for the optimised observables from the method of moments}
\label{sec:appendix:bootstrap:correlation:optimised}

Correlation matrices between the optimised asymmetries  in the different \qsq bins are provided in Tables~\ref{appendix:moments:correlation:optimised:1}--\ref{appendix:moments:correlation:optimised:15} for the moment analysis. The correlations are determined by a bootstrapping technique.

\begin{table}[!htb]
\caption{
Correlation matrix for the optimised angular observables obtained for the method of moments in the bin $0.10<q^2<0.98\gevgevcccc$. 
\label{appendix:moments:correlation:optimised:1}
}
\centering
\begin{tabular}{l|rrrrrrrr}
 & $F_{\rm L}$ & $P_1$ & $P_2$ & $P_3$ & $P'_4$ & $P'_5$ & $P'_6$ & $P'_8$ \\
\hline
$F_{\rm L}$ & $1.00 $ &         $0.00 $ &       $0.12 $ &       $0.09 $ &       $-0.04 $ &      $-0.23 $ &      $-0.04 $ &      $-0.08 $ \\
$P_1$ &  &        $1.00 $ &        $-0.02 $ &      $-0.05 $ &      $0.04 $ &       $0.08 $ &       $0.01 $ &       $-0.04 $ \\    
$P_2$ &  &  &     $1.00 $ &        $0.06 $ &       $-0.05 $ &      $0.11 $ &       $0.08 $ &       $-0.08 $ \\    
$P_3$ &  &  & &   $1.00 $ &        $-0.04 $ &      $-0.01 $ &      $-0.10 $ &       $-0.03 $ \\    
$P'_4$ & &  & & & $1.00 $ &        $-0.22 $ &      $-0.07 $ &      $0.07 $ \\     
$P'_5$ & & & & & &  $1.00 $ &        $0.01 $ &       $-0.07 $ \\    
$P'_6$ & & & & & & &  $1.00 $ &        $-0.08 $ \\    
$P'_8$ & & & & & & & &      $1.00 $ \\      
\end{tabular}
\end{table}

\begin{table}[!htb]
\caption{
Correlation matrix for the optimised angular observables obtained for the method of moments in the bin $1.1<q^2<2.0\gevgevcccc$. 
\label{appendix:moments:correlation:optimised:2}
}
\centering
\begin{tabular}{l|rrrrrrrr}
 & $F_{\rm L}$ & $P_1$ & $P_2$ & $P_3$ & $P'_4$ & $P'_5$ & $P'_6$ & $P'_8$ \\
\hline
$F_{\rm L}$ & $1.00 $ &         $0.02 $ &       $-0.09 $ &      $0.03 $ &       $0.21 $ &       $0.44 $ &       $-0.35 $ &      $-0.17 $ \\    
$P_1$ &  &        $1.00 $ &        $-0.17 $ &      $0.03 $ &       $0.01 $ &       $0.08 $ &       $0.04 $ &       $-0.04 $ \\    
$P_2$ &  & &      $1.00 $ &        $-0.31 $ &      $-0.17 $ &      $-0.40 $ &       $0.31 $ &       $0.11 $ \\     
$P_3$ &  & & &     $1.00 $ &        $0.10 $ &        $0.18 $ &       $-0.10 $ &       $-0.04 $ \\    
$P'_4$ & & & & &    $1.00 $ &        $0.16 $ &       $-0.24 $ &      $-0.07 $ \\    
$P'_5$ & & & & & &    $1.00 $ &        $-0.32 $ &      $-0.21 $ \\    
$P'_6$ & & & & & & &      $1.00 $ &        $0.09 $ \\     
$P'_8$ & & & & & & & &    $1.00 $ \\      
\end{tabular}
\end{table}

\begin{table}[!htb]
\caption{
Correlation matrix for the optimised angular observables obtained for the method of moments in the bin $2.0<q^2<3.0\gevgevcccc$. 
\label{appendix:moments:correlation:optimised:3}
}
\centering
\begin{tabular}{l|rrrrrrrr}
 & $F_{\rm L}$ & $P_1$ & $P_2$ & $P_3$ & $P'_4$ & $P'_5$ & $P'_6$ & $P'_8$ \\
\hline
$F_{\rm L}$ & $1.00 $ &         $-0.09 $ &      $-0.39 $ &      $0.00 $ &       $-0.45 $ &      $0.20 $ &        $-0.40 $ &       $-0.12 $ \\   
$P_1$ & &   $1.00 $ &        $0.11 $ &       $0.01 $ &       $-0.03 $ &      $0.02 $ &       $0.11 $ &       $-0.01 $ \\   
$P_2$ & & &       $1.00 $ &        $-0.03 $ &      $0.35 $ &       $-0.24 $ &      $0.31 $ &       $0.04 $ \\    
$P_3$ & & & &      $1.00 $ &        $0.04 $ &       $0.07 $ &       $-0.01 $ &      $0.06 $ \\    
$P'_4$ & & & & &       $1.00 $ &        $-0.23 $ &      $0.30 $ &        $0.08 $ \\    
$P'_5$ & & & & & &      $1.00 $ &        $-0.23 $ &      $-0.03 $ \\   
$P'_6$ & & & & & & &    $1.00 $ &        $0.00 $ \\     
$P'_8$ & & & & & & & &    $1.00 $ \\     
\end{tabular}
\end{table}

\begin{table}[!htb]
\caption{
Correlation matrix for the optimised angular observables obtained for the method of moments in the bin $3.0<q^2<4.0\gevgevcccc$. 
\label{appendix:moments:correlation:optimised:4}
}
\centering
\begin{tabular}{l|rrrrrrrr}
 & $F_{\rm L}$ & $P_1$ & $P_2$ & $P_3$ & $P'_4$ & $P'_5$ & $P'_6$ & $P'_8$ \\
\hline
$F_{\rm L}$ & $1.00 $ &         $-0.23 $ &      $0.08 $ &       $-0.33 $ &      $-0.08 $ &      $-0.23 $ &      $0.30 $ &        $0.07 $ \\    
$P_1$ & &       $1.00 $ &        $-0.11 $ &      $0.38 $ &       $-0.01 $ &      $-0.05 $ &      $-0.04 $ &      $-0.03 $ \\   
$P_2$ & & &      $1.00 $ &        $-0.28 $ &      $0.06 $ &       $0.07 $ &       $-0.05 $ &      $-0.05 $ \\   
$P_3$ & & & &      $1.00 $ &        $-0.03 $ &      $-0.11 $ &      $0.18 $ &       $0.04 $ \\    
$P'_4$ & & & & &    $1.00 $ &        $0.18 $ &       $-0.03 $ &      $-0.06 $ \\   
$P'_5$ & & & & & &    $1.00 $ &    $-0.14 $ &      $-0.03 $ \\   
$P'_6$ & & & & & & &      $1.00 $ &        $0.21 $ \\    
$P'_8$ & & & & & & & &  $1.00 $ \\     
\end{tabular}
\end{table}

\begin{table}[!htb]
\caption{
Correlation matrix for the optimised angular observables obtained for the method of moments in the bin $4.0<q^2<5.0\gevgevcccc$. 
\label{appendix:moments:correlation:optimised:5}
}
\centering
\begin{tabular}{l|rrrrrrrr}
 & $F_{\rm L}$ & $P_1$ & $P_2$ & $P_3$ & $P'_4$ & $P'_5$ & $P'_6$ & $P'_8$ \\
\hline
$F_{\rm L}$ & $1.00 $ &         $-0.34 $ &      $-0.08 $ &      $0.24 $ &       $-0.27 $ &      $-0.55 $ &      $-0.24 $ &      $0.25 $ \\     
$P_1$ & &       $1.00 $ &        $0.31 $ &       $-0.64 $ &      $-0.14 $ &      $-0.24 $ &      $-0.03 $ &      $0.00 $ \\     
$P_2$ & & &       $1.00 $ &        $-0.37 $ &      $-0.13 $ &      $-0.23 $ &      $-0.01 $ &      $-0.01 $ \\    
$P_3$ & & & &      $1.00 $ &        $0.13 $ &       $0.38 $ &       $0.17 $ &       $-0.13 $ \\    
$P'_4$ & & & & &     $1.00 $ &        $0.36 $ &       $0.07 $ &       $0.03 $ \\     
$P'_5$ & & & & & &     $1.00 $ &        $0.28 $ &       $-0.25 $ \\    
$P'_6$ & & & & & & &       $1.00 $ &        $-0.05 $ \\    
$P'_8$ & & & & & & & &   $1.00 $ \\      
\end{tabular}
\end{table}

\begin{table}[!htb]
\caption{
Correlation matrix for the optimised angular observables obtained for the method of moments in the bin $5.0<q^2<6.0\gevgevcccc$. 
\label{appendix:moments:correlation:optimised:6}
}
\centering
\begin{tabular}{l|rrrrrrrr}
 & $F_{\rm L}$ & $P_1$ & $P_2$ & $P_3$ & $P'_4$ & $P'_5$ & $P'_6$ & $P'_8$ \\
\hline
$F_{\rm L}$ & $1.00 $ &         $-0.28 $ &      $0.14 $ &       $0.19 $ &       $-0.25 $ &      $-0.14 $ &      $0.01 $ &       $-0.13 $ \\ 
$P_1$ & &  $1.00 $ &        $-0.18 $ &      $-0.21 $ &      $0.08 $ &       $0.00 $ &       $-0.04 $ &      $0.06 $ \\  
$P_2$ & & &  $1.00 $ &        $0.13 $ &       $-0.14 $ &      $-0.11 $ &      $0.01 $ &       $-0.04 $ \\ 
$P_3$ & & & &    $1.00 $ &        $-0.11 $ &      $-0.11 $ &      $0.08 $ &       $0.07 $ \\  
$P'_4$ & & & & &    $1.00 $ &        $0.17 $ &       $0.06 $ &       $0.06 $ \\  
$P'_5$ & & & & & &     $1.00 $ &        $-0.04 $ &      $0.07 $ \\  
$P'_6$ & & & & & & &     $1.00 $ &        $0.07 $ \\  
$P'_8$ & & & & & & & &      $1.00 $ \\   
\end{tabular}
\end{table}

\begin{table}[!htb]
\caption{
Correlation matrix for the optimised angular observables obtained for the method of moments in the bin $6.0<q^2<7.0\gevgevcccc$. 
\label{appendix:moments:correlation:optimised:7}
}
\centering
\begin{tabular}{l|rrrrrrrr}
 & $F_{\rm L}$ & $P_1$ & $P_2$ & $P_3$ & $P'_4$ & $P'_5$ & $P'_6$ & $P'_8$ \\
\hline
$F_{\rm L}$ & $1.00 $ &         $-0.21 $ &      $0.20 $ &        $-0.15 $ &      $-0.41 $ &      $-0.33 $ &      $0.09 $ &       $0.12 $ \\   
$P_1$ & &     $1.00 $ &        $-0.13 $ &      $0.06 $ &       $0.02 $ &       $-0.07 $ &      $0.00 $ &        $-0.07 $ \\  
$P_2$ & & &      $1.00 $ &        $0.00 $ &        $-0.22 $ &      $-0.16 $ &      $0.05 $ &       $0.02 $ \\   
$P_3$ & & & &     $1.00 $ &        $0.12 $ &       $0.10 $ &        $0.09 $ &       $0.04 $ \\   
$P'_4$ & & & & &      $1.00 $ &        $0.25 $ &       $-0.04 $ &      $-0.05 $ \\  
$P'_5$ & & & & & &       $1.00 $ &        $-0.03 $ &      $-0.04 $ \\  
$P'_6$ & & & & & & &     $1.00 $ &        $0.21 $ \\   
$P'_8$ & & & & & & & &      $1.00 $ \\    
\end{tabular}
\end{table}

\begin{table}[!htb]
\caption{
Correlation matrix for the optimised angular observables obtained for the method of moments in the bin $7.0<q^2<8.0\gevgevcccc$. 
\label{appendix:moments:correlation:optimised:8}
}
\centering
\begin{tabular}{l|rrrrrrrr}
 & $F_{\rm L}$ & $P_1$ & $P_2$ & $P_3$ & $P'_4$ & $P'_5$ & $P'_6$ & $P'_8$ \\
\hline
$F_{\rm L}$ & $1.00 $ &         $-0.08 $ &      $0.58 $ &       $-0.18 $ &      $-0.26 $ &      $-0.41 $ &      $-0.13 $ &      $-0.16 $ \\ 
$P_1$ & &       $1.00 $ &        $-0.01 $ &      $0.03 $ &       $-0.08 $ &      $-0.09 $ &      $0.06 $ &       $0.03 $ \\  
$P_2$ & & &      $1.00 $ &        $-0.19 $ &      $-0.28 $ &      $-0.45 $ &      $-0.12 $ &      $-0.26 $ \\ 
$P_3$ & & & &     $1.00 $ &        $0.05 $ &       $0.06 $ &       $0.12 $ &       $0.11 $ \\  
$P'_4$ & & & & &       $1.00 $ &        $0.26 $ &       $-0.01 $ &      $0.13 $ \\  
$P'_5$ & & & & & &       $1.00 $ &        $0.17 $ &       $0.11 $ \\  
$P'_6$ & & & & & & &       $1.00 $ &        $0.10 $ \\   
$P'_8$ & & & & & & & &        $1.00 $ \\   
\end{tabular}
\end{table}

\begin{table}[!htb]
\caption{
Correlation matrix for the optimised angular observables obtained for the method of moments in the bin $11.00 <q^2<11.75\gevgevcccc$. 
\label{appendix:moments:correlation:optimised:9}
}
\centering
\begin{tabular}{l|rrrrrrrr}
 & $F_{\rm L}$ & $P_1$ & $P_2$ & $P_3$ & $P'_4$ & $P'_5$ & $P'_6$ & $P'_8$ \\
\hline
$F_{\rm L}$ & $1.00 $ &         $-0.32 $ &      $0.60 $ &        $0.17 $ &       $0.15 $ &       $0.02 $ &       $-0.06 $ &      $-0.13 $ \\   
$P_1$ & &       $1.00 $ &        $-0.24 $ &      $-0.05 $ &      $-0.12 $ &      $-0.19 $ &      $0.06 $ &       $0.06 $ \\    
$P_2$ & & &      $1.00 $ &        $0.16 $ &       $-0.03 $ &      $-0.09 $ &      $-0.03 $ &      $-0.14 $ \\   
$P_3$ & & & &       $1.00 $ &        $0.05 $ &       $0.09 $ &       $0.01 $ &       $0.07 $ \\    
$P'_4$ & & & & &      $1.00 $ &        $0.19 $ &       $-0.11 $ &      $-0.14 $ \\   
$P'_5$ & & & & & &        $1.00 $ &        $-0.12 $ &      $-0.10 $ \\    
$P'_6$ & & & & & & &      $1.00 $ &        $0.24 $ \\    
$P'_8$ & & & & & & & &      $1.00 $ \\     
\end{tabular}
\end{table}

\begin{table}[!htb]
\caption{
Correlation matrix for the optimised angular observables obtained for the method of moments in the bin $11.75 <q^2<12.50\gevgevcccc$. 
\label{appendix:moments:correlation:optimised:10}
}
\centering
\begin{tabular}{l|rrrrrrrr}
 & $F_{\rm L}$ & $P_1$ & $P_2$ & $P_3$ & $P'_4$ & $P'_5$ & $P'_6$ & $P'_8$ \\
\hline
$F_{\rm L}$ & $1.00 $ &         $-0.41 $ &      $0.55 $ &       $-0.10 $ &       $-0.46 $ &      $-0.47 $ &      $-0.37 $ &      $-0.10 $ \\    
$P_1$ & &     $1.00 $ &        $-0.48 $ &      $0.05 $ &       $0.15 $ &       $0.16 $ &       $0.20 $ &        $0.04 $ \\    
$P_2$ & & &      $1.00 $ &        $-0.19 $ &      $-0.55 $ &      $-0.57 $ &      $-0.40 $ &       $-0.13 $ \\   
$P_3$ & & & &     $1.00 $ &        $0.11 $ &       $0.18 $ &       $0.20 $ &        $0.09 $ \\    
$P'_4$ & & & & &       $1.00 $ &        $0.52 $ &       $0.40 $ &        $0.23 $ \\    
$P'_5$ & & & & & &       $1.00 $ &        $0.45 $ &       $0.16 $ \\    
$P'_6$ & & & & & & &      $1.00 $ &        $0.22 $ \\    
$P'_8$ & & & & & & & &       $1.00 $ \\     
\end{tabular}
\end{table}

\begin{table}[!htb]
\caption{
Correlation matrix for the optimised angular observables obtained for the method of moments in the bin $15.0 <q^2<16.0\gevgevcccc$. 
\label{appendix:moments:correlation:optimised:11}
}
\centering
\begin{tabular}{l|rrrrrrrr}
 & $F_{\rm L}$ & $P_1$ & $P_2$ & $P_3$ & $P'_4$ & $P'_5$ & $P'_6$ & $P'_8$ \\
\hline
$F_{\rm L}$ & $1.00 $ &         $-0.03 $ &      $0.41 $ &       $0.09 $ &       $0.12 $ &       $0.06 $ &       $0.00 $ &       $0.05 $ \\    
$P_1$ & &       $1.00 $ &        $0.07 $ &       $0.09 $ &       $-0.15 $ &      $-0.28 $ &      $-0.03 $ &      $0.01 $ \\    
$P_2$ & & &      $1.00 $ &        $0.08 $ &       $-0.02 $ &      $-0.11 $ &      $-0.01 $ &      $-0.01 $ \\   
$P_3$ & & & &       $1.00 $ &        $0.02 $ &       $0.06 $ &       $0.10 $ &        $0.12 $ \\    
$P'_4$ & & & & &       $1.00 $ &        $0.34 $ &       $-0.02 $ &      $-0.16 $ \\   
$P'_5$ & & & & & &       $1.00 $ &        $-0.13 $ &      $-0.02 $ \\   
$P'_6$ & & & & & & &      $1.00 $ &        $0.12 $ \\    
$P'_8$ & & & & & & & &   $1.00 $ \\     
\end{tabular}
\end{table}

\begin{table}[!htb]
\caption{
Correlation matrix for the optimised angular observables obtained for the method of moments in the bin $16.0 <q^2<17.0\gevgevcccc$. 
\label{appendix:moments:correlation:optimised:12}
}
\centering
\begin{tabular}{l|rrrrrrrr}
 & $F_{\rm L}$ & $P_1$ & $P_2$ & $P_3$ & $P'_4$ & $P'_5$ & $P'_6$ & $P'_8$ \\
\hline
$F_{\rm L}$ & $1.00 $ &         $-0.09 $ &      $0.33 $ &       $0.03 $ &       $0.17 $ &       $0.23 $ &       $0.06 $ &       $-0.01 $ \\  
$P_1$ & &       $1.00 $ &        $0.07 $ &       $0.03 $ &       $-0.12 $ &      $-0.14 $ &      $0.01 $ &       $-0.02 $ \\  
$P_2$ & & &       $1.00 $ &        $0.04 $ &       $-0.15 $ &      $-0.06 $ &      $0.02 $ &       $0.02 $ \\   
$P_3$ & & & &        $1.00 $ &        $-0.06 $ &      $-0.18 $ &      $0.13 $ &       $0.08 $ \\   
$P'_4$ & & & & &     $1.00 $ &        $0.25 $ &       $0.09 $ &       $-0.02 $ \\  
$P'_5$ & & & & & &       $1.00 $ &        $0.02 $ &       $0.06 $ \\   
$P'_6$ & & & & & & &     $1.00 $ &        $0.14 $ \\   
$P'_8$ & & & & & & & &      $1.00 $ \\    
\end{tabular}
\end{table}

\begin{table}[!htb]
\caption{
Correlation matrix for the optimised angular observables obtained for the method of moments in the bin $17.0 <q^2<18.0\gevgevcccc$. 
\label{appendix:moments:correlation:optimised:13}
}
\centering
\begin{tabular}{l|rrrrrrrr}
 & $F_{\rm L}$ & $P_1$ & $P_2$ & $P_3$ & $P'_4$ & $P'_5$ & $P'_6$ & $P'_8$ \\
\hline
$F_{\rm L}$ & $1.00 $ &         $-0.07 $ &      $0.28 $ &       $0.11 $ &       $0.04 $ &       $0.21 $ &       $-0.06 $ &      $0.06 $ \\  
$P_1$ & &      $1.00 $ &        $0.02 $ &       $0.07 $ &       $-0.12 $ &      $-0.20 $ &       $0.10 $ &        $0.00 $ \\  
$P_2$ & & &   $1.00 $ &        $0.00 $ &        $-0.07 $ &      $0.01 $ &       $-0.01 $ &      $0.06 $ \\  
$P_3$ & & & &  $1.00 $ &        $0.00 $ &        $-0.03 $ &      $0.19 $ &       $0.05 $ \\  
$P'_4$ & & & & &       $1.00 $ &        $0.15 $ &       $0.04 $ &       $-0.12 $ \\ 
$P'_5$ & & & & & &       $1.00 $ &        $-0.17 $ &      $0.08 $ \\  
$P'_6$ & & & & & & &      $1.00 $ &        $0.11 $ \\  
$P'_8$ & & & & & & & &       $1.00 $ \\   
\end{tabular}
\end{table}

\begin{table}[!htb]
\caption{
Correlation matrix for the optimised angular observables obtained for the method of moments in the bin $18.0 <q^2<19.0\gevgevcccc$. 
\label{appendix:moments:correlation:optimised:14}
}
\centering
\begin{tabular}{l|rrrrrrrr}
 & $F_{\rm L}$ & $P_1$ & $P_2$ & $P_3$ & $P'_4$ & $P'_5$ & $P'_6$ & $P'_8$ \\
\hline
$F_{\rm L}$ & $1.00 $ &         $0.03 $ &       $0.39 $ &       $0.10 $ &        $-0.06 $ &      $-0.01 $ &      $0.00 $ &       $0.06 $ \\   
$P_1$ & &        $1.00 $ &        $0.02 $ &       $0.03 $ &       $-0.14 $ &      $-0.18 $ &      $0.00 $ &       $0.03 $ \\   
$P_2$ & & &  $1.00 $ &        $0.00 $ &        $-0.27 $ &      $-0.29 $ &      $-0.03 $ &      $0.07 $ \\   
$P_3$ & & & &        $1.00 $ &        $-0.01 $ &      $-0.03 $ &      $0.17 $ &       $0.02 $ \\   
$P'_4$ & & & & &      $1.00 $ &        $0.38 $ &       $0.03 $ &       $0.01 $ \\   
$P'_5$ & & & & & &       $1.00 $ &        $-0.01 $ &      $-0.03 $ \\  
$P'_6$ & & & & & & &      $1.00 $ &        $0.19 $ \\   
$P'_8$ & & & & & & & &     $1.00 $ \\    
\end{tabular}
\end{table}

\begin{table}[!htb]
\caption{
Correlation matrix for the optimised angular observables obtained for the method of moments in the bin $15.0 <q^2<19.0\gevgevcccc$. 
\label{appendix:moments:correlation:optimised:15}
}
\centering
\begin{tabular}{l|rrrrrrrr}
 & $F_{\rm L}$ & $P_1$ & $P_2$ & $P_3$ & $P'_4$ & $P'_5$ & $P'_6$ & $P'_8$ \\
\hline
$F_{\rm L}$ &  $1.00 $ &   $-0.05 $ &   $0.33 $ &   $0.08 $ &   $0.08 $ &   $0.13 $ &   $0.01 $ &   $0.03 $ \\     
$P_1$ & &   $1.00 $ &   $0.05 $ &   $0.05 $ &   $-0.14 $ &   $-0.21 $ &   $0.02 $ &   $0.01 $ \\   
$P_2$ & & &   $1.00 $ &   $0.03 $ &   $-0.10 $ &   $-0.07 $ &   $-0.01 $ &   $0.02 $ \\    
$P_3$ & & & &  $1.00 $ &   $-0.02 $ &   $-0.06 $ &   $0.15 $ &   $0.07 $ \\    
$P'_4$ & & & & &   $1.00 $ &   $0.24 $ &   $0.04 $ &   $-0.07 $ \\   
$P'_5$ &  &  &  & & &   $1.00 $ &   $-0.07 $ &   $0.03 $ \\  
$P'_6$ &  &  & & &  & &  $1.00 $ &   $0.14 $ \\    
$P'_8$ & &  & & &  &  & &   $1.00 $ \\     
\end{tabular}
\end{table}

\clearpage

\newpage
\centerline{\large\bf LHCb collaboration}
\begin{flushleft}
\small
R.~Aaij$^{39}$, 
C.~Abell\'{a}n~Beteta$^{41}$, 
B.~Adeva$^{38}$, 
M.~Adinolfi$^{47}$, 
A.~Affolder$^{53}$, 
Z.~Ajaltouni$^{5}$, 
S.~Akar$^{6}$, 
J.~Albrecht$^{10}$, 
F.~Alessio$^{39}$, 
M.~Alexander$^{52}$
% ...

{\footnotesize \it
% ...
$ ^{5}$Clermont Universit\'{e}, Universit\'{e} Blaise Pascal, CNRS/IN2P3, LPC, Clermont-Ferrand, France\\
$ ^{6}$CPPM, Aix-Marseille Universit\'{e}, CNRS/IN2P3, Marseille, France\\
% ...
$ ^{10}$Fakult\"{a}t Physik, Technische Universit\"{a}t Dortmund, Dortmund, Germany\\
% ...
$ ^{38}$Universidad de Santiago de Compostela, Santiago de Compostela, Spain\\
$ ^{39}$European Organization for Nuclear Research (CERN), Geneva, Switzerland\\
% ...
$ ^{41}$Physik-Institut, Universit\"{a}t Z\"{u}rich, Z\"{u}rich, Switzerland\\
% ...
$ ^{47}$H.H. Wills Physics Laboratory, University of Bristol, Bristol, United Kingdom\\
% ...
$ ^{52}$School of Physics and Astronomy, University of Glasgow, Glasgow, United Kingdom\\
$ ^{53}$Oliver Lodge Laboratory, University of Liverpool, Liverpool, United Kingdom\\
% ...
}
\end{flushleft}




%

\end{document}